\documentclass[11pt,a4paper]{article}
\pdfoutput=1
\usepackage[pdftex]{graphics}
\usepackage{jheppub}
\usepackage{amsmath,amssymb,amsfonts}
\usepackage{array,booktabs}
\usepackage{slashed}
\usepackage{tabularray}

\let\oldc\c
\let\oldi\i
\let\oslash\o

\usepackage[dvipsnames]{xcolor}
\def\e{\epsilon}
\def\ve{\varepsilon}
\def\k{\kappa}
\def\l{\lambda}
\def\i{\iota}
\def\m{\mu}
\def\n{\nu}
\def\r{\rho}
\def\s{\sigma}
\def\t{\tau}
\def\L{\Lambda}

\def\a{{\alpha}}
\def\b{{\beta}}
\def\c{{\gamma}}
\def\d{{\delta}}
\def\da{{\dot{\a}}}
\def\db{{\dot{\b}}}
\def\dc{{\dot{\c}}}

\def\bs{{\bar{\sigma}}}

\definecolor{labelcolor}{RGB}{194, 175, 116}
\usepackage[final]{showlabels} %

\def\mem{\hspace{0.1em}}

\def\rambda{{\bar{\lambda}}}

\def\bz{\bar{z}}
\def\bZ{{\bar{Z}}}

\def\rmA{{\mathrm{A}}}
\def\rmB{{\mathrm{B}}}
\def\rmC{{\mathrm{C}}}

\usepackage{enumitem}
\setlist[itemize]{
    label=\adjustbox{scale=0.8}{$\bullet$}, 
    itemsep=0.5pt,
    topsep=3px,
    leftmargin=25pt,
    rightmargin=10pt
}
\setlist[enumerate]{
    label=(\alph*),
    itemsep=0.5pt,
    topsep=3px,
    rightmargin=10pt
}

\newcommand{\hla}[1]{{\color[RGB]{220,10,10}{}#1}}
\newcommand{\hlx}[1]{{\color[RGB]{10,110,244}{}#1}}

\newcommand{\hlg}[1]{{\color[RGB]{0,136,186}{}#1}}

\def\g{\mathfrak{g}}

\def\SU{\mathrm{SU}}
\def\GL{\mathrm{GL}}
\def\SL{\mathrm{SL}}
\def\U{\mathrm{U}}

\newcommand{\BB}[1]{\Big(\,{#1}\,\Big)}
\newcommand{\bb}[1]{\bigg(\,{#1}\,\bigg)}
\newcommand{\BBBB}[1]{\Bigg(\,{#1}\,\Bigg)}

\newcommand{\bbsq}[1]{\bigg[\,{#1}\,\bigg]}

\newcommand{\bigbig}[1]{\big(\mem{#1}\mem\big)}

\newcommand{\pr}[1]{(\hhem{#1}\hem)}

\newcommand{\lrp}[1]{\left(\,{#1}\,\right)}
\newcommand{\lrsq}[1]{\left[\,{#1}\,\right]}

\def\mem{\hspace{0.1em}}
\def\hem{\hspace{0.05em}}
\def\nem{\hspace{-0.1em}}
\def\hnem{\hspace{-0.05em}}
\def\hhem{\hspace{0.025em}}
\def\hhnem{\hspace{-0.025em}}
\def\hhhem{\hspace{0.0125em}}

\def\blank{{\,\,\,\,\,}}

\usepackage[export]{adjustbox}
\usepackage{mathtools}

\newcommand{\fref}[1]{Fig.\,\ref{#1}}
\renewcommand{\eqref}[1]{Eq.\,(\ref{#1})}
\newcommand{\eqrefs}[2]{Eqs.\,(\ref{#1}) and (\ref{#2})}
\newcommand{\eqrefss}[3]{Eqs.\,(\ref{#1}), (\ref{#2}), and (\ref{#3})}

\newcommand{\eqrefsor}[2]{Eqs.\,(\ref{#1}) or (\ref{#2})}
\newcommand{\Sec}[1]{Sec.\,\ref{#1}}
\newcommand{\Secs}[2]{Secs.\,\ref{#1} and \ref{#2}}
\newcommand{\Secss}[3]{Secs.\,\ref{#1}, \ref{#2}, and \ref{#3}}
\newcommand{\Secsto}[2]{Secs.\,\ref{#1} to \ref{#2}}

\newcommand{\App}[1]{App.\,\ref{#1}}
\newcommand{\rcite}[1]{Ref.\,\cite{#1}}
\newcommand{\rrcite}[1]{Refs.\,\cite{#1}}
\newcommand{\transition}[1]{\qquad\adjustbox{scale=0.95}{\text{#1}}\qquad}
\newcommand{\transit}[1]{\quad\adjustbox{scale=0.95}{\text{#1}}\quad}

\def\lb{\{\kern-0.15em\{}
\def\rb{\}\kern-0.15em\}}

\newcommand{\pb}[2]{\{\hem{#1},{#2}\hem\}}

\newcommand{\comm}[2]{[\hem{#1},{#2}\hem]}

\newcommand{\act}[1]{[\,{#1}\,]}

\newcommand{\pif}[2]{{\Pi^\circ\hnem
    \bigbig{
        \hem{d#1},
        {d#2}\hem
    }
}}
\newcommand{\pie}[2]{{\Pi^\circ\hnem
    \bigbig{
        \hem{#1},
        {#2}\hem
    }
}}
\newcommand{\pib}[2]{{\Pi^\bullet\hnem
    \bigbig{
        \hem{#1},
        {#2}\hem
    }
}}
\newcommand{\pia}[2]{{\Pi\hhhem
    \bigbig{
        \hem{#1},
        {#2}\hem
    }
}}

\def\fbullet{{\color{fakecolor}\bullet}}
\newcommand{\pifb}[2]{{\Pi^{\fbullet}\hnem
    \bigbig{
        \hem{#1},
        {#2}\hem
    }
}}

\newcommand{\cont}[2]{\big\langle\hem{#1},\hem{#2}\hem\big\rangle}
 
\def\O{\mathcal{O}}

\def\GG{\mathscr{G}}

\newcommand{\Exp}[2]{
    \mathrm{Exp}\BB{{#1}
        ,\mem
    {#2}}
}

\def\qfq{{\quad\iff\quad}}
\def\qiq{{\quad\implies\quad}}

\def\Kerr{{\smash{\text{$\kern-0.075em\sqrt{\text{Kerr\hem}}$}}}}

\def\VO{{\smash{\text{VO}}}}
\def\SO{{\smash{\text{SO}}}}

\def\ST{{\smash{\text{ST}}}}
\def\MT{{\smash{\text{MT}}}}

\def\BMT{{\smash{\text{BMT}}}}

\def\QMPD{{\smash{\text{QMPD}}}}

\def\mass{{\smash{\text{mass}}}}
\def\spin{{\smash{\text{spin}}}}

\def\ba{\bar{\alpha}}

\def\bPsi{\bar{\Psi}}
\def\bzeta{\bar{\zeta}}
\def\bxi{\bar{\xi}}
\def\bmu{\bar{\mu}}
\def\bR{\bar{R}}
\def\bF{\bar{F}}
\def\bkappa{\bar{\kappa}}
\def\bk{\bar{\kappa}}
\def\bgamma{\bar{\gamma}}
\def\bpsi{\bar{\psi}}

\def\rambda{\bar{\lambda}}

\def\tE{\tilde{E}}

\def\lsq{{
    \kern-0.037em
    \adjustbox{scale=0.99,valign=c}{$
        {\lfloor \llap{\reflectbox{\rotatebox[origin=c]{180}{$\lfloor$}}}}
    $}
    \kern-0.04em
}}
\def\rsq{{
    \kern-0.04em
    \adjustbox{scale=0.99,valign=c}{$
        {\rlap{\reflectbox{\rotatebox[origin=c]{180}{$\rfloor$}}} \rfloor}
    $}
    \kern-0.037em
}}

\def\Pexp{\mathrm{P}\kern-0.1em\exp}

\def\Pexp{\mathrm{P}\kern-0.1em\exp}

\def\mathe{{\scalebox{1.025}[1]{$\mathrm{e}$}}}

\def\mtimes{{\mem\times\mem}}
\def\mdot{{\mem\cdot\mem}}
\def\mplus{{\mem+\mem}}
\def\mminus{{\mem-\mem}}

\def\md{{\mem\cdot}}

\def\sprime{{\mathrlap{\smash{{}^\prime}}{\hspace{0.05em}}}}

\def\swedge{{\mem{\wedge}\,}}
\let\oldcap\cap
\renewcommand{\cap}{{\,\oldcap\,}}

\def\C{\mathbb{C}}
\def\R{\mathbb{R}}
\def\Z{\mathbb{Z}}

\newcommand{\wrap}[1]{{\smash{#1}\vphantom{\b}}}

\def\hx{\hat{x}}
\def\hy{\hat{y}}
\def\hp{\hat{p}}

\def\hlambda{\hat{\lambda}}

\def\hdelta{\hat{\delta}}

\usepackage{dsfont}
\def\ups{\mathbf{P}\kern-0.0025em}
\def\ps{\mathds{P}}

\def\mflat{\mathbb{M}}

\def\P{\mathcal{P}}
\def\M{\mathcal{M}}

\def\SO{\mathrm{SO}}

\def\Sm{\mathscr{S}^-_{\smash{\hnem\nem\times}}}
\def\SS{\mathscr{S}_2}

\definecolor{linkcolor}{RGB}{27,54,126}
\hypersetup{
    citecolor=linkcolor,
    linkcolor=linkcolor,
    urlcolor=linkcolor
}

\usepackage{hyphenat}
\definecolor{paracolor}{RGB}{27,54,126}

\newcommand{\para}[1]{\paragraph{\color{paracolor}#1}}

\definecolor{fakecolor}{RGB}{224,41,135}

\newcommand{\fake}[1]{{%
    \color{fakecolor}\hem\mathbf{#1}%
}}

\def\fx{\fake{x}}

\def\fL{\fake{\L}}
\def\fS{\fake{S}}

\def\fp{\fake{p}}

\def\fu{\fake{u}}
\def\fv{\fake{v}}

\usepackage{bm}
\def\fa{\fake{\bm{\a}}}
\def\fba{\fake{\bm{\ba}}}

\def\fpsi{\fake{\bm{\psi}}}
\def\fbpsi{\fake{\bm{\bpsi}}}

\def\fchi{\fake{\bm{\chi}}}

\newcommand{\fakeit}[1]{{%
    \color{fakecolor}\bm{#1}\hhhem%
}}

\def\fb{\fakeit{b}}
\def\fR{\fakeit{R}}

\usepackage[outline]{contour}
\contourlength{0.15pt}
\def\vbet{{\color{fakecolor}\hhem{\contour{fakecolor}{$\bm{\vec{\beta}}$}}\hem}}
\def\bet{{\color{fakecolor}\hhem{\contour{fakecolor}{$\bm{{\beta}}$}}\hhem}}

\def\vs{\vec{\sigma}}

\def\fth{{\fakeit{\vartheta'}\kern-4pt}^{\vphantom{\prime}}}
\def\fom{{\fakeit{\varpi'}\kern-4pt}^{\vphantom{\prime}}}

\contourlength{0.2pt}
\newcommand{\textfake}[1]{{%
    \color{fakecolor}{\contour{fakecolor}{#1}}%
}}

\def\Delt{\mathit{\Delta}}

\def\acX{\smash{\acute{X}}}
\def\acY{\smash{\acute{Y}}}

\usepackage{mathrsfs}

\def\ss{{\kern0.05em/\kern-0.25em/\kern0.10em}}

\def\inn{{\:\in\:}}
\def\eqq{{\:=\:}}
\def\too{{\:\to\:}}

\def\dX{{{\delta}X}}

\def\Cinfty{C^\infty\hnem}

\def\vex{\vec{x}}
\def\vea{\vec{a}}

\def\mlra{{\mem\leftrightarrow\mem}}

\def\V{\mathcal{V}}

\newcommand{\expval}[1]{
    \big\langle\hem{
        #1
    }\hem\big\rangle
}

\DeclareMathOperator{\sinc}{sinc}

\title{Universality in Relativistic Spinning Particle Models}

\author[a]{Joon-Hwi Kim}
\author[b]{Sangmin Lee}

\affiliation[a]{Walter Burke Institute for Theoretical Physics,\\
California Institute of Technology, Pasadena, CA 91125, U.S.A.}
\affiliation[b]{School of Physics, Korea Institute for Advanced Study,\\
85 Hoegi-ro, Dongdaemun-Gu, Seoul 02455, Korea} 

\abstract{
    We establish an equivalence between
    massive spinning particle models in four spacetime dimensions
    coupled to electromagnetism or gravity,
    within the spin-magnitude-preserving sector.
    Four representative models in the literature
    are shown to describe
    exactly the same physics
    in their free and interacting theories:
    vector oscillator,
    spinor oscillator,
    spherical top,
    and
    massive twistor.
    The Bargmann-Michel-Telegdi (BMT) and 
    quadrupolar Mathisson-Papapetrou-Dixon (QMPD) equations 
    are derived in a model-independent fashion.
    This universal framework allows for incorporating higher spin multipole interactions as well.
    We establish the rigorous construction of the interacting theory of the spherical top model
    with emphasis on spin gauge invariance.
    Applications to
    black hole physics, conserved charges,
    and post-Newtonian or post-Minkowskian frameworks
    are discussed.
}

\emailAdd{joonhwi@caltech.edu, sangminlee@kias.re.kr}

\begin{document}
\begin{flushright}
    \texttt{CALT-TH 2026-008}\\
    \texttt{KIAS-P26006}
\end{flushright}
\maketitle

\newpage
\section{Introduction}

The construction of
Lagrangian or Hamiltonian models of relativistic spinning particles
is an old yet foundational subject
in physics.
Early approaches by Frenkel \cite{Frenkel:1926zz} and Thomas
\cite{Thomas:1927yu}
date back a whole century.
A vast variety of models have appeared in the literature since then,
employing
bosonic
\cite{Mathisson:1937zz,Papapetrou:1951pa,Dixon:1970zza,Hanson:1974qy,Bailey:1975fe,%
souriau1970structure,souriau:1970b,Souriau:1974ahp,Kunzle:1972uk,%
Grassberger:1977tn,Rempel:2015foa,Deriglazov:2015bqa,Basile:2023vyg,Haddad:2024ebn%
},
fermionic
\cite{Berezin:1976eg,Casalbuoni:1975hx,Casalbuoni:1976tz,Brink:1976sz,Brink:1976uf},
or 
spinorial/twistorial
\cite{woodhouse1997geometric,Lyakhovich:1996we,Rempel:2016jbn,%
Zima:1995db,%
Mezincescu:2015apa,%
Penrose:1974di,Perjes:1974ra,tod1977some,Bette:1989zt,Bette:2004ip,Fedoruk:2003td,Fedoruk:2007dd,Fedoruk:2014vqa,deAzcarraga:2014hda,Deguchi:2015iuw,Kim:2021rda,ambikerr1}
variables.
See also \rrcite{Frydryszak:1996mu,Rivas:2002}
for an overview
and \rrcite{Plyushchay:1990eq,Plyushchay:1992ga,Bars:2005ze,Bars:2005ax,Deriglazov:2017jub,Jakobsen:2023tvm} for more related works.
A systematic geometric perspective on the subject
was pioneered by Souriau
\cite{souriau1970structure,souriau:1970b,Souriau:1974ahp},
the significance of which is difficult to overstate;
see \rcite{Damour:2024mzo}
for a recent account.

The abundance of spinning particle models 
is both a blessing and a complication.
On the one hand,
model-specific features or intuitions
may be advantageous,
rendering
certain properties of 
the free or interacting theories of spinning particles
more transparent.
On the other hand, they 
might hinder appreciating the
underlying physical structures that are common to all implementations.
Typical models involve redundant variables encoding unphysical (i.e., gauge) degrees of freedom,
in which case model-specific calculations and derivations tend to be inefficient and less intuitive.
Translating between different models
is generically 
a nontrivial task
and may involve 
computing
Dirac brackets.

The latter standpoint might have been implied in the recent article \cite{Damour:2024mzo} by Damour and Iglesias-Zemmour,
where Souriau \cite{souriau1970structure,souriau:1970b,Souriau:1974ahp}'s
(pre)symplectic formulation of
an elementary 
spinning particle
is identified as 
the most economical framework
employing a minimal number of variables
whereas other formulations
are viewed as rather redundant.

Recently, \rrcite{sst-asym,Kim:2024grz}
have made useful observations
regarding
the universal, model-independent structures
of massive spinning particles
in four spacetime dimensions.
Firstly, 
\rcite{sst-asym}
provided a constructive proof that
a set of physical variables $(x^\m,\hy^\m,p_\m)$
exists in 
any phase space of a free massive spinning particle
as a direct implication of global Poincar\'e symmetry,
realizing a universal Poisson bracket relation.\footnote{
    In \rcite{sst-asym},
    interactions are added from the perspective of scattering amplitudes at asymptotia.
}
Secondly,
\rcite{Kim:2024grz}
noticed that
the bulk dynamics
of a massive spinning particle model
coupled to electromagnetism
can effectively be characterized by
knowing only
the Poisson brackets
between the physical variables $(x^\m,\hy^\m,p_\m)$
while forgetting the extra details
of the model.
The lesson is that
the interacting theories of different spinning particle models
can describe the same physics
via such consistent truncations.

By taking
these observations due to \rrcite{sst-asym,Kim:2024grz}
as input,
this paper thoroughly
establishes an equivalence between massive spinning particle models in four spacetime dimensions
with or without electromagnetic and gravitational interactions.
This identifies
a universality in spinning particle models
as a direct consequence of physical principles
such as Poincar\'e symmetry,
gauge invariance,
and equivalence principle.
This universality implies that
the
Bargmann-Michel-Telegdi (BMT) \cite{Bargmann:1959gz} equations,
the 
Mathisson-Papapetrou-Dixon (MPD) \cite{Mathisson:1937zz,Papapetrou:1951pa,Dixon:1970zza}
equations,
and their arbitrary higher multipole extensions
admit unified derivations within any massive spinning particle model.
We provide a consistent presentation of these ideas
in the Hamiltonian formulation
while speaking in the
languages of symplectic/Poisson geometries.

In \Sec{FREE},
we reproduce \rcite{sst-asym}'s construction
to explicate the universality at the level of free theory,
reviewing the universal Poisson bracket relation between the physical variables 
$x^\m$\:(position),
$\hy^\m$\:(spin length),
$p_\m$\:(momentum).
We provide an additional geometric analysis
that clarifies the connection to
Souriau \cite{souriau1970structure}'s
construction
in the context of
the coadjoint orbit
\cite{kirillov1975elements,kostant1970orbits,%
bacry1967space,arens1971classical,Carinena:1989uw} method.
These establish
the existence of
ten- or eight-dimensional
universal phase spaces
in any massive spinning particle model
as symplectic submanifolds.

\begin{figure}[t]
    \centering
    \includegraphics[scale=0.49,
        trim = 0pt 40pt 0pt 0pt
    ]{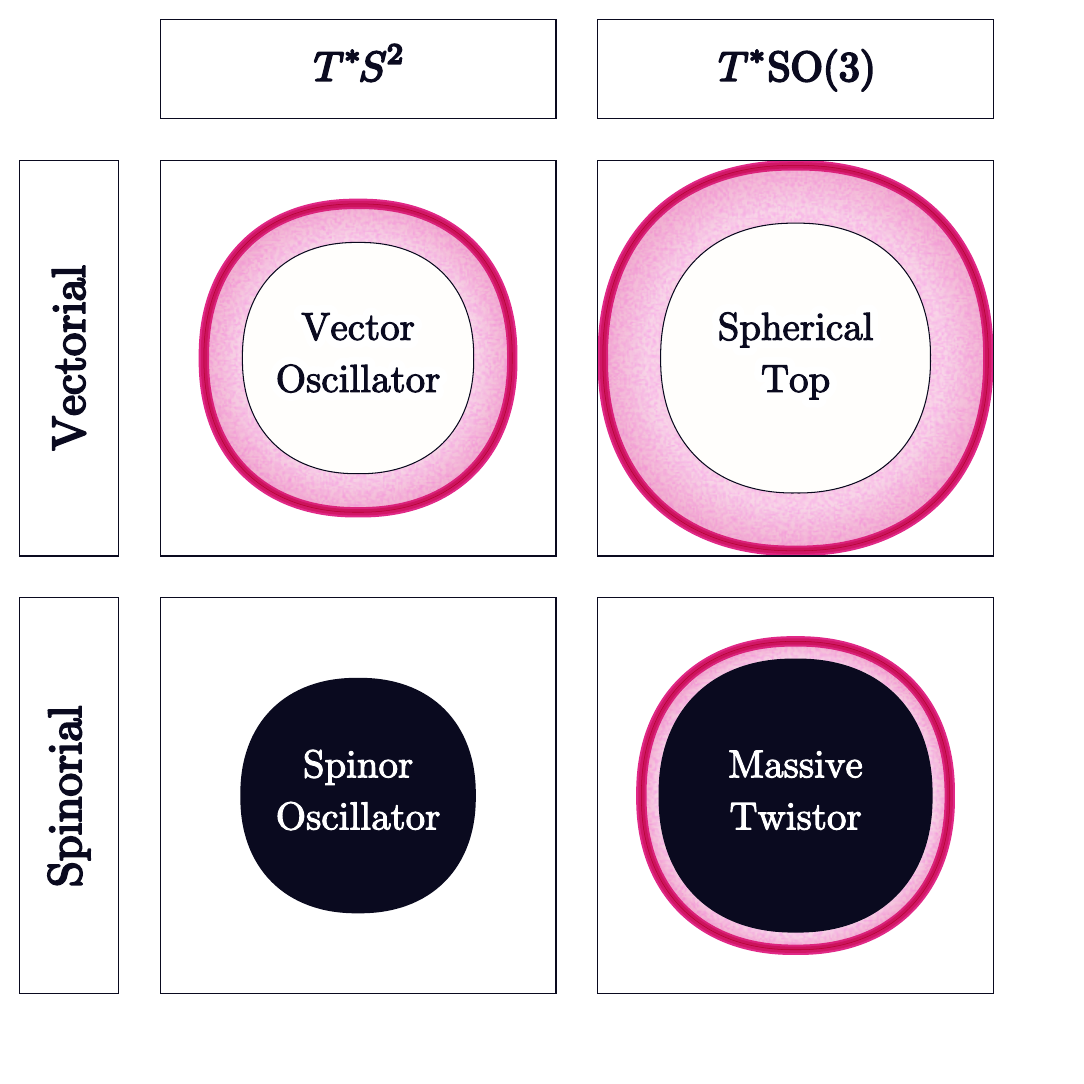}
    \caption{
        The four models are
        classified by their spin phase spaces:
        $T^*S^2$ or $T^*\SO(3)$
        up to double cover.
        In this figure,
        the sizes of 
        the white/black cores
        visualize
        the dimensions of
        the reduced phase spaces
        that involve no
        unphysical spinning degrees of freedom.
        The pink areas
        visualize
        the surplus dimensions
        (number of unphysical variables)
        employed by
        the defining phase spaces.
    }
    \label{overview}
\end{figure}

Next, \Sec{FOUR}
explicitly demonstrates this universality
with four concrete example systems
gathered from the literature:
vector oscillator
\cite{Grassberger:1977tn,Rempel:2015foa,Deriglazov:2015bqa,Basile:2023vyg,Haddad:2024ebn},
spinor oscillator
\cite{woodhouse1997geometric,Lyakhovich:1996we,Rempel:2016jbn},
spherical top
\cite{Hanson:1974qy,Bailey:1975fe},
and
massive twistor
\cite{Penrose:1974di,Perjes:1974ra,tod1977some,Bette:1989zt,Bette:2004ip,Fedoruk:2007dd,Deguchi:2015iuw,Kim:2021rda,ambikerr1}
models.
Despite the full generality of
our claims on the universality,
these four instances
represent
a notable subset
in the vast zoo of models
that is worth spotlighting
in a modern context.
We presume isotropy,
meaning that
an internal $\SO(3)$ (``little group'') symmetry action
is present in the particle's rest frame.

\fref{overview}
provides an overview of these four models.
Hamiltonian systems exhibiting an $\SO(3)$ symmetry
are classified 
in terms of the rotational (spin) phase spaces they employ.
The sensible options are
$T^*\SO(3)$,
$T^*S^2$,
and
$S^2$
in the order of decreasing dimension,
with the understanding that
their double cover counterparts are also 
considered
for spinorial implementations.
The models in the first column
employ 
the four-dimensional spin phase space
$T^*S^2$.
The models in the second column
employ
the six-dimensional spin phase space
$T^*\SO(3)$.
The 
spinor oscillator and massive twistor models
are the spinorial reformulations of
vector oscillator and spherical top models,
respectively.

In order to maintain relativistic covariance,
these models typically employ more variables than are strictly necessary.
\fref{overview} represents
the number of unphysical spin variables 
by
the size of the pink margins, 
demonstrating
the visual grammar of this paper:
\begin{align*}
    \textfake{Pink}
    \,\,=\,\,
        \text{Unphysical,{\mem} Gauge,{\mem} Spurious,{\mem} Fake,{\mem} Redundant}
    \,.
\end{align*}
One sees that
the vectorial implementations are more redundant than spinorial ones.
For example, the spherical top 
involves six unphysical variables
reflecting the notorious issue of
spin supplementary conditions (SSCs)
\cite{pryce1948mass,Newton:1949cq,moller1949dynamique,moller1949definition,fleming1965covariant,Dixon:1970zz,%
Costa:2014nta,Mikoczi:2016fiy,Witzany:2018ahb%
},
whose precise mathematical formulation 
in the standard frameworks of gauge redundancies
has been achieved recently
\cite{Steinhoff:2015ksa,Kim:2021rda}.

Advocates of the Souriau \cite{souriau1970structure,souriau:1970b,Souriau:1974ahp}
formulation
highlight that
the smallest possible spin phase space is
the two-dimensional $S^2$.
The universal geometrical analysis
of \Sec{FREE}
implies that
all of the four models in \fref{overview}
achieve such $S^2$
via appropriate reductions.
This point is explicitly substantiated in
\Sec{HIER}
by establishing the hierarchy
$T^*\SO(3) {\:\supset\:} T^*S^2 {\:\supset\:} S^2$
of spin phase spaces
in the Dirac bracket framework.

\begin{figure}
    \centering
    \includegraphics[width=0.77\linewidth,
        trim = 0pt 10pt 0pt 0pt
    ]{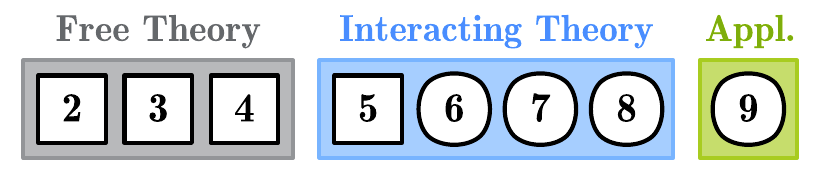}
    \caption{
        The structure of this paper.
        \Secsto{FREE}{HIER} concern free theory.
        \Secsto{INT1}{COVPS} concern interacting theory.
        \Sec{APPL} showcase applications.
        In \Secsto{FREE}{INT1}, we work in flat spacetime (special relativity).
        In \Secsto{INT2}{APPL}, our discussions involve curved spacetime (general relativity).
    }
    \label{chapitres}
\end{figure}

Having established universality in the free theory,
we then consider 
interacting theories:
electromagnetism and gravity.
See \fref{chapitres} for an overview of our plan.

\Sec{INT1} establishes
universality in electromagnetic backgrounds.
It is shown that the BMT equations
\cite{Bargmann:1959gz}
are directly implied by
the universal Poisson bracket relation between
the physical variables $(x,\hy,p)$
and thus
arise in any spinning particle model.

\Sec{INT2} then
makes an important conceptual leap
that proposes
the existence of 
the variables $(x,\hy,p)$
and their universal Poisson bracket relation
in \textit{general relativity}.
Based on this crucial generalization,
it is shown that the MPD equations 
\cite{Mathisson:1937zz,Papapetrou:1951pa,Dixon:1970zza}
and its quadrupolar extension,
which we refer to as the
quadrupolar MPD (QMPD) equations
\cite{Steinhoff:2009tk,Harte:2011ku,Vines:2016unv,Compere:2023alp,Ramond:2026fpi,%
khriplovich1989particle,Yee:1993ya,Khriplovich:1997ni,Thorne:1984mz},
are direct implications of the
general-relativistic universal variables $(x,\hy,p)$
and their Poisson algebra.

Two theoretical frameworks play critical roles
in these discussions.
The first is
\textit{Souriau's method of minimal coupling}
\cite{souriau1970structure,souriau:1970b,Souriau:1974ahp},
where 
interactions are implemented by
modifications of the symplectic structure
instead of the Hamiltonian.
This approach has been also known as
the ``Feynman bracket'' method in the literature
\cite{dyson1990feynman,lee1990feynman,tanimura1992relativistic,stern1993deformed,chou1994dynamical,berard1999dirac}
and has been systematized and generalized under the name ``symplectic perturbation theory'':
see \rcite{csg} and also \rcite{ambikerr1}.
Brief reviews of this method are provided
at the start of \Secs{INT1}{INT2}.

The second is
\textit{the covariant Poisson bracket} \cite{csg},
which characterizes Poisson structures
in a manifestly covariant fashion
that never writes down bare Christoffel symbols or connection coefficients
at every intermediate step.
To our best knowledge,
this idea is due to \rcite{csg}
while its precursors may trace back to
\rrcite{souriau1970structure,dewitt1965dynamical,DeWitt:2011nnj,vanHolten:1992bu,Gibbons:1993ap,dAmbrosi:2015ndl}.
This paper does not assume prior familiarity with \rcite{csg},
where a comprehensive formalization is provided
in terms of
almost-symplectic geometry
and Ehresmann connections.

To explicitly prove the existence of the general-relativistic universal variables postulated by \Sec{INT2},
we perform a thorough investigation on
the covariant phase space geometries of 
the four models
in \Secs{HR}{COVPS}.
The spherical top model exhibits an exceptional degree of complexity
due to its \textit{spin gauge redundancies};
hence it is discussed separately in \Sec{HR}.
 
Since Dixon \cite{Dixon:1970zza},
it has been well-known that
the definition of center coordinates
for extended relativistic bodies
is intrinsically tied to the problem of
separating the total angular momentum
into orbital and spin parts.
This point gives rise to 
the aforementioned issue of SSCs,
addressed during the derivations of the MPD equations
\cite{Mathisson:1937zz,Papapetrou:1951pa,Dixon:1970zza,tulczyjew1959motion,dixon1964covariant}
and investigated throughout
\rrcite{pryce1948mass,Newton:1949cq,moller1949dynamique,moller1949definition,fleming1965covariant,Dixon:1970zz};
see
\rrcite{Costa:2014nta,Mikoczi:2016fiy,Witzany:2018ahb}
for recent reviews.
Steinhoff \cite{Steinhoff:2015ksa}
has established
the modern reformulation of SSCs 
as gauge redundancies
in the spherical top model
at the level of free theory,
while \rcite{Kim:2021rda} provided
a subtle yet crucial refinement.

To our knowledge, however,
the status of this reformulation has not been established at the level of interacting theory.
Importantly, the interaction action must be spin gauge invariant
for physical observables to be independent of one's customary choice of a SSC.
\Sec{HR} of this paper,
for the first time,
establishes such spin gauge invariant couplings 
and explicitly identifies the spin gauge-invariant center coordinates in general relativity
by exploiting the all-orders geodesic deviation formalism of \rcite{gde}
that builds upon the Synge framework \cite{Ruse:1931ht,Synge:1931zz,Poisson:2011nh,Vines:2014oba}.
This rigorous formulation of the interacting spherical top
reveals that the addition of ``fictitious terms''
in the Lagrangian
is mandatory in noncovariant SSCs.

The subject of relativistic spinning particle models
has gained revived interest
in the recent gravitational physics community.
The context is the point-particle effective theory of astrophysical objects
\cite{Goldberger:2004jt,Porto:2005ac,Levi:2015msa,Porto:2016pyg,Levi:2018nxp},
the applications of which
range from post-Newtonian (PN) or post-Minkowskian (PM) approximations 
\cite{Will:2011nz,Blanchet:2013haa,Levi:2018nxp,%
Damour:2016gwp,Cheung:2018wkq,Cristofoli:2019neg,Bjerrum-Bohr:2021din,Kalin:2020mvi,Kalin:2020fhe,Mogull:2020sak,Jakobsen:2021smu,Bjerrum-Bohr:2022blt,Vines:2017hyw}
to 
extreme mass-ratio inspirals \cite{Cheung:2023lnj,Cui:2025bgu,Skoupy:2025nie,Chen:2025ncm}.
The most intriguing yet challenging problem
may have been
the construction of worldline models
that can exactly represent spinning black holes
to all orders.
The all-orders effective actions of 
{\Kerr},
Kerr, and Kerr-Newman probes
are established by
\rrcite{gmoov,probe-nj,njmagic.1,njmagic.11}.

With this context in mind,
we end by showcasing the applications of our constructions
in \Sec{APPL}.
\Sec{APPL>Q}
approaches dynamical symmetries and conserved charges
directly from the universal variables,
in light of a recent surge of interest on the topic
\cite{Compere:2021kjz,Compere:2023alp,Gonzo:2024zxo,Ramond:2024ozy,Akpinar:2025tct,Ramond:2026fpi}.
\Secs{APPL>PM}{APPL>PN}
discuss implications on
the PM and PN expansions
while commenting on
the necessary inclusion of the ``fictitious vertices.''
\Sec{APPL>POLES}
shows how higher multipolar couplings are implemented in our framework.
\Sec{APPL>BH}
explains
how the universal variables $(x,\hy,p)$ and their Poisson algebra
admit 
a strikingly simpler
complex-geometrical description
via
the spinspacetime program of \rcite{sst-asym},
realizing
the Newman-Janis \cite{Newman:1965tw-janis} 
property
in terms of complex coordinates ``{\mem$z \eqq x {\:+\,} i\hy$\,}''
\cite{sst-asym,gmoov,probe-nj,njmagic.1,njmagic.11}.
Keywords include exact {\Kerr} equations of motion (EoM)
and
dynamical Newman-Janis shift of conserved charges.

Overall, this work should provide a solid foundation for relativistic spinning particle models and 
offer clear guidance for future research in point-particle effective theory.

\newpage
\section{Universality in Free Theory}
\label{FREE}

This section reviews a set of crucial facts shown by \rcite{sst-asym},
which enables direct identifications of
the universal geometrical structures
of massive spinning-particle phase spaces.
An additional analysis on symplectic leaves is provided.

\subsection{Universal Poisson Bracket Relation from Poincar\'e Symmetry}
\label{FREE>FULL}

Suppose any Hamiltonian formulation of
a massive physical system
that enjoys global Poincar\'e symmetry.
\rcite{sst-asym} shows that
one can always construct
functions 
$\hx^\m$, $\hy^\m$, $p_\m$
on its phase space $\ps$
such that 
$p \mdot \hx = 0 = p \mdot \hy$
and
\begin{align}
\begin{split}
\label{M3*}
    \pb{\hx^\m}{\hx^\n}
    \,\,&=\,\,
        \frac{1}{-p^2}\, 
        \BB{\mem
            \hx^\m p^\n {-\,} p^\m \hx^\n
            + \ve^{\m\n\r\s} \hy_\r\mem p_\s
        }
    \,,\\
    \pb{\hx^\m}{\hy^\n}
    \,\,&=\,\,
        \frac{1}{-p^2}\,
            \hy^\m p^\n
    \,,\quad
    \pb{\hy^\m}{\hy^\n}
    \,=\,
        \frac{1}{-p^2}\, 
            \ve^{\m\n\r\s} \hy_\r\mem p_\s
    \,,\\
    \pb{\hx^\m}{p_\n}
    \,\,&=\,\,
        \delta^\m{}_\n - \frac{p^\m\hem p_\n}{p^2}
    \,,\quad
    \pb{\hy^\m}{p_\n}
    \,=\,0
    \,,\quad
    \pb{p_\m}{p_\n}
    \,=\, 0
    \,.
\end{split}
\end{align}
Here, $\pb{f}{g}$ denotes the Poisson bracket
between functions $f$ and $g$ on the phase space $\ps$.
This result applies to not only discrete systems but also continuous systems:
a compact mass distribution on flat spacetime
with a conserved stress-energy $T^{\m\n}$, for instance.
Yet, the interest of this paper will be restricted specifically to one-particle systems.

Firstly,
Poincar\'e symmetry
implies that
one can construct
the Poincar\'e charges
as functions on the phase space $\ps$:
momentum $p_\m$ and total angular momentum $J^{\m\n}$.
They must satisfy the following Poisson brackets
as per the definition of the Poincar\'e algebra:
\begin{align}
\begin{split}
\label{poincare}
    \pb{J^{\m\n}}{J^{\r\s}}
    \,\,&=\,\,
          J^{\m\r}\mem \eta^{\n\s}
        - J^{\m\s}\mem \eta^{\n\r}
        + J^{\n\s}\mem \eta^{\m\r}
        - J^{\n\r}\mem \eta^{\m\s}
    \,,\\[0.15\baselineskip]
    \pb{J^{\m\n}}{p_\r}
    \,\,&=\,\,  
          \delta^\m{}_\r\mem p^\n 
        - \delta^\n{}_\r\mem p^\m 
    \,,\\[0.15\baselineskip]
    \pb{p_\m}{p_\n} \,\,&=\,\, 0
    \,.
\end{split}
\end{align}
Here, $\eta^{\m\n}$ is the flat inverse metric
of signature $(-,+,+,+)$.
The system being massive means that its classical states are not allowed to approach the zero locus of $p^2$
in $\ps$.

Secondly,
\textit{define} $\hx^\m$ and $\hy^\n$ as
\begin{align}
    \label{hxy-def}
    \hx^\m
    \,:=\,
        \frac{1}{p^2}\, J^{\m\n}\mem p_\n
    \,,\quad
    \hy^\m
    \,:=\,
        \frac{1}{-p^2}\, {*}J^{\m\n}\mem p_\n
    \,,
\end{align}
where ${*}J^{\m\n} = \frac{1}{2}\mem \ve^{\m\n\r\s} J_{\r\s}$
describes the Hodge star
in the convention $\ve_{0123} = +1$.
By direct computation,
one shows that
the Poincar\'e algebra in \eqref{poincare}
implies
the Poisson brackets in \eqref{M3*}
between $\hx^\m$, $\hy^\m$, and $p_\m$.
Also,
$p \mdot \hx = 0 = p \mdot \hy$
by the antisymmetry of $J^{\m\n}$ and ${*}J^{\m\n}$.
This concludes the proof.

On account of the well-known definition of the Pauli-Lubanski pseudovector,
the variable $\hy^\m$
in \eqref{hxy-def}
describes the spin pseudovector
normalized in units of length.
In this paper,
we refer to this $\hy^\m$
as \textit{spin length pseudovector}.

In fact,
$\hy^\m$ can evaluate to zero in some systems;
take a free relativistic scalar particle
formulated in the cotangent bundle of Minkowski space,
for instance.
In this paper,
we concern systems with nontrivial $\hy^\m$ only.

\subsection{Partly Reduced Formulation}
\label{FREE>PART}

Physically,
the notion of mass dimension must exist in the phase space $\ps$.
Especially, the momentum $p_\m$
carries mass dimension $+1$
while the angular momentum $J^{\m\n}$
carries mass dimension $0$.
This implies that the dilatation charge $D$
would also exist
as a function on the phase space $\ps$,
such that the Poisson bracket with $D$
measures the mass dimension:
\begin{align}
    \label{dilatation}
    \pb{p_\m}{D}
    \,=\, 
        +p_\m
    \,,\quad
    \pb{J^{\m\n}}{D}
    \,=\,
        0
    \,.
\end{align}

\rcite{sst-asym} shows that
\eqrefs{poincare}{dilatation}
together implies the existence of 
functions $x^\m$, $\hy^\m$, and $p_\m$
on $\ps$
such that
$p \mdot \hy = 0$
and
\begin{align}
\begin{split}
\label{M3}
    \pb{x^\m}{x^\n}
    \,\,&=\,\,
        \frac{1}{-p^2}\, 
            \ve^{\m\n\r\s} \hy_\r\mem p_\s
    \,=\,
    \pb{\hy^\m}{\hy^\n}
    \,,\\
    \pb{x^\m}{\hy^\n}
    \,\,&=\,\,
        \frac{1}{-p^2}\,\BB{
            \hy^\m p^\n + p^\m \hy^\n 
        }
    \,,\\[0.15\baselineskip]
    \pb{x^\m}{p_\n}
    \,\,&=\,\,
        \delta^\m{}_\n
    \,,\quad
    \pb{\hy^\m}{p_\n}
    \,=\,0
    \,,\quad
    \pb{p_\m}{p_\n}
    \,=\, 0
    \,.
\end{split}
\end{align}

The proof is straightforward.
Define
\begin{align}
    \label{x-def}
    x^\m
    \,:=\,
        \frac{1}{p^2}\, 
        \BB{
            J^{\m\n}\mem p_\n
            - D\mem p^\m
        }
    \,,
\end{align}
which adds an additional term to $\hx^\m$.
Then direct computation verifies \eqref{M3}.

In terms of $x^\m$, $\hy^\m$, and $p_\m$,
the total angular momentum $J^{\m\n}$
and the dilatation charge $D$
are expressed as
\begin{align}
    \label{JD}
    J^{\m\n}
    \,=\,
        x^\m p^\n {-\,} p^\m x^\n 
        + \ve^{\m\n\r\s} \hy_\r\mem p_\s
    \,,\quad
    D
    \,=\,
        - p \mdot x
    \,.
\end{align}
From \eqref{JD}
and the Poisson bracket $\pb{x^\m}{p_\n} = \delta^\m{}_\n$ in 
\eqref{M3},
it is evident that
the interpretation of $x^\m$
is a reconstructed notion of
flat spacetime coordinates.

Again, the above result applies to any massive system.
In the context of
effective point-particle description of
compact bodies,
$x^\m$ in \eqref{x-def}
describes the Poincar\'e-covariant center coordinates.
Accordingly,
\eqref{JD}
describes the covariant split
of the total angular momentum $J^{\m\n}$
to orbital
($x^\m p^\n {-\,} p^\m x^\n$)
and
spin
($\ve^{\m\n\r\s} \hy_\r\mem p_\s$)
parts.

\subsection{Symplectic Leaves}
\label{FREE>LEAVES}

Lastly, let us comprehend the above results
from a mathematical perspective.

A Poisson manifold is a manifold equipped with a Poisson bracket
$\pb{f}{g} = \Pi(df,dg)$
between functions $f$ and $g$,
which arises from a bivector $\Pi$.
A symplectic manifold is a manifold equipped with a symplectic form $\omega$,
which is a closed nondegenerate two-form.
Symplectic manifolds are Poisson manifolds, but the converse is not true.
This is because the Poisson bivector $\Pi$ can be degenerate,
unlike the symplectic form $\omega$.

Yet,
any Poisson manifold is foliated by 
\textit{symplectic leaves},
which are
(connected)
symplectic submanifolds of maximal dimension.
The dimension of symplectic leaves
is determined by the rank of the Poisson bivector $\Pi$.

Provided that $\hy^\m$ is nontrivial,
$\hx^\m$, $\hy^\m$, and $p_\m$
in \eqref{M3*}
together describes
a $10$-dimensional Poisson submanifold embedded in the phase space $\ps$.
Its symplectic leaves are $8$-dimensional.
To see this,
regard 
the Poisson brackets in \eqref{M3*}
as components of an antisymmetric matrix
and compute its rank: $8$.
Furthermore, examining its kernel shows that
each of such symplectic leaves
is the locus of 
\begin{align}
    -p^2 \,=\, m^2
    \,,\quad
    -p^2\mem \hy^2 \,=\, w^2
    \,,
\end{align}
where $m$ and $w$ are constants.
Physically,
the first equation computes the \textit{rest mass} $m$
whereas
the second equation computes the \textit{spin magnitude} $w$.

In this way, it follows that
one can always identify an $8$-dimensional symplectic submanifold $\ups_{8}(m,w)$
inside any phase space $\ps$
of a massive system
with nontrivial $\hy^\m$.
This symplectic submanifold
will be referred to as the \textit{8-dimensional universal phase space}
of the massive spinning system,
on which the rest mass and the spin magnitude
are fixed to constant values $m$ and $w$:
\begin{align}
    \label{ups8}
    \ps
    \,\,\supset\,\,
    \ups_{8}(m,w)
    \,.
\end{align}

In turn, $8$ sets the minimal bound
for the dimension of a massive spinning phase space.
It achieves three translational degrees of freedom and one spinning degree of freedom.
It can be seen that 
$\ups_8(m,w)$ is homeomorphic to $\R^6 {\,\times\,} S^2$,
which
describes the coadjoint orbit
\cite{kirillov1975elements,kostant1970orbits,souriau1970structure}
of the Poincar\'e group
\cite{souriau1970structure,bacry1967space,arens1971classical,Carinena:1989uw}.
It follows that 
$\ups_8(m,w)$ is equivalent to
Souriau's ``$8$-dimensional space of motion''
reviewed in \rcite{Damour:2024mzo}.

Similarly,
$x^\m$, $\hy^\m$, and $p_\m$ 
in \eqref{M3}
together describes a $11$-dimensional
Poisson submanifold in $\ps$
whose symplectic leaves are $10$-dimensional.
Computation shows that
each symplectic leaf is the locus of 
$-p^2\mem \hy^2 = w^2$.
This identifies the \textit{10-dimensional universal phase space} $\ups_{10}(w)$,
on which the spin magnitude
is fixed to a constant value $w$:
\begin{align}
    \label{ups10}
    \ps
    \,\,\supset\,\,
    \ups_{10}(w)
    \,.
\end{align}
It can be seen that $\ups_{10}(m,w)$
is homeomorphic to $\R^8 {\,\times\,} S^2$.

It is possible to reproduce
the $8$-dimensional universal phase space in \eqref{ups8}
from the $10$-dimensional universal phase space in \eqref{ups10}:
\begin{align}
    \label{quotient}
    \ups_{8}(m,w)
    \,\,=\,\,
        \Big\{\mem{
            (x,\hy,p)
            \in
                \ups_{10}(w)
        \,\Big|\,\hem
            p^2 + m^2 = 0
        }\,\Big\}
        \hem\Big/\nem\hnem
            \sim
    \,.
\end{align}
This describes a quotient along
one-parameter orbits
in $\ups_{10}(w)$,
\begin{align}
    \label{reparam}
    \bigbig{
        x^\m
        ,\mem
        \hy^\m 
        ,\mem
        p_\m
    }
    \,\,\sim\,\,
    \bigbig{
        x^\m + k\mem p^\m
        ,\mem
        \hy^\m 
        ,\mem
        p_\m
    }
    \,,
\end{align}
which is the Hamiltonian flow of $p^2 + m^2$.
Physically, \eqref{reparam} implements a translation of the reconstructed spacetime coordinates in \eqref{x-def}
along the momentum direction.

Mathematically,
quotients of the form \eqref{quotient}
could be called 
\textit{symplectic quotient}:
impose a constraint \textit{and} quotient by its Hamiltonian flow
\cite{marsden1974reduction,meyer1973symmetries}.
The process of reducing a symplectic manifold to a smaller one
via symplectic quotient
will be referred to as \textit{symplectic reduction}.
Clearly,
symplectic reduction always reduces the dimension
by an even number.

\newpage

The physicists' toolkit for symplectic reduction
is known as 
\textit{Dirac bracket} 
\cite{Dirac:1950pj,dirac1964lectures,Henneaux:1992ig}.
In this paper,
we use Dirac brackets to 
explicitly realize symplectic quotients
as symplectic submanifolds.
See \App{DIRAC} for a review of symplectic reduction and Dirac bracket.

It is a simple exercise to show that
the universal phase space 
$\ups_8(m,w)$ in \eqref{ups8}
is a symplectic submanifold of
$\ups_{10}(w)$ in \eqref{ups10},
by constructing the Dirac bracket
that fixes $p^2 + m^2$ and $p \mdot x$ to zero
within $\ups_{10}(w)$:
\begin{align}
    \label{upss}
    \ps
    \,\,\supset\,\,
    \ups_{10}(w)
    \,\,\supset\,\,
    \ups_8(m,w)
    \,.
\end{align}
Surely, this reduction recipe
applies universally in every massive spinning system.

\section{The Four Models}
\label{FOUR}

In \Sec{FREE},
we have employed results due to \rcite{sst-asym}
to establish
the universal existence of
physical variables $x^\m$, $\hx^\m$, $\hy^\m$, and $p_\m$,
satisfying universal Poisson bracket relations
dictated by Poincar\'e symmetry.
In turn, the symplectic submanifolds 
$\ups_{10}(w)$ and $\ups_{8}(m,w)$
arise
universally,
characterized by 
mass $m$ or spin magnitude $w$.

The objective of this section is to demonstrate
this universal construction
for the four concrete example systems in \fref{overview}:
\begin{align*}
\begin{split}
    1.\,\,\,\text{
        Vector Oscillator
        (VO)
    }
    &\,,\quad
    3.\,\,\,\text{
        Spherical Top
        (ST)
    }
    ,\\
    2.\,\,\,\text{
        Spinor Oscillator
        (SO)
    }
    &\,,\quad
    4.\,\,\,\text{
        Massive Twistor
        (MT)
    }
    .
\end{split}
\end{align*}
Firstly, the formulae in \eqrefs{hxy-def}{x-def}
are applied to 
validate the universal Poisson bracket relations.
Secondly,
the universal Poisson bracket relations are reproduced as Dirac brackets
via concrete identifications of constraints.
We restrict our attention to free theory,
and wait to add on interactions in later sections.

The four models above 
are defined on symplectic manifolds
with varying dimensions:
\begin{align}
    \label{fantastic-four}
    \ps_{16}^\VO
    \,,\quad
    \ps_{12}^\SO
    \,,\quad
    \ps_{20}^\ST
    \,,\quad
    \ps_{16}^\MT
    \,.
\end{align}
These models tend to involve unphysical spinning degrees of freedom
for the sake of manifest relativistic covariance,
except in the case of spinor oscillator.
After appropriate reduction procedures,
the phase spaces in \eqref{fantastic-four} are boiled down to
\begin{align}
    \label{fantastic-four-dia}
    \ps_{12}^\VO
    \,,\quad
    \ps_{12}^\SO
    \,,\quad
    \ps_{14}^\ST
    \,,\quad
    \ps_{14}^\MT
    \,,
\end{align}
respectively.
It then follows that
\begin{align}
    \label{double}
    \ps_{12}^\VO
    \,\,\cong\,\,
    \ps_{12}^\SO \big/\mem \Z_2
    \,,\quad
    \ps_{14}^\ST
    \,\,\cong\,\,
    \ps_{14}^\MT \big/\mem \Z_2
    \,,
\end{align}
establishing equivalences up to double cover.

It is worth mentioning that
\textit{none} of the physical phase spaces in \eqref{fantastic-four-dia}
necessitate fixed spin magnitude.
On a related note,
we will assume strictly constant mass $m$
when imposing the mass-shell constraint,
although
one could incorporate the Regge trajectory \cite{Hanson:1974qy}
in principle;
see \App{REGGE} for a further exploration on this point.

For the reader's sake,
we recapitulate the key formulae in \Sec{FREE}.
It suffices to identify
\begin{align}
    \label{sum.xy}
    x^\m
    \,:=\,
        \frac{1}{p^2}\, 
        \BB{
            J^{\m\n}\mem p_\n
            - D\mem p^\m
        }
    \,,\quad
    \hy^\m
    \,:=\,
        - \frac{1}{p^2}\, {*}J^{\m\n}\mem p_\n
    \,,\quad
    p_\m
    \,,
\end{align}
as $\hx^\m = x^\m - p^\m\mem p\mdot x/ p^2$
is automatic by projection of $x^\m$ along $p^\m$.

\subsection{Vector Oscillator}
\label{FOUR>VO}

Our first example is
the vector oscillator model,
which has been discovered multiple times over the years 
\cite{Grassberger:1977tn,Rempel:2015foa,Deriglazov:2015bqa,Basile:2023vyg,Haddad:2024ebn}.
The idea of this model is
to implement the spin angular momentum
as a composite variable
arising from a pair of vectors,
just like orbital angular momentum.

\para{Definition}
Let $\mflat = (\R^4,\eta)$
denote the Minkowski space
equipped with $(-,+,+,+)$-signature flat metric $\eta_{\m\n}$.
The phase space of the vector oscillator model is
\begin{align}
    \label{VO.psr}
    \ps^\VO_{16}
    \,\,=\,\,
    \nem
        \BB{
            T^* \oplus T \oplus T^*
        }\mem \mflat
    \,,
\end{align}
which has dimension $4 + 4 + 4 + 4 = 16$.
This is a direct sum bundle
with real coordinates
\begin{align}
    \label{VO.coordsr}
    \bigbig{
        \fx^\m
        ,\mem
        p_\m
        ,\mem
        \fu^\m
        ,\mem
        \fv_\m
    }
    \,.
\end{align}
Here, $\fx^\m$ coordinatize the base manifold $\mflat$
while $p_\m$, $\fu^\m$, and $\fv_\m$ coordinatize the fibers.
In this precise geometric sense, $\fx^\m$ are external degrees of freedom
whereas $p_\m$, $\fu^\m$, and $\fv_\m$ are internal degrees of freedom.
The symplectic potential is given by
\begin{align}
    \label{VO.thetar}
    \theta
    \,=\,
        p_\m\mem d\fx^\m
        + \fv_\m\mem d\fu^\m
    \,,
\end{align}
so the nonvanishing components of the Poisson bracket are
\begin{align}
    \label{VO.pbr}
    \pb{\fx^\m}{p_\n}
    \,=\,
        \delta^\m{}_\n
    \,,\quad
    \pb{\fu^\m}{\fv_\n}
    \,=\,
        \delta^\m{}_\n
    \,.
\end{align}
A symplectic potential is a one-form whose exterior derivative is the symplectic form $\omega = d\theta$.
The pointwise inverse
of the symplectic form
is the Poisson bivector.
The components of the Poisson bivector are the Poisson brackets.
In this way, \eqref{VO.thetar} derives \eqref{VO.pbr}.

The spin sector is more conveniently described
in terms of complex variables
$\fa^\m = (\fu^\m + i\fv^\m)/\nem\sqrt{2}$,
in which case
the phase space is recast as
\begin{align}
    \label{VO.psc}
    \ps^\VO_{16}
    \,\,\cong\,\,
    \nem
        \BB{
            T^* \oplus T^\C
        }\mem \mflat
    \,.
\end{align}
Here, $T^\C\mflat$ denotes the complexified tangent bundle of $\mflat$.
The coordinates are now
\begin{align}
    \label{VO.coordsc}
    \bigbig{
        \fx^\m
        ,\mem
        p_\m
        ,\mem
        \fa^\m
    }
    \,,
\end{align}
while the symplectic potential reads 
\begin{align}
    \label{VO.thetac}
    \theta
    \,=\,
        p_\m\mem d\fx^\m
        + \frac{i}{2}\mem
        \BB{
            \fba_\m\mem d\hnem\fa^\m
            - d\hnem\fba_\m\mem \fa^\m
        }
    \,.
\end{align}
Now the nonvanishing components of the Poisson bracket are identified as
\begin{align}
    \label{VO.pbc}
    \pb{\fx^\m}{p_\n}
    \,=\,
        \delta^\m{}_\n
    \,,\quad
    \pb{\fa^\m}{\fba_\n}
    \,=\,
        -i\mem \delta^\m{}_\n
    \,,
\end{align}
so $\fa^\m$ and $\fba_\m$ could serve as
lowering and raising operators in the quantum theory.

\para{Universality}
Firstly, we apply \eqref{sum.xy}
to construct the physical variables
$(x^\m,\hy^\m,p_\m)$.

The Poincar\'e and dilatation charges
are readily derived by
implementing Noether theorem
in the Hamiltonian framework.
Due to the geometrical identification in \eqref{VO.psc},
spacetime translations act on the phase space $\ps^\VO_{16}$ as
$\d_\xi(\fx^\m,p_\m,\fa^\m) = (\fx^\m \mplus \xi^\m,p_\m,\fa^\m)$.
The momentum is the generator of this transformation,
which is simply $p_\m$.
Lorentz transformations, on the other hand,
rotate all variables as
$\d_\ve(\fx^\m,p_\m,\fa^\m) = (\ve^\m{}_\n\mem \fx^\m,-p_\n\mem \ve^\n{}_\m,$ $\ve^\m{}_\n\mem\fa^\n)$,
where
$\ve^{\m\n} = -\ve^{\n\m}$
is infinitesimal.
From the symplectic potential in \eqref{VO.thetac},
it follows that this transformation is generated by
$-\frac{1}{2}\, \ve_{\m\n}\mem J^{\m\n}$,
where
\begin{align}
    \label{VO.J}
    J^{\m\n}
    \,=\,
        \BB{
            \fx^\m p^\n - p^\m \fx^\n
        }
        + i\mem \BB{
            \fa^\m \fba^\n - \fba^\m \fa^\n
        }
    \,.
\end{align}
Meanwhile, the mass dimensions of $\fx^\m$, $p_\m$, and $\fa^\m$ are $-1$, $1$, and $0$, respectively.
Thus, the dilatation charge is
$D = -p\mdot \fx$.

From the above identification of the Poincar\'e and dilatation charges,
\eqref{sum.xy} implies
\begin{align}
    \label{VO.xy}
    x^\m
    \,=\,
        \fx^\m
        + \frac{i}{p^2}\, \BB{
            \fa^\m \fba^\n - \fba^\m \fa^\n
        }\mem p_\n
    \,,\quad
    \hy^\m
    \,=\,
        \frac{i}{-p^2}\,
            \ve^{\m\n\r\s} p_\n\mem \fa_\r\mem \fba_\s
    \,.
\end{align}
Direct computation using \eqrefs{VO.pbc}{VO.xy} verifies that 
\eqrefs{M3*}{M3} are reproduced in the vector oscillator model,
substantiating the claims in \Sec{FREE}.

\eqref{VO.xy} shows that
the basic spacetime variable $\fx^\m$ 
provided by the model
is mismatched from the physical center coordinates $x^\m$.
Indeed, $\pb{\fx^\m}{\fx^\n} = 0$ while $\pb{x^\m}{x^\n} \neq 0$.
Below, we refer to $\fx^\m$ as spurious center coordinates.

\para{Dirac Brackets}

Secondly,
we perform the Dirac bracket analysis.
The question here is whether imposing appropriate constraints on the phase space $\ps^\VO_{16}$
reproduces Dirac brackets that agree with
the universal Poisson brackets in \eqref{M3}.

Six constraints have been identified
in the literature
\cite{Rempel:2015foa,Rempel:2016jbn}:
\begin{align}
\begin{split}
    \label{VO.constraints}
    \Psi_1
    \,=\,
        p\mdot\hnem \fa 
    \,,&\quad 
    \Psi_2
    \,=\,
        \frac{1}{2}\, \fa^2
    \,,\quad 
    \Psi_3
    \,=\,
        \fba \mdot\hnem \fa - w
    \,,\\
    \bPsi_1
    \,=\,
        p\mdot\hnem \fba 
    \,,&\quad 
    \bPsi_2
    \,=\,
        \frac{1}{2}\, \fba^2
    \,,\quad 
    \Psi_4
    \,=\,
        \frac{1}{2i} \log\hnem \left( \frac{\eta\mdot\hnem \fa}{\eta\mdot\hnem \fba} \right)
    \,.
\end{split}
\end{align} 
Here, $\eta^\mu$ is a constant spacelike reference vector.
It is known that
\smash{$(\Psi_2,\bPsi_2,\Psi_3)$} form an $\mathrm{sl}(2,\R)$ algebra under the Poisson bracket \cite{Haddad:2024ebn}.
The constraints $(\Psi_1,\bPsi_1)$
shift the spurious center $\fx^\m$
by their Hamiltonian actions,
while the other constraints cannot.

The geometry of the internal phase space starts from
the unconstrained $T^*\mflat \cong \mflat^\C$
and is reduced as
$T^*\R^3 \supset T^*S^2 \supset S^2$
via successive impositions of 
$(\Psi_1,\bPsi_1)$,
$(\Psi_2,\bPsi_2)$,
and $(\Psi_3,\Psi_4)$.
The imposition of $(\Psi_1,\bPsi_1)$ is a strict necessity,
as it sets $\fx^\m = x^\m$
such that the ambiguities in defining the center position is removed.
Meanwhile, from physical grounds, the imposition of $(\Psi_2,\bPsi_2)$
would be also obligatory
since it kills
translational modes
in $T^*\R^3$
which apparently bear no clear physical significance.
The internal phase spaces $T^*S^2$ and $S^2$, on the other hand,
nicely qualify as rotational phase spaces.

\newpage

Consequently, the following series of reductions arises
for the vector oscillator model:
\begin{align}
    \label{VO.series}
    \ps^\VO_{16}
    \,\supset\,
    \ps^\VO_{12}
    \,\supset\,
    \ps^\VO_{10}(w)
    \,\supset\,
    \ps^\VO_{8}(m,w)
    \,.
\end{align}
Here, $\ps^\VO_{12}$ imposes $(\Psi_1,\bPsi_1,\Psi_2,\bPsi_2)$
while $\ps^\VO_{10}(w)$ imposes all constraints in \eqref{VO.constraints}.
The last reduction is via the constraint pair
$(p^2 \mplus m^2, p \mdot x)$
as discussed around \eqref{upss}.

The derivation of the Dirac brackets
is not difficult.
For instance,
on $\ps^\VO_{12}$ one finds
\begin{align}
\begin{split}
    \label{VO.db12}
    \pb{\fx^\m}{\fx^\n}_{12}
    \,=\,
        \frac{1}{-p^2}\,
            \fS^{\m\n}
    &\,,\quad
    \pb{\fx^\m}{p_\n}_{12}
    \,=\,
        \delta^\m{}_\n
    \,,\\
    \pb{\fx^\m}{\fa^\n}_{12}
    \,=\,
        \frac{1}{-p^2}\,
            \fa^\m\mem p^\n
    &\,,\quad
    \pb{\fa^\m}{\fba_\n}_{12}
    \,=\,
        -i\,\bb{
            \hdelta^\m{}_\n
            - \frac{\fba^\m \fa_\n}{\fba\mdot\fa}
        }
    \,,
\end{split}
\end{align}
where
$\fS^{\m\n} := i\mem (\fa^\m \fba^\n - \fba^\m \fa^\n)$.

It follows that
the Dirac brackets between 
$(\fx^\m,\hy^\m,p_\m)$
take the same form as
the Poisson brackets between $(x^\m,\hy^\m,p_\m)$
in \eqref{M3}
on $\ps_{12}^\VO$ or $\ps_{10}^\VO(w)$,
for instance.
Similarly,
the Dirac brackets between 
$(\fx^\m,\hy^\m,p_\m)$
take the same form as
the Poisson brackets between $(\hx^\m,\hy^\m,p_\m)$
in \eqref{M3}
on the fully reduced $\ps_8^\VO(m,w)$.
This establishes that 
the universal phase spaces
in \Sec{FREE}
are concretely realized in the vector oscillator model as
\begin{align}
    \ps_{10}^\VO(w)
    \,\cong\,
        \ups_{10}(w)
    \,,\quad
    \ps_8^\VO(m,w)
    \,\cong\,
        \ups_8(m,w)
    \,.
\end{align}

\subsection{Spinor Oscillator}
\label{FOUR>SO}

Our second example is 
the spinor oscillator model.
This model is explicitly formulated in 
Woodhouse \cite{woodhouse1997geometric}
(see pp.\:117-119),
for instance,
and appears also in \rrcite{Zima:1995db,Lyakhovich:1996we,Rempel:2016jbn}.
The idea is to realize the spin phase space $S^2$
as the Riemann sphere $\mathbb{CP}^1$,
in terms of an internal two-spinor variable $\zeta_\a$.
This is essentially the 
relativistic generalization of the construction
\cite{jordan1935zusammenhang,Schwinger1952NYO3071}
widely known under the name of
``Schwinger's oscillator model of spin.''

\para{Definition}
The phase space of
the spinor oscillator model
is
\begin{align}
    \label{SO.ps}
    \ps^\SO_{12}
    \,\,=\,\,
        \nem\BB{
            T^* \oplus \Sm
        }\mem \mflat
    \,,
\end{align}
where $\Sm\mflat$
is the left-handed spinor bundle over
the flat spacetime $\mflat$
with zero section removed.
That is, the typical fiber of $\Sm\mflat$
is $\C^2_\times := \C^2 \setminus\nem \{0\}$.
The phase space in \eqref{SO.ps} has $4 + 4 + 4 = 12$ real dimensions
and is coordinatized by
\begin{align}
    \label{SO.coords}
    \bigbig{
        x^\m
        ,\mem
        p_\m
        ,\mem
        \zeta_\a
    }
    \,,
\end{align}
where $\zeta_\a \in \C^2_\times$.
The symplectic potential is
\begin{align}
\begin{split}
    \label{SO.theta}
    \theta
    \,=\,
        p_\mu\mem dx^\mu 
        + 
        \frac{1}{2i}\mem
        \BB{\mem
            \bzeta_\da\, \hat{p}^{\da\a} d\zeta_\a 
            - d\bzeta_\da\mem \hat{p}^{\da\a}\hem \zeta_\a
        }
    \,,
\end{split}
\end{align}
where $\hat{p}_\m {\,:=\,} p_\m/|p|$
and $|p| {\,:=\,} (-p^2)^{1/2}$.
The nonzero components of the Poisson bracket are 
\begin{align}
\begin{split}
    \label{SO.pb}
    \pb{x^\m}{x^\n}
    \,=\,
        \frac{1}{-p^2}\,
            S^{\m\n}
    &\,,\quad
    \pb{x^\m}{\zeta_\a}
    \,=\,
        \frac{
            (\sigma^{\m\n}\zeta)_\a
        }{2p^2}\,    
            p_\n
    \,,\\
    \pb{x^\m}{p_\n}
    \,=\,
        \delta^\m{}_\n
    &\,,\quad
    \pb{\zeta_\a}{\bzeta_\da}
    \,=\,
        i\mem \hat{p}_{\a\da}
    \,.
\end{split}
\end{align}
Our spinor conventions are identical to \rcite{Kim:2021rda}.
$S^{\m\n}$ is a composite variable
defined as
\begin{align}
    \label{SO.S}
    S^{\m\n} 
    \,:=\,
        \frac{1}{4i}\,\mem
        \bzeta_\da\mem
        \BB{\mem
            (\bar{\sigma}^{\m\n})^\da{}_\db\,
            \hat{p}^{\db\a}
            \mem-\mem
            \hat{p}^{\da\b}\mem 
            (\sigma^{\m\n})_\b{}^\a
        \mem}\mem
        \zeta_\a
    \,.
\end{align}


\para{As a Reformulation}

The spinor oscillator model
is essentially a reformulation
of the vector oscillator model in \Sec{FOUR>VO}
(cf. \rcite{Rempel:2016jbn}).
Recall the four constraints
$(\Psi_1, \bPsi_1, \Psi_2, \bPsi_2)$
in \eqref{VO.constraints}
that achieved the $12$-dimensional $\ps_{12}^\VO$ in \eqref{VO.series}.
The constraints $\Psi_2 {\:=\:} 0$ and $\bPsi_2 {\:=\:} 0$
are explicitly solved by 
a spinorial parametrization of the form
\begin{align}
    \fa^{\da\a} 
    \,=\,
        \frac{1}{\sqrt{2w}}\,
            \xi^\da \zeta^\a 
    \,.
    \label{v2s}
\end{align}
To impose
$\Psi_1 {\:=\:} 0$ and $\bPsi_1 {\:=\:} 0$
as well,
one stipulates the relation\footnote{
    This declares a Euclidean conjugation relationship
    between $\xi^\da$ and $\zeta_\a$
    in the rest frame of $\hat{p}^\m$.
    The sign choice in \eqref{Econj} sets $-\hat{p}^{\da\a} \doteq \delta^{\da\a}$ in the rest frame
    in our conventions.
    Also, $\zeta^\a = + \xi_\db\mem \hat{p}^{\db\a}$.
}
\begin{align}
    \xi^\da \,=\, - \hat{p}^{\da\a}\mem \zeta_\a 
    \,.
    \label{Econj}
\end{align}
At this stage,
we are free to take any one of the spinor variables
$\zeta_\a$ and $\xi^\da$
as fundamental
and regard the other as dependent.

By plugging in \eqref{v2s}
to $\fa^\m$ given in \eqref{VO.thetac},
renaming $\fx^\m$ to $x^\m$,
and discarding a total derivative,
one finds
\begin{align}
\begin{split}
    \label{SO.thetasolved}
    \theta
    \,=\,
    p_\m\hem dx^\m
    + \frac{i}{2}\mem\BB{
        \bxi^\a\mem d\zeta_\a -\hem d\bzeta_\da\mem \xi^\da 
    }
    \,.
\end{split}
\end{align}
This readily reproduces \eqref{SO.theta}
upon eliminating $\xi^\da$ via \eqref{Econj}.
Mathematically, this implements an explicit embedding $\ps_{12}^\SO \to \ps_{16}^\VO$
and its pullback.

In this sense,
the phase space $\ps^\SO_{12}$ of the spinor oscillator model
in \eqref{SO.ps}
provides a spinorial reformulation of the
constrained phase space $\ps_{12}^\VO$
of the vector oscillator model
in \eqref{VO.series}
by solving
$(\Psi_1, \bPsi_1, \Psi_2, \bPsi_2)$
explicitly.
Their precise relation is given in \eqref{double},
which arises from
the double cover relationship between
the spin phase spaces
$\C^2_\times$ and $T^*S^2 {\mem\setminus\mem} S^2$;
consider the representation of a two-spinor as a flag
\cite{penrose:1985spinors1},
for instance.

This equivalence up to double cover
is established again by the fact that
the Poisson brackets 
between $(x^\m, p_\m, \a^\m, \ba_\m)$
in the spinor oscillator model
reproduces the Dirac brackets
of the vector oscillator model
in \eqref{VO.db12}
via \eqrefs{v2s}{Econj}.

As a side note,
we may want to remark on
a chiral approach
that can be useful in some calculations.
The idea is to take
the left-handed (anti-self-dual) spinors
$\zeta_\a$ and $\bxi^\a$
as fundamental variables,
which yet sacrifices
manifest reality of the symplectic structure.
Then the symplectic potential in \eqref{SO.thetasolved}
becomes
\begin{subequations}
\label{SO.thetac}
\begin{align}
    \label{SO.theta-}
    \theta
    \,=\,
        p_{\a\da}\mem\hem
        d\bb{
            x^{\da\a}
            \mem-\mem
            \frac{i}{4|p|}\, \BB{
                \xi^\da \bar{\xi}^\a -\bar{\zeta}^\da \zeta^\a
            } 
        \nem}
        + i\mem \bxi^\a\mem d\zeta_\a
    \,.
\end{align}
If instead the right-handed (self-dual) spinors
$\xi^\da$ and $\bzeta_\da$
are taken as fundamental,
the symplectic potential in \eqref{SO.thetasolved}
becomes
\begin{align}
    \label{SO.theta+}
    \theta
    \,=\,
        p_{\a\da}\mem\hem
        d\bb{
            x^{\da\a}
            \mem+\mem
            \frac{i}{4|p|}\, \BB{
                \xi^\da \bar{\xi}^\a -\bar{\zeta}^\da \zeta^\a
            } 
        \nem}
        - i\mem d\bzeta_\da\mem \xi^\da
    \,.
\end{align}
\end{subequations}
By using any of \eqrefsor{SO.theta-}{SO.theta+},
the symplectic form can be 
presented in the Darboux format
at the expense of manifest reality.

\para{Universality}

Now
we apply \eqref{sum.xy}
to identify
the physical variables
$(x^\m,\hy^\m,p_\m)$.

Working similarly as in \Sec{FOUR>VO},
the Poincar\'e and dilatation charges
are found as
\begin{align}
    \label{charges.SO}
    p_\m
    \,,\quad
    J^{\m\n}
    \,=\,
        \BB{
            x^\m p^\n - x^\n p^\m
        }
        + S^{\m\n}
    \,,\quad
    D
    \,=\,
        -p\mdot x
    \,,
\end{align}
where $S^{\m\n}$ is given in \eqref{SO.S}.
From \eqrefs{sum.xy}{charges.SO},
it follows that
the default position variable $x^\m$
of the spinor oscillator model
is exactly the physical center coordinates,
as the notation has already suggested.
The spin length pseudovector is
\begin{align}
    \label{SO.y}
    \hy^{\da\a}
    \,=\,
        -
        \frac{1}{4|p|}\, \BB{
            \xi^\da \bar{\xi}^\a -\bar{\zeta}^\da \zeta^\a
        }
    \,,
\end{align}
so \eqref{SO.S} describes
$S^{\m\n} = \ve^{\m\n\r\s}\mem \hy_\r\mem p_\s$.

Direct computation using \eqrefs{SO.pb}{SO.y}
verifies that 
\eqrefs{M3*}{M3}
are reproduced in the spinor oscillator model,
substantiating the claims in \Sec{FREE}.

\para{Dirac Brackets}

Next,
we perform the Dirac bracket analysis.

The constraint pair relevant for fixing the spin magnitude can be identified as
\begin{align}
    \label{SO.constraints}
    \phi_1
    \,=\,
        \bar{\xi}^\a \zeta_\a - 2w
    \,,\quad
    \chi^1
    \,=\,
        \frac{i}{2}\mem
            \log\bb{
                \frac{
                    \langle \eta \zeta \rangle
                }{
                    \langle \bxi \eta \rangle
                }
            }
    \,,
\end{align}
with the aid of a constant reference spinor $\eta^\a$.
Here, $\phi_1$ generates $\U(1)$ orbits 
in the space of $\zeta_\a$.
Consequently, the following series of reductions arises for the spinor oscillator model:
\begin{align}
    \label{SO.series}
    \ps_{12}^\SO
    \,\supset\,
    \ps_{10}^\SO(w)
    \,\supset\,
    \ps_8^\SO(m,w)
    \,.
\end{align}
For instance, the Dirac bracket on $\ps_{10}^\SO(m,w)$ is found as
\begin{align}
\begin{split}
    \label{SO.db10}
    \pb{x^\m}{x^\n}_{10}
    \,=\,
        \frac{1}{-p^2}\,
            S^{\m\n}
    &\,,\quad
    \pb{x^\m}{\zeta_\a}_{10}
    \,=\,
        \frac{
            (\sigma^{\m\n}\zeta)_\a
        }{2p^2}\,    
            p_\n
    \,,\quad
    \pb{\zeta_\a}{\zeta_\b}_{10}
    \,=\,
        \frac{i}{4}\, \e_{\a\b}
    \,,\\
    \pb{x^\m}{p_\n}_{10}
    \,=\,
        \delta^\m{}_\n
    &\,,\quad
    \pb{\zeta_\a}{\bzeta_\da}_{10}
    \,=\,
        i\mem 
        \bb{
            \hat{p}_{\a\da}
            - \frac{1}{2}\, 
                \frac{\zeta_\a\mem \eta^\b \hat{p}_{\b\da}}{\lsq\bzeta|\hat{p}|\eta\rangle}
            - \frac{1}{2}\,
                \frac{\eta_\a\mem \bzeta_\da}{\langle\eta\zeta\rangle}
        }
    \,.
\end{split}
\end{align}

It is left as an exercise to verify that
the Poisson brackets between
$(x^\m,\hy^\m,p_\m)$
reproduce \eqref{M3}
in $\ps_{12}^\SO$.
Since $(x^\m,\hy^\m,p_\m)$ are invariant under the $\U(1)$ rephasing of $\zeta_\a$,
this implies that
the Dirac brackets between 
$(x^\m,\hy^\m,p_\m)$
on $\ps_{10}^\SO$
also reproduces \eqref{M3}.
Similarly,
the Dirac brackets between 
$(x^\m,\hy^\m,p_\m)$
reproduce \eqref{M3*}
on $\ps_{10}^\SO(m,w)$.
Therefore,
the universal phase spaces
in \Sec{FREE}
are concretely realized in the spinor oscillator model as
\begin{align}
    \label{SO.get}
    \ps_{10}^\SO(w)
    \,\cong\,
        \ups_{10}(w)
    \,,\quad
    \ps_8^\SO(m,w)
    \,\cong\,
        \ups_8(m,w)
    \,.
\end{align}
Note the absence of $\Z_2$ quotients here.

\subsection{Spherical Top}
\label{FOUR>ST}

Our third example is
the spherical top model,
introduced by Hanson and Regge \cite{Hanson:1974qy}
in 1974.
The idea of this model is to pursue a smooth conceptual extension of
rigid body dynamics in Newtonian mechanics
by generalizing the Eulerian angles---%
as coordinates on the group manifold $\SO(3)$---%
to a Lorentz frame.


Concretely, let $\SO^+\hnem(1,3)$ denote the restricted Lorentz group,
and
let $\mflat \rtimes \SO^+\hnem(1,3)$ be the restricted Poincar\'e group.
The phase space of the spherical top model is
the cotangent bundle of the restricted Poincar\'e group,
which has dimension $2 \times (4+6) = 20$:
\begin{align}
    \label{ST.ps}
    \ps^\ST_{20}
    \,\,=\,\,
    \nem
        T^*(
            \mflat \rtimes \SO^+\hnem(1,3)
        )
    \,.
\end{align}
The coordinates are
\begin{align}
    \label{ST.coords}
    \bigbig{
        \fx^\m
        ,\mem
        \fL^\m{}_A
        ,\mem
        p_\m
        ,\mem
        \fS_{\m\n}
    }
    \,,
\end{align}
where $(\fx^\m,\fL^\m{}_A) \in \mflat {\,\rtimes\,} \SO(1,3)$
coordinatize the base group manifold
while $(p_\m,\fS_{\m\n}) \in \mathrm{Lie}(\mflat {\,\rtimes\,} \SO^+\hnem(1,3))$
coordinatize the cotangent fibers.
$\fL^\m{}_A {\:\in\:} \SO^+\hnem(1,3)$ is a restricted Lorentz group element,
dubbed body frame.
$\fS_{\m\n} {\:=\:} {-\fS_{\n\m}}$ is a Lorentz algebra element.

\begin{figure}[t]
    \centering
    \includegraphics[scale=0.4]{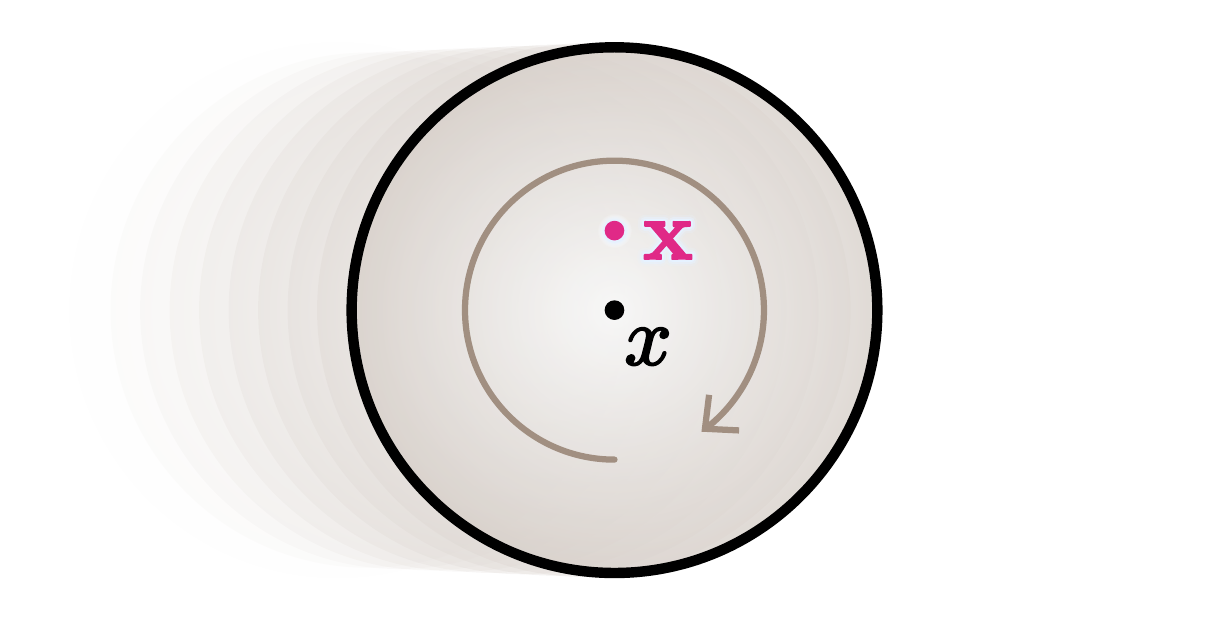}
    \caption{
        A demonstration of a Lorentz-violating SSC,
        due to Fleming \cite{fleming1965covariant}.
        Suppose a perfectly symmetric rotating sphere
        whose geometric center is at rest
        in an inertial frame.
        Suppose an observer
        that moves with a constant relative velocity.
        In the observer's frame,
        the apparent center of energy
        (defined in terms of the energy density $T^{00}$),
        $\fx$,
        is displaced from the geometric center.
        Namely, the Lorentz factor
        describes that
        the fast-moving parts of the sphere
        appear heavy,
        while the slow-moving parts appear light.
        However,
        the physical center $x$ 
        has always been
        the geometric center.
        The mismatch between $\fx$ and $x$
        is
        because the observer has
        defined $\fx$
        in a way that involves their own four-velocity vector $u^\m_\text{obs}$,
        which is a Lorentz-violating reference:
        $T^{00} = T_{\m\n}\hem u_\text{obs}^\m u_\text{obs}^\n$.
        This demonstrates the fakeness or arbitrariness of the spurious center $\fx$
        in noncovariant SSCs.
    }
    \label{fleming}
\end{figure}

The symplectic potential reads
\begin{align}
    \label{ST.theta}
    \theta
    \,=\,
        p_\m\mem d\fx^\m
        + \frac{1}{2}\, 
            \fS_{\m\n}\mem \fL^\m{}_A\mem d\hnem\fL^\n{}_B
            \mem \eta^{AB}
    \,,
\end{align}
so the nonvanishing components of the Poisson bracket are
\begin{align}
\begin{split}
    \label{ST.pb}
    \pb{\fx^\m}{p_\n}
    \,\,&=\,\,
        \delta^\m{}_\n
    \,,\\
    \pb{\fL^\m{}_A}{\fS_{\r\s}}
    \,\,&=\,\,
        \BB{
            -2\mem \delta^\m{}_{[\r} \eta_{\s]\n}
        }\mem
        \fL^\n{}_A
    \,,\\
    \pb{\fS_{\m\n}}{\fS_{\r\s}}
    \,\,&=\,\,
        \BB{
            -4\mem \delta^{[\k}{}_{[\m}\mem
            \eta_{\n][\r}\mem
            \delta^{\l]}{}_{\s]}
        }\mem \fS_{\k\l}
    \,.
\end{split}
\end{align}

The spherical top model notoriously
suffers from
unphysical redundancies,
often discussed in the context of
SSCs
\cite{pryce1948mass,Newton:1949cq,moller1949dynamique,moller1949definition,fleming1965covariant,Dixon:1970zz,Steinhoff:2015ksa}; 
see \fref{fleming}.
Crucially,
the generalization from $\SO(3)$ to $\SO^+\hnem(1,3)$ 
doubles the number of spinning variables
for the sake of relativistic covariance,
introducing fake \textit{boost} degrees of freedom.
Three of the six components of
$\fL^\m{}_A$
are unphysical;
three of the six components of
$\fS_{\m\n}$
are also unphysical.
In particular, $\fS_{\m\n}$
does not faithfully represent the physical spin angular momentum of the system,
the intricacy of which leads to
ambiguities in
the definition of the center position, $\fx^\m$.
Below, we refer to 
$\fx^\m$, $\fL^\m{}_A$, and $\fS_{\m\n}$
respectively as
spurious position,
body frame,
and
spin tensor.

\para{Universality}

Firstly,
we apply \eqref{sum.xy}
to construct the physical variables
$(x^\m,\hy^\m,p_\m)$.

The Poincar\'e charges
are uniquely determined from
the natural Poincar\'e group action
(left action)
on the cotangent bundle
$\ps^\ST_{20} = T^*( \mflat \rtimes \SO^+\hnem(1,3) )$.
Meanwhile, the dilatation charge simply
follows from the mass dimensions:
$-1$, $0$, $+1$, and $0$
for $\fx^\m$, $\fL^\m{}_A$, $p_\m$, and $\fS_{\m\n}$.
The results are
\begin{align}
    \label{ST.JD}
    p_\m
    \,,\quad
    J^{\m\n}
    \,=\,
        \BB{
            \fx^\m p^\n - \fx^\n p^\m
        }
        + \fS^{\m\n}
    \,,\quad
    D \,=\, -p \mdot \fx
    \,.
\end{align}

From the above identification of the Poincar\'e and dilatation charges,
\eqref{sum.xy} implies
\begin{align}
    \label{ST.xy}
    x^\m
    \,=\,
        \fx^\m
        + \frac{1}{p^2}\, \fS^{\m\n}\mem p_\n
    \,,\quad
    \hy^\m
    \,=\,
        \frac{1}{-p^2}\,
            {*}\fS^{\m\n}\mem p_\n
    \,.
\end{align}
Direct computation using \eqrefs{ST.pb}{ST.xy} verifies that
\eqrefs{M3*}{M3} are reproduced in the spherical top model,
substantiating the claims in \Sec{FREE}.

\para{Dirac Brackets}

Secondly,
we perform the Dirac bracket analysis.

The three spurious degrees of freedom
in the spherical top model
have been addressed since
the original work \cite{Hanson:1974qy}.
However,
their precise mathematical formulation
within the well-established framework of gauge redundancies
was only facilitated by
Steinhoff \cite{Steinhoff:2015ksa}.
Upon a slight yet important refinement \cite{Kim:2021rda},
the three gauge generators are identified as
\begin{align}
    \label{phia}
    \phi_a
    \,=\,
        \frac{1}{2}\,
            \BB{
                \hat{p}^\m +\hem \fL^\m{}_0
            }\mem
                \fS_{\m\n}
            \mem \fL^\n{}_a
    \qiq
    \pb{\phi_a}{\phi_b}
    \,=\,
        0
    \,,
\end{align}
where $a \in \{1,2,3\}$
describes the spatial part of the internal index $A \in \{0,1,2,3\}$.
The vanishing Poisson bracket in \eqref{phia}
establishes that the gauge algebra is abelian,
so the gauge orbits form $\R^3$
\cite{Kim:2021rda}.
To remove all spurious spinning degrees of freedom,
one takes the symplectic quotient
\begin{align}
    \label{ST.quotient14}
    \ps_{14}^\ST
    \,\,=\,\,
        \Big\{\mem{
            (\fx,\fL,p,\fS)
            \in
                \ps_{20}^\ST
        \,\Big|\,\hem
            \phi_a = 0
        }\,\Big\}
        \hem\Big/\mem
            \R^3
    \,.
\end{align}

In this framework,
SSCs
are elegantly reformulated as gauge-fixing conditions
\cite{Steinhoff:2015ksa}.
For example, consider
the gauge-fixing functions
\begin{align}
    \label{chia}
    \text{Covariant Gauge}\,
    :\quad
    \fchi^a
    \,=\,
        \hat{p}_\m\hem \fL^{\m a}
    \,.
\end{align}
Imposing both $\phi_a {\:=\:} 0$ and $\fchi^a {\:=\:} 0$
not only
implements the covariant SSC
$\fS_{\m\n}\mem p^\n {\:=\:} 0$
but also aligns
the spurious body frame 
with a rest frame of the momentum:
$\fL^\m{}_0 {\:=\:} \hat{p}^\m$.
Hence
three boost components are removed 
from both $\fS_{\m\n}$ and $\fL^\m{}_A$
as desired.
In this gauge fixing,
the spurious center coincides with the physical center.

\newpage

For another example,
consider the gauge-fixing functions
\begin{align}
    \label{PNW}
    \text{Canonical Gauge}\,
    :\quad
    \fchi^a
    \,=\,
        l_\m\hem \fL^{\m a}
    \,,
\end{align}
where $l^\m$ is
a constant vector that is
unit and timelike: $l^2 {\:=\:} {-1}$.
Imposing both $\phi_a {\:=\:} 0$ and $\fchi^a {\:=\:} 0$
now implements
$\fS_{\m\n}\hem ( \hat{p}^\n {\:+\:} l^\n ) = 0$,
which is known as the Pryce-Newton-Wigner
\cite{pryce1948mass,Newton:1949cq}
SSC.
As will be revisited in \Sec{APPL>PN},
this is the SSC
canonically
utilized in post-Newtonian calculations:
the constant unit timelike vector
is naturally interpreted as
the four-velocity
of an inertial observer
so that
$\fL^\m{}_0 = l^\m = u^\m_\text{obs}$.
Evidently, this observer four-velocity
is an explicit Lorentz-violating reference structure.
Thus, in this gauge choice,
Lorentz symmetry 
is spuriously broken down
to the spatial $\SO(3)$ isotropy in the observer's rest frame.
See also the appendix of \rcite{sst-asym}
for an insightful ``bootstrap'' derivation of this gauge-fixing condition. 

Any choice
for the three gauge-fixing functions
establishes
the following series of symplectic reductions arises for the spherical top model:
\begin{align}
    \label{ST.series}
    \ps_{20}^\ST
    \,\supset\,
    \ps_{14}^\ST
    \,\supset\,
    \ps_{12}^\ST(m)
    \,.
\end{align}

The Dirac brackets on $\ps_{14}^\ST$ and $\ps_{12}^\ST(m)$
have been well-described
in various gauges
\cite{Hanson:1974qy,Kim:2021rda,Lee:2023nkx}.
If one adopts the covariant gauge condition in \eqref{chia},
the Dirac bracket
on $\ps_{14}^\ST$
describes
\begin{align}
\begin{split}
\label{ST.db14@}
    \pb{\fx^\mu}{\fx^\nu}_{14}
    \,=\, 
        \frac{1}{-p^2}\,
            \fS^{\m\n}
    &\,,\quad
    \pb{\fx^\m}{\fS_{\r\s}}_{14} 
    \,=\, 
        \frac{1}{p^2}\,
            2\mem \fS^\m{}_\wrap{[\r}\mem p_\wrap{\s]}
    \,,\\
    \pb{\fx^\mu}{p_\nu}_{14} 
    \,=\, 
        \delta^\m{}_\n 
    &\,,\quad
    \pb{\fS_{\m\n}}{\fS_{\r\s}}_{14}
    \,=\,
        \BB{
            -4\mem \hdelta^{[\k}{}_{[\m}\mem
            \hat{\eta}_{\n][\r}\mem
            \hdelta^{\l]}{}_{\s]}
        }\mem \fS_{\k\l}
    \,,\\
    \pb{ \fL^\m{}_a }{ \fx^\n }_{14} 
    \,=\,
        \frac{1}{p^2}\,
            p^\m\hem
            \fL^\n{}_a
    &\,,\quad
    \pb{ \fL^\m{}_a }{ \fS_{\r\s} }_{14} 
    \,=\,
        \BB{
            -2\mem \hdelta^\m{}_{[\r} \hat{\eta}_{\s]\n}
        }\mem
        \fL^\n{}_a
    \,,
\end{split}
\end{align}
where $\hdelta^\m{}_\n {\::=\:} \delta^\m{}_\n - p^\m p_\n/p^2$
and $\hat{\eta}_{\m\n} {\::=\:} \eta_{\m\n} - p_\m p_\n /p^2$.
Computation shows that
the Dirac brackets between $(\fx^\m,\hy^\m,p_\m)$
due to \eqref{ST.db14@}
take the same form as
the universal Poisson brackets in \eqref{M3}.
Similarly,
one can verify that
the Dirac brackets between $(\fx^\m,\hy^\m,p_\m)$
on $\ps_{12}^\ST$
take the same form as
\eqref{M3*}.

To elaborate,
the first two rows in \eqref{ST.db14@}
directly reproduce the universal Poisson algebra.
The last row in \eqref{ST.db14@},
on the other hand,
encodes extra information 
only available for $T^*\SO(3)$ models.
Also, note that $\pb{\fL^\m{}_a}{\fS_{\r\s}}_{14}$ in \eqref{ST.db14@}
boils down to
\begin{align}
    \label{ST.ly}
    \pb{\fL^\m{}_a}{\hy^\n}_{14}
    \,&=\,
        \frac{1}{p^2}\,
            \ve^{\m\n}{}_{\r\s}\mem
            p^\r \fL^\s{}_a
    \,.
\end{align}

The spin magnitude
has not been fixed so far.
By further reductions of the spin phase space
described in \Sec{HIER},
one constructs submanifolds of $\ps_{14}^\ST$ and $\ps_{12}^\ST(m)$,
such as
\begin{align}
    \ps_{10}^\ST(w)
    \,\cong\,
        \ups_{10}(w)
    \,,\quad
    \ps_8^\ST(m,w)
    \,\cong\,
        \ups_8(m,w)
\end{align}
that provide explicit embeddings of the universal phase spaces in \Sec{FREE}.

\subsection{Massive Twistor} 
\label{FOUR>MT}

Our fourth example is 
the massive twistor model,
which traces back to 
the '70s twistor particle program
of Penrose and Perj\'es
\cite{Penrose:1974di,Perjes:1974ra,tod1977some}.
While this model
has been revisited several times
in \rrcite{Bette:1989zt,Bette:2004ip,Fedoruk:2007dd,Deguchi:2015iuw},
we follow the modern formulation put forward by \rrcite{Kim:2021rda,ambikerr1},
which has provided a proper interpretation of the internal symmetries.

It is now widely known that
the configurations of a massive spinning particle
can be represented by the massive spinor-helicity variable $\lambda_\a{}^I$ \cite{ahh2017}.
The idea of the massive twistor model
is to
take the space of massive spinor-helicity variables
as the configuration space
so that its cotangent bundle
serves as the phase space
of a massive spinning particle.

\para{Definition}
The precise definition of
the phase space of the massive twistor model is
\begin{align}
    \label{MT.ps}
    \ps^\MT_{16}
    \,\,=\,\,
        \C^8
        \setminus\nem
        \Delta
    \,,
\end{align}
where $\Delta$ is a quadric in $\C^8$.
This phase space has $16$ real dimensions
and is coordinatized by $8$ complex variables,
\begin{align}
    \label{MT.coords}
    \bigbig{
        \lambda_\a{}^I
        ,\mem
        \mu^{\da I}
    }
    \,.
\end{align}
Here, $\a {\:\in\:} \{0,1\}$ and $\da {\:\in\:} \{0,1\}$ are $\mathrm{SL}(2,\C)$ spinor indices,
while $I {\:\in\:} \{0,1\}$ is an $\mathrm{SU}(2)$ spinor index.
The quadric $\Delta$ removes the degenerate locus 
$\det(\lambda) = 0$, so
\begin{align}
    \lambda_\a{}^I
    \,\in\,
    \GL(2,\C)
    \,,\quad
    \mu^{\da I}
    \,\in\,
    \C^4
    \,.
\end{align}

The symplectic potential is
\begin{align}
\begin{split}
    \label{MT.theta}
	\theta
	\,=\,
    \frac{1}{2}\,\BB{
        \bmu_I{}^\a\mem d\lambda_\a{}^I
        - d\bmu_I{}^\a\mem \lambda_\a{}^I
    }
    +
    \frac{1}{2}\,\BB{
        d\rambda_{I\da}\mem \mu^{\da I}
        - \rambda_{I\da}\mem d\mu^{\da I}
    }
    \,,
\end{split}
\end{align}
which describes that $(\lambda_\a{}^I,\mu^{\da I})$
are canonical (Darboux) coordinates
in the complex sense.
Thus,
the nonzero components of the Poisson bracket are simply
\begin{align}
\begin{split}
    \label{MT.pb}
    \pb{\lambda_\a{}^I}{\bmu_J{}^\b}
    \,=\,
        \delta_I{}^J\mem \delta_\a{}^\b
    \,.
\end{split}
\end{align}
As usual,
we omit components related by taking complex conjugations:
$[\lambda_\a{}^I]^* {\:=\:} \rambda_{I\da}$
and
$[\mu^{\da I}]^* {\:=\:} \bmu_I{}^\a$.\footnote{
    The $\SU(2)$ spinor representation is pseudoreal,
    meaning that an invariant tensor $\delta_{I\bar{I}}$
    is involved in the complex conjugation.
}
It should be clear from \eqrefs{MT.theta}{MT.pb}
that $\ps^\MT_{16}$ can be viewed as
the cotangent bundle of the space $\GL(2,\C)$ of massive spinor-helicity variables.

At first sight, the massive twistor model
seems disparate from
any of the three spinning particle models
discussed in \Secss{FOUR>VO}{FOUR>SO}{FOUR>ST}.
Especially,
\eqrefs{MT.ps}{MT.pb} are purely based on spinorial variables
so that
the notion of spacetime seems completely absent.
Hence one could wonder how this model can even 
achieve a conventional picture of
a particle in spacetime.
Notably, however, \rcite{sst-asym}'s construction
shown in \eqref{sum.xy}
reconstructs and identifies spacetime
in any phase space with Poincar\'e symmetry,
as is demonstrated below.

\para{Symmetries}

We start by releasing more details
on the phase space
in terms of its symmetries,
although the reader is encouraged to skip the following
and directly jump to \eqref{MT.charges}.
The twistor space
could be 
defined as the representation space $\C^{2,2}$ of the conformal group $\SU(2,2)$,
which is a K\"ahler vector space
\cite{penrose:maccallum,tod1977some}.
The complex linear space $\C^8$ in \eqref{MT.ps}
has denoted 
the product $(\C^{2,2})^2$ of two twistor spaces.
To see this,
consider
\begin{align}
    \label{blocks}
    Z_\rmA{}^I
    \,=\,
        \begin{pmatrix}
            \lambda_\a{}^I
            \\
            i\hhem \mu^{\da I}
        \end{pmatrix}
    \,,\quad
    \bZ_I{}^\rmA
	\,=\,
	\BB{
		-i\hem \bmu_I{}^\a	
		\,\,\,
		\rambda_{I\da}
	}
    \,,
\end{align}
which identifies
two twistors (Dirac spinors) $Z_\rmA{}^0$ and $Z_\rmA{}^1$.
In \eqref{blocks},
the $(2,2)$-signature metric $A^{\bar{\rmB}\rmA}$ of the twistor space
facilitates the conjugation
$\bZ_I{}^\rmA = [Z_\rmB{}^I]^*\mem A^{\bar{\rmB}\rmA}$.
Note that the Poisson bracket relation in \eqref{MT.pb}
translates to $i\mem \pb{Z_\rmA{}^I}{\bZ_J{}^\rmB} = \delta_\rmA{}^\rmB\mem \delta_J{}^I$.

On the linear space of $Z_\rmA{}^I$,
the conformal group $\SU(2,2)$
acts from the left
(i.e., on the Dirac spinor index $\rmA$)
while
an $\SU(2)$ internal symmetry
acts from the right
(i.e., on the $\SU(2)$ spinor index $I$).
There is also
a total $\U(1)$ rephasing transformation
$Z_\rmA{}^I \mapsto e^{i\psi/2}\hem Z_\rmA{}^I$.
The interpretation established by \rrcite{Kim:2021rda,ambikerr1}
states that
the $\SU(2)$ is global 
(massive little group)
while the $\U(1)$ is merely gauge.
Meanwhile,
the fifteen Noether charges of
the $\SU(2,2)$ group action
arise from
the traceless bilinear
$Z_\rmA{}^I \bZ_I{}^\rmB - \frac{1}{4}\mem \delta_\rmA{}^\rmB\mem (\bZ_K{}^\rmC Z_\rmC{}^K)$.
These encode spacetime transformations
and derive \eqref{MT.charges}.

\para{Universality}

Let us apply \eqref{sum.xy}
to construct
the universal variables $(x^\m,\hy^\m,p_\m)$.
This analysis was performed by \rcite{sst-asym}.


The conformal group action $\SU(2,2)$ on $\ps^\MT_{16}$
includes Poincar\'e and dilatation
transformations.
This uniquely stipulates
the Poincar\'e and dilatation charges as
\cite{Kim:2021rda,ambikerr1}
\begin{align}
\begin{split}
    \label{MT.charges}
    p_{\a\da}
    \mem=\mem
        -\lambda_\a{}^I\mem \rambda_{I\da}
    \,,\,\,\,\,
    J^\da{}_\db
    \mem=\mem
		\mu^{\da I}\hem \rambda_\wrap{I\db}
		- \tfrac{1}{2}\mem \delta^\da{}_\db\,
			\mu^{\dc I}\hem \rambda_\wrap{I\dc}
    \,,\,\,\,\,
    D 
    \mem=\mem
        \frac{1}{2}\,\BB{
            \rambda_{I\da}\mem \mu^{\da I}
            + \bmu_I{}^\a\mem \lambda_\a{}^I
        }
    \,.
\end{split}
\end{align}
Meanwhile, in the spinor notation,
\eqref{sum.xy}
boils down to
\begin{align}
    \label{Gpformula}
    x^{\da\a} + i\hem \hy^{\da\a}
    \,=\,
        - \BB{
            J^\da{}_\db + \tfrac{1}{2}\, \delta^\da{}_\db\mem D
        }\mem 
        (p^{-1})^{\db\a}
    \,.
\end{align}
By plugging in \eqref{MT.charges},
one identifies the equation
\begin{align}
    \label{MT.zcoords}
    x^{\da\a} + i\hem y^{\da\a}
    \,=\,
        \mu^{\da I}\mem (\lambda^{-1})_I{}^\a
    \,,
\end{align}
which describes
$x^\m + iy^\m$
in the vector notation.
It follows that the real part $x^\m$ is the physical spacetime coordinates,
while the imaginary part $y^\m$ yields the physical spin length pseudovector $\hy^\m$
via the projection $\hy^\m = y^\m - p^\m\mem  p\mdot y/p^2$.
\eqrefs{MT.pb}{MT.zcoords}
verify
\eqrefs{M3*}{M3}
in the massive twistor model,
substantiating the claims in \Sec{FREE}.

Note that \eqref{MT.zcoords} can be written as
\begin{align}
    \label{incidence}
    \mu^{\da I}
    \,=\,
        \BB{
            x^{\da\a} + i\hem y^{\da\a}
        }\mem
        \lambda_\a{}^I
    \,,
\end{align}
which is known as the massive generalization of
incidence relation.
The incidence relation is a central equation in twistor theory that establishes the relationship between
twistor space and (complex) spacetime
\cite{penrose1967twistoralgebra}.
Amusingly,
\rcite{sst-asym}'s
derivation of \eqref{MT.zcoords}
rediscovers the massive incidence relation,
verifying that
the complexified spacetime in twistor particle theory
\cite{penrose:maccallum,newman1974curiosity,Shirafuji:1983zd}
unifies spacetime and spin as real and imaginary parts.

\para{As a Reformulation}

The massive twistor model is essentially a reformulation of the spherical top model
in \Sec{FOUR>ST}.
To see this,
recall the constraints $\phi_a$ and $\fchi^a$
in \eqrefs{phia}{chia}
that derived the constrained phase space $\ps_{14}^\ST {\:\subset\:} \ps_{20}^\ST$.
Notably, these constraints
are explicitly solved by the following parametrization: 
\begin{align}
\begin{split}
    \label{ST.solve}
    p_{\a\da}
    \mem=\mem
        -\lambda_\a{}^I\mem \rambda_{I\da}
    \,,\,\,\,\,\,\,
    \fL^{\da\a}{}_a
    \mem\propto\mem
        (\rambda^{-1})^{\da I}\mem
        (\s_a)_I{}^J\hem
        (\lambda^{-1})_J{}^\a
    \,,\,\,\,\,\,\,
    \fS_{\m\n}
    \mem=\mem
        \ve_{\m\n\r\s}\mem y^\r\hem p^\s
    \,.
\end{split}
\end{align}
The first two spinorial parametrizations in \eqref{ST.solve}
trivialize 
the orthogonality $p_\m\hem \L^\m{}_a = 0$.
In fact, they precisely describe the well-known
parametrization of 
massive momentum $p_\m$
and polarization vectors $\L^\m{}_{a=1,2,3}$
in terms of massive spinor-helicity variables $\lambda_\a{}^I$.
In the meantime,
the last parametrization in \eqref{ST.solve}
trivializes the covariant SSC $S_{\m\n}\hem p^\n = 0$
in terms of the pseudovector variable $y^\m$.

Yet, these are at the expense of two additional redundancies.
First,
\eqref{ST.solve} is invariant under
the $\U(1)$ rephasing transformation
$\lambda_\a{}^I \sim e^{i\psi/2}\mem \lambda_\a{}^I$.
Second,
\eqref{ST.solve} is invariant under
the shift transformation
$y^\m \sim y^\m + k\mem p^\m$
for $k {\:\in\:} \R$.

By plugging in 
\eqref{ST.solve}
to \eqref{ST.theta}
and renaming $\fx^\m$ to $x^\m$, 
one obtains
\begin{align}
    \label{K.theta}
	\theta
	\,=\,
	- \lambda_\a{}^I\mem \rambda_{I\da}\, dx^{\da\a}
	+ i\mem y^{\da\a}\mem
	\BB{
		\lambda_\a{}^I\hem d\rambda_{I\da}
		{\,-\,}
		d\lambda_\a{}^I\mem \rambda_{I\da}
	}
    \,,
\end{align}
defined on the 
$16$-dimensional space of
$x^\m {\:\in\:} \R^4$, $y^\m {\:\in\:} \R^4$, and $\lambda_\a{}^I {\:\in\:} \GL(2,\C)$:
$4 + 4 + 8$.
It is easy to show that \eqref{K.theta}
is equivalent to the symplectic potential in \eqref{MT.theta}
via the massive incidence relation in \eqref{incidence}.
It should be clear that
\eqref{MT.theta} has computed 
a pullback due to 
an embedding
$\ps_{14}^\MT \to \ps_{20}^\ST$.

In sum, 
the massive twistor model can be regarded as
a reformulation of the spherical top
that arises by solving the constraints
for the covariant SSC,
explicitly in a spinorial fashion.
Their precise relation is given in \eqref{double},
which arises from
the double cover relationship between
the spin phase spaces
$T^*\SU(2)$ and $T^*\SO(3)$.
It could be helpful to note that
$\GL(2,\C) \supset \SL(2,\C) \supset \SU(2)$.
In \rrcite{Kim:2021rda,ambikerr1},
this equivalence up to double cover
is established by
matching between the Dirac brackets
in the spherical top and massive twistor models
via \eqref{ST.solve}.

This realization provides an alternative view on
the massive twistor model
that emphasizes the spacetime picture.
Namely,
there exists
an invertible diffeomorphism
that maps 
$\C^8 \setminus\nem \Delta$ in \eqref{MT.ps}
to a direct sum bundle over Minkowski space:
\begin{align}
    \label{K.ps}
    \ps^\MT_{16}
    \,\,\cong\,\,
        \BB{
            T \oplus \SS
        }\mem \mflat
    \,.
\end{align}
Here, 
$\SS\mflat$ denotes a fiber bundle over $\mflat$
whose typical fiber is $\GL(2,\C)$:
massive spinor-helicity bundle over spacetime.
That is, \eqref{K.ps}
is coordinatized by
\begin{align}
    \label{K.coords}
    \bigbig{
        x^\m
        ,\mem
        y^\m
        ,\mem
        \lambda_\a{}^I
    }
    \,.
\end{align}

Concretely, the invertible diffeomorphism
from $(T \oplus \SS)\mem \mflat$
to $\C^8 \setminus\nem \Delta$
is $(x^\m,y^\m,\lambda_\a{}^I) \mapsto {(\lambda_\a{}^I,(x \mplus iy)^{\da\a}\mem \lambda_\a{}^I)}$,
which provides a mathematical formalization 
for the massive incidence relation in \eqref{incidence}.\footnote{
    In twistor-theoretic terms,
    \eqref{K.ps}
    describes the massive analog of the correspondence space.
    Physically, working in this massive correspondence space
    is what \rrcite{Shirafuji:1983zd,Fedoruk:2007dd}
    mean by the ``hybrid'' description.
}
The one-form in \eqref{K.theta}
is the pullback of the one-form in \eqref{MT.theta}
by this diffeomorphism.


\para{Dirac Brackets}

The above analysis leads to a clear identification of the constraints of massive twistor model.
The unphysical redundancy is the $\U(1)$ redundancy,
generated by
\begin{align}
    \label{MT.phi1}
    \phi_1
    \,=\,
        - \frac{1}{2}\,
            \bZ_I{}^\rmA Z_\rmA{}^I
    \,=\,
        \frac{1}{2i}\,\BB{\hem
            \rambda_{I\da}\mem \mu^{\da I}
            - \bmu_I{}^\a\mem \lambda_\a{}^I
        }
    \,=\,
        -p \mdot y
    \,.
\end{align}
The resulting symplectic quotient is
\begin{align}
    \label{MT.quotient14}
    \ps_{14}^\MT
    \,\,=\,\,
        \Big\{\mem{
            (\lambda_\a{}^I,\mu^{\da I})
            \in
                \ps_{16}^\MT
        \,\Big|\,\hem
            \phi_1 = 0
        }\,\Big\}
        \hem\Big/\mem
            \U(1)
    \,,
\end{align}
which is free from
both the rephasing and shift redundancies
identified earlier.
A particular slice on the $\U(1)$ orbit may be chosen in terms of the gauge-fixing function
\begin{align}
    \label{MT.chi1}
    \chi^1 
    \,=\,
        \frac{1}{2i}\mem
        \log\bb{
            \frac{\det(\lambda)}{\det(\rambda)}
        }
    \,.
\end{align}
Consequently, the following series of reductions arises for the massive twistor model:
\begin{align}
    \label{MT.series}
    \ps_{16}^\MT
    \,\supset\,
    \ps_{14}^\MT
    \,\supset\,
    \ps_{12}^\MT(m)
    \,.
\end{align}

Computation shows that the Dirac brackets between
$(x^\m,y^\m,p_\m)$ on $\ps_{14}^\MT$
take the same form as the universal Poisson brackets in \eqref{M3}:
see \rcite{Kim:2024grz}.
Again, the Dirac brackets split into universal and model-specific parts: 
see \eqref{MT.db14}.
Similarly,
the Dirac brackets between
$(x^\m,y^\m,p_\m)$ on $\ps_{12}^\MT$
take the same form as 
\eqref{M3*}.

By further reductions of the spin phase space
described in \Sec{HIER},
one constructs submanifolds of $\ps_{14}^\MT$ and $\ps_{12}^\MT(m)$
providing explicit embeddings of the universal phase spaces:
\begin{align}
    \ps_{10}^\MT(w)
    \,\cong\,
        \ups_{10}(w)
    \,,\quad
    \ps_8^\MT(m,w)
    \,\cong\,
        \ups_8(m,w)
    \,.
\end{align}

\section{Hierarchy of Spin Phase Spaces}
\label{HIER}

In \Sec{FOUR},
we have reviewed four spinning particle models
in Hamiltonian formulation
and clarified their constraints.
After appropriate removal of unphysical degrees of freedom,
the oscillator class of models
describes $12$-dimensional reduced phase spaces
while
the spherical top and massive twistor models
describe $14$-dimensional reduced phase spaces.
It can be seen that they are respectively isomorphic to
$T^*\mflat \times T^*S^2$
and
$T^*\mflat \times T^*\SO(3)$
up to double cover,
which describe
spin (rotational) phase spaces
$T^*S^2$ and $T^*\SO(3)$.

In this section, 
we establish a hierarchy of spin phase spaces:
\begin{align}
    \label{hierarchy}
    T^*\SO(3)
    \,\supset\,
    T^*S^2
    \,\supset\,
    S^2
    \,.
\end{align}
Via double cover, it also holds that
$T^*\SU(2) \supset \C^2_\times \supset \mathbb{CP}^1 \cong S^2$.
This establishes that 
all of the four models
are physically equivalent up to symplectic quotients,
in the free theory.
In the interacting theory,
this fact will imply that
the physics of the four models
are the same
as long as their interactions with external fields
are compatible with the quotients.

\newpage

\subsection{Definitions}
\label{HIER>DEF}

For the sake of clarity,
we shall start by stating the definitions of each spin phase space in \eqref{hierarchy}.
It suffices to work in the nonrelativistic notation.

\para{$\bm{T^*\SO(3)}$}

The six-dimensional spin phase space $T^*\SO(3)$
is the cotangent bundle of the group manifold $\SO(3)$.
It is coordinatized by
\begin{align}
    \label{spin6.coords}
    \bigbig{
        \Lambda^i{}_a
        ,\mem
        S_i
    }
    \,,
\end{align}
where the indices $i$ and $a$
run through $\{1,2,3\}$.
The group element $\L^i{}_a \in \SO(3)$
encodes the three Eulerian angles.
The fiber coordinates $S_i$
describe the spin angular momentum as a pseudovector.
Physically, this phase space describes
the nonrelativistic spherical top
with fixed center.
The Poisson structure is given by
\begin{align}
    \label{pb6}
    \pb{\L^i{}_a}{\L^j{}_b}_6
    \,=\,
        0
    \,,\quad
    \pb{\L^i{}_a}{S_k}_6
    \,=\,
        \ve^i{}_{kj}\, \L^j{}_a
    \,,\quad
    \pb{S_i}{S_j}_6
    \,=\,
        \ve^k{}_{ij}\mem S_k
    \,.
\end{align}

There are two $\SO(3)$ symmetries acting on this phase space:
the left symmetry that rotates the external index $i$
and
the right symmetry that rotates the internal index $a$.
The Poisson brackets in \eqref{pb6}
are invariant under these symmetry actions.

\para{$\bm{T^*S^2}$}

The four-dimensional spin phase space $T^*S^2$
is the cotangent bundle of the two-sphere $S^2$.
It is coordinatized by 
$(\theta,\phi,p_\theta,p_\phi)$,
where $(\theta,\phi)$ coordinatizes the base $S^2$.
The Poisson structure is given such that
$(p_\theta,p_\phi)$ are canonically conjugate
to $(\theta,\phi)$.\footnote{
    A discussion of $T^*S^2$ as a spin phase space
    may be an uncommon practice in the literature.
    For instance, \rcite{Pauli:1927qhd} briefly introduces four variables $(s,\chi, s_z,\varphi)$
    but then quickly discards $(s,\chi)$ to descend to $S^2$.
    A similar approach is taken by \rcite{Kuzenko:1994ju} as well.
}

Physically, $T^*S^2$ is the phase space of 
a familiar example in constrained Hamiltonian mechanics:
particle on a sphere.
A point particle in $\R^3$ describes the canonical brackets
\begin{align}
    \label{UV.pb}
    \pb{U^i}{U^j}
    \,=\, 0
    \,,\quad
    \pb{U^i}{V^j}
    \,=\, \delta^{ij}
    \,,\quad
    \pb{V^i}{V^j}
     \,=\, 0
    \,,
\end{align}
where $\vec{U}$ and $\vec{V}$ 
denote the position and momentum, respectively.
The set of constraints that restricts this particle on
a sphere $S^2$ embedded in the three-space
is given by
\begin{align}
    \label{UV.constraints}
    \vec{U}^2 - \rho^2
    \,=\, 0
    \,,\quad
    \vec{V} \mdot\mem \vec{U}
    \,=\, 0
    \,.
\end{align}
Here, the radius of the sphere is characterized by a dimensionful constant $\rho$
while $\vec{V} \mdot\mem \vec{U} = 0$ ensures the tangency of the particle's momentum to the sphere.
Imposing the constraints in \eqref{UV.constraints}
derives the Dirac bracket,
\begin{align}
\begin{split}
    \label{pb4.UV}
    \pb{U^i}{U^j}_4
    \,&=\,
        0
    \,,\\[0.05\baselineskip]
    \pb{U^i}{V^j}_4
    \,&=\,
        \delta^{ij}
        - \frac{1}{\rho^2}\,
            U^i\hem U^j
    \,,\\
    \pb{V^i}{V^j}_4
    \,&=\,
        - \frac{1}{\rho^2}\,
        \BB{
            U^i\hem V^j - V^i\hem U^j
        }
    \,.
\end{split}
\end{align}
It can be shown that
\eqref{pb4.UV} precisely describes the canonical Poisson structure on $T^*S^2$.
Mathematically, this establishes that 
$T^*S^2$ is embedded in $T^*\R^3$.
The explicit embedding map 
describes $U^i = \rho\mem (\sin\theta\cos\phi,\sin\theta\sin\phi,\cos\theta)$,
for instance.

\newpage

Notably, 
there exists an alternative embedding of
$T^*S^2$ in $T^*\R^3$.
Suppose three-vector variables $\vec{u}, \vec{v} \in \R^3$
that satisfy 
\begin{align}
    \label{uv.pb}
    \pb{u^i}{u^j}
    \,=\, 0
    \,,\quad
    \pb{u^i}{v^j}
    \,=\, \delta^{ij}
    \,,\quad
    \pb{v^i}{v^j}
    \,=\, 0
    \,.
\end{align}
To clarify,
this posits a new $T^*\R^3$
that is independent from the earlier $T^*\R^3$ in \eqref{UV.pb}.
With this understanding, suppose constraints
\begin{align}
    \label{uv.constraints}
    \vec{u}^2 - \vec{v}^2 
    \,=\, 0
    \,,\quad
    \vec{u} \mdot \vec{v} 
    \,=\, 0
    \,.
\end{align}
The resulting Dirac bracket describes
\begin{align}
\begin{split}
    \label{pb4}
    \pb{u^i}{u^j}_4
    \,&=\,
        - \frac{1}{2|\vec{u}||\vec{v}|}\,
        \BB{
            u^i\hem v^j - v^i\hem u^j
        }
    \,=\,
    \pb{v^i}{v^j}_4
    \,,\\
    \pb{u^i}{v^j}_4
    \,&=\,
        \delta^{ij}
        - \frac{1}{2|\vec{u}||\vec{v}|}\,
        \BB{
            u^i\hem u^j + v^i\hem v^j
        }
    \,,
\end{split}
\end{align}
which describes the Poisson structure of a four-dimensional submanifold.
It turns out that this submanifold is exactly $T^*S^2$,
as \eqref{pb4} precisely reproduces \eqref{pb4.UV}
via
\begin{align}
    \vec{u}
    \,=\,
        \frac{\rho}{|\vec{U}|}\,
        \vec{U}
    \,,\quad
    \vec{v}
    \,=\,
        \frac{|\vec{U}|}{\rho}\,
        \vec{V}
    \qiq
    \vec{u} \times \vec{v}
    \,=\,
        \vec{U} \times \vec{V}
    \,.
\end{align}
In this light,
the constant $\rho$ would be considered spurious:
it completely disappears from the formulation in \eqref{pb4}.

Physically, this alternative embedding of $T^*S^2$
in $T^*\R^3$
arises from the nonrelativistic vector oscillator model.
The constraints in
\eqref{uv.constraints}
correspond to
$(\Psi_2,\bPsi_2)$
in \eqref{VO.constraints}.

\para{$\bm{S^2}$}

The two-dimensional spin phase space $S^2$
is the two-sphere with coordinates $(\theta,\phi)$
in which the area element,
oriented as $\omega = w\mem \sin\theta\, d\phi \swedge d\theta$,
serves as the symplectic form
\cite{shankar2012principles}.
This means to have the Poisson bracket 
\begin{align}
    \label{pb2.angles}
    \pb{\theta}{\phi}_2
    \,=\, 
        \frac{1}{w\mem \sin\theta}
    \,.
\end{align}

Alternatively,
the two-dimensional spin phase space $S^2$
can be realized as a coadjoint orbit
\cite{kirillov1975elements,kostant1970orbits,souriau1970structure,woodhouse1997geometric,kirillov2004lectures}.
Consider the three-dimensional Poisson manifold $\R^3 \hhnem\nem\setminus\nem\nem \{0\}$,
whose coordinates $S_i$ satisfy the Poisson bracket relation
$\pb{S_i}{S_j} = \ve^k{}_{ij}\mem S_k$.
The rank of $\ve^k{}_{ij}\mem S_k$
as an antisymmetric matrix,
described by row and column indices $i$ and $j$, 
is $2$.
Explicitly, the Casimir $\vec{S}^2$ Poisson-commutes with any function of $\vec{S}$.
Thus, the symplectic leaves are characterized by the equation
\begin{align}
    \label{S2leaf}
    \vec{S}^2 
    - w^2
    \,=\,
        0
    \,,
\end{align}
where $w > 0$ is a constant.
The symplectic leaf characterized by \eqref{S2leaf}
is a two-sphere $S^2 \subset \R^3$ of radius $w$
and exhibits the Poisson structure
\begin{align}
    \label{pb2}
    \pb{S_i}{S_j}_2
    \,=\,
        \ve^k{}_{ij}\mem S_k
    \,.
\end{align}

The above two definitions are equivalent.
The explicit embedding $S^2 \to \R^3$ is
\begin{align}
    \label{angle-to-S}
    S_i
    \,=\,
        w\mem 
        \bigbig{
            \sin\theta\cos\phi
            ,\mem
            \sin\theta\sin\phi
            ,\mem
            \cos\theta
        }
    \,.
\end{align}
Combining
\eqrefs{pb2.angles}{angle-to-S}
reproduces \eqref{pb2}.

\newpage

\subsection{Successive Reductions}
\label{HIER>RED}

Provided the characterizations of the spin phase spaces in \Sec{HIER>DEF},
we can now readily establish their hierarchy
via successive implementations of symplectic reduction.

\para{From 6 to 4}
First, we establish the reduction from
the six-dimensional
$T^*\SO(3)$ to 
the four-dimensional
$T^*S^2$.
This reduction is relevant for the spherical top and massive twistor models in \Secs{FOUR>ST}{FOUR>MT}.
It suffices to work in the nonrelativistic notation.

Take the definition of $T^*\SO(3)$
by \eqref{pb6}.
Let $R^a{}_b$ be an arbitrary constant $\SO(3)$ matrix.
Imposing $S_i\, \L^i{}_a\mem R^a{}_1 {\:=\:} 0 {\:=\:} S_i\, \L^i{}_a\mem R^a{}_2$
makes $\L^i{}_a\hem R^a{}_3$ parallel to $S^i$.
Without loss of generality,
we can perform a right symmetry transformation
$\L^i{}_a \mapsto \L^i{}_b\mem (R^{-1})^b{}_a$,
$S_i \mapsto S_i$
on $T^*\SO(3)$
to have
\begin{align}
    \label{constraints64}
    \Psi_1
    \,=\,
        S_i\, \L^i{}_1
    \,,\quad
    \Psi_2
    \,=\,
        S_i\, \L^i{}_2
    \,.
\end{align}
Computation shows that the Dirac bracket
setting $\Psi_1 = 0 = \Psi_2$
is given by
\begin{align}
\label{db4}
{\renewcommand{\arraystretch}{1.9}
\begin{array}{lll}
    \displaystyle
    \pb{\L^i{}_1}{\L^j{}_2}_4
    \,=\,
        \frac{\L^i{}_3\mem \L^j{}_3}{|\vec{S}|}
    &\,,\quad
    \displaystyle
    \pb{\L^i{}_1}{\L^j{}_3}_4
    \,=\,
        -\frac{\L^i{}_3\mem \L^j{}_2}{|\vec{S}|}
    &\,,\quad
    \displaystyle
    \pb{\L^i{}_a}{S_k}_4
    \,=\,
        \ve^i{}_{kj}\, \L^j{}_a
    \,,\\
    \displaystyle
    \pb{\L^i{}_2}{\L^j{}_3}_4
    \,=\,
        \frac{\L^i{}_3\hem \L^j{}_1}{|\vec{S}|}
    &\,,\quad
    \displaystyle
    \pb{\L^i{}_3}{\L^j{}_3}_4
    \,=\,
        \frac{\ve^{ij}{}_k\mem \L^k{}_3}{|\vec{S}|}
    &\,,\quad
    \displaystyle
    \pb{S_i}{S_j}_4
    \,=\,
        \ve^k{}_{ij}\, S_k
    \,.
\end{array}}
\end{align}

The four-dimensional constrained phase space which \eqref{db4} describes is $T^*S^2$.
The explicit embedding
$T^*S^2 \to T^*\SO(3)$
is given by
\begin{align}
    \label{4to6}
    \bigbig{
        \L^i{}_1
        ,\mem
        \L^i{}_2
        ,\mem
        \L^i{}_3
    }
    \,=\,
        \bb{
            \frac{u^i}{|\vec{u}|}
            ,\mem
            \frac{v^i}{|\vec{v}|}
            ,\mem
            \frac{
                \ve^i{}_{jk}\hem u^j v^k
            }{|\vec{u}||\vec{v}|}
        }
    \,,\quad
    S^i
    \,=\,
        \ve^i{}_{jk}\hem u^j v^k
    \,.
\end{align}
\eqref{pb4} implies \eqref{db4}
via \eqref{4to6}.

\para{From 4 to 2}
Next, we establish the reduction from the four-dimensional $T^*S^2$ to the two-dimensional $S^2$.
This reduction is relevant for all models
in our setup,
as $T^*S^2$ serves as a bridge between $T^*\SO(3)$ and $S^2$.

Take the definition of $T^*S^2$ by \eqref{pb4}.
Suppose constraints
\begin{align}
    \label{constraints42}
    \Psi_3
    \,=\,
        |\vec{u}||\vec{v}|
        -
        w
    \,,\quad
    \Psi_4
    \,=\,
        \tan^{-1}\nem\bb{
            \frac{
                \vec{\eta}\mdot\vec{v}
            }{
                \vec{\eta}\mdot\vec{u}
            }
        }
    \,,
\end{align}
where $\vec{\eta}$ is a constant reference.
These are the nonrelativistic analogs of
$(\Psi_3,\Psi_4)$
in \eqref{VO.constraints}.
$\Psi_3$ generates $\SO(2)$ rotations on the two-plane
perpendicular to $\vec{S} = \vec{u}\times\vec{v}$
while fixing $|\vec{S}|$;
$\Psi_4$ is the angle such that 
$\pb{\Psi_3}{\Psi_4}_4 = -1$.
The resulting Dirac bracket 
exhibits dependencies in $\vec{\eta}$
and may describe subtleties
regarding topological obstructions
in promoting $\Psi_4$ to
a globally defined function.
Regardless, $S^i = \ve^i{}_{jk}\mem u^j\hem v^k$
serve as good coordinates on the two-dimensional constraint surface.
Their Dirac brackets are precisely \eqref{pb2}
since $\pb{S_i}{\Psi_3} = 0$,
establishing that
imposing \eqref{constraints42}
on $T^*S^2$
derives $S^2$.

\para{From 6 to 2}

Lastly, it should be clear that 
a direct reduction from the six-dimensional $T^*\SO(3)$ to the two-dimensional $S^2$
is also viable,
in which case one imposes
$(\Psi_1,\Psi_2,\Psi_3,\Psi_4)$
at once
with 
$(\Psi_1,\Psi_2)$ in \eqref{constraints64}
and
$\Psi_3 = |\vec{S}| - w$,
$\Psi_4 = \tan^{-1}( \eta_i\mem \Lambda^i{}_2 / \eta_j\mem \Lambda^j{}_1 )$.

\newpage

\section{Universality in Interacting Theory: Electromagnetism}
\label{INT1}

Through \Secs{FREE}{FOUR},
we have explicitly established that
any model of a massive spinning particle
describes
physical variables
$(x^\m, \hy^\m, p_\m)$
that exhibit the following
universal Poisson bracket relation
in the \textit{free theory}:
\begin{align}
\begin{split}
\label{xyp}
    \pb{x^\m}{x^\n}^\circ
    \,\,&=\,\,
        \frac{1}{-p^2}\, 
            \ve^{\m\n\r\s} \hy_\r\mem p_\s
    \,=\,
    \pb{\hy^\m}{\hy^\n}^\circ
    \,,\\
    \pb{x^\m}{\hy^\n}^\circ
    \,\,&=\,\,
        \frac{1}{-p^2}\,\BB{
            \hy^\m p^\n + p^\m \hy^\n 
        }
    \,,\\[0.15\baselineskip]
    \pb{x^\m}{p_\n}^\circ
    \,\,&=\,\,
        \delta^\m{}_\n
    \,.
\end{split}
\end{align}
Our objective now
is to
generalize such universal structures
in the \textit{interacting theory}.
The output is model-agnostic derivations of
the physical EoM of massive spinning particles
coupled to electromagnetism and gravity.
For clarity of exposition, we first discuss electromagnetic interactions in this section and defer gravitational interactions to \Sec{INT2}.

\subsection{Couplings as Symplectic Perturbations}
\label{INT>SPT}

Our point of departure is
the celebrated fact in symplectic geometry
that
a manifestly gauge-invariant coupling
of particles to electromagnetic fields
is viable by
modifying the symplectic structure
in terms of the field strength.
This construction is due to Souriau \cite{souriau1970structure}
and also traces back to Feynman \cite{dyson1990feynman}.
Historically, this manifestly gauge-invariant method has served as an important milestone in the study of particle mechanics in the symplectic framework
\cite{woodhouse1997geometric,torrence1973gauge,guillemin1978equations,sternberg1978classical,guillemin1990symplectic}.
Unfortunately,
it appears to be less recognized in recent literature,
so we shall provide a brief review
(see also \rcite{csg}).

For the sake of concreteness,
let us suppose a simple example:
a massive scalar particle in Minkowski space.
This is a Hamiltonian system defined on
the cotangent bundle $T^*\mflat$.
In the free theory,
this phase space is
equipped with the symplectic form and Hamiltonian
\begin{align}
    \label{scalar.omega0}
    \omega^\circ \,=\,
        dp_\m \swedge dx^\m
    \,,\quad
    \phi_0 \,=\,
        \frac{1}{2m}\mem
        \BB{
            p^2 + m^2
        }
    \,.
\end{align}
We have started to put the accent $^\circ$
to explicitly indicate free-theory objects.
The canonical Poisson bracket reads
\begin{align}
    \label{scalar.pb0}
    \pb{x^\m}{x^\n}^\circ
    \,=\,
        0
    \,,\quad
    \pb{x^\m}{p_\n}^\circ
    \,=\,
        \delta^\m{}_\n
    \,,\quad
    \pb{p_\m}{p_\n}^\circ
    \,=\,
        0
    \,.
\end{align}

The Souriau-Feynman method of electromagnetic coupling
modifies the geometric data in \eqref{scalar.omega0} as
\begin{align}
    \label{scalar.omega}
    \omega \,=\,
        dp_\m \swedge dx^\m 
        \,+\,
        qF
    \,,\quad
    \phi_0 \,=\,
        \frac{1}{2m}\mem
        \BB{
            p^2 + m^2
        }
    \,,
\end{align}
where $F = \frac{1}{2}\mem F_{\m\n}(x)\mem dx^\m \swedge dx^\n$ is the field strength two-form,
and $q$ is the electric charge.
Crucially, the Hamiltonian is left unchanged
while the symplectic form is perturbed as
\begin{align}
    \label{omega-split}
    \omega \,=\,
        \omega^\circ \mem+\, \omega'
    \,,
\end{align}
with $\omega' = qF$.
As a result, the Poisson bracket in \eqref{scalar.pb0} is altered as
\begin{align}
    \label{scalar.pb}
    \pb{x^\m}{x^\n}
    \,=\,
        0
    \,,\quad
    \pb{x^\m}{p_\n}
    \,=\,
        \delta^\m{}_\n
    \,,\quad
    \pb{p_\m}{p_\n}
    \,=\,
        qF_{\m\n}(x)
    \,.
\end{align}

\newpage

To see why \eqref{scalar.omega} achieves the electromagnetic coupling,
one can derive the EoM:
\begin{align}
\begin{split}
    \label{lorforce}
    \dot{x}^\m
    \mem=\mem
        \pb{x^\m}{\phi_0}
    \mem=\mem
        \frac{p^\m}{m}
    \,,\quad
    \dot{p}_\m
    \mem=\mem
        \pb{p_\m}{\phi_0}
    \mem=\mem
        \frac{q}{m}\,
        F_{\m\n}(x)\mem p^\n
    \,.
\end{split}
\end{align}
\eqref{lorforce} exactly describes 
the EoM of a charged scalar particle in electromagnetic background.
The modified $\pb{p_\m}{p_\n}$ bracket in \eqref{scalar.pb}
implements the Lorentz force law
in \eqref{lorforce}.
As a result, the intuition for noncanonical bracket $\pb{p_\m}{p_\n} \neq 0$
is the cyclotron precession exhibited by the momentum $p_\m$
in the external electromagnetic field.

Crucially,
this derivation of the Lorentz force law
does not invoke the electromagnetic gauge potential at all,
as is emphasized by both 
Souriau \cite{souriau1970structure}
and Feynman \cite{dyson1990feynman}.
As a result, $p_\m$
is the gauge-invariant, physical momentum
(also known as kinetic momentum).

The only consistency requirement in this framework
is the closure of the modified symplectic form in \eqref{scalar.omega}
\cite{souriau1970structure},
which holds via the Bianchi-type field equations $dF = 0$ on the background \cite{dyson1990feynman}.
Physically, this closure
amounts to the \textit{Liouville property} of classical time evolution:
conservation of classical probability
(as phase space measure) \cite{Kim:2025sey,csg}.

If one insists on keeping the canonical Poisson brackets,
one can perform
a gauge-dependent coordinate transformation 
in the phase space
(a change of worldline field basis):
\begin{align}
    \label{kincan}
    p_\m
    \,=\,
        \fp_\m - qA_\m(x)
    \,.
\end{align}
Here, $A {\:=\:} A_\m(x)\mem dx^\m$ is the electromagnetic gauge potential
such that $F {\:=\:} dA$.
As the symplectic form in \eqref{scalar.omega}
is the exterior derivative of
$p_\m\mem dx^\m + qA = \fp_\m\mem dx^\m$,
the geometric data in
\eqref{scalar.omega} now appears as
\begin{align}
    \label{scalar.omega.can}
    \omega \,=\,
        d\fp_\m \wedge dx^\m 
    \,,\quad
    \phi_0 \,=\,
        \frac{1}{2m}\mem
        \BB{
            \fp^2 + m^2
        }
        - \frac{q}{m}\mem A_\m(x)\mem \fp^\m
        + \frac{q^2}{2m}\mem
            A_\m(x)\mem A^\m(x)
    \,.
\end{align}
As a result, the Hamiltonian is perturbed and solely encodes the electromagnetic interactions,
while the symplectic structure is brought back to the free-theory form:
\begin{align}
    \label{scalar.pb.can}
    \pb{x^\m}{x^\n}
    \,=\,
        0
    \,,\quad
    \pb{x^\m}{\fp_\n}
    \,=\,
        \delta^\m{}_\n
    \,,\quad
    \pb{\fp_\m}{\fp_\n}
    \,=\,
        0
    \,.
\end{align}

This formulation, however, exhibits evident disadvantages
in both conceptual and practical aspects.
First, gauge invariance is grossly obscured.
The canonical momentum $\fp_\m$ is not a gauge-invariant observable.
Second, the coupling is nonlinear.
In perturbation theory,
the $(q^2/2)\mem A_\m(x)\mem A^\m(x)$ term in \eqref{scalar.omega.can}
translates to the spurious quartic vertex in scalar quantum electrodynamics.
Third, the derivation of the Lorentz force law
from the Hamiltonian EoM
becomes not so simple.

In this paper,
we always fix the Hamiltonian in the free-theory form
such that the external field couplings are 
implemented solely by modifications of the symplectic structure
(declaration of worldline field basis).
When the symplectic form is modified in the form of \eqref{omega-split},
we say that the two-form $\omega'$
is a \textit{symplectic perturbation}.

We end with a useful formula.
The pointwise inverses of $\omega^\circ$ and $\omega$
define
the Poisson bivectors 
$\Pi^\circ$ and $\Pi$
of the free and interacting theories,
respectively.
Their relation is
\begin{align}
    \label{ppert}
    \omega \,=\,
        \omega^\circ \mem+\, \omega'
    \qiq
    \Pi
    \,=\,
        \Pi^\circ
        - \Pi^\circ\mem \omega'\, \Pi
    \,.
\end{align}
In writing down \eqref{ppert}, we have treated
$\omega$, $\omega^\circ$, $\omega'$, $\Pi$, $\Pi^\circ$
like antisymmetric matrices.

\newpage

\subsection{Model-Agnostic Derivation of BMT Equations}
\label{INT>BMT}

By combining the ideas in \Secs{FREE}{INT>SPT},
we establish
universal recipes
of coupling masssive spinning particles to electromagnetism.

\para{Universal Recipe for Minimal Coupling}

Suppose a free massive spinning particle
formulated in a symplectic manifold $\ps$.
The reader may pick any of the explicit examples in \Sec{FOUR}.
Let $\omega^\circ$ be the symplectic form given to $\ps$.
Time evolution describes a Hamiltonian vector field on $\ps$,
which we denote as $T^\circ$.

The result in \Sec{FREE>PART}
implies that
there exist functions $(x^\m, \hy^\m, p_\m)$
on $\ps$
that satisfy the universal Poisson brackets in \eqref{xyp}.
Without loss of generality, we assume that
these functions evolve under the free-theory time evolution as
\begin{align}
    \label{T0}
    T^\circ\act{
        x^\m
    }
    \,=\, \frac{p^\m}{m}
    \,,\quad
    T^\circ\act{
        \hy^\m
    }
    \,=\, 0
    \,,\quad
    T^\circ\act{
        p_\m
    }
    \,=\, 0
    \,.
\end{align}
Physically, this means that 
the free particle follows a straight-line trajectory
while carrying a constant spin pseudovector.

As an application of the Souriau-Feynman framework in \Sec{INT>SPT},
we suppose the following symplectic perturbation on $\ps$:
\begin{align}
    \label{EM.sp0}
    \omega'
    \,\,=\,\,
        qF
    \,\,=\,\,
        \frac{1}{2}\,
            qF_{\r\s}(x)\,
            dx^\r \swedge dx^\s
    \,.
\end{align}
The formula in \eqref{ppert} then implies that
the Poisson bivector is modified as
\begin{align}
\begin{split}
    \label{ppert.EM}
    \Pi(df,dg)
    \,=\,
        \Pi^\circ(df,dg)
        \mem-\mem
        \Pi^\circ(df,dx^\r)\,
            qF_{\r\s}(x)\,
        \Pi(dx^\s,dg)
    \,.
\end{split}
\end{align}
Iterating \eqref{ppert.EM} will provide an infinite series expansion for $\pb{f}{g} = \Pi(df,dg)$,
which governs how the Poisson bracket is modified under the symplectic perturbation:
$
    \pb{f}{g}
    =
        \pb{f}{g}^\circ
        \mem-\mem
        \pb{f}{x^\r}^\circ\,
            qF_{\r\s}(x)\,
        \pb{x^\s}{g}^\circ
        \mem+\mem
        \pb{f}{x^\r}^\circ\,
            qF_{\r\s}(x)\,
        \pb{x^\s}{x^\k}^\circ\,
            qF_{\k\l}(x)\,
        \pb{x^\l}{g}^\circ
        \mem-\mem
        \cdots
$.

Let $T$ be the time-evolution Hamiltonian vector field 
in the interacting theory
such that the Hamiltonian EoM reads $\dot{f} = T\act{f}$.
\eqrefs{T0}{ppert.EM} together implies that
\begin{align}
    \label{ppert.EM0}
    \dot{f}
    \,=\,
        T^\circ\act{
            f
        }
        \mem-\mem
        \pb{f}{x^\r}^\circ\,
            qF_{\r\s}(x)\,
        \dot{x}^\s
    \,.
\end{align}
Plugging in the physical variables $(x^\m,\hy^\m,p_\m)$ to $f$ in \eqref{ppert.EM0},
we find
\begin{align}
\begin{split}
    \label{uBMT}
    \dot{x}^\m
    \,&=\,
            \frac{p^\m}{m} 
        \mem-\mem
        \frac{1}{m^2}\mem
        \BB{
            \ve^{\m\r\k\l} \hy_\k\mem p_\l
        }\mem
        \BB{
            qF_{\r\s}(x)\mem \dot{x}^\s
        }
    \,,\\
    \dot{\hy}^\m
    \,&=\,
        \frac{1}{m^2}\mem
        \BB{
            \hy^\m\mem p^\r
            + p^\m\mem \hy^\r
        }\mem
        \BB{
            qF_{\r\s}(x)\mem 
            \dot{x}^\s
        }
    \,,\\[0.12\baselineskip]
    \dot{p}^\m
    \,&=\,
        qF^\m{}_\n(x)\mem \dot{x}^\n
    \,,
\end{split}
\end{align}
where we have used \eqref{xyp}
to evaluate 
$\pb{x^\m}{x^\r}^\circ$,
$\pb{\hy^\m}{x^\r}^\circ$,
and
$\pb{p_\m}{x^\r}^\circ$.

\eqref{uBMT} defines the classical EoM of a consistent Hamiltonian system,
as the modified symplectic form due to \eqref{EM.sp0} is closed.
What is its physical interpretation, then?


\para{BMT Equations}

The BMT \cite{Bargmann:1959gz} equations 
dictate the dynamics of charged massive spinning particles in external electromagnetic fields
at the linear order in electric charge $q$,
in the regime where derivative effects can be completely ignored:
$F(x) \gg \hy\mem \partial F(x), \hy^2\mem \partial^2 F(x), \cdots$.
In our conventions, they are
\begin{subequations}
\label{BMT}
\begin{align}
\label{BMT.x}
    \dot{x}^\m
    \,&=\,
        \frac{p^\m}{m}
        \mem-\mem
        \frac{q}{m}\,
        \bbsq{    
            \frac{g}{2}\,
                {*}\hnem F^\m{}_\n(x)
            + \bb{1 - \frac{g}{2}}\,
                \hdelta^\m{}_\k\,
                {*}\hnem F^\k{}_\n(x)
        }\, \hy^\n
    + \cdots
    \,,\\
\label{BMT.y}
    \dot{\hy}^\m
    \,&=\,
        \frac{q}{m}\,
        \bbsq{    
            F^\m{}_\n(x)
            - \bb{1 - \frac{g}{2}}\,
                \hdelta^\m{}_\k\mem
                F^\k{}_\n(x)
        }\, \hy^\n
    + \cdots
    \,,\\[0.12\baselineskip]
\label{BMT.p}
    \dot{p}^\m
    \,&=\,
        \frac{q}{m}\,
            F^\m{}_\n(x)\mem p^\n
    + \cdots
    \,,
\end{align}
\end{subequations}
where the ellipses signify that
terms of $\O(q^{1+n})$ or $\O(\partial^n\nem F)$
are discarded from the right-hand sides
for $n \geq 1$.
See \App{REVIEW>BMT} for more details.

Many references present just \eqrefs{BMT.y}{BMT.p}
without explicitly stating the relation between $p^\m$ and $\dot{x}^\m$.
However, it should be emphasized that
\eqref{BMT.x}
must be specified 
together with \eqrefs{BMT.y}{BMT.p}
to fully define the time evolution
and ensure its consistency,
especially when working in the Hamiltonian formulation.

In \eqref{BMT}, the constant parameter $g$
is the gyromagnetic ratio,
also known as the $g$-factor.
The gyromagnetic ratio controls the spin-induced magnetic dipole moment.
The specific value $g{\:=\:}0$
has been identified as the minimal coupling
in the traditional sense,
in which case
the BMT equations in \eqref{BMT} read
\begin{subequations}
\label{BMT0}
\begin{align}
\label{BMT0.x}
    \dot{x}^\m
    \,&=\,
        \frac{p^\m}{m}
        \mem-\mem
        \frac{q}{m}\,
            \hdelta^\m{}_\k\mem
            {*}\hnem F^\k{}_\n(x)\mem \hy^\n
    + \cdots
    \,,\\
\label{BMT0.y}
    \dot{\hy}^\m
    \,&=\,
        - \frac{q}{m}\,
            \hp^\m\mem \hp_\k\hhem
            F^\k{}_\n(x)\mem  \hy^\n
    + \cdots
    \,,\\[0.12\baselineskip]
\label{BMT0.p}
    \dot{p}^\m
    \,&=\,
        \frac{q}{m}\,
            F^\m{}_\n(x)\mem p^\n
    + \cdots
    \,.
\end{align}
\end{subequations}

Notably,
the EoM derived in \eqref{uBMT} 
yield the $g {\:=\:} 0$ BMT equations in \eqref{BMT0}
through
plugging in $\dot{x}^\m = p^\m\nem/m + \O(q^1)$
and using an identity involving two epsilon tensors.
Therefore, 
the $g{\:=\:}0$ BMT equations
arise in any massive spinning particle model
by prescribing
the symplectic perturbation
in \eqref{EM.sp0}
realized at the physical center $x^\m$.

\para{Universal Recipe for Dipolar Coupling}

To implement the dipolar coupling in our universal framework,
consider the following symplectic perturbation:
\begin{align}
    \label{EM.sp1}
    \omega'
    \,\,=\,\,
        \frac{1}{2}\,
            qF_{\m\n}(x)\,
            dx^\m \swedge dx^\n
        \,+\,
        q\mem 
        d\,\bb{
            c_1\,
            \hy^\r\,
            {*}F_{\r\s}(x)\,
            dx^\s
        }
    \,.
\end{align}
Here, $c_1$ is a constant.
\eqref{EM.sp1} is the most general symplectic perturbation
that can contribute to the $\O(q^1\hy^1)$ part of the EoM.
It is \textit{bootstrapped} by 
the physical principles of
\begin{align}
\label{bootstrap.EM}
\begin{split}
    \text{(a)}\,\,&\text{
        Liouville property%
    }
    \,,\\
    \text{(b)}\,\,&\text{
        Poincar\'e invariance%
    }
    \,,\\
    \text{(c)}\,\,&\text{
        Gauge invariance%
    }
    \,,\\
    \text{(d)}\,\,&\text{
        Parity invariance%
    }
    \,.
\end{split}
\end{align}

By applying the method demonstrated in \eqref{ppert.EM0},
one finds that 
the $\O(q^1y^1)$ contribution to the resulting EoM
arises from the combination
\begin{align}
    \label{EM.q1y1}
    - \frac{q}{m}\,
    \bbsq{    
        \pb{f}{x^\r}^\circ\,
        F_{\r\s}(x)\mem p^\s
        +
        c_1\,
        \pb{f}{\hy^\r}^\circ\,
        {*}F_{\r\s}(x)\mem p^\s
    }
    \,.
\end{align}
By examining \eqref{EM.q1y1} for $f = (x^\m,\hy^\m,p_\m)$,
one exactly reproduces the BMT equations in \eqref{BMT}
provided the identification
\begin{align}
    c_1
    \,=\,
        g/2
    \,.
\end{align}

Therefore, the BMT equations with the generic gyromagnetic ratio $g$
arise in any massive spinning particle model
by prescribing the symplectic perturbation in \eqref{EM.sp1}
in terms of
the physical center $x^\m$ and spin length pseudovector $\hy^\m$.

\subsection{The Four Models with Electromagnetic Couplings}
\label{INT>FOUR1}

It remains to substantiate and reproduce
the universal derivation of BMT equations 
in \Sec{INT>BMT}
from each of our four example models.

To be fully instructive,
this subsection wishes to
derive the EoM
by varying first-order actions
with Lagrange multipliers,
instead of retaining the sophisticated geometrical languages of symplectic forms and Dirac brackets.
Our first-order actions
take the form
\begin{align}
    \label{psaction}
    \int\,
    \bbsq{\,
        \theta 
        \,-\,
            \k^0\mem \phi_0\, d\t
        \,-\,
            \BB{
                \text{extra constraint terms}
            }\, d\t
    \mem}
    \,,
\end{align}
which defines sigma models from $\R$ to
phase spaces with constraints.
Here,
$\t$ is the worldline parameter,
$\theta$ is (the pullback of) the symplectic potential,
$\phi_0$ is the mass-shell constraint,
and $\kappa^0$ is a Lagrange multiplier.
Furthermore,
there will be extra constraints
so that
more Lagrange multipliers are employed,
except for the spinor oscillator model.
We start from the spinorial models,
as they do not involve the intricacies regarding the spurious center.

\para{Spinor Oscillator}

The special-relativistic spinor oscillator
is a sigma model $\R {\:\to\:} \ps^\SO_{12}$,
whose target space is
coordinatized by $(x^\m,p_\m,\zeta_\a)$.

The electromagnetic minimal coupling
is implemented by the action
\begin{align}
    \label{SOEM.action}
    \int\,
        \bbsq{\,
            p_\m\mem \dot{x}^\m
            +
            \frac{1}{2i}\mem
            \BB{\mem
                \bzeta_\da\, \hat{p}^{\da\a} \dot{\zeta}_\a 
                - \dot{\bzeta}_\da\mem \hat{p}^{\da\a}\hem \zeta_\a
            }
            + qA_\m(x)\mem \dot{x}^\m
            - \frac{\k^0}{2m}\mem\BB{
                p^2 + m^2
            }
        }\, d\t
    \,,
\end{align}
which is a functional of
$(x^\m,p_\m,\zeta_\a)$ and $\kappa^0$.
It is easy that
the symplectic potential due to \eqref{SOEM.action} reproduces
the symplectic form in \eqref{EM.sp0}.
Variations of \eqref{SOEM.action}
give
\begin{align}
\begin{split}
    \label{SOEM.eom1}
    \dot{x}^\m
    \,&=\,
        \frac{\k^0\hem p^\m}{m}
        - \frac{q}{m^2}\,
            S^{\m\r}\mem F_{\r\s}(x)\mem \dot{x}^\s
    \,,\\
    \dot{p}_\m
    \,&=\,
        qF_{\m\n}(x)\mem \dot{x}^\n
    \,,\\
    \dot{\zeta}_\a
    \,&=\,
            -\frac{q}{2m^2}\,
            F_\wrap{\r\s}(x)\mem \dot{x}^\s
            \mem p_\n\mem
            (\s^{\r\n})_\a{}^\b\mem \zeta_\b
    \,,
\end{split}
\end{align}
and $p^2 + m^2 = 0$.
Here, $S^{\m\n} := \ve^{\m\n\r\s} \hy_\r\hem p_\s$.

\newpage

The worldline parametrization can be fixed as $\k^0 = 1$.
Then \eqref{SOEM.eom1} readily reproduces the $g = 0$ BMT equations.
In particular, the spin precession is governed by
\begin{align}
\begin{split}
    \label{SOEM.eom2}
    \dot{\zeta}_\a
    \,&=\,
        \frac{q}{m}\mem
        \bb{
            \bF_\a{}^\b(x)
            + \hat{p}_{\a\da}\,
                F^\da{}_\db(x)\,
            \hat{p}^{\db\b}
        }\mem \zeta_\b
        + \O(q^2)
    \,.
\end{split}
\end{align}

Similarly, the BMT equations with generic $g$-factor
arise by
adding $(g/2)\mem q\mem {*}\hnem F_{\m\n}(x)\mem \hy^\m\hhem \dot{x}^\n$
in the Lagrangian in \eqref{SOEM.action}
with the definition of $\hy^\m$ in \eqref{SO.y}.
In this case,
one finds that
the spin precession is governed by
$\dot{\zeta}_\a = (\Omega_\BMT)_\a{}^\b\mem \zeta_\b$
with \cite{ambikerr1}
\begin{align}
    \label{prec1}
    (\Omega_\BMT)_\a{}^\b
    \,=\,
        \frac{q}{m}\mem
        \bb{
            \frac{1{\,+\,}g/2}{2}\,
            \bF_\a{}^\b(x)
            + 
            \frac{1{\,-\,}g/2}{2}\,
            \hat{p}_{\a\da}\,
                F^\da{}_\db(x)\,
            \hat{p}^{\db\b}
        }
    \,.
\end{align}

\para{Massive Twistor}

The special-relativistic massive twistor
is a sigma model $\R {\:\to\:} \ps^\MT_{16}$,
whose target space can be
coordinatized by $(x^\m,y^\m,\lambda_\a{}^I)$.

The electromagnetic minimal coupling
is implemented by the action
\begin{align}
    \label{MTEM.action}
    \int\,
        \lrsq{
        \begin{aligned}[c]
        &
        	{- \lambda_\a{}^I\mem \rambda_{I\da}\, \dot{x}^{\da\a}}
        	+ i\mem y^{\da\a}\mem
        	\BB{
        		\lambda_\a{}^I\hem \dot{\rambda}_{I\da}
        		{\,-\,}
        		\dot{\lambda}_\a{}^I\mem \rambda_{I\da}
        	}
            + qA_\m(x)\mem \dot{x}^\m
        \\
        &
            - \frac{\k^0}{2m}\mem\BB{
                m^2 - \det(\lambda)\mem \det(\rambda)
            \hnem}
            - \k^1\,
                \rambda_{I\da}\mem y^{\da\a}\hem \lambda_\a{}^I
        \end{aligned}
        }\, d\t
    \,,
\end{align}
which is a functional of
$(x^\m,y^\m,\lambda_\a{}^I)$, and $(\kappa^0, \kappa^1)$.
It is easy that the symplectic potential due to \eqref{MTEM.action} reproduces
the symplectic form in \eqref{EM.sp0}.
Variations of \eqref{MTEM.action} give
\begin{align}
\begin{split}
    \label{MTEM.eom1}
    \dot{x}^\m
    \,&=\,
        \frac{\k^0\hem p^\m}{m}
        - \frac{q}{m^2}\,
            \BB{
                \ve^{\m\r}{}_{\k\l}\mem y^\k p^\l
            }\mem F_{\r\s}(x)\mem \dot{x}^\s
    \,,\\
    \dot{y}^\m
    \,&=\,
        \frac{q}{m^2}\,
            \BB{
                y^\m p^\r + p^\m y^\r
            }\mem F_{\r\s}(x)\mem \dot{x}^\s
    \,,\\
    \dot{\lambda}_\a{}^I
    \,&=\,
        - \frac{1}{2}\,
            (\s^\r)_{\a\da}\mem (\rambda^{-1})^{\da I}\mem
            qF_{\r\s}(x)\mem \dot{x}^\s
        - \frac{i\hhem\kappa^1}{2}\,
            \lambda_\a{}^I
    \,,
\end{split}
\end{align}
together with $p^2 + m^2 = 0$ and $p\mdot y = 0$.

The Lagrange multiplier $\k^1$
will be fixed by imposing
an explicit gauge-fixing condition
such as the one in \eqref{MT.chi1}.
Physical observables are nonetheless $\U(1)$ invariant,
so their time evolution is uniquely determined
even if $\k^1$ is left agnostic.
With this understanding,
it is easy to reproduce the $g{\:=\:}0$ BMT equations
from \eqref{MTEM.eom1}.

The massive twistor model exhibits some nice features.
First of all,
the universal derivation in \Sec{INT>BMT}
reincarnates
in an almost identical fashion
when obtaining
the first two lines in \eqref{MTEM.eom1},
as
the model naturally features the variables
$x^\m$ and $y^\m$.
Second of all,
the EoM of the spin frame $\lambda_\a{}^I$
encapsulate the EoM of both the momentum and spin,
thanks to the fact that
the body-frame components of spin
are dynamically conserved:
$\hy^{\da\a} \propto (\rambda^{-1})^{\da I}\mem (\vec{W}\mdot\vec{\s})_I{}^J\mem (\lambda^{-1})_J{}^\a$
for some constant $\vec{W}$.

Similarly, it is easy to derive the BMT equations with generic $g$,
in which case
the spin frame's EoM are
$\dot{\lambda}_\a{}^I = (\Omega_\BMT)_\a{}^\b\mem \lambda_\b{}^I - (i\k^1/2)\mem \lambda_\a{}^I$
with $(\Omega_\BMT)_\a{}^\b$ given in \eqref{prec1}.

\para{Vector Oscillator}

The special-relativistic vector oscillator
is a sigma model $\R {\:\to\:} \ps^\VO_{16}$,
whose target space $\ps^\VO_{16}$ is
coordinatized by $(\fx^\m,p_\m,\fa^\m)$.
The reduced phase space $\ps^\VO_{12}$
with no unphysical spinning degrees of freedom
arises by imposing four spin constraints.

\newpage

The electromagnetic minimal coupling is implemented by the action
\begin{align}
    \label{VOEM.action}
    \int\,
        \lrsq{\,
            \begin{aligned}[c]
                &
                    p_\m\mem \dot{\fx}^\m 
                    + \frac{i}{2}\mem
                    \BB{
                        \fba_\m\mem \dot{\fa}^\m
                        - \dot{\fba}_\m\mem \fa^\m
                    }
                \\
                &
                    + qA_\m\hnem\hhnem\bb{
                        \fx + i\mem \frac{(\fa\swedge\fba) p}{p^2}
                    }\,
                    \frac{d}{d\t}\bb{
                        \fx^\m + i\mem \frac{(\fa\swedge\fba)^{\m\n} p_\n}{p^2}
                    }
                \\
                &
                    - \frac{\k^0}{2m}\mem\BB{
                        p^2 + m^2
                    }
                    - \kappa^1\mem p\mdot\hnem\fa
                    - \bar{\kappa}^1\mem p\mdot\hnem\fba
                    - \frac{\kappa^2}{2}\mem \fa^2
                    - \frac{\bkappa^2}{2}\mem \fba^2
            \end{aligned}
        \mem}\, d\t
    \,,
\end{align}
which is a functional of $(\fx^\m,p_\m,\fa^\m)$ and $(\k^0,\k^1,\k^2)$.
The complex Lagrange multipliers $\k^1$ and $\k^2$
impose the four necessary spin constraints,
$p\mdot\hnem\fa {\:=\:} 0 {\:=\:} p\mdot\hnem\fba$
and
$\fa^2 {\:=\:} 0 {\:=\:} \fba^2 $.

It should be clear that
\eqref{VOEM.action}
is uniquely determined by
\eqrefs{EM.sp0}{VO.xy}.
Unfortunately, it describes a bulky formula 
due to the very mismatch between $x$ and $\fx$.
However, two simplified formulations are viable.

The first approach
is based on the observation that
the constraints
$p\mdot\hnem\fa {\:=\:} 0 {\:=\:} p\mdot\hnem\fba$
are being explicitly imposed by Lagrange multipliers
in \eqref{VOEM.action},
on the support of which
the spurious and physical centers coincide.
Thus, an equivalent, simplified action is
\begin{align}
    \label{VOEM.action.approx}
    \int\,
        \lrsq{\,
            \begin{aligned}[c]
                &
                    p_\m\mem \dot{\fx}^\m 
                    + \frac{i}{2}\mem
                    \BB{
                        \fba_\m\mem \dot{\fa}^\m
                        - \dot{\fba}_\m\mem \fa^\m
                    }
                    + qA_\m(\fx)\mem
                        \dot{\fx}^\m
                \\
                &
                    - \frac{\k^0}{2m}\mem\BB{
                        p^2 + m^2
                    }
                    - \kappa^1\mem p\mdot\hnem\fa
                    - \bar{\kappa}^1\mem p\mdot\hnem\fba
                    - \frac{\kappa^2}{2}\mem \fa^2
                    - \frac{\bkappa^2}{2}\mem \fba^2
            \end{aligned}
        \mem}\, d\t
    \,.
\end{align}
In this particular sense,
the vector oscillator model
happens to allow for
realizing the interaction at the unphysical center $\fx^\m$,
\textit{at the level of action}.

Computation shows that
the saddle of either \eqrefs{VOEM.action}{VOEM.action.approx} describes
\begin{align}
\begin{split}
\label{VOEM.eom2}
    \dot{\fx}^\m
    \,&=\,
        \frac{p^\m}{m}
        - \frac{i\hem (\fa\swedge\fba)^{\m\r}}{m^2}\mem
            qF_{\r\s}(\fx)\mem
        \dot{\fx}^\s
    \,,\\
    \dot{\fa}^\m
    \,&=\,
        -
        \bb{
            \frac{p^\m\dot{\fx}^\r}{m^2}\mem
                qF_{\r\s}(\fx)
        \hnem}\mem
        \fa^\s
    \,,\\
    \dot{p}_\m
    \,&=\,
        qF_{\m\n}(\fx)\mem \dot{\fx}^\n
    \,,
\end{split}
\end{align}
where we have set $\k^0 = 1$
and fixed $\k^1$ and $\k^2$
by demanding the preservation of
the four spin constraints.
When using \eqref{VOEM.action.approx},
the unique solution is
(assuming $\fba\mdot\fa \neq 0$)
\begin{align}
    \label{VOEM.kappasol}
    \bk^1
    \,=\,
        -\frac{i}{m^2}\,
            qF_{\r\s}(\fx)\mem \dot{\fx}^\r\mem \fa^\s
    \,,\quad
    \bk^2
    \,=\,
        0
    \,.
\end{align}
It is easy to see that
\eqref{VOEM.eom2} implies the $g {\:=\:} 0$ BMT equations.
Similarly,
one incorporates generic $g$
by simply adding
$(g/2)\mem q\mem {*}\hnem F_{\m\n}(\fx)\mem \hy^\m\hhem \dot{\fx}^\n$
to the Lagrangian in \eqref{VOEM.action.approx}.

The second approach is based on
an alternative coordinatization of the phase space $\ps_{16}^\VO$
that arises by studying the Hamiltonian flows of
the constraints $p\mdot\hnem\fa$ and $p\mdot\hnem\fba$:
\begin{align}
    \label{VOalt.coords}
    \bigbig{
        x^\m
        ,\mem
        p_\m
        ,\mem
        \a^\m
        ,\mem
        \fpsi_1
    }
    \,,
\end{align}
where $p\mdot\a = 0$ and $\psi_1{\:\in\:}\C$.
Specifically, we have defined
\begin{align}
    \label{VO.bparam}
    \a^\m 
    \mem:=\mem
        \hdelta^\m{}_\n\mem \fa^\n
    \,,\quad
    \fpsi_1
    \mem:=\mem
        -i\mem \hp\mdot\hnem\fba
    \qiq
    \fx^\m
    \,=\,
        x^\m
        - \frac{1}{|p|}\,\BB{
            \a^\m\mem \fpsi_1
            + 
            \ba^\m\mem \fbpsi_1
        }
    \,.
\end{align}
The free-theory symplectic potential then boils down to
\begin{align}
\begin{split}
    \label{VO.theta0.split}
    \theta^\circ
    \,&=\,
        p_\m\mem dx^\m
        + \frac{i}{2}\mem
        \BB{
            \ba_\m\mem d\a^\m
            - d\ba_\m\mem \a^\m
        }
        + \frac{i}{2}\mem
        \BB{
            \fbpsi_1\mem d\fpsi_1
            - d\fbpsi_1\mem \fpsi_1
        }
    \,.
\end{split}
\end{align}
As a result, the action in \eqref{VOEM.action}
can be alternatively formulated
as a functional of $(x^\m,p_\m,\a^\m,\fpsi_1)$ and $(\k^0,\k^1,\k^2)$.
By integrating out the $(\k^1,\fpsi_1)$ sector, one obtains\footnote{
    When varying these types of actions, one should be careful about the fact that not all four complex components of $\a^\m$
    are independent.
    The same comment applies to 
    \eqref{VOGR.action.correct} as well.
}
\begin{align}
    \label{VOEM.action.gi}
    \int\,
        \lrsq{\,
            \begin{aligned}[c]
                &
                    p_\m\mem \dot{x}^\m 
                    + \frac{i}{2}\mem
                    \BB{
                        \ba_\m\mem \dot{\a}^\m
                        - \dot{\ba}_\m\mem \a^\m
                    }
                    + qA_\m(x)\mem
                        \dot{x}^\m
                \\
                &
                    - \frac{\k^0}{2m}\mem\BB{
                        p^2 + m^2
                    }\mem d\t
                    - \frac{\kappa^2}{2}\mem \a^2
                    - \frac{\bkappa^2}{2}\mem \ba^2
            \end{aligned}
        \mem}\, d\t
    \,.
\end{align}
The $g{\:=\:}0$ BMT equations can be derived from \eqref{VOEM.action.gi}.
Incorporating generic $g$ is easy.

\para{Spherical Top}

The spherical top
is a sigma model $\R {\:\to\:} \ps^\ST_{20}$,
whose target space $\ps^\ST_{20}$ is coordinatized by $(\fx^\m,\fL^\m{}_A,p_\m,\fS_{\m\n})$.
The reduced phase space $\ps^\ST_{14}$
with no fake spinning degrees of freedom
arises by imposing the gauge generator constraints $\phi_a {\:=\:} 0$
that pair up with
the gauge-fixing condition $\fchi^a {\:=\:} 0$.
The imposition of $\phi_a {\:=\:} 0$ is a must,
while different choices exist for the gauge-fixing function $\fchi^a$.

\eqrefs{EM.sp0}{ST.xy}
together stipulates that the action
of the minimally coupled charged spherical top is
\begin{align}
    \label{STEM.action}
    \int\,
        \lrsq{\,
            \begin{aligned}[c]
                &
                    p_\mu\mem \dot{\fx}^\mu 
                    + \frac{1}{2}\, 
                        \fS_{\m\n}\mem \fL^\m{}_A\mem \dot{\fL}{}^\n{}^A
                +
                    qA_\m\hnem\hhnem\bigbig{
                        \fx
                        + \fS p / p^2
                    }\,
                    \frac{d}{d\t}\BB{
                        \fx^\m
                        + \fS^{\m\n}\hem p_\n / p^2
                    }
                \\
                &
                    - \frac{\k^0}{2m}\mem\BB{
                        p^2 + m^2
                    }
                    - \kappa^a\mem \phi_a
            \end{aligned}
        \mem}\, d\t
    \,,
\end{align}
which is a functional of
$(\fx^\m,\fL^\m{}_A,p_\m,\fS_{\m\n})$ and $(\kappa^0, \kappa^a)$.

Importantly, the action in \eqref{STEM.action}
allows for adopting one's favorite SSC.
While the gauge generators $\phi_a$ are explicitly incorporated in the action,
the gauge-fixing function $\fchi^a$ will be only specified
\textit{a posteriori} during the process of determining $\k^a$.
The action in \eqref{STEM.action}
itself is
agnostic of the choice of the SSC
and is spin gauge-invariant.
Hence, the situation differs from 
the vector oscillator:
one cannot proactively replace
$\fx + \fS p/ p^2$ in \eqref{STEM.action}
with $\fx$.
Note that
$\fS^{\m\n} p_\n = 0$
iff one adopts the covariant gauge-fixing condition via \eqref{chia}.

A correct understanding on this point
facilitates
a consistent derivation of 
the BMT equations
with \textit{any} choice of the SSC.
For example,
the saddle of \eqref{STEM.action} is found 
in the covariant spin gauge as
\begin{align}
\begin{split}
\label{STEM.eom}
    \dot{\fx}^\m
    \,=\,
        \frac{p^\m}{m}
        - \frac{\fS^{\m\r}}{m^2}\mem
            qF_{\r\s}(\fx)\mem
        \dot{\fx}^\s
    &\,,\quad
    \dot{\fL}{}^\m{}_a
    \,=\,
        -
        \bb{
            \frac{p^\m\dot{\fx}^\r}{m^2}\mem
                qF_{\r\s}(\fx)
        \hnem}\mem
        \fL^\s{}_a
    \,,\\
    \dot{p}_\m
    \,=\,
        qF_{\m\n}(\fx)\mem \dot{\fx}^\n
    &\,,\quad
    \dot{\fS}^{\m\n}
    \,=\,
        -2\mem \bb{
            \frac{p^{[\m|}\dot{\fx}^\r}{m^2}\mem
                qF_{\r\s}(\fx)
        \hnem}\mem
        \fS^{\s|\n]}
    \,.
\end{split}
\end{align}
Compare this with \eqref{VOEM.eom2}.
\eqref{STEM.eom} readily derives the $g {\:=\:} 0$ BMT equations.
It is left as an exercise 
to derive the BMT equations
in noncovariant spin gauges as well,
such as the canonical gauge in \eqref{PNW}. 
Incorporating generic $g$ is straightforward.

\section{Universality in Interacting Theory: Gravity}
\label{INT2}

\subsection{Couplings as Covariant Symplectic Perturbations}
\label{INT>COVSPT}

Let us now move on to gravitational interactions.
As a preliminary analysis,
we begin by describing
how the Souriau-Feynman method in \Sec{INT>SPT}
generalizes to gravity \cite{csg}.

\para{Flat Spacetime}

We revisit the free scalar particle in \eqref{scalar.omega0}.
The symplectic form $\omega^\circ$ is described in 
the coordinate one-form basis
$(dx^\m,dp_\m)$,
where $\m$
is a \textit{global Lorentz} index:
\begin{align}
    \label{scalarfree}
    \omega^\circ \,=\,
        dp_\m \wedge dx^\m
    \,.
\end{align}

As an intermediate step toward general relativity,
it is always instructive to
investigate
``flat spacetime in noninertial frames.''
The particle's position variable $x^\m$
becomes curvilinear coordinates
on the Minkowski space $\mflat = (\R^4,\eta)$,
which behave as $x^\m \mapsto x'^\m {\:=\:} f^\m(x)$
under coordinate transformations.
The particle's momentum variable is $p_\m$,
which transforms as $p_\m \mapsto p_\n\mem (\partial x^\n \nem/ \partial x'^\m)$.

Crucially,
the coordinate basis $(dx^\m,dp_\m)$ of one-forms
is not generally covariant,
since $dp_\m$ is not transformed to
$dp_\n\mem (\partial x^\n \nem/ \partial x'^\m)$
due to a term
$p_\n\mem (\partial^2 x^\n \nem/ \partial x'^\m \partial x'^\r)\mem dx'^\r$
that involves a second derivative.
As is well-known,
such second-derivative terms are typically canceled by
exploiting an affine connection.
In particular, let $\nabla$ be the Levi-Civita connection of the flat metric $\eta$.
In curvilinear coordinates, it describes nonzero Christoffel symbols $\Gamma^\m{}_{\n\r}(x)$.
Consider the covariant exterior derivative
$Dp_\m {\:=\:} dp_\m {\:-\:} p_\n\mem \Gamma^\n{}_{\m\r}(x)\mem dx^\r$.
Then $(dx^\m,Dp_\m)$ is a generally covariant basis of one-forms on $T^*\mflat$,
as 
$dx^\m \mapsto (\partial x'^\m \nem/\partial x^\n)\mem dx^\n$
and
$Dp_\m \mapsto Dp_\n\mem (\partial x^\n \nem/\partial x'^\m)$
under coordinate transformations.

In this covariant basis,
the free-theory symplectic form in 
\eqref{scalarfree}
is represented as
\begin{align}
    \label{scalargencov}
    \omega^\circ \,=\,
        Dp_\m \wedge dx^\m
    \,,
\end{align}
which is manifestly under coordinate transformations.

In fact,
a more sophisticated treatment
will employ the tetrad formalism.
Let $e^m {\:=\:} $ $ e^m{}_\m(x)\mem dx^\m$
be the orthonormal coframe
such that $\eta_{\m\n}(x) {\:=\:} \eta_{mn}\mem e^m{}_\m(x)\mem e^n{}_\n(x)$.
Here, $m,n,$ $r,s,\cdots \in \{0,1,2,3\}$
are \textit{local Lorentz} indices.
The components of the particle's momentum
are $p_m$,
where $p_m\mem e^m{}_\m(x) = p_\m$.
In this case,
the free-theory symplectic form is written as
\begin{align}
    \label{scalarlorcov}
    \omega^\circ \,=\,
        Dp_m \wedge e^m
    \,,
\end{align}
which manifests invariances under 
coordinate and local Lorentz transformations.
Here, the covariant exterior derivative $D$
utilizes the spin connection.

The advantage of employing the tetrad formalism
is that the Hamiltonian in \eqref{scalar.omega}
remains in the free-theory form:
$\phi_0 {\:=\:} (\eta^{mn}\hem p_m\hhem p_n \mplus m^2)/2m$,
where $\eta^{mn}$ describes a \textit{constant} matrix.
In our view,
this refinement is necessary
for fully succeeding
the philosophy of the Souriau-Feynman method in \Sec{INT>SPT}:
manifest covariance
\textit{and}
fixed Hamiltonian.

\para{Curved Spacetime}

We are now ready to proceed to general relativity.
Spacetime is a smooth pseudo-Riemannian four-manifold $\M {\:=\:} (\R^4,g)$.
Let $\nabla$ be the Levi-Civita connection of the metric $g$,
whose connection coefficients are the Christoffel symbols
$\Gamma^\m{}_{\n\r}(x)$.
Let $e^m {\:=\:} e^m{}_\m(x)\mem dx^\m$ be an orthonormal coframe such that
$g_{\m\n}(x) = $ $ \eta_{mn}\mem e^m{}_\m(x)\mem e^n{}_\n(x)$.
Let $\gamma^m{}_n {\:=\:} \gamma^m{}_{n\r}(x)\mem dx^\r$ describe the spin connection.
The covariant exterior derivative is denoted as $D$.
As is well-known,
the Riemann curvature and torsion two-forms
are defined as
$R^m{}_n = \frac{1}{2}\mem R^m{}_{nrs}(x)\mem e^r \swedge e^s = d\gamma^m{}_n + \gamma^m{}_r \swedge \gamma^r{}_n$
and
$T^m = De^m = de^m + \gamma^m{}_n \swedge e^n$,
the latter of which vanishes for the Levi-Civita connection.

The Hamiltonian formulation of
a scalar particle in $\M$
takes the cotangent bundle $T^*\nem\M$
as the phase space.
This phase space
will be endowed with
a nondegenerate two-form,
\begin{align}
    \label{scalar.omega-cov}
    \omega^\bullet
    \,=\,
        Dp_m \wedge e^m
    \,.
\end{align}
Physically speaking, \eqref{scalar.omega-cov}
implements the free-theory symplectic form 
$dp_\m \wedge dx^\m$
in local inertial frames.
This describes the covariantization map \cite{csg}
\begin{align}
    \label{covmap.xpform}
    \bigbig{
        dx^\m
        ,\mem
        dp_\m
    }
    \,\,\mapsto\,\mem
    \bigbig{
        e^m
        ,\mem
        Dp_m
    }
    \,,
\end{align}
which promotes \textit{global Lorentz} indices to \textit{local Lorentz} indices.

With this understanding,
the Einstein equivalence principle
implies that
the particle's symplectic form
$\omega$
must take the form
\begin{align}
    \label{covsplit}
    \omega
    \,\,=\,\,
        \omega^\bullet
        \mem+\mem
        \omega'
    \transition{where} 
    \lim_{R\to0} \omega'
    \,=\, 0
    \,.
\end{align}
This means that 
$\omega$
approaches to 
$\omega^\bullet$ in \eqref{scalar.omega-cov}
when curvature effects are ignored.
Said in another way,
physics in curved spacetime
should approximately look like
physics in flat spacetime
when examined in local Lorentz frames,
in the limit of negligible tidal effects.

Crucially,
$\omega'$ in \eqref{covsplit} 
must be invariant under
coordinate and local Lorentz transformations,
as $\omega^\bullet$ and $\omega$ are.
Hence \eqref{covsplit}
describes a covariant split of the gravity-coupled symplectic form $\omega$.
It understands $\omega$
as a curvature-induced perturbation around
the covariantization $\omega^\bullet$
of the free theory's $\omega^\circ$.
In this sense,
we refer to
such $\omega'$
as a \textit{covariant symplectic perturbation}.\footnote{
    See \rcite{csg} for its precise mathematical definition and formalization.
}

It should be clear that
this construction parallels \Sec{INT>SPT},
where the electromagnetism-coupled symplectic form $\omega$
arises as a curvature-induced
(meaning the curvature $F$ of the $\mathrm{U}(1)$ gauge bundle)
perturbation around
the free theory's $\omega^\circ$.

Again, an important consistency condition
is the Liouville property of classical time evolution,
which ensures the conservation of classical probability in a sense \cite{Kim:2025sey,csg}.
The Liouville property is mathematically equivalent to the closure of the symplectic form,
$d\omega = 0$.
Imposing this closure on $\omega$ in \eqref{covsplit},
one finds that
$\omega'$ must satisfy
\begin{align}
    \label{bianchiR}
    d\omega'
    \,=\,
        -d\omega^\bullet
    \,=\,
        - D^2 p_m \wedge e^m
        + Dp_m \swedge De^m
    \,=\,
        p_m\mem \BB{
            R^m{}_n \swedge e^n
        }
    \,.
\end{align}
Here, we have used the fact that $\omega^\bullet$ carries no free indices, so
$d\omega^\bullet = D\omega^\bullet$.
We have also used the definitions of the Riemann curvature and torsion two-forms.

The bracketed term in \eqref{bianchiR}
identically vanishes 
by the algebraic Bianchi identity of the Riemann tensor.
Therefore, the simplest gravitational coupling
consistent with 
gauge invariances,
equivalence principle,
and Liouville property
is achieved by
$\omega' = 0$.
It turns out that 
$\omega' = 0$ precisely realizes
the minimally coupled scalar particle
whose EoM are the geodesic equations,
as shown below.

\para{Covariant Poisson Bracket Relation}

Note that
$(e^m,Dp_m)$ provides a complete,
yet noncoordinate,
basis of one-forms on the phase space $T^*\M$.

The pointwise inverse of $\omega^\bullet$
is the unique bivector $\Pi^\bullet$
on $T^*\M$
such that
\begin{align}
    \label{scalar.Pi-cov.components}
    \pib{e^m}{e^n}
    \mem=\mem
        0
    \,,\quad
    \pib{e^m}{Dp_n}
    \mem=\mem
        \delta^m{}_n
    \,,\quad
    \pib{Dp_m}{Dp_n}
    \mem=\mem
        0
    \,.
\end{align}
\eqref{scalar.Pi-cov.components}
is nothing but the covariantization of
the free theory's Poisson bracket relation in \eqref{scalar.pb0}
via the replacements
$\Pi^\circ \mapsto \Pi^\bullet$
and
$(dx^\m,dp_\m) \mapsto (e^m,Dp_m)$:
recall \eqref{covmap.xpform}.

Let $\Pi$ be the pointwise inverse of the full symplectic form $\omega$.
Suppose the simplest covariant symplectic perturbation,
$\omega' = 0$.
In this case, we have
$\Pi = \Pi^\bullet$.
Notably,
\eqref{scalar.Pi-cov.components}
facilitates a \textit{manifestly covariant derivation}
of the Hamiltonian EoM:
\begin{align}
\begin{split}
    \label{grav0.eom}
    &
    e^m{}_\m(x)\mem \dot{x}^\m
    \,=\,
        \pia{e^m}{d\phi_0}
    \,=\,
        \pia{e^m}{D\phi_0}
    \,=\,
        \pia{e^m}{Dp_n}
        \mem p^n
    \,=\,
        \frac{p^m}{m}
    \,,\\
    &
    \frac{Dp_m}{d\t}
    \,=\,
        \pia{Dp_m}{d\phi_0}
    \,=\,
        \pia{Dp_m}{D\phi_0}
    \,=\,
        \pia{Dp_m}{Dp_n}
        \mem p^n
    \,=\,
        0
    \,.
\end{split}
\end{align}
Namely, we insert the covariant one-forms $e^m$ and $Dp_m$
to the Poisson bivector $\Pi$.
We also use the fact that the Hamiltonian is a scalar function,
so $d\phi_0 = D\phi_0$.
It is easy to check that \eqref{grav0.eom}
reproduces the geodesic equation.

More explicitly,
let $E_m {\:=\:} E^\m{}_m(x)\mem \partial_\m$
be the frame vector fields (tetrad)
on the spacetime
that is dual to the coframe $e^m = e^m{}_\m(x)\mem dx^\m$.
Namely,
$e^m{}_\m(x)\mem E^\m{}_n(x) = \delta^m{}_n$
and
$E^\m{}_m(x)\mem e^m{}_\n(x) = \delta^\m{}_\n$.
It is not difficult to see that
$(\tilde{E}_m, \partial/\partial p_m)$
provides a covariant basis of vector fields
on the phase space,
where 
\begin{align}
    \label{scalar.tE}
    \tilde{E}_r
    \,=\,
        E^\r{}_r(x)\,
        \bb{
            \frac{\partial}{\partial x^\r}
            + p_n\mem \gamma^n{}_{mr}(x)\,
                \frac{\partial}{\partial p_m}  
        }
    \,.
\end{align}
In the mathematical jargon,
$\tE_r$ in
\eqref{scalar.Pi-cov} is the \textit{horizontal lift}
\cite{ehresmann1948connexions,Mason:2013sva,gde} of the spacetime tetrad $E_r$ 
to the phase space
with respect to the Levi-Civita connection.

It follows that
$(\tE_m,\partial/\partial p_m)$
is the dual basis to $(e^m,Dp_m)$.
As a result,
the bivector
$\Pi^\bullet$ in
\eqref{scalar.Pi-cov.components}
is explicitly found as
\begin{align}
    \label{scalar.Pi-cov}
    \Pi^\bullet
    \,=\,
        \tE_m \wedge \frac{\partial}{\partial p_m}
    \,.
\end{align}
It is left as an exercise to
examine the components of \eqref{scalar.Pi-cov}
in any basis
to reproduce \eqref{grav0.eom}.

We end with 
a useful formula.
Let $\omega = \omega^\bullet + \omega'$
in a generic phase space.
Let $\Pi^\bullet$ and $\Pi$ be
the pointwise inverses of $\omega^\bullet$ and $\omega$.
Recalling \eqref{ppert},
it is easy to see that
\begin{align}
    \label{ppertc}
    \omega \,=\,
        \omega^\bullet \mem+\, \omega'
    \qiq
    \Pi
    \,=\,
        \Pi^\bullet
        - \Pi^\bullet\mem \omega'\, \Pi
    \,.
\end{align}
In writing down \eqref{ppertc}, we have treated
$\omega$, $\omega^\bullet$, $\omega'$, $\Pi$, $\Pi^\bullet$
like antisymmetric matrices.

\newpage

\subsection{Universal Covariant Poisson Bracket Relation}
\label{INT>CUPB}

Next,
we formulate
a generalization of
the universal Poisson bracket relation in \eqref{xyp}
to describe spinning particles
with curvilinear coordinates.

Suppose a free massive spinning particle in Minkowski space $\mflat$,
formulated on a symplectic manifold $\ps$ as a phase space.
As reviewed in \Sec{FREE>PART},
Poincar\'e symmetry implies
the existence of functions $(x^\m,\hy^\m,p_\m)$ on $\ps$
that satisfy
the universal Poisson bracket relation in \eqref{xyp}
\cite{sst-asym}.
Mathematically, \eqref{xyp} 
plugs in the one-forms
$(dx^\m,d\hy^\m,dp_\m)$
to the free-theory Poisson bivector $\Pi^\circ$:
\begin{align}
\begin{split}
\label{xyp.pi0}
    \pif{x^\m}{x^\n}
    \,&=\,
        \frac{1}{-p^2}\, 
            \ve^{\m\n\r\s} \hy_\r\mem p_\s
    \,=\,
    \pif{\hy^\m}{\hy^\n}
    \,,\\
    \pif{x^\m}{\hy^\n}
    \,&=\,
        \frac{1}{-p^2}\,\BB{
            \hy^\m p^\n + p^\m \hy^\n 
        }
    \,,\\[0.15\baselineskip]
    \pif{x^\m}{p_\n}
    \,&=\,
        \delta^\m{}_\n
    \,,\\[0.215\baselineskip]
    \pif{\hy^\m}{p_\n} 
    \,&=\, 0
    \,,\\[0.15\baselineskip]
    \pif{p_\m}{p_\n} 
    \,&=\, 0
    \,.
\end{split}
\end{align}

Now suppose the same particle
is put in a curved spacetime $\M = (\R^4,g)$,
in which case
the phase space will be a symplectic manifold $\P$
such that
$\dim \P = \dim \ps$.
Let $\omega$ be the symplectic form of $\P$.
Based on the Einstein equivalence principle,
we postulate the local existence of 
functions $(x^\m,\hy^\m,p_\m)$ 
and a nondegenerate bivector $\Pi^\bullet$
on $\P$
such that
\begin{align}
\begin{split}
\label{xyp.cov}
    \pib{e^m}{e^n}
    \,&=\,
        \frac{1}{-p^2}\, 
            \ve^{mnrs} \hy_r\mem p_s
    \,=\,
    \pib{D\hy^m}{D\hy^n}
    \,,\\
    \pib{e^m}{D\hy^n}
    \,&=\,
        \frac{1}{-p^2}\,\BB{
            \hy^m p^n + p^m \hy^n 
        }
    \,,\\[0.15\baselineskip]
    \pib{e^m}{Dp_n}
    \,&=\,
        \delta^m{}_n
    \,,\\[0.215\baselineskip]
    \pib{D\hy^m}{Dp_n} 
    \,&=\, 0
    \,,\\[0.15\baselineskip]
    \pib{Dp_m}{Dp_n} 
    \,&=\, 0
    \,,
\end{split}
\end{align}
where
$e^m = e^m{}_\m(x)\mem dx^\m$,
$D\hy^m = d\hy^m + \gamma^m{}_{n\r}(x)\mem \hy^n \mem dx^\r$,
and
$Dp_\m = dp_m - p_n\mem \gamma^n{}_{m\r}(x)\mem dx^\r$
utilizes the coframe and spin connection at $x$.
It should be clear that \eqref{xyp.cov}
arises from \eqref{xyp}
by the following replacements
(covariantization map),
which promotes global Lorentz indices to local Lorentz indices:\footnote{
    It should be clear that \eqref{xyp.cov}
    can be also stated in terms of
    $(dx^\m, D\hy^\m, Dp_\m)$:
    $e^m {\:=\:} e^m{}_\m(x)\mem dx^\m$,
    $D\hy^m {\:=\:} e^m{}_\m(x)\mem D\hy^\m$,
    and
    $Dp_m {\:=\:} Dp_\m\mem E^\m{}_m(x)$
    are identities.
    The sophisticated tetrad language
    is employed here
    to emphasize the conceptual aspect of the equivalence principle.
    Furthermore, \eqref{xyp.cov}
    can be also stated in terms of the coordinate differentials
    $(dx^\m, d\hy^\m, dp_\m)$,
    in which case one separates out bare Christoffel symbols on the right-hand sides
    (which is against the covariant Poisson bracket philosophy \cite{csg}).
}
\begin{align}
    \label{covmap-xyp}
    \Pi^\circ
    \,\,\mapsto\,\mem
    \Pi^\bullet
    \,,\quad
    \bigbig{
        dx^\m
        ,\mem
        d\hy^\m
        ,\mem
        dp_\m
    }
    \,\,\mapsto\,\mem
    \bigbig{
        e^m
        ,\mem
        D\hy^m
        ,\mem
        Dp_m
    }
    \,.
\end{align}
We refer to \eqref{xyp.cov}
as the \textit{universal covariant Poisson bracket relation}
of massive spinning particles,
which reincarnates \eqref{xyp.pi0}
in local inertial frames.

Physically, 
the universal covariant Poisson bracket relation
\eqref{xyp.cov}
will be regarded as
a consequence of local Poincar\'e covariance:
diffeomorphism and local Lorentz.

Yet,
any formula deduced from the equivalence principle
could ignore tidal (curvature) effects.
To formulate this idea in a precise fashion,
let $\omega^\bullet$ be
the pointwise inverse of $\Pi^\bullet$,
which defines a nondegenerate two-form on $\P$.
Then the equivalence principle implies that
the particle's symplectic form
will arise by a curvature correction on $\omega^\bullet$:
\begin{align}
    \label{covsplit.re}
    \omega
    \,\,=\,\,
        \omega^\bullet
        \mem+\mem
        \omega'
    \transition{where}
    \lim_{R\to0} \omega'
    \,=\,
        0
    \,.
\end{align}
Surely,
$\omega^\bullet$ and $\omega'$
must be separately invariant under
coordinate and local Lorentz transformations.
The two-form $\omega'$ in \eqref{covsplit.re}
is the \textit{covariant symplectic perturbation}
for massive spinning particles.

\subsection{Model-Agnostic Derivation of QMPD Equations}
\label{INT>MPD}

By combining the ideas in \Secs{INT>COVSPT}{INT>CUPB},
we establish
universal recipes
of coupling massive spinning particles to gravity.

\para{Universal Recipe for Minimal Coupling}

Again, an important consistency condition
that must be imposed on \eqref{covsplit.re}
is the Liouville property,
i.e.,
the closure $d\omega = 0$.
Computation shows that
$\omega^\bullet$ is not closed
iff the spacetime exhibits nonzero curvature.
It can be shown that
the simplest possible covariant symplectic perturbation 
that makes $\omega = \omega^\bullet + \omega'$ closed is\footnote{
    The proofs of these facts
    may check
    the Jacobi identities
    with the functions $(x^\m,\hy^\m,p_\m)$,
    for $\Pi^\bullet$ and $\Pi$
    (cf. covariant Jacobi identity \cite{csg}).
    The computations are tedious but doable.
    Another method is to utilize symplectic realizations,
    which is the exploration in 
    \Sec{COVPS} in essence.
}
\begin{align}
    \label{GR.sp0}
    \omega'
    \,\,=\,\,
        p_m\mem {\star}R^m{}_n\mem \hy^n
    \,\,=\,\,
        \frac{1}{2}\,
            p_m\mem
            {\star}R^m{}_{nrs}(x)\,
                \hy^n\mem
                e^r \swedge e^s
    \,,
\end{align}
where $\star$ is the Hodge star
acting on local Lorentz indices.

Plugging in
\eqref{GR.sp0} to \eqref{ppertc}
yields
\begin{align}
    \label{ppert.ab}
    \pia{\a}{\b}
    \,=\,
        \pib{\a}{\b}
        - 
        \pib{\a}{e^r}\,
        \BB{
            p_m\mem 
            {\star}R^m{}_{nrs}(x)
            \mem \hy^n
        \nem}
        \,\pia{e^s}{\b}
    \,,
\end{align}
where $\a$ and $\b$ are arbitrary one-forms on the phase space $\P$.

On account of the equivalence principle,
we postulate the local existence of
a vector field $T^\bullet$ on $\P$
such that
\begin{align}
    \label{Tcov}
    \cont{e^m}{T^\bullet}
    \,=\, \frac{p^m}{m}
    \,,\quad
    \cont{D\hy^m}{T^\bullet}
    \,=\, 0
    \,,\quad
    \cont{Dp_m}{T^\bullet}
    \,=\, 0
    \,,
\end{align}
where $\langle \blank, \blank \rangle$
denotes the contraction between one-forms and vector fields.
\eqref{Tcov} is the covariantization of \eqref{T0}
via \eqref{covmap-xyp}.

Let $T$ be the time-evolution Hamiltonian vector field 
in the interacting theory
such that the Hamiltonian EoM reads $\dot{f} = T\act{f}$.
\eqrefs{Tcov}{ppert.ab} together implies that
\begin{align}
    \label{ppert.GR0}
    \cont{\a}{T}
    \,=\,
        \cont{\a}{T^\bullet}
        \mem-\mem
        \pib{\a}{e^r}\,
        \BB{
            p_m\mem 
            {\star}R^m{}_{nrs}(x)
            \mem \hy^n
        \nem}\,
        e^s{}_\s(x)\mem \dot{x}^\s
    \,.
\end{align}
Plugging in the 
noncoordinate one-forms
$(e^m,D\hy^m,Dp_m)$
to $\a$ in \eqref{ppert.GR0}
readily derives the covariant EoM
in the tetrad frame.
The final result is then 
presented in
the coordinate frame as
\begin{align}
\begin{split}
    \label{uMPD}
    \dot{x}^\m
    \,&=\,
            \frac{p^\m}{m} 
        \mem-\mem
        \frac{1}{m^2}\mem
        \BB{
            \ve^{\m\r\k\l} \hy_\k\mem p_\l
        }\mem
        \BB{
            p_\zeta\mem 
            {\star}R^\zeta{}_{\xi\r\s}(x)
            \mem \hy^\xi
        \hnem}\, \dot{x}^\s
    \,,\\
    \frac{D\hy^\m}{d\t}
    \,&=\,
        \frac{1}{m^2}\mem
        \BB{
            \hy^\m\mem p^\r
            + p^\m\mem \hy^\r
        }\mem
        \BB{
            p_\zeta\mem 
            {\star}R^\zeta{}_{\xi\r\s}(x)
            \mem \hy^\xi
        \hnem}\, \dot{x}^\s
    \,,\\[0.12\baselineskip]
    \frac{Dp_\m}{d\t}
    \,&=\,
        \BB{
            p_\zeta\mem 
            {\star}R^\zeta{}_{\xi\m\n}(x)
            \mem \hy^\xi
        \hnem}\, \dot{x}^\n
    \,.
\end{split}
\end{align}

\eqref{uMPD} defines the classical EoM of a consistent Hamiltonian system,
as the modified symplectic form due to \eqref{GR.sp0} is closed.
What is its physical interpretation, then?

\para{MPD Equations}

The MPD \cite{Mathisson:1937zz,Papapetrou:1951pa,Dixon:1970zza} equations 
dictate the dynamics of massive spinning particles in curved spacetime
at the leading order in curvature $R$
and up to linear coupling in spin $\hy$,
in the regime where derivative effects 
are completely suppressed:
$R(x) \gg \hy\mem DR(x), \hy^2\mem D^2 R(x), \cdots$.
In other words,
the MPD equations capture the universal behavior
of massive spinning particles
in the leading pole-dipole, small-curvature approximations.
In this paper,
the MPD equations refer to
\begin{subequations}
\label{MPD}
\begin{align}
\label{MPD.x}
    \dot{x}^\m
    \,&=\,
        \frac{p^\m}{m}
        \mem-\mem
        \frac{1}{m}\mem
        \bbsq{\mem
            \hdelta^\m{}_\k\mem
            \BB{
                R^\k{}_{\n\r\s}(x)\mem 
                \hy^\r p^\s
            }
        }\mem
        \hy^\n
        + \cdots
    \,,\\
\label{MPD.y}
    \frac{D\hy^\m}{d\t}
    \,&=\,
        -\frac{1}{m}\mem
        \bbsq{
            \hp^\m \hp_\k\,
            \BB{
                - {*}\hnem R^\k{}_{\n\r\s}(x)\mem 
                \hy^\r p^\s
            }
        }\mem
        \hy^\n
        + \cdots
    \,,\\[0.12\baselineskip]
\label{MPD.p}
    \frac{Dp^\m}{d\t}
    \,&=\,
        \frac{1}{m}\mem
        \BB{
            - {*}\hnem R^\m{}_{\n\r\s}(x)\mem
            \hy^\r p^\s
        }\mem 
        p^\n
        + \cdots
    \,,
\end{align}
\end{subequations}
where the ellipses signify that terms of $\O(R^{1+n})$, $\O(\hy^{1+n})$, or $\O(D^n\nem R)$ are discarded from the right-hand sides
for $n \geq 1$.

The above definition is a refinement of the original result by \rcite{Mathisson:1937zz,Papapetrou:1951pa,Dixon:1970zza};
see \App{REVIEW>MPD} for details.
First,
the original references 
\cite{Mathisson:1937zz,Papapetrou:1951pa,Dixon:1970zza}
utilize the spin tensor $S^{\m\n}$
instead of the pseudovector
$\hy^\m = - {*}S^{\m\n} p_\n/p^2$.
Second,
the original references
\cite{Mathisson:1937zz,Papapetrou:1951pa,Dixon:1970zza}
do not state \eqref{MPD.x},
despite its necessity
for well-definedness and consistency
as remarked in \Sec{INT>BMT}.

Straightforward algebra shows that
the EoM obtained in \eqref{uMPD}
are equivalent to \eqref{MPD}.
Therefore,
in any general-relativistic massive spinning particle model,
the minimal (universal) coupling \`a la MPD equations
arises if
the covariant symplectic perturbation is given by \eqref{GR.sp0}
in terms of
the physical center position $x^\m$ 
and spin length pseudovector $\hy^m$
defined in the covariant SSC.

\para{Covariant Double Copy}

It is useful to note that
the MPD equations in \eqref{MPD}
are isomorphic to
the $g = 0$ BMT equations in
\eqref{uBMT}
via the replacement
\begin{align}
    \label{cov-dc}
    qF_{\m\n}(x)
    \quad\xleftrightarrow[]{\,\,\,\,\,\,}\quad
    p_\k\mem {*}\hnem R^\k{}_{\l\m\n}(x)\mem \hy^\l
    \,=\,
        - {*}\hnem R_{\m\n\r\s}(x)\mem \hy^\r\hem p^\s
    \,.
\end{align}
In our systematic framework,
this ``covariant double copy'' replacement rule
is nicely explained by
the correspondence between
the (covariant) symplectic perturbations
of electromagnetism and gravity
in \eqrefs{EM.sp0}{GR.sp0}
(to be more precise, see \rcite{csg}).

\newpage

\para{Universal Recipe for Quadrupolar Coupling}

It is possible to incorporate nonminimal couplings in our universal framework.
For instance,
the most general covariant symplectic perturbation
that contributes 
to the $\O(R^1y^2)$ part of the EoM
for $\hy^m$ and $p_m$
is uniquely determined as
\begin{align}
    \label{GR.sp2}
    \omega'
    \,\,=\,\,
        \frac{1}{2}\,
            p_m\mem
            {\star}R^m{}_{nrs}(x)\,
                \hy^n\mem
                e^r \swedge e^s
        \,+\,
        d\,\bb{
            - 
            \frac{C_2}{2!}\,
            p_m\hem R^m{}_{nrs}(x)\mem \hy^n\hem
                \hy^r e^s
        }
    \,,
\end{align}
where $C_2$ is a constant.
\eqref{GR.sp2} is bootstrapped by
the physical principles of
\begin{align}
\label{bootstrap.GR}
\begin{split}
    \text{(a)}\,\,&\text{
        Liouville property%
    }
    \,,\\
    \text{(b)}\,\,&\text{
        Diffeomorphism invariance%
    }
    \,,\\
    \text{(c)}\,\,&\text{
        Local Lorentz invariance%
    }
    \,,\\
    \text{(d)}\,\,&\text{
        Parity invariance%
    }
    \,.
\end{split}
\end{align}

By applying the method demonstrated in \eqref{ppert.GR0},
one finds that 
the $\O(R^1\hy^1)$ contribution to the resulting covariant EoM
arises from the combination
\begin{align}
    \label{Grav.R1y2}
    - \frac{1}{m}\,
    \lrp{\,
    \begin{aligned}[c]
        &
        \pib{\a}{e^r}\,
        p_m\mem
        {\star}R^m{}_{nrs}(x)\mem 
        \hy^n\mem
        p^s
        \\
        &+
        C_2\,
        \pib{\a}{D\hy^r}\,
        p_m\mem
        R^m{}_\wrap{(nr)s}(x)\mem 
        \hy^n\mem
        p^s
        \\
        &+
        \frac{C_2}{2}\,
        \pib{\a}{Dp_m}\,
        R^m{}_\wrap{nrs}(x)\mem 
        \hy^n
        \hy^r\hem
        p^s
    \end{aligned}
    }
    \,.
\end{align}
By inserting 
the covariant one-forms
$(e^m,D\hy^m,Dp_m)$
to $\a$ in \eqref{Grav.R1y2},
we obtain
\begin{align}
    &
    \dot{x}^\m
    \,=\,
        \frac{p^\m}{m}
        \mem-\mem
        \frac{1}{m}\mem
        \bbsq{\mem
            \frac{C_2}{2}\,
            \BB{
                R^\m{}_{\n\r\s}(x)\mem 
                \hy^\r p^\s
            }
            +
            \bigbig{1-C_2}\,
            \hdelta^\m{}_\k\mem
            \BB{
                R^\k{}_{\n\r\s}(x)\mem 
                \hy^\r p^\s
            }
        }\mem
        \hy^\n
        + \cdots
    \nonumber
    \,,\\
    &
    \frac{D\hy^\m}{d\t}
    \,=\, -
        \frac{1}{m}\mem
        \bbsq{
            \BB{
                {*}\hnem R^\m{}_{\n\r\s}(x)\mem 
                \hy^\r p^\s
            }
            - 
            \bigbig{1-C_2}\,
            \hdelta^\m{}_\k\,
            \BB{
                {*}\hnem R^\k{}_{\n\r\s}(x)\mem 
                \hy^\r p^\s
            }
        }\mem
        \hy^\n
        + \cdots
    \nonumber
    \,,\\[0.12\baselineskip]
    &
    \frac{Dp^\m}{d\t}
    \,=\, -
        \frac{1}{m}\mem
        \BB{
            {*}\hnem R^\m{}_{\n\r\s}(x)\mem
            \hy^\r p^\s
        }\mem 
        p^\n
        + \cdots
    \,,
    \label{QMPD}
\end{align}
where the ellipses signify that terms of $\O(R^{1+n})$, $\O(\hy^{2+n})$, or $\O(D^n\nem R)$ are discarded from the right-hand sides
for $n \geq 1$.\footnote{
    The last two equations in \eqref{QMPD}
    are isomorphic to \eqrefs{BMT.y}{BMT.p}
    via
    the covariant double copy replacement rule in \eqref{cov-dc},
    upon replacing $c_1 {\:=\:} g/2$ with $C_2$.
    The first equation in \eqref{QMPD}
    exhibits a minor mismatch in coefficient
    due to the nonabelian nature of gravitational interactions:
    the ``gravitational charge'' $S_{\m\n} {\,=\,} \ve_{\m\n\r\s} \hy^\r p^\s$ is a variable
    instead of a constant $q$.
    See \App{REVIEW>QMPD}.
}

\para{QMPD Equations}

\eqref{QMPD} is a generalization of \eqref{MPD}
by a new parameter $C_2$
characterizing quadrupolar spin coupling.
In this paper, we refer to \eqref{QMPD}
as the QMPD equations.
\App{REVIEW}  
explicates how it agrees
with previous works \cite{Steinhoff:2009tk,Harte:2011ku,Vines:2016unv,Compere:2023alp,Ramond:2026fpi}.

Our analysis establishes that
the QMPD equations in \eqref{QMPD} provide
the unique and complete
extension of the MPD equations
up to $\O(R^1y^2)$,
when the derivative effects are suppressed.
Our analysis also shows that
the QMPD equations with 
the arbitrary quadrupole coefficient $C_2$
arises in any massive spinning particle model
if the covariant symplectic
perturbation is given as \eqref{GR.sp2},
in terms of
the center position $x^\m$ 
and spin length pseudovector $\hy^m$
defined in the covariant SSC.

Notably, the effect of a nonzero $C_2$ on the $\O(y^2)$ EoM
is just as strong as that of the minimal coupling.
The QMPD equations are thus of a phenomenological importance,
as spinning black holes carry a large value $C_2 {\:=\:} 1$
\cite{Khriplovich:1997ni,Thorne:1984mz,Hansen:1974zz,Newman:1965tw-janis,Hernandez:1967zza,Thorne:1980ru}.
Any analysis on Kerr black hole orbits
based on the MPD ($C_2 {\:=\:} 0$) equations
is bound to fail, 
producing \textit{qualitatively wrong} answers.

\subsection{The Four Models with Gravitational Couplings}
\label{INT>FOUR2}

In \Sec{INT>CUPB},
we have postulated
the universal existence of functions $(x^\m,\hy^m,p_m)$
in the general-relativistic phase spaces of massive spinning particles,
satisfying the covariant Poisson bracket relation in \eqref{xyp.cov}.
We have also claimed that
the condition in \eqref{covsplit.re} holds
such that the minimal covariant symplectic perturbation is given by \eqref{GR.sp0}.
As a result,
\Sec{INT>MPD} has derived
the MPD and QMPD equations
in a universal fashion.

The main mission of 
the remaining parts of this paper
is to prove the above postulate and claim
in each of our four concrete models.
The procedure is straightforward for
the spinorial models
because their basic position variables are physical.
The vectorial models,
however,
requires much more work.

We start by providing
the gravitational counterpart of the analysis given in \Sec{INT>FOUR1}.
The general-relativistic phase spaces will be identified
as per the equivalence principle.
The general-relativistic worldline actions
and their resulting EoM
will be explored.

\para{Spinor Oscillator}

The general-relativistic spinor oscillator
is a sigma model $\R {\:\to\:} \P^\SO_{12}$,
whose target space is
\begin{align}
    \label{SOGR.ps}
\smash{
    \P^\SO_{12}
    \,\,=\,\,
        \nem\BB{
            T^* \oplus \Sm
        }\mem \M
    \,.
}
\end{align}
Here, $\M {\:=\:} (\R^4,g)$ is the curved spacetime.
The orthonormal coframe $e^m$ defines a local trivialization of $T\M$,
which decomposes into the right-handed and left-handed spinor bundles,
$\mathscr{S}^\pm\nem\M$.
Namely,
local Lorentz indices are traded off with spinor indices as
$e^m \leftrightarrow e^{\da\a}$.
With this understanding,
the phase space in \eqref{SOGR.ps}
is coordinatized by
\begin{align}
    \label{SOGR.coords}
\smash{
    \bigbig{
        x^\m
        ,\mem
        p_m
        ,\mem
        \zeta_\a
    }
    \,.
}
\end{align}

The gravitational minimal coupling
is implemented by the action
\begin{align}
    \label{SOGR.action}
    \int\,
        \bbsq{\,
            p_{\a\da}\mem e^{\da\a}{}_\m(x)\mem \dot{x}^\m
            +
            \frac{1}{2i}\mem
            \bb{\mem
                \bzeta_\da\, \hat{p}^{\da\a} 
                \mem
                \frac{D\zeta_\a}{d\t}
                - \frac{D\bzeta_\da}{d\t}\, \hat{p}^{\da\a}\hem \zeta_\a
            }
            - \frac{\k^0}{2m}\mem\BB{
                p^2 + m^2
            }
        }\, d\t
    \,,
\end{align}
which is a functional of
$(x^\m,p_m,\zeta_\a)$ and $\kappa^0$.
Variations of \eqref{SOGR.action}
derive
\begin{align}
\begin{split}
    \label{SOGR.eom1}
    \dot{x}^\m
    \,&=\,
        \frac{\k^0\hem p^\m}{m}
        + \frac{1}{2m^2}\,
            S^{\m\r}\mem S^{\k\l} R_{\k\l\r\s}(x)\mem \dot{x}^\s
    \,,\\
    \frac{Dp_\m}{d\t}
    \,&=\,
        -\frac{1}{2}\,
            S^{\k\l} R_{\k\l\m\n}(x)\mem \dot{x}^\n
    \,,\\
    \frac{D\zeta_\a}{d\t}
    \,&=\,
            \frac{q}{4m^2}\,
            S^{\k\l}
            R_\wrap{\k\l\r\s}(x)\mem \dot{x}^\s
            \mem p_\n\mem
            (\s^{\r\n})_\a{}^\b\mem \zeta_\b
    \,,
\end{split}
\end{align}
\newpage\noindent
and $p^2 + m^2 = 0$.
Here, we have kept the definitions of 
$S$, $\xi$, $\hy$
in \eqrefss{SO.S}{Econj}{SO.y}
in local Lorentz frames.
Setting $\k^0 {\:=\:} 1$,
\eqref{SOGR.eom1} is isomorphic to \eqref{SOEM.eom1}
via covariantization
and the replacement in \eqref{cov-dc}.
Thus it reproduces the MPD equations.

Similarly, the QMPD equations arise by
adding 
$
    - (C_2/2)\,
    p_m\hem R^m{}_{nrs}(x)\mem \hy^n\hem
        \hy^r e^s{}_\s(x)\mem \dot{x}^\s
$
in the Lagrangian in \eqref{SOGR.action}.
The spin precession is covariantly characterized by
$D\zeta_\a/d\t = (\Omega_\QMPD)_\a{}^\b\mem \zeta_\b$,
where \cite{ambikerr1}
\begin{align}
    \label{prec2}
    (\Omega_\QMPD)_\a{}^\b
    \,=\,
        -\frac{S^{\k\l}}{2m}\mem
        \bb{
            \frac{1{\,+\,}C_2}{2}\,
            \bR_\a{}^\b{}_{\k\l}(x)
            + 
            \frac{1{\,-\,}C_2}{2}\,
            \hat{p}_{\a\da}\,
                R^\da{}_\wrap{\db\k\l}(x)\,
            \hat{p}^{\db\b}
        }
    \,.
\end{align}
Again,
the isomorphism between \eqrefs{prec1}{prec2} is obvious
and describes \eqref{cov-dc},
which was observed by \rcite{ambikerr1}.

\para{Massive Twistor}

The general-relativistic massive twistor
is a sigma model $\R {\:\to\:} \P^\MT_{16}$,
whose target space is
\begin{align}
    \label{MTGR.ps}
    \P^\MT_{16}
    \,\,=\,\,
        \nem\BB{
            T \oplus \SS
        }\mem \M
    \,.
\end{align}
By local trivializations,
$\P^\MT_{16}$
is coordinatized by
\begin{align}
    \label{MTGR.coords}
    \bigbig{
        x^\m
        ,\mem
        y^m
        ,\mem
        \lambda_\a{}^I
    }
    \,.
\end{align}

The gravitational minimal coupling
is implemented by the action
\begin{align}
    \label{MTGR.action}
    \int\,
        \lrsq{
        \begin{aligned}[c]
        &
        	{- \lambda_\a{}^I\mem \rambda_{I\da}\, e^{\da\a}{}_\m(x)\mem \dot{x}^\m}
        	+ i\mem y^{\da\a}\mem
        	\bb{
        		\lambda_\a{}^I\mem
                    \frac{D\rambda_{I\da}}{d\t}
        		{\,-\,}
        		\frac{D\lambda_\a{}^I}{d\t}\mem \rambda_{I\da}
        	}
        \\[-0.1\baselineskip]
        &
            - \frac{\k^0}{2m}\mem\BB{
                m^2 - \det(\lambda)\mem \det(\rambda)
            \hnem}
            - \k^1\,
                \rambda_{I\da}\mem y^{\da\a}\hem \lambda_\a{}^I
        \end{aligned}
        }\, d\t
    \,,
\end{align}
which is a functional of
$x^\m,y^m,\lambda_\a{}^I$, and $\kappa^0, \kappa^1$.
Variations of \eqref{SOGR.action}
derive
\eqref{MTEM.eom1}
with $qF_{\m\n}(x)$ replaced with
$p_\k\mem {*}\hnem R^\k{}_{\l\m\n}(x)\mem y^\l$,
so
it is easy to reproduce the MPD equations.

Similarly, it is easy to derive
the QMPD equations with generic $C_2$,
in which case
one finds
\smash{$D\lambda_\a{}^I \nem/d\t = (\Omega_\QMPD)_\a{}^\b\mem \lambda_\b{}^I - (i\k^1/2)\mem \lambda_\a{}^I$}
with $(\Omega_\QMPD)_\a{}^\b$ given in \eqref{prec2}.

\para{Vector Oscillator (First Attempt)}
The general-relativistic vector oscillator
is the sigma model $\R \to \P_{16}^\VO$,
whose target space is
\begin{align}
    \label{VOGR.ps}
    \P^\VO_{16}
    \,\,=\,\,
        \nem\BB{
            T^* \oplus T^\C
        }\mem \M
    \,.
\end{align}
Via local trivializations,
the phase space in \eqref{VOGR.ps}
is coordinatized by
\begin{align}
    \label{VOGR.coords}
    \bigbig{
        \fx^\m
        ,\mem
        p_m
        ,\mem
        \fa^m
    }
    \,.
\end{align}

Of course, \eqrefs{VOGR.ps}{VOGR.coords}
covariantize \eqrefs{VO.psc}{VO.coordsc}.
Accordingly, 
one might naively conclude that
the covariantization of \eqref{VO.thetac}
will define the gravitational minimal coupling.
The first-order action \`a la \eqref{psaction}
reads
\begin{align}
    \label{VOGR.action}
    \int\,
        \lrsq{\,
            \begin{aligned}[c]
                &
                    p_m\mem e^m{}_\m(\fx)\mem \dot{\fx}^\m
                    + \frac{i}{2}\mem
                    \bb{
                        \fba_m\mem \frac{D\hnem\fa^m}{d\t}
                        - \frac{D\hnem\fba_m}{d\t}\mem \fa^m
                    }
                \\
                &
                    - \frac{\k^0}{2m}\mem\BB{
                        p^2 + m^2
                    }\mem d\t
                    - \kappa^1\mem p\mdot\hnem\fa
                    - \bar{\kappa}^1\mem p\mdot\hnem\fba
                    - \frac{\kappa^2}{2}\mem \fa^2
                    - \frac{\bkappa^2}{2}\mem \fba^2
            \end{aligned}
        \mem}\, d\t
    \,,
\end{align}
which is a functional of $(\fx^\m,p_m,\fa^m)$ and $(\k^0,\k^1,\k^2)$.
\newpage

\eqref{VOGR.action}
is an action localized on the spurious worldline $\t \mapsto \fx^\m(\t)$ in curved spacetime $\M$.
For instance,
local Lorentz transformations act on it
as $p_m \mapsto p_n\mem (\Omega^{-1}\hnem(\fx))^n{}_m$,
$e^m{}_\m(\fx) \mapsto \Omega^m{}_n(\fx)\mem e^n{}_\m(\fx)$, etc.
The covariant derivatives encode spin connections on the spurious worldline:
$D\hnem\fa^m = d\hnem\fa^m + \gamma^m{}_{n\r}(\fx)\mem \fa^n\mem d\fx^\r$.

This could have been a dangerous approach.
First of all, the gravitational fields
$e^m{}_\m(\fx)$ and $\gamma^m{}_{\n\r}(\fx)$
are coupled to the unphysical center
$\fx^\m$.
Moreover,
the precise relation between the fake center $\fx^\m$ and the physical center $x^\m$
is even lost (yet to be identified) in the general-relativistic case.
Concretely, what is the explicit formula
that expresses $x^\m$
as a composite variable
in the defining worldline field basis
$(\fx^\m,p_m,\fa^m)$?

Fortunately,
the vector oscillator model
might allow for using the action in \eqref{VOGR.action}
since it
mandates the explicit imposition of
$p\mdot\hnem\fa {\:=\:} 0 = p\mdot\hnem\fba$
by Lagrange multipliers,
on the support of which
one might argue that
$\fx$ still coincides with the physical center $x$.

Indeed,
the saddle of \eqref{VOGR.action}
is given in the form of
\eqref{VOEM.eom2}
via replacing
$qF_{\r\s}(\fx)$
with
$i\hem \fba_m\hem R^m{}_{n\r\s}(\fx)\hem \fa^n$.
Amusingly,
this replacement rule applies to
the solution to the Lagrange multipliers in 
\eqref{VOEM.kappasol} as well.
Consequently,
one finds that
\eqref{VOGR.action}
correctly derives
the MPD equations.
Similarly, 
the QMPD equations 
can be derived by
adding the term
$-(C_2/2)\,
    p_m\hem R^m{}_{nrs}(\fx)\mem \hy^n \hy^r\mem e^s{}_\s(\fx)\mem \dot{\fx}^\s
$
to the Lagrangian in \eqref{VOGR.action}.

Note also that a proper refinement of \eqref{VOGR.action}
will be facilitated later by \eqref{VOGR.theta}.

\para{Vector Oscillator (Second Attempt)}
The astute reader will point out that
\eqref{VOGR.action}
describes the analog of the first simplified formulation in \eqref{VOEM.action.approx}.
The correct way to couple the vector oscillator to the gravitational background
is to parallel
the second simplified formulation in \eqref{VOEM.action.gi},
in fact.

First,
switch to the alternative coordinate system in \eqref{VOalt.coords} 
already 
at the free theory level.
Second, covariantize 
the symplectic potential in \eqref{VO.theta0.split}.
Third, consider integrating out the \textit{scalars} $(\k^1,\fpsi_1)$.
The resulting action is
\begin{align}
    \label{VOGR.action.correct}
    \int\,
        \lrsq{\,
            \begin{aligned}[c]
                &
                    p_m\mem e^m{}_\m(x)\mem \dot{x}^\m
                    + \frac{i}{2}\mem
                    \bb{
                        \ba_m\mem \frac{D\hnem\a^m}{d\t}
                        - \frac{D\hnem\ba_m}{d\t}\mem \a^m
                    }
                \\
                &
                    - \frac{\k^0}{2m}\mem\BB{
                        p^2 + m^2
                    }\mem d\t
                    - \frac{\kappa^2}{2}\mem \a^2
                    - \frac{\bkappa^2}{2}\mem \ba^2
            \end{aligned}
        \mem}\, d\t
    \,,
\end{align}
which is a functional of $(x^\m,p_m,\a^m)$ and $(\k^0,\k^2)$
and is localized on 
the physical worldline
$\t \mapsto x^\m(\t)$.
\eqref{VOGR.action.correct} is a sensible action
for the minimally coupled general-relativistic vector oscillator.

\para{Spherical Top}

The general-relativistic spherical top is the sigma model $\R \to \P_{20}^\ST$,
whose target space is
\begin{align}
    \label{STGR.ps}
    \P^\ST_{20}
    \,\,=\,\,
        \nem
        T^*\bigbig{
            F_{\mathrm{\SO^+\hnem(1,3)}}(T\hnem\M)
        }
    \,.
\end{align}
Here,
$F_{\mathrm{\SO^+\hnem(1,3)}}(T\hnem\M)$
is the restricted orthonormal frame bundle
for the pseudo-Riemannian four-manifold
$\M {\:=\:} (\R^4,g)$.
Via local trivializations,
this phase space
is coordinatized by
\begin{align}
    \label{STGR.coords}
    \bigbig{
        \fx^\m
        ,\mem
        p_m
        ,\mem
        \fL^m{}_A
        ,\mem
        \fS_{mn}
    }
    \,.
\end{align}
\newpage

To implement the gravitational minimal coupling
in the spherical top model,
typical references simply
suppose the covariantized action
\begin{align}
    \label{STGR.action}
    \textfake{\scshape Wrong:}\quad
    \int\,
        \lrsq{\,
        \begin{aligned}[c]
            &
                p_m\mem e^m{}_\m(\fx)\mem \dot{\fx}^\mu 
                + \frac{1}{2}\, 
                    \fS_{mn}\mem \fL^m{}_A\mem \frac{D\nem\fL^n{}_B}{d\t}
                    \mem \eta^{AB}
            \\
            &
                - \frac{\k^0}{2m}\mem\BB{
                    p^2 + m^2
                }\mem d\t
                - \kappa^a\mem \phi_a
        \end{aligned}
        \mem}\, d\t
    \,,
\end{align}
which is a functional of
$(\fx^\m,\fL^m{}_A,p_m,\fS_{mn})$ and $(\kappa^0, \kappa^a)$.

\eqref{STGR.action}
is a worldline action localized on
$\t \mapsto \fx^\m(\t)$.
For instance,
the covariant derivative is
$D\nem\fL^m{}_A = d\hnem\fL^m{}_A + \gamma^m{}_{n\r}(\fx)\mem \fL^n{}_A\mem d\fx^\r$.

The theory which \eqref{STGR.action} defines
is \textit{seriously ill}
due to the spin gauge dependence of the spurious center $\fx^\m$,
on which the gravitational fields
$e^m{}_\m(\fx)$ and $\gamma^m{}_{n\r}(\fx)$ are coupled to:
the value of the action in \eqref{STGR.action}
changes when
one changes the SSC from one to another!
The gravitational fields must couple to a physical center $x^\m$,
while
the curved-spacetime generalization 
of the formula in \eqref{ST.xy}
is missing.
However, we do not have the definition of the 
physical center $x^\m$
in terms of 
the defining variables
$(\fx^\m,\fL^m{}_A,p_m,\fS_{mn})$,
such that $\pb{x^\m}{\phi_a} = 0$.
In fact, we do not even know
if such a gauge-invariant center exists at all!

Due to the utterly nonsensical status of \eqref{STGR.action},
we shall avoid paying attention to its EoM.
The correct answer will be
given in \eqref{STGR.action.correct}.
We have checked that
\eqref{STGR.action.correct} 
consistently derives
the MPD and QMPD equations
in \textit{any} SSC
just like in \Sec{INT>FOUR1}.

\section{The Interacting Theory of Spherical Top} 
\label{HR}

In \Sec{INT>FOUR2},
we have observed that
the interacting theory of the spherical top
involves subtleties
regarding
the spin gauge invariance
of the background-coupled action.

To reiterate,
the interacting theory develops fatal inconsistencies
if the action is 
coupled to the spin gauge-dependent center $\fx$.
Namely,
one's customary choice of the SSC
\textit{changes the predictions on physical observables}.
It is a strict necessity
to realize the 
external field couplings
at a spin gauge-invariant center $x$.

It is the very mission of this section to establish
the interacting theory of the spherical top model
in a rigorous fashion
and prove the existence of a spin gauge-invariant center $x$.
This mandates a nontrivial analysis,
which must explicitly identify
\begin{enumerate}
    \item 
        the spin gauge-invariant center coordinates and actions,
    \item
        their explicit formulae
        in the defining worldline field basis,
    \item
        the spin gauge algebra and flow in the presence of interactions,
    \item
        accurate realizations of the universal covariant structures postulated in \Sec{INT>CUPB}. 
\end{enumerate}

To reiterate,
the most pressing problem is that
the explicit formula for the physical center in \eqref{ST.xy}
is lost in curved spacetimes,
since
it is nonsensical to add
the \textit{coordinates} $\fx^\m$
with (the components of) a \textit{vector} $\fS^{\m\n} p_\n /p^2$.
To this end, we again utilize the strategy taken in \Sec{INT>COVSPT}:
``flat spacetime in noninertial frames''
as a bridge toward general relativity.

\subsection{Explicit Parametrization of Spin Gauge Redundancy}
\label{HR>ORB}

Our point of departure
is a deeper understanding on the spin gauge transformations
at the level of free theory.

\para{Spin Gauge-Invariant Observables}

The free spherical top describes a $20$-dimensional phase space $\ps^\ST_{20}$
whose coordinates are $(\fx^\m,\fL^\m{}_A,p_\m,\fS_{\m\n})$,
equipped with the three spin gauge generators $\phi_a$ in \eqref{phia}
that define the symplectic quotient
$\ps^\ST_{14}$ in \eqref{ST.quotient14}.

A function $f$ on $\ps^\ST_{20}$
is \textit{spin gauge invariant}
iff it descends to the quotient $\ps^\ST_{14}$:
\begin{align}
    \pb{f}{\phi_a}
    \,=\,
        0
    \,.
\end{align}
For example,
the universal variables $(x^\m,\hy^\m,p_\m)$
are spin gauge invariant:
\begin{subequations}
\label{sgi}
\begin{align}
    \label{sgi.xyp}
    x^\m
    \,=\,
        \fx^\m
        + \frac{1}{p^2}\,
            \fS^{\m\n}\hem p_\n
    \,,\quad
    \hy^\m
    \,=\,
        - \frac{1}{p^2}\,
            {*}\fS^{\m\n}\hem p_\n
    \,,\quad
    p_\m
    \,.
\end{align}

The $14$-dimensional quotient $\ps^\ST_{14}$ 
can be explicitly coordinatized by finding $14$ spin gauge-invariant variables
that are functionally independent.
As noted in \Sec{FREE>PART},
$(x^\m,\hy^\m,p_\m)$ together describes an $11$-dimensional space.
It is known \cite{Steinhoff:2015ksa}
that
the remaining three spin gauge-invariant variables are
\begin{align}
    \label{sgi.L}
    \L^\m{}_a
    \,:=\,
    \bb{
        \delta^\m{}_\n
        - \frac{\hem{
            (\hp {\,+} \fL_0)^\m\mem \hp_\n
        \mem}}{
            \hp \mdot (\hp {\,+} \fL_0)
        }
    }\mem \fL^\n{}_a
    \,.
\end{align}
\end{subequations}
Direct computation shows that
$\pb{\L^\m{}_a}{\phi_c} = 0$
and $p_\m\hem \L^\m{}_a = 0$.
In sum, $(x^\m,\hy^\m,p_\m,\L^\m{}_a)$
in \eqref{sgi}
provides a well-behaved,
manifestly spin gauge-invariant
coordinate chart on 
the physical phase space
$\ps_{14}^\ST$.

Geometrically, the formula in \eqref{sgi.L}
originates from the minimal Lorentz boost 
that aligns the spurious body frame 
with the momentum
\cite{Steinhoff:2015ksa}:
\begin{align}
    \bb{
        \delta^\m{}_\n
        - \frac{\hem{
            (\hp {\,+} \fL_0)^\m
            \mem
            (\hp {\,+} \fL_0)_\n
        \mem}}{
            \hp \mdot (\hp {\,+} \fL_0)
        }
        - 2\mem \hp^\m\hnem \fL_{\n0}
    }\mem 
    \begin{pmatrix}
        \fL^\n{}_0
        \\
        \fL^\n{}_a
    \end{pmatrix}
    \,=\,
    \begin{pmatrix}
        \hp^\m
        \\
        \L^\m{}_a
    \end{pmatrix}
    \,.
\end{align}
We will refer to $\L^\m{}_a$---%
more accurately the Lorentz matrix $\L^\m{}_A = (\hp^\m,\L^\m{}_a)$---%
as the \textit{physical body frame}.
The physicalness of $\L^\m{}_a$
is expected from 
the perspective of
\eqref{ST.solve}.

\para{Decomposition into Physical and Gauge Parts}

Given the clear identifications of the $14$ physical variables
$(x^\m,\hy^\m,p_\m,\L^\m{}_a)$
characterizing the physical phase space $\ps^\ST_{14}$,
we now investigate
how the unphysical redundancies are wired in $\ps^\ST_{20}$.

First of all, let us define
\begin{align}
    \label{fb}
    \fb^\m
    \,:=\,
        \frac{1}{-p^2}\,
            \fS^{\m\n}\hem p_\n
    \,,\quad
    S_{\m\n}
    \,:=\,
        \ve_{\m\n\r\s}\hem \hy^\r p^\s 
    \,,
\end{align}
so $p \mdot \fb = 0$ and $S_{\m\n}\hem p^\n = 0$.
The interpretation of $S_{\m\n}$ is the physical spin tensor.
The vector $\fb^\m$ describes the
separation between the physical and spurious centers:
\begin{align}
    \label{fcen.flat}
    \fx^\m 
    \,=\,
        x^\m \hem\hhem+\mem \fb^\m
    \,.
\end{align}
It can be shown that
\begin{subequations}
\label{bcollect}
\begin{align}
    \label{bparam}
    \fS_{\m\n}\nem
    \,=\,
        S_{\m\n}
        - (\fb \swedge p)_{\m\n}
    \,,\quad
    \fL^\m{}_A
    \,=\,
        \L^\m{}_B\mem \fR^B{}_A
    \,,
\end{align}
where $\fR^A{}_B$ is a Lorentz group element defined by
\begin{align}
\begin{split}
    \label{Rmatrix}
    \fR^A{}_B(\vbet) 
    \,&=\, 
    \frac{1}{1\mminus\vbet^2\nem/4}\,
    \BBBB{\,\hnem
    {\renewcommand{\arraystretch}{1.2}
    \begin{matrix}
        {1\mplus\vbet^2\nem/4} & -\bet_b
        \\
        -\bet^a & 
            ({1\mminus\vbet^2\nem/4})\mem \delta^a{}_b 
            + \bet^a \bet_b /2
    \end{matrix}}
    \,}
    \,.
\end{split}
\end{align}
Here, $\vbet$ is a three-vector
that exhibits a nonlinear relationship with $\fb^\m$:
\begin{align}
    \label{b-beta}
    \fb^\m\hhnem(\vbet)
    \,=\,
        \L^{\m a}\,
            \frac{1}{|p|}
            \bbsq{
                \bigbig{
                    S_{\r\s}\mem \L^\r{}_a\mem \L^\s{}_b
                }\mem \frac{\bet^b}{2}
                -
                \bb{
                    1 - \frac{\vbet^2}{4}
                }\mem \phi_a
            }
    \,.
\end{align}
\end{subequations}
In this way,
the spurious variables
$(\fx^\m,\fL^\m{}_A,\fS_{\m\n})$
are found as functions of the physical variables
$(x^\m,\L^\m{}_A,S_{\m\n})$
and a three-vector $\vbet$.
That is, we have shown that
a single three-vector variable $\vbet$
parametrizes the entire gauge content of the spurious variables.

\para{Symplectic Structure Revisited}

Further computation
based on \eqref{bcollect}
shows that
\begin{align}
    \label{RdR}
            \L^\m{}_A\mem
            \L^\n{}_B\mem
            \fR^A{}_C\mem d\fR^B{}_D
            \mem\eta^{CD}
    \,=\,
            \BB{
                \hat{p}^{[\m} + \fL^{[\m}{}_0
            }\mem
            \fL^{\n]}{}_a
        \mem
        d\bet^a
    \,.
\end{align}
From
Eqs.\:(\ref{fb}), (\ref{bparam}), (\ref{RdR}), and the definition of $\phi_a$ in (\ref{phia}),
one can derive
\begin{subequations}
\label{thetapieces}
\begin{align}
    \label{thetapiece0}
    p_\m\hem d\fb^\m
    \,&=\,
        -dp_\m \fb^\m
    \,,\\
    \label{thetapiece1}
    \frac{1}{2}\,
        \fS_{\m\n}\mem
            \L^\m{}_A\mem
            d\L^\n{}_B\mem
            \fR^A{}_C\mem \fR^B{}_D
            \mem\eta^{CD}
    \,&=\,
    dp_\m \fb^\m
    + \frac{1}{2}\,
        S_{\m\n}\mem
            \L^\m{}_a\mem
            d\L^{\n a}
    \,,\\
    \label{thetapiece2}
    \frac{1}{2}\,
        \fS_{\m\n}\mem
            \L^\m{}_A\mem
            \L^\n{}_B\mem
            \fR^A{}_C\mem d\fR^B{}_D
            \mem\eta^{CD}
    \,&=\,
        \phi_a\mem d\bet^a
    \,.
\end{align}
\end{subequations}
where \eqref{thetapiece1} uses
$d(p_\m\hem \L^\m{}_a) {\:=\:} 0$.
The sum of 
the above three equations
yields
\begin{align}
\begin{split}
    \label{piece.theta}
    \theta^\circ
    \,&=\,
    p_\m\mem d\fx^\m
    +
    \frac{1}{2}\,
        \fS_{\m\n}\mem
            \fL^\m{}_A\mem
            d\fL^{\n A} 
    \,=\,
        p_\m\mem dx^\m
        + \frac{1}{2}\,
            S_{\m\n}\mem
                \L^\m{}_a\mem
                d\L^{\n a}
        + \phi_a\mem d\bet^a
    \,.
\end{split}
\end{align}
Notably, \eqref{piece.theta} establishes a clean decomposition of the symplectic potential into physical and gauge parts.
Consequently, the symplectic form is also cleanly split as
\begin{align}
    \label{piece.omega}
    \omega^\circ
    \mem\,=\,\mem
        dp_\m \swedge dx^\m
        + \frac{1}{2}\,
            dS_{\m\n}
            \swedge
                (\L^\m{}_a\mem
                d\L^{\n a})
        + \frac{1}{2}\,
            S_{\m\n}\mem
                d\L^\m{}_a
                \swedge
                d\L^{\n a}
        + d\phi_a \swedge d\bet^a
    \,.
\end{align} 

\para{Gauge Orbit}

Consistently with \eqref{piece.omega},
one finds that
the spin gauge flow describes
\begin{align}
\begin{array}{rrr}
    \pb{
        x^\m
    }{\phi_c}
    \,=\,
        0
    \,,&\quad
    \pb{
        \L^\m{}_a
    }{\phi_c}
    \,=\,
        0
    \,,&\quad
    \pb{
        \bet^a
    }{\phi_c}
    \,=\,
        \delta^a{}_c
    \,,\\
    \pb{
        p_\m
    }{\phi_c}
    \,=\,
        0
    \,,&\quad
    \pb{
        S_{\m\n}
    }{\phi_c}
    \,=\,
        0
    \,,&\quad
    \pb{
        \phi_a
    }{\phi_c}
    \,=\,
        \mathrlap{0\,.}\phantom{\delta^a{}_c\,,}
\end{array}
\end{align}
As a result, the $\R^3$ gauge orbits in $\ps^\ST_{20}$
are explicitly found as
\begin{align}
    \label{orbit}
    \exp\nem\BB{
        \pb{\blank}{\phi_c}\mem 
        \sigma^c
    }
    \,\,\,:\,\,\,\,
    \left\{\mem
    {\renewcommand{\arraystretch}{1.1}
    \begin{array}{rl}
        \fx
        &\,\,\,\mapsto\,\,\,
            \fx
            - \fb(\vbet)
            + \fb(\vbet \mplus\vs)
        \\
        p
        &\,\,\,\mapsto\,\,\,
            p
        \\
        \fL
        &\,\,\,\mapsto\,\,\,
            \fL\,
                \fR^{-1}\hnem(\vbet)\,
                \fR(\vbet \mplus\vs)
        \\
        \fS
        &\,\,\,\mapsto\,\,\,
            \fS
            + \fb(\vbet)
                \wedge p
            - \fb(\vbet \mplus\vs)
                \wedge p
    \end{array}}
    \,\mem
    \right\}
    \,,
\end{align}
where we have made indices implicit
to avoid clutter.
The three-vector variable
$\vs\in\R^3$ explicitly parametrizes the gauge orbits.
The submanifold $\vbet = 0$
sets a special ``origin'' in \eqref{orbit}
so that the gauge algebra is abelian
despite the nonlinearities in 
$\fb(\vbet)$ and $\fR(\vbet)$.

\newpage

It should be clear that
the gauge orbit in \eqref{orbit}
solves
the following set of partial differential equations:
\begin{align}
\begin{split}
    \label{flow-pde}
    \frac{\partial
        \fx^\m
    }{\partial \bet^a}
    \,=\,
        \pb{
            \fx^\m
        }{\phi_a}
    &\,,\quad
    \frac{\partial\nem
        \fL^\m{}_A
    }{\partial \bet^a}
    \,=\,
        \pb{
            \fL^\m{}_A
        }{\phi_a}
    \,,\\
    \frac{\partial
        p_\m
    }{\partial \bet^a}
    \,=\,
        \pb{
            p_\m
        }{\phi_a}
    &\,,\quad
    \frac{\partial
        \fS_{\m\n}
    }{\partial \bet^a}
    \,=\,
        \pb{
            \fS_{\m\n}
        }{\phi_a}
    \,.
\end{split}
\end{align}
Previously,
the flow equations in \eqref{flow-pde}
were solved
on the $17$-dimensional submanifold $\phi_a {\:=\:} 0$
in $\ps^\ST_{20}$
\cite{Kim:2021rda},
in which case $\bet^a$ and
$\fchi^a {\:=\:} \hat{p}_\m\hhem \fL^{\m a}$
exhibit the relations
$\vec{\fchi} = \vbet / (1 {\,-\,} \vbet^2\hnem/4)$
and
$\vbet = 2\vec{\fchi} \mem/ ( 1 {\,+\,} \sqrt{1\mplus\vec{\fchi}^2} )$.
The new contribution of this paper
is the general solution in \eqref{orbit}
that applies to any point in $\ps^\ST_{20}$
without submanifold restrictions.

\subsection{The Physical Center in General Relativity}
\label{HR>CENTER}

Our next mission is to explicitly construct the physical center $x^\m$
in the general-relativistic theory of spherical top.

\para{Flat Spacetime Revisited}

We begin with revisiting the free spherical top
in curvilinear coordinates.
\eqref{fcen.flat} states that
the fake and physical centers
are related by $\fx^\m = x^\m + \fb^\m$.
Crucially,
this equation can only make sense in \textit{Cartesian coordinates}
in which the concept of ``position vector'' is valid.

What is the coordinate-free content of \eqref{fcen.flat}?
Geometrically, it states that the spurious center $\fx$ is reached from the physical center $x$ by moving along the straight-line path in the direction of the vector $\fb$.
A well-known fact is that straight lines are the geodesics of flat space.
Therefore, \eqref{fcen.flat} states that the spurious center $\fx$ is obtained by the \textit{geodesic flow} from the physical center $x$
generated by the vector $\fb$.

Based on this understanding,
\eqref{fcen.flat} appears in curvilinear coordinates as
\begin{align}
    \label{fcen.curved}
    \fx^{\m\sprime}
    \,=\,
    \delta^{\m\sprime}{}_\m\mem
    \bb{\,
        x^\m \hem\hhem+\mem \fb^\m
        \hem\hhem-\mem \frac{1}{2}\, \Gamma^\m{}_{\r\s}(x)\mem \fb^\r\hem \fb^\s
        + \O(\fb^3)
    \,}
    \,,
\end{align}
where $\Gamma^\m{}_{\r\s}(x)$ are the Christoffel symbols
of the flat metric $\eta_{\m\n}(x)$.
\eqref{fcen.curved} is easily extended to all orders in $\fb$ by solving the geodesic equation
with the initial condition $(x^\m,\fb^\m)$.
A pedagogical exercise is to deduce \eqref{fcen.curved}
by directly performing a Cartesian-to-curvilinear coordinate transformation on \eqref{fcen.flat}.

\begin{figure}
    \centering
    \includegraphics[scale=0.5,
        trim= 0pt 10pt 0pt 20pt
    ]{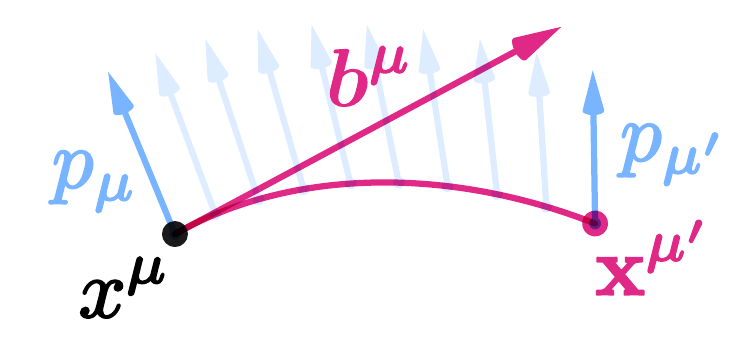}
    \caption{
        A geodesic connects between
        the physical center $x^\m$ and the fake center $\fx^{\m\sprime}$,
        along which the parallel propagator $W^\m{}_{\m'}$ is constructed
        so that local tensor degrees of freedom 
        are transported like
        $p_\m\hem W^\m{}_{\m'} = p_{\m'}$. 
    }
    \label{gdex}
\end{figure}

In \eqref{fcen.curved},
we have started employing the Synge bitensor notation \cite{Ruse:1931ht,Synge:1931zz,Poisson:2011nh,gde}.
The idea of this notation is to employ
separate index sets---%
primed and unprimed---%
when studying \textit{bilocal objects}
in differential geometry.
As illustrated in \fref{gdex},
the central object of investigation
in our case
is a spacelike geodesic segment,
whose endpoints provide two spacetime points $x$ and $\fx$.
Unprimed indices $\m,\n,\r,\s,\cdots$ are assigned to the physical point $x$.
Primed indices $\m',\n',\r',\s',\cdots$ are assigned to the spurious point $\fx$.
Tensors based at $x$ carry unprimed indices.
Tensors based at $\fx$ carry primed indices.
The former transform with the Jacobian factors at $x$.
The latter transform with the Jacobian factors at $\fx$.

In this fashion, one practices the very philosophy of index notation: 
the transformation behavior of an object
is completely specified by---%
and is evident from---%
its index form.

As shown in \fref{gdex},
$\fb^\m$ is the tangent vector of the geodesic segment
at the point $x$;
hence it carries an unprimed index. 
Namely, $\fb^\m$ 
is an element of
the tangent space at $x$.

\para{Parallel Propagator}

To compare between
tensors at $x$ and $\fx$,
a law of parallel propagation is necessary:
the very mathematical idea of a connection.

The parallel propagator is an object of the index form
$W^{\m\sprime}{}_\n$,
which describes an isomorphism between
the tangent spaces at
$x$ and $\fx$.
In the Synge framework
\cite{Ruse:1931ht,Synge:1931zz,Poisson:2011nh,gde},
it is a common practice to
consider the parallel propagator
due to the Levi-Civita connection
about the geodesic contour.
The explicit power series expansion for
$W^{\m\sprime}{}_\n$
in this case reads
\begin{align}
    \label{Wformula} 
        \delta^{\m\sprime}{}_\m
        \mem\bb{
            \delta^\m{}_\n
            - \Gamma^\m{}_{\n\r}(x)\mem \fb^\r
            - \frac{1}{2}\,
                \BB{\hhem
                    \Gamma^\m{}_{\n\r,\s}(x)
                    - 2\mem \Gamma^\m{}_{\l\r}(x)\,
                    \Gamma^\l{}_{\n\s}(x)
                \hnem}\mem
                \fb^\r \fb^\s
            + \O(\fb^3)
        }
    \,.
\end{align}
It a nice exercise to derive
\eqref{Wformula}
from
a path-ordered exponential
constructed on the geodesic contour
from $x$ to $\fx$.

The parallel propagator facilitates
transporting tensors at point $x$ to point $\fx$
(and vice versa).
For example, a vector $v^\m$ at $x$ is parallel-transported to $\fx$ as $v^{\m\sprime} = W^{\m\sprime}{}_\m\mem v^\m$.
Similarly, a covector $p_\m$ at $x$ is parallel-transported to $\fx$ as
$p_{\m'} = p_\m\mem W^\m{}_{\m'}$,
as is visualized in \fref{gdex}.
In general,
a rank-$(p,q)$ tensor at $x$ is parallel-transported to $\fx$ as
\begin{align}
    T^{\m\sprime_1 \cdots \m\sprime_p}{}_{\n\sprime_1 \cdots \n\sprime_q}
    \,=\,
        W^{\m\sprime_1}{}_{\m_1}
        {\cdots\mem}
        W^{\m\sprime_p}{}_{\m_p}
        \,
            T^{\m_1 \cdots \m_p}{}_{\n_1 \cdots \n_q}
        \,
        W^{\n_1}{}_{\n'_1}
        {\cdots\mem}
        W^{\n_q}{}_{\n'_q}
    \,,
\end{align}
where our notation is such that
\begin{align}
    W^{\m\sprime}{}_\k\mem W^\k{}_{\n'}
    \,=\,
    \delta^{\m\sprime}{}_{\n'}
    \,,\quad
    W^\m{}_{\k\sprime}\,\hem W^{\k\sprime}{}_\n 
    \,=\,
    \delta^\m{}_\n
    \,.
\end{align}

In fact, this gymnastics can be applied to tensor-valued differential forms as well.
Especially, it is left as an exercise to check the following identities
regarding vector-valued differential one-forms,
in flat spacetime:
\begin{subequations}
\label{Wdiffs}
\begin{align}
    \label{Wdiff.x}
    D^2 =\mem 0
    &\qiq
    d\fx^{\m\sprime}
    \,=\,
        W^{\m\sprime}{}_\m\mem
        \BB{
            dx^\m + D\fb^\m
        }
    \,,\\
    \label{Wdiff.v}
    D^2 =\mem 0
    &\qiq
    Dv^{\m\sprime}
    \,=\,
        W^{\m\sprime}{}_\m\mem
            Dv^\m
    \,.
\end{align}
\end{subequations}
The second identity assumes the relation
$v^{\m\sprime} = W^{\m\sprime}{}_\m\mem v^\m$.
Also,
$D\fb^\m = d\fb^\m + \Gamma^\m{}_{\n\r}(x)\mem \fb^\n\mem dx^\r$,
$Dv^{\m\sprime} = dv^{\m\sprime} + \Gamma^{\m\sprime}{}_{\n'\r'}(\fx)\mem v^{\n\sprime}\mem d\fx^{\r\sprime}$,
and
$Dv^\m = dv^\m + \Gamma^\m{}_{\n\r}(x)\mem v^\n\mem dx^\r$.
Our index notation
is supposed to unambiguously clarify
the position which the connection coefficients inside the covariant derivatives
couple to.

\para{Symplectic Structure Revisited}

Finally,
let us apply the Synge formalism to the free spherical top
to rederive \eqref{piece.theta}
in curvilinear coordinates.
The first line now reads
\begin{align}
\label{gentheta.1}
    \theta^\circ
    \,&=\,
    p_{\m'}\mem d\fx^{\m\sprime}
    +
    \frac{1}{2}\,
        \fS_{\m\sprime\mem\n'}
            \fL^{\m\sprime}{}_A\mem
            D\nem\fL^{\n\sprime A}
    \,.
\end{align}
To proceed,
we not only use \eqref{fcen.curved}
but also
stipulate that the local tensor degrees of freedom
of the spherical top
are parallel-transported 
to the physical center 
as
\begin{align}
    \label{Wlocal}
    p_{\m\sprime}
    \,=\,
        p_\m\mem W^\m{}_{\m\sprime}\mem
    \,,\quad
    \fL^{\m\sprime}{}_A
    \,=\,
        W^{\m\sprime}{}_\m\mem \fL^\m{}_A
    \,,\quad
    \fS_{\m\sprime\mem\n\sprime}
    \,=\,
        \fS_{\m\n}\mem W^\m{}_{\m'}\hem W^\n{}_{\n\sprime}\mem
    \,.
\end{align}
By plugging in \eqrefs{fcen.curved}{Wlocal}
to \eqref{gentheta.1},
we find
\begin{align}
\begin{split}
    \label{gentheta.2}
    \theta^\circ 
    \,&=\,
    p_\m\mem dx^\m
    +
        p_\m\hem D\fb^\m
    +
    \frac{1}{2}\,
        \fS_{\m\n}\mem
            \fL^\m{}_A\mem
            D\nem\fL^{\n A}
    \,,
\end{split}
\end{align}
through using the identities in \eqref{Wdiffs}.
Since \eqref{gentheta.2} involves only first-order differentials,
the computation in \eqref{thetapieces}
readily generalizes to yield
\begin{align}
    \label{gentheta.3}
    \theta^\circ
    \,=\,
        p_\m\mem dx^\m
        + \frac{1}{2}\,
            S_{\m\n}\mem
                \L^\m{}_a\mem
                D\hhnem\L^{\n a}
        + \phi_a\mem d\bet^a
    \,,
\end{align}
provided the definitions in
\eqrefs{fb}{bcollect}
are kept the same.

\subsection{The Spin Gauge-Invariant Action}
\label{HR>ACTION}

We may now transition to curved spacetime
$\M = (\R^4,g)$.
Suppose there exists a physical center
in the presence of curvature.
Then its relation with the fake center
must smoothly approach \eqref{fcen.curved}
in the flat limit.
With hindsight,
let us simply postulate that
the relation in \eqref{fcen.curved}
does not develop curvature corrections.

Similarly,
we would like to
again utilize the geodesic parallel propagator due to the Levi-Civita connection
of the curved metric $g$,
to transport local tensor degrees of freedom.
Importantly, this implies
that our parallel propagator
will be metric-preserving:
\begin{align}
    \label{metric-preserving}
    g_{\m\n}(x)\,
        W^\m{}_{\m'}\mem W^\n{}_{\n'}\nem
    \,=\,
        g_{\m'\nem\n'}(\fx)
    \,,\quad
    g_{\m'\nem\n'}(\fx)\,
        W^{\m\sprime}{}_\m\mem W^{\n\sprime}{}_\n
    \,=\,
        g_{\m\n}(x)
    \,.
\end{align}

\para{Minimal Coupling Done Right}

Next, we recall from \Sec{HR>CENTER}
that
two equivalent formulations of the free spherical top
are provided by
\eqrefs{gentheta.1}{gentheta.3}
in the limit of vanishing curvature.
Let us reproduce them for convenience:
\begin{subequations}
\label{genthetas}
\begin{align}
    \label{gentheta'}
    \theta^\circ
    \,&=\,
    p_{\m'}\mem d\fx^{\m\sprime}
    +
    \frac{1}{2}\,
        \fS_{\m\sprime\mem\n'}
            \fL^{\m\sprime}{}_A\mem
            D\nem\fL^{\n\sprime A}
    \,,\\
    \label{gentheta}
    \,&=\,
        p_\m\mem dx^\m
        + \frac{1}{2}\,
            S_{\m\n}\mem
                \L^\m{}_a\mem
                D\hhnem\L^{\n a}
        + \phi_a\mem d\bet^a
    \,.
\end{align}
\end{subequations}
Both of \eqrefs{gentheta'}{gentheta}
are generally covariant.
Thus,
following the strategy that has been repeatedly practiced in this paper,
we might transplant
either of the symplectic potentials in
\eqrefs{gentheta'}{gentheta}
to curved spacetime.

However,
the astute reader will point out that
one of \eqrefs{gentheta'}{gentheta}
is not sensible
in view of spin gauge invariance.
Moreover,
the mathematical equivalence between \eqrefs{gentheta'}{gentheta}
breaks down
in the presence of spacetime curvature,
as the identities in  \eqref{Wdiffs}
gain curvature corrections.

\newpage

Based on our preliminary explorations in \Sec{INT>FOUR2},
the answer should be obvious:
the correct symplectic potential is \eqref{gentheta}.
Thus, we declare that
the symplectic potential
of the general-relativistic spherical top
in curved spacetime
is
\begin{align}
    \label{HRGR.theta}
    \theta
    \,=\,
        p_\m\mem dx^\m
        + \frac{1}{2}\,
            S_{\m\n}\mem
                \L^\m{}_a\mem
                D\hhnem\L^{\n a}
        + \phi_a\mem d\bet^a
    \,.
\end{align}
Here,
it is important that
$\bet^a$ describes a set of \textit{scalars},
carrying no spacetime indices,
so $d\beta^a = D\hnem\bet^a$:
the index $a \in \{1,2,3\}$ is a global index,
not gauged.

\begin{figure}
    \centering
    \includegraphics[scale=0.5,
        trim= 0pt 10pt 0pt 20pt
    ]{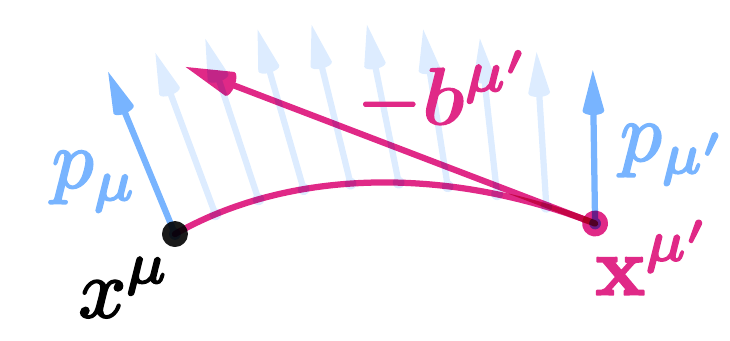}
    \caption{
        Flipside view on \fref{gdex}.
        The physical center $x^\m$
        is reached from the fake center $\fx^{\m\sprime}$
        through deviating by the tangent vector $-\fb^{\m\sprime}$.
    }
    \label{gdefx}
\end{figure}

\para{Action In the Original Worldline Field Basis}

To return to the primed variables
$(\fx^{\m\sprime},p_{\m'},\fL^{\m\sprime}{}_A,\fS_{\m'\nem\n'})$,
we reverse the derivation in \eqrefss{gentheta.1}{gentheta.2}{gentheta.3}
while taking curvature effects into account.
We find that
\begin{align}
    \label{varths}
    \theta
    \,=\,
    p_{\m'}\mem d\fx^{\m\sprime}
    +
    \frac{1}{2}\,
        \fS_{\m\sprime\mem\n'}
            \fL^{\m\sprime}{}_A\mem
            D\nem\fL^{\n\sprime A}
    +
        \fth_\text{mass}
    +
        \fth_\text{spin} 
    \,,
\end{align}
where $\vartheta'_{(0)}$ and $\vartheta'_{(1)}$
are curvature corrections
such that
\begin{align}  
    \label{thetafict.limits}
    \lim_{R\to0}
    \fth_\text{mass}
    \,=\,
        0
    \,,\quad
    \lim_{R\to0}
    \fth_\text{spin}
    \,=\,
        0
    \,,\quad
    \lim_{\fx\to x}
    \fth_\text{mass}
    \,=\,
        0
    \,,\quad
    \lim_{\fx\to x}
    \fth_\text{spin}
    \,=\,
        0
    \,.
\end{align}
Let us elaborate on the details in four steps.

First,
we retain the definitions in
\eqrefs{fb}{bcollect}
to bring \eqref{HRGR.theta} to
\begin{align}
    \label{HRGR.theta.2}
    \theta
    \,=\,
        p_\m\mem dx^\m
        +
            p_\m\hem D\fb^\m
        +
        \frac{1}{2}\,
            \fS_{\m\n}\mem
                \fL^\m{}_A\mem
                D\nem\fL^{\n A}
    \,,
\end{align}
which merges the physical and gauge parts
for the rotational sector.

Second, \eqref{fcen.curved} states that 
$\fx$ is obtained by the unit-time geodesic flow from $x$
by $\fb^\m$.
Its inverse relation reads
\begin{align}
    \label{fcen.curved.inv}
    x^\m
    \,=\,
    \delta^\m{}_{\m'}\hem
    \bb{\,
        \fx^{\m\sprime} \hem\hhem-\mem \fb^{\m\sprime}
        \hem\hhem-\mem \frac{1}{2}\, \Gamma^{\m\sprime}{}_{\r'\nem\s'}(\fx)\mem \fb^{\r\sprime}\hem \fb^{\s\sprime}
        + \O(\fb^3)
    \,}
    \,,
\end{align}
meaning that
the physical center $x^\m$
is obtained by
the unit-time geodesic flow
from $\fx^{\m\sprime}$
by $-\fb^{\m\sprime}$.
Here, 
$\fb^{\m\sprime} = W^{\m\sprime}{}_\m\mem \fb^\m$
is the tangent vector of the geodesic segment
at $\fx$
that points outwards,
so $-\fb^{\m\sprime}$
is the tangent vector pointing inwards.
As mentioned before,
we stipulate that
the local tensor variables in \eqref{HRGR.theta.2},
$p_\m$, $\fL^\m{}_A$, and $\fS_{\m\n}$,
are parallel-transported
to the spurious point $\fx$ 
along the geodesic path
to become
$p_{\m'}$, $\fL^{\m\sprime}{}_A$, and $\fS_{\m'\nem\n'}$,
respectively.

Third,
we generalize
the identities in \eqref{Wdiffs}
in the presence of curvature.
Due to general covariances at both $x$ and $\fx$,
they will be generalized in the following form:
\begin{subequations}
\label{RWdiffs}
\begin{align}
    \label{RWdiff.x}
    W^{\m\sprime}{}_\m\mem dx^\m
    \,&=\,
        X^{\m\sprime}{}_{\s'}\mem
            d\fx^{\s\sprime}
        -
        Y^{\m\sprime}{}_{\s'}\hem
            D\fb^{\s\sprime}
    \,,\\
    \label{RWdiff.v}
    W^{\m\sprime}{}_\m\mem Dv^\m
    \,&=\,
    Dv^{\m\sprime}
    \mem+
    \BB{
        \acX^{\m\sprime}{}_{\n'\nem\s'}\mem
            d\fx^{\s\sprime}
        -
        \acY^{\m\sprime}{}_{\n'\nem\s'}\hem
            D\fb^{\s\sprime}
    \mem}\, v^{\n\sprime}
    \,.
\end{align}
\end{subequations}
Here, 
$X^{\m\sprime}{}_{\s'}$,
$Y^{\m\sprime}{}_{\s'}$,
$\acX^{\m\sprime}{}_{\n'\nem\s'}$,
and
$\acY^{\m\sprime}{}_{\n'\nem\s'}$
are all tensorial at $\fx$.
Therefore,
they must arise by combining
the Riemann tensor $R^{\m\sprime}{}_{\n'\nem\r'\nem\s'}(\fx)$,
their covariant derivatives,
and the vector $\fb^{\m\sprime}$.
In particular, 
$X^{\m\sprime}{}_{\s'}$
and
$Y^{\m\sprime}{}_{\s'}$
are known as the
Jacobi propagators
\cite{Dixon:1970zza,Dixon:1974xoz},
whose all-orders formulae
were studied by \rcite{Vines:2014oba}.
Meanwhile, the recent work \cite{gde} 
has provided
a versatile and powerful formalism
that systematically generates
all-order formulae
for
$\Delt X^{\m\sprime}{}_{\s'}$,
$\Delt Y^{\m\sprime}{}_{\s'}$,
$\acX^{\m\sprime}{}_{\n'\nem\s'}$,
and
$\acY^{\m\sprime}{}_{\n'\nem\s'}$.

Finally,
using the identities in \eqref{RWdiffs}
brings \eqref{HRGR.theta.2}
into the form in \eqref{varths},
with
the curvature corrections being
\begin{subequations}
\label{vths}
\begin{align}
    \label{vth0}
    \fth_\text{mass}
    \,&=\,
        p_{\m'}\mem
        \BB{
                \Delt X^{\m\sprime}{}_{\s'}\mem
                d\fx^{\s\sprime}
            -
                \Delt Y^{\m\sprime}{}_{\s'}\hem
                D\fb^{\s\sprime}
        \,}
    \,,\\
    \label{vth1}
    \fth_\text{spin}
    \,&=\,
        - \frac{1}{2}\,
            S_{\m'\nem\n'}\mem
            \BB{
                \acX^{\m'\nem\n\sprime}{}_{\s'}\mem
                    d\fx^{\s\sprime}
                -
                \acY^{\m'\nem\n\sprime}{}_{\s'}\hem
                    D\fb^{\s\sprime}
            \,}
    \,.
\end{align}
\end{subequations}
Here, we have defined 
$\Delt X^{\m\sprime}{}_{\s'} := 
    X^{\m\sprime}{}_{\s'}
    - \delta^{\m\sprime}{}_{\s'}
$
and
$\Delt Y^{\m\sprime}{}_{\s'} := 
    Y^{\m\sprime}{}_{\s'}
    - \delta^{\m\sprime}{}_{\s'}
$.
It should be clear that
the relations in \eqrefs{fb}{bcollect}
imply
\begin{align}
    -\fb^{\m\sprime}
    \,=\,
        \frac{1}{p^2}\,
            \fS^{\m'\nem\n\sprime} p_{\n'}
    \,,\quad
    S_{\m'\nem\n'}\nem
    \,=\,
        \fS_{\m'\nem\n'}
        + (\fb \swedge p)_{\m'\nem\n'}
    \,=\,
        \fS_{\k'\nem\l'}\mem
        \hdelta^{\k\sprime}{}_{\m'}\hem
        \hdelta^{\l\sprime}{}_{\n'}
    \,.
\end{align}
In conclusion,
the spin gauge-invariant action
of the general-relativistic spherical top model
is explicitly found as
\begin{align}
    \label{STGR.action.correct}
    \int\,
        \lrsq{\,
        \begin{aligned}[c]
            &
                p_{\m'}\hem \dot{\fx}^{\m\sprime} 
                + \frac{1}{2}\,
                    \fS_{\m\sprime\mem\n'}
                        \fL^{\m\sprime}{}_A\mem
                        \frac{D\nem\fL^{\n\sprime A}}{d\t}
            \\
            &
                + 
                p_{\m'}\mem
                    \Delt X^{\m\sprime}{}_{\s'}\mem
                    \dot{\fx}^{\s\sprime}
                +
                p_{\m'}\mem
                    \Delt Y^{\m\sprime}{}_{\s'}
                    \mem
                    \frac{D}{d\t}\mem
                    \bb{
                        \frac{
                            \fS^{\s'\nem\xi\sprime} p_{\xi'}
                        }{p^2}
                    }
            \\
            &
                - \frac{1}{2}\,
                    \fS_{\k'\nem\l'}\mem
                    \hdelta^{\k\sprime}{}_{\m'}\hem
                    \hdelta^{\l\sprime}{}_{\n'}
                    \bb{
                        \acX^{\m'\nem\n\sprime}{}_{\s'}\mem
                            \dot{\fx}^{\s\sprime}
                        +
                        \acY^{\m'\nem\n\sprime}{}_{\s'}\hem
                        \mem
                        \frac{D}{d\t}\mem
                        \bb{
                            \frac{
                                \fS^{\s'\nem\xi\sprime} p_{\xi'}
                            }{p^2}
                        }
                    }
            \\
            &
                - \frac{\k^0}{2m}\mem\BB{
                    g^{\m'\nem\n\sprime}(\fx)\,
                        p_{\m'}\hhem p_{\n\sprime}
                    + m^2
                }\mem d\t
                - \frac{\kappa^a}{2}\,
                    \BB{
                        \hat{p}^{\m\sprime} +\hem \fL^{\m\sprime}{}_0
                    }\mem
                        \fS_{\m'\nem\n'}
                    \fL^{\n\sprime}{}_a
        \end{aligned}
        \mem}\, d\t
    \,,
\end{align}
which is a functional of
$(\fx^{\m\sprime},\fL^{\m\sprime}{}_A,p_{\m'},\fS_{\m'\nem\n'})$ and $(\kappa^0, \kappa^a)$.
The all-orders formulae due to \rcite{gde}
state that
\begin{subequations}
\label{gdelow}
\begin{align}
    \Delt X^{\m\sprime}{}_{\s'}
    \,&=\,
        \frac{(-1)^2}{2}\,
            R^{\m\sprime}{}_{\n'\nem\r'\nem\s'}(\fx)\,
            \fb^{\n\sprime}
            \fb^{\r\sprime}
        + \frac{(-1)^3}{6}\,
            R^{\m\sprime}{}_{\n'\nem\r'\nem\s';\k'}(\fx)\,
            \fb^{\n\sprime}
            \fb^{\r\sprime}
            \fb^{\k\sprime}
        + \O(\fb^4)
    \,,\\[-0.05\baselineskip]
    \Delt Y^{\m\sprime}{}_{\s'}
    \,&=\,
        \frac{(-1)^2}{6}\,
            R^{\m\sprime}{}_{\n'\nem\r'\nem\s'}(\fx)\,
            \fb^{\n\sprime}
            \fb^{\r\sprime}
        + \frac{(-1)^3}{12}\,
            R^{\m\sprime}{}_{\n'\nem\r'\nem\s';\k'}(\fx)\,
            \fb^{\n\sprime}
            \fb^{\r\sprime}
            \fb^{\k\sprime}
        + \O(\fb^4)
    \,,\\[-0.05\baselineskip]
    \acX^{\m\sprime}{}_{\n'\nem\s'}
    \,&=\,
        \frac{(-1)^1}{1}
            R^{\m\sprime}{}_{\n'\nem\r'\nem\s'}(\fx)\,
            \fb^{\r\sprime}
        + \frac{(-1)^2}{2}\,
            R^{\m\sprime}{}_{\n'\nem\r'\nem\s';\k'}(\fx)\,
            \fb^{\r\sprime}
            \fb^{\k\sprime}
        + \O(\fb^3)
    \,,\\[-0.05\baselineskip]
    \acY^{\m\sprime}{}_{\n'\nem\s'}
    \,&=\,
        \frac{(-1)^1}{2}\,
            R^{\m\sprime}{}_{\n'\nem\r'\nem\s'}(\fx)\,
            \fb^{\r\sprime}
        + \frac{(-1)^2}{3}\,
            R^{\m\sprime}{}_{\n'\nem\r'\nem\s';\k'}(\fx)\,
            \fb^{\r\sprime}
            \fb^{\k\sprime}
        + \O(\fb^3)
    \,,
\end{align}
\end{subequations}
and there is no difficulty in
proceeding to higher orders
or even
writing down
a closed-form expression (finite sum) at each order in $\fb$.

\newpage

We should highlight the fact that
nonlinear-in-Riemann terms start to appear
from the first omitted spin orders in \eqref{gdelow}.
For instance,
the $\O(\fb^4)$ part of $\Delt X^{\m\sprime}{}_{\s'}$ is
\begin{subequations}
\begin{align}
    \frac{(-1)^4}{4!}\,
    \BB{
        R^{\m\sprime}{}_{\n'\nem\r'\nem\s';\k';\l'}(\fx)
        + R^{\m\sprime}{}_{\n'\nem\r'\nem\xi'}(\fx)\mem
        R^{\xi\sprime}{}_{\k'\nem\l'\nem\s'}(\fx)
    }\,
        \fb^{\n\sprime}
        \fb^{\r\sprime}
        \fb^{\k\sprime}
        \fb^{\l\sprime}
    \,.
\end{align}
The $\O(\fb^6)$ part of $\Delt X^{\m\sprime}{}_{\s'}$ is, schematically,
\begin{align}
    \frac{(-1)^6}{6!}\,
    \BB{
        {(D^4\nem R)}
        + 6\, {(D^2\nem R)}\mem R
        + 4\, {(D\hnem R)}\hem {(D\hnem R)}
        +     R\mem {(D^4\nem R)}
        + R\mem R\mem R
    }\,
        \fb\mem
        \fb\mem
        \fb\mem
        \fb\mem
        \fb\mem
        \fb
    \,.
\end{align}
\end{subequations}
Note that the power of Riemann tensors
is unbounded.

Crucially, 
the constraints do not change their forms
in \eqref{STGR.action.correct}
because
they are \textit{scalars} and
the parallel propagator is metric-preserving.
This demonstrates the importance of \rcite{Kim:2021rda}'s refinement
on the spin gauge generators
in the gravity-coupled theory.
The metric-preserving property also ensures that
the geodesic flow from $\fx$ to $x$
commutes with
merging the physical and gauge parts
in the above derivation.
Lastly,
it should be also clear that our methodology here
readily applies to nonminimal couplings.

\para{Fictitious Interaction Lagrangian}

Clearly,
working in noncovariant gauges
costs us infinitely many extra interaction vertices;
their implications for 
post-Newtonian expansions
will be commented in \Sec{APPL}. 
We may want to describe such extra terms
with the adjective ``fictitious,''
as their physical meaning is reminiscent of the
fictitious forces:
\vspace{-4pt}
\begin{align}
    \label{genGR.theta}
    \theta
    \,\,=\,\,
        p\mem dx
        \,+\, \tfrac{1}{2}\,
            S\mem
                \L\mem
                D\hhnem\L
        \,+\, \phi_a\mem d\bet^a
    \,\,=\,\,
        p\mem d\fx
        \,+\, \tfrac{1}{2}\,
            \fS\mem
                \fL\mem
                D\nem\fL
    \,+\,
    \overbrace{
        \fth_\text{mass}
        +
    \mathrlap{\vphantom{\Big|}}   
        \fth_\text{spin}
    }^{\adjustbox{raise=2pt}{\textfake{
        \scshape\footnotesize Fictitious
    }}}
    \,.
\end{align}

The perspectives toward gauge redundancies
may be twofold.
On the one hand,
gauge redundancies are a useful utility
that allows one to
describe the same physics
in different ``reference frames.''
A typical example is 
a free particle in special relativity
described in curvilinear coordinates:
changing coordinates
generates fictitious forces
in the equations of motion,
but the physical observables are the same.
This perspective
observes the beauty of gauge ``symmetries''
and their constraining power in writing down interaction Lagrangians.

On the other hand,
gauge redundancies are
an absurdity in our description,
sometimes serving as sheer sources of confusion.
A typical argument
appeals to the Stueckelberg trick
\cite{stuckelberg1938wechselwirkungskrafte},
which explicates that
gauge redundancies can be arbitrarily introduced or not.
This perspective emphasizes the fakeness of gauge redundancies
and their null physical content.

In our case,
the spin gauge ``symmetry''
has granted us the power of
changing the SSC freely
by the Hamiltonian flows of $\phi_a$.
Once the 
time evolution is
known in one SSC,
one readily determines 
the time evolution in other SSCs by spin gauge transformations.

At the same time,
the spurious description
on the fake worldline $\t \mapsto \fx^{\m\sprime}(\t)$
of one's choice
has generated
an infinite tower of fictitious terms
in the interaction Lagrangian,
which are needed to
keep the observables on the physical worldline the same.\footnote{
    In fact, the transition from \eqref{HRGR.theta}
    to \eqref{varths},
    summarized in \eqref{genGR.theta},
    may be thought of as a Stueckelberg trick
    that introduces $\fx^{\m\sprime}$
    as a reference point.
}
For instance,
suppose one arbitrarily picks a worldline 
$\t \mapsto \fx^{\m\sprime}(\t)$
that exhibits a certain wiggly behavior.
To generate the newly introduced
accelerations of this worldline
and
precessions of the accordingly parallel-transported local tensor variables,
the force law---EoM---must change.
This is precisely the meaning and role
of the fictitious terms
found in \eqref{STGR.action.correct}:
the fictitious forces
due to changing the SSC (as a ``frame'') from covariant to noncovariant.

\subsection{Preservation of Spin Gauge Flow and Algebra}
\label{HR>PRES}

Lastly,
we provide some vital consistency checks
on our constructions.

\para{Spin Gauge Flow with Electromagnetic Couplings}

We begin with the electromagnetic coupling.
To this end, we apply our geometric framework in \Sec{INT1} to the spherical top model.
Let $\omega^\circ$
be the free-theory symplectic form
given to $\ps_{20}^\ST$,
so $d\omega^\circ = 0$.
Suppose 
a symplectic perturbation
$\omega'$ on $\ps_{20}^\ST$
such that
$d(\omega^\circ + \omega') = 0$.

The spin gauge transformations
are generated by
the Hamiltonian vector fields
\begin{align}
    \GG_a^\circ
    \,:=\,
        \pie{\blank}{d\phi_a}
    \,,\quad
    \GG_a
    \,:=\,
        \pia{\blank}{d\phi_a}
\end{align}
in the free and interacting theories,
respectively.
Are these vector fields the same?

We identify a sufficient condition for
a spin gauge-invariant symplectic perturbation:
\begin{align}
    \label{spt.suff}
    \i_{\GG_a^\circ}\mem \omega' 
    \mem=\mem 0
    \qiq
    \pounds_{\GG^\circ_a}\mem \omega'
    \,=\,
        d\mem \i_{\GG^\circ_a}\mem \omega' 
        +
        \i_{\GG^\circ_a}\mem d\omega'
    \,=\,
        0
    \,.
\end{align}
Provided \eqref{spt.suff},
we recall \eqref{ppert}
to find
\begin{align}
    \label{spt.pres}
    \GG_a
    \,\,=\,\,
        \GG_a^\circ
        \,-\,
        \pia{
            \i_{\GG_a^\circ}\mem \omega'
        }{\blank}
    \,\,=\,\,
        \GG_a^\circ
    \,.
\end{align}
Therefore, the spin gauge flow
is completely ignorant of the symplectic perturbation $\omega'$.
In the Poisson bracket notation, \eqref{spt.pres} 
boils down to
\begin{align}
    \label{spt.pres.pb}
    \pb{f}{\phi_a}
    \,=\,
        \pb{f}{\phi_a}^\circ
    \,,
\end{align}
for any function $f$ on $\ps_{20}^\ST$.
Thus the gauge orbit 
in the interacting theory
is identical
to the $\R^3$ found in \eqref{orbit},
and the abelian gauge algebra is left intact:
\begin{align}
    \label{spt.abelian}
    \pb{\phi_a}{\phi_b}
    \,=\,
        \pb{\phi_a}{\phi_b}^\circ
    \,=\,
        0
    \,.
\end{align}

All of the electromagnetic couplings considered in this paper satisfy the sufficient condition in \eqref{spt.suff}.
For instance, the minimal coupling is given by
$
\omega' =
\frac{1}{2}\, F_{\m\n}(x)\,
    dx^\m \swedge dx^\n
$.
Since $\pb{x^\n}{\phi_a}^\circ = 0$,
this implies
\begin{align}
    \pb{f}{\phi_a}
    \,\,=\,\,
        \pb{f}{\phi_a}^\circ
        -
        \pb{f}{x^\m}\,
        F_{\m\n}(x)\,
        \pb{x^\n}{\phi_a}^\circ
    \,\,=\,\,
        \pb{f}{\phi_a}^\circ
    \,.
\end{align}

\para{Spin Gauge Flow with Gravitational Couplings}

Next, we consider the gravitational coupling.
To this end, we apply our geometric framework in \Sec{INT2} to the spherical top model.
By the postulate in \Sec{INT>CUPB},
the general-relativistic phase space $\P_{20}^\ST$
is equipped by
a nondegenerate two-form
$\omega^\bullet$.
As will be shown shortly in \Sec{COVPS>ST},
the explicit construction of $\omega^\bullet$
shows that
$d\omega^\bullet = -d(p_\m\hem {*\hnem}R^\m{}_\n\mem \hy^\n)$,
so the claim in \Sec{INT>MPD} is verified.
Suppose a covariant symplectic perturbation
$\omega'$ on $\P_{20}^\ST$
such that
$d(\omega^\bullet + \omega') = 0$.

\newpage

Let us define
\begin{align}
    \label{covspt.GG}
    \GG_a^\bullet
    \,:=\,
        \pib{\blank}{d\phi_a}
    \,,\quad
    \GG_a
    \,:=\,
        \pia{\blank}{d\phi_a}
\end{align}
in the free and interacting theories,
respectively.
Our postulates imply that
$\lim_{R\to0} \GG_a = \GG^\bullet_a$.
Are these vector fields the same?

We identify a sufficient condition for
a spin gauge-invariant covariant symplectic perturbation:
\begin{align}
    \label{covspt.suff}
    \i_{\GG_a^\bullet}\mem \omega' 
    \mem=\mem 0
    \qiq
    \pounds_{\GG^\bullet_a}\mem \omega'
    \,=\,
        d\mem \i_{\GG^\bullet_a}\mem \omega' 
        +
        \i_{\GG^\bullet_a}\mem d\omega'
    \,=\,
        - \i_{\GG^\bullet_a}\mem d\omega^\bullet
    \,=\,
        0
    \,.
\end{align}
In \eqref{covspt.suff},
the last equality follows from 
the condition
$d\omega^\bullet = -d(p_\m\hem {*\hnem}R^\m{}_\n\mem \hy^\n)$
and
\begin{align}
    \label{xyp.invariance}
    \pib{
        dx^\m
    }{d\phi_a}
    \,=\,0
    \,,\quad
    \pib{
        D\hy^\m
    }{d\phi_a}
    \,=\,0
    \,,\quad
    \pib{
        Dp_\m
    }{d\phi_a}
    \,=\,0
    \,.
\end{align}
\eqref{xyp.invariance}
arises by covariantizing the Poisson brackets in the free theory:
see \eqref{HRGR.pib},
for instance.
Provided \eqref{covspt.suff},
we recall \eqref{ppertc}
to find
\begin{align}
    \label{covspt.pres}
    \GG_a
    \,\,=\,\,
        \GG_a^\bullet
        \,-\,
        \pia{
            \i_{\GG_a^\bullet}\mem \omega'
        }{\blank}
    \,\,=\,\,
        \GG_a^\bullet
    \,.
\end{align}
Therefore, the spin gauge flow
is completely ignorant of the covariant symplectic perturbation $\omega'$.
To clarify,
this means that tidal effects can be ignored
when computing Poisson brackets with 
the gauge generators $\phi_a$.
Clearly, the abelian gauge algebra is left intact:
\begin{align}
    \label{covspt.abelian}
    \pb{\phi_a}{\phi_b}
    \,=\,
        \pb{\phi_a}{\phi_b}^\bullet
    \,=\,
        0
    \,.
\end{align}
Again, the last equality in \eqref{covspt.abelian}
follows by covariantization; see \Sec{COVPS>ST}.

Crucially, the definitions and identities used in the above derivation
count heavily on the fact that
$d\phi_a = D\phi_a$.
It cannot be overemphasized that
it is significantly important to
formulate the spin gauge generators
as spacetime scalars.

All of the gravitational couplings considered in this paper satisfy the sufficient condition in \eqref{covspt.suff}.
For instance, the minimal coupling descirbes
$
\omega' =
    p_\m\hem {*\hnem}R^\m{}_\n\mem \hy^\n
$.
Via \eqref{xyp.invariance},
this implies
\begin{align}
\begin{split}
    \pb{f}{\phi_a}
    \,\,&=\,\,
        \pib{df}{d\phi_a}
        -
        \pib{df}{dx^\r}\,
        \BB{
            p_\m\hem {*\hnem}R^\m{}_{\n\r\s}(x)\mem \hy^\n
        }\,\hem
        \pib{dx^\s}{d\phi_a}
    \,,\\
    \,\,&=\,\,
        \pib{df}{d\phi_a}
    \,.
\end{split}
\end{align}
In contrast,
the Poisson algebra of
spin gauge-invariant observables
is clearly sensitive to the covariant symplectic perturbation.

Finally, the $\R^3$ gauge orbit can be explicitly identified.
The action of
$\exp\bigbig{
    \pb{\blank}{\phi_c}\mem \sigma^c
}$ on $\P^\MT_{20}$
is found as 
\begin{align}
    \label{orbit.GR}
    \left\{\mem
    {\renewcommand{\arraystretch}{1.45}
    \begin{array}{rl}
        \fx
        &\,\,\,\mapsto\,\,\,
            \Exp{
                \fb(\vbet \mplus\vs)
            }{
                \Exp{-\fb(\vbet)}{\fx}
            }
        \\
        p_{\m\sprime}
        &\,\,\,\mapsto\,\,\,
            p_{\m'}\hem
            W^{\m\sprime}{}_{\m}\hem
            W^\m{}_{\m\sprime\mem\sprime}
        \\
        \fL^{\m\sprime}{}_A
        &\,\,\,\mapsto\,\,\,
            W^{\m\sprime\mem\sprime}{}_\m\hem
            W^\m{}_{\m'}
            \fL^{\m\sprime}{}_B\mem
                \bigbig{\fR^{-1}\hnem(\vbet)\,
                \fR(\vbet\mplus\vs)
                }\hnem{}^B{}_A
        \\
        \fS_{\m\sprime\mem\n\sprime}
        &\,\,\,\mapsto\,\,\,
\begin{aligned}[t]
        &
            \fS_{\m\sprime\mem\n\sprime}
            \,
            W^{\m\sprime}{}_{\m}\hem
            W^{\n\sprime}{}_{\n}\hem
            W^\m{}_{\m\sprime\mem\sprime}\,
            W^\n{}_{\n\sprime\mem\sprime}
        \\
        &
        \vphantom{\frac{0}{0}}
            + 
            \bigbig{
                \fb(\vbet) \swedge p
            }{}_{\m\sprime\mem\n'}
            W^{\m\sprime}{}_{\m}\hem
            W^{\n\sprime}{}_{\n}\hem
            W^\m{}_{\m\sprime\mem\sprime}\,
            W^\n{}_{\n\sprime\mem\sprime}
            -
            \bigbig{
                \fb(\vbet\mplus\vs) \swedge p
            }{}_{\m\sprime\mem\sprime}{}_{\n\sprime\mem\sprime}
\end{aligned}
    \end{array}}
    \,\mem
    \right\}
    \,,
\end{align}
where the unprimed, single-primed, and double-primed
indices are associated with
the gauge representatives at
$0$,
$\vbet$,
and
$\vbet\mplus\vs$,
respectively.
In particular,
$W^{\m\sprime}{}_\m$
describes the geodesic contour
from the physical center to the original fake center,
while 
$W^{\m\sprime\mem\sprime}{}_\m$
describes the geodesic contour
from the physical center to the transformed fake center.
Note that 
$
    W^{\m\sprime\mem\sprime}{}_\m\hem
    W^\m{}_{\m'}
    \neq
    W^{\m\sprime\mem\sprime}{}_{\m'}
$
in the presence of spacetime curvature.
The definitions in \eqref{bcollect}
are kept at each point,
while $\mathrm{Exp}$
denotes the exponential map:
geodesic deviation by
$v {\:\in\:} T_x\M$
from $x {\:\in\:} \M$
yields the point $\mathrm{Exp}(v,x) {\:\in\:} \M$.
To put it simply,
\eqref{orbit.GR} reincarnates \eqref{orbit}
in the Riemann normal coordinate system
based at the physical center.
\eqref{orbit.GR} implements
the spin gauge transformations
by always referencing
the physical center $x$
as the point of a special status,
consistently with the fact that the gauge algebra is abelian.

\subsection{Bottom-Up Approach to Spin Gauge-Invariant Coupling}
\label{HR>BOOTS}

In the above analysis,
we have provided a top-down construction
of the spin gauge-invariant gravitational coupling.
This top-down approach
first insists on the existence of a physical center
(\Sec{HR>CENTER})
and \textit{by construction} makes it physical afterwards
(\Sec{HR>PRES}).
It identifies a working proposal
for the spin gauge-invariant coupling,
from which
the action in the original field basis is completely determined as well
(\Sec{HR>ACTION}).

The reader could question
whether every aspect of this procedure was inevitable.
For instance,
one might ask if
the formulae in \eqrefs{fcen.curved}{Wformula}
can be generalized to
more complicated forms
that possibly involve the local tensor degrees of freedom,
although
there exist arguments why the geodesic deviation in \eqref{fcen.curved}
and the parallel propagation with respect to the Levi-Civita connection in \eqref{Wformula}
are natural options in Einstein gravity where metric dictates the entire geometry.

To present a complementary perspective,
this subsection provides a bottom-up construction of the spin gauge-invariant gravitational coupling.
We demonstrate that the spin gauge-invariant action,
in the original field basis,
can be perturbatively constructed
by imposing the preservation of the gauge algebra and flow
as bootstrap conditions.
We sketch this analysis
at a few low orders in $\fb$,
providing a brief justification for our top-down results.

\para{Electromagnetism}

The top-down derivation of 
the spin gauge-invariant
electromagnetic interaction symplectic potential
has been given in \eqref{STEM.action}:
$qA_\m(x)\mem dx^\m$ with $x^\m = \fx^\m - \fb^\m$.
It can be shown \cite{gde} that
$qA_\m(x)\mem dx^\m$ is equivalent to
$qA_\m(\fx)\mem d\fx^\m + \fth_\text{charge}$
modulo total derivative,
where the fictitious symplectic potential is given by
\begin{align}
\begin{split}
\label{fth.EM}
    \fth_\text{charge}
    \,=\,
        q\, \sum_{\ell=1}^\infty\mem
            \frac{(-1)^\ell}{\ell!}
            \lrp{
            \begin{aligned}[c]
            &
                F_{\r_1\m,\r_2\cdots\r_\ell}(\fx)\mem 
                \fb^{\r_1} \fb^{\r_2} {\cdots} \fb^{\r_\ell}\,
                d\fx^\m
            \\
            &
                + (\ell{\,-\,}1)\,
                F_{\r_1\m,\r_2\cdots\r_{\ell-1}}(\fx)\mem 
                \fb^{\r_1} \fb^{\r_2} {\cdots} \fb^{\r_{\ell-1}}\,
                d\fb^\m
            \end{aligned}
            }
\end{split}
\end{align}
so that $\lim_{\fx \to x} \fth_\text{charge} = 0$.
\eqref{fth.EM} can be viewed as providing the corrections to the spurious minimal coupling $qA_\m(\fx)\mem d\fx^\m$
that restore spin gauge invariance
at every order in $\fb$.
The interaction symplectic form is given by
\begin{align}
    \label{fom.EM}
    \frac{1}{2}\, qF_{\m\n}(\fx)\mem d\fx^\m \swedge d\fx^\n
    \mem+\mem
        \fom_\text{charge}
    \,=\,
        \frac{1}{2}\, qF_{\m\n}(\fx\mminus\fb)\mem d(\fx\mminus\fb)^\m \swedge d(\fx\mminus\fb)^\n
    \,,
\end{align}
where the right-hand side is readily Taylor-expanded in the orders of $\fb$.

\newpage

Our message here is that the fictitious symplectic form in \eqref{fom.EM}
can be discovered \textit{from scratch} in the following fashion.
The point of departure is the free spherical top described in \Sec{FOUR>ST}.
Its symplectic form is
\begin{align}
    \label{ST.omega0}
    \omega^\circ
    \,&=\, 
        dp_\m \swedge d\fx^\m
        +
        \frac{1}{2}\,
            d\fS_{\m\n} \swedge
                (\fL^\m{}_A\mem
                d\fL^{\n A})
        +
        \frac{1}{2}\,
            \fS_{\m\n}\mem
                d\fL^\m{}_A \swedge
                d\fL^{\n A}
    \,,
\end{align}
whose pointwise inverse
derives the Poisson brackets in \eqref{ST.pb}.
It follows that
\begin{align}
\label{wfl}
    &
    \pb{\fx^\m}{\phi_a}^\circ
    \,=\,
        w^\m{}_a
    \,=\,
    \pb{\fb^\m}{\phi_a}^\circ 
    \,,\quad
    \pb{\fS^{\m\n}}{\phi_a}^\circ = -\bigbig{w_a \wedge p}{}^{\m\n}
    \,,
\end{align}
where we have introduced a notation
\begin{align}
    \label{wa}
    w^\m{}_a
    \,:=\,
        \frac{1}{2|p|}\,
            \hat{\eta}^{\m\n}
            \fS_{\n\r} \fL^\r{}_a 
    \,.
\end{align}

A naive attempt to endow a charge
supposes the spurious minimal coupling
$\omega = 
    \omega^\circ
    + \frac{1}{2}\,
        q\hem F_{\m\n}(\fx)\,
            d\fx^\m \swedge d\fx^\n
$.
By utilizing a geometric series expansion,
one finds that
the resulting Poisson bracket describes
\begin{align}
\begin{split}
    \label{xphi.EM0}
    \pb{\phi_a}{\phi_b}
    \,&=\,
    \begin{aligned}[t]
        &
        \pb{\phi_a}{\phi_b}^\circ
        -
            \pb{\phi_a}{\fx^\m}^\circ
            \mem qF_{\m\n}(\fx)\,
            \pb{\fx^\n}{\phi_b}^\circ
        \\
        &
        +
            \pb{\phi_a}{\fx^\m}^\circ
            \mem qF_{\m\n}(\fx)\,
            \pb{\fx^\n}{\fx^\r}^\circ
            \mem qF_{\r\s}(\fx)\,
            \pb{\fx^\s}{\phi_b}^\circ
        -
            \cdots
        \,,
    \end{aligned}
    \\
    \,&=\,
        qF_{\m\n}(\fx)\, w^\m{}_a\hem w^\n{}_b
    \,,
\end{split}
\end{align}
which is nonvanishing.
Apparently, the gauge algebra is deformed
and is not closed.

To fix this problem, the most general modification of the symplectic form is
\begin{align}
    \label{EM.ansatz}
    \omega 
    \,=\,
    \omega^\circ
    \mem+\mem
    \frac{1}{2}\,
    qF_{\m\n}(\fx)\,
        d\fx^\m \swedge d\fx^\n
    \mem+\mem
    \fom
    \transition{where}
    d\fom \,=\, 0
    \transit{and}
    \lim_{\fb\to0} \fom \,=\, 0
    \,.
\end{align}
Due to the condition 
$\lim_{\fb\to0} \fom = 0$,
the viable ansatz seems to be
\begin{align}
\begin{split}
    \fom
    \mem\,=\,\mem
    \sum_{\ell=0}^\infty\,
    \frac{(-1)^\ell}{\ell!}\mem
    \bb{
        \frac{1}{2}\,
            f_{\m\n}^{(\ell,0)}\,
                d\fx^\m \swedge d\fx^\n
        +
            f_{\m\n}^{(\ell,1)}\,
                d\fb^\m \swedge d\fx^\n
        +
        \frac{1}{2}\,
            f_{\m\n}^{(\ell,2)}\,
                d\fb^\m \swedge d\fb^\n
    }
    \,,
\end{split}
\end{align}
where $f_{\m\n}^{(\ell,j)}$ 
is of $\O(\fb^{\ell-j})$ for $\ell {\:\geq\:} j$
and $f_{\m\n}^{(\ell,j)} = 0$ for $\ell {\:<\:} j$.
It suffices to set $f_{\m\n}^{(0,0)} = 0$.

With this ansatz,
we examine the implication of the condition
\begin{align}
    \label{boots.EM}
    0
    \,=\,
        \pb{\phi_a}{\phi_b}
    \,.
\end{align}
At the leading order in $\fb$,
using \eqref{wfl} gives
\begin{align}
    \label{boots.EMab.1}
    0
    \,&=\,
        \BB{
            qF_{\m\n}(\fx)
            - 2\mem f^{(1,1)}_{[\m\n]}
            + f^{(2,2)}_{\m\n}
        }\mem
        w^\m{}_a\hem w^\n{}_b
    \,,
\end{align}
so $f^{(1,1)}_{\m\n} = f^{(2,2)}_{\m\n} = qF_{\m\n}(\fx)$
qualifies as a solution.
Proceeding to higher orders, it can be checked that
the following qualifies as a solution:
\begin{align}
\begin{split}
    \label{fom.EM.series}
    \fom
    \mem\,=\,\mem 
    {}&{}
        - \frac{1}{2}\,
            qF_{\m\n,\r}(\fx)\,
            \fb^\r\,
                d\fx^\m \swedge d\fx^\n
        -
        qF_{\r\n}(\fx)\,
            d\fb^\r \swedge d\fx^\n
    \\
    {}&{}
        +
        q\,
        \sum_{\ell=2}^\infty\,
            \frac{(-1)^\ell}{\ell!}\,
            \fb^{\r_3} {\cdots} \fb^{\r_\ell}
            \mem\lrp{
            \begin{aligned}[c]
            &
                \tfrac{1}{2}\,
                F_{\m\n,\r_1\cdots\r_\ell}\hnem(\fx)\,
                \fb^{\r_1} \fb^{\r_2}
                \,
                    d\fx^\m \swedge d\fx^\n
            \\&
                +
                F_{\r_1\m,\r_2\cdots\r_\ell}\nem(\fx)\,
                \fb^{\r_1}
                    d\fb^{\r_2} \swedge d\fx^\n
            \\&
                +
                \tfrac{1}{2}\,
                F_{\r_1\r_2,\r_3\cdots\r_\ell}\nem(\fx)\,
                    d\fb^{\r_1} \swedge d\fb^{\r_2}
            \end{aligned}
            }
    \,.
\end{split}
\end{align}
Certainly, \eqref{fom.EM.series}
reproduces 
the interaction symplectic form described in \eqref{fom.EM}.

\newpage

\para{Gravity}

For gravity,
the point of departure
of the bottom-up construction
is the covariantization of \eqref{ST.omega0}:
\begin{align}
    \label{ST.omegafb}
    \omega^\fbullet
    \,&=\, 
        Dp_{\m'} \swedge d\fx^{\m\sprime}
        +
        \frac{1}{2}\,
            D\fS_{\m\sprime\mem\n'} \swedge
                (\fL^{\m\sprime}{}_A\mem
                D\nem\fL^{\n\sprime A})
        +
        \frac{1}{2}\,
            \fS_{\m\sprime\mem\n'}\mem
                D\nem\fL^{\m\sprime}{}_A \swedge
                D\nem\fL^{\n\sprime A}
    \,.
\end{align}
The resulting covariant Poisson brackets
simply covariantize those in \eqref{ST.pb}.
That is,
the nonvanishing components of
the pointwise inverse $\Pi^\fbullet$
of \eqref{ST.omegafb}
in the generally covariant (yet not spin gauge-invariant) basis
$(d\fx^{\m\sprime},Dp_{\m'},D\nem\fL^{\m\sprime}{}_A,D\fS_{\m\sprime\mem\n'}\hnem)$
of one-forms
are
\begin{align}
\begin{split}
    \label{ST.pb.fcov}
    \pifb{d\fx^{\m\sprime}}{Dp_{\n'}}
    \,\,&=\,\,
        \delta^{\m\sprime}{}_{\n'}
    \,,\\
    \pifb{D\nem\fL^{\m\sprime}{}_A}{D\fS_{\r\sprime\mem\s'}}
    \,\,&=\,\,
        \BB{
            -2\mem \delta^{\m\sprime}{}_{[\r'} \eta_{\s']\n'}
        }\mem
        \fL^{\n\sprime}{}_A
    \,,\\
    \pifb{D\fS_{\m\sprime\mem\n'}}{D\fS_{\r\sprime\mem\s'}}
    \,\,&=\,\,
        \BB{
            -4\mem \delta^{[\k\sprime}{}_{[\m'}\mem
            \eta_{\n'][\r'}\mem
            \delta^{\l']}{}_{\s']}
        }\mem \fS_{\k\sprime\mem\l'}
    \,.
\end{split}
\end{align}
It follows that
\begin{align}
\label{wfl.fcov}
    &
    \pifb{d\fx^{\m\sprime}}{d\phi_a}
    \,=\,
        w^{\m\sprime}{}_a
    \,=\,
    \pifb{D\fb^{\m\sprime}}{d\phi_a}
    \,,\quad
    \pifb{D\fS^{\m\sprime\mem\n\sprime}}{d\phi_a}
    \,=\,
        - \bigbig{w_a \wedge p}{}^{\m\sprime\mem\n\sprime}
    \,,
\end{align}
where
\begin{align}
    \label{fwa}
    w^{\m\sprime}{}_a
    \,:=\,
        \frac{1}{2|p|}\,
            \hat{\eta}^{\m\sprime\mem\n'}
            \fS_{\n\sprime\mem\r'} \fL^{\r\sprime}{}_a
    \,.
\end{align}

It is easily seen that \eqref{ST.omegafb} is not symplectic:
$d\omega^\fbullet \neq 0$.
The minimal correction is
$\omega = 
    \omega^\fbullet
    - \frac{1}{4}\,
        \fS_{\k\sprime\mem\l'}
        R^{\k\sprime\mem\l\sprime}{}_{\m\sprime\mem\n'\hhnem}(\fx)\,
            d\fx^{\m\sprime} \swedge d\fx^{\n\sprime}
$.
By utilizing a geometric series expansion,
one finds that
the resulting Poisson bracket describes
\begin{align}
\begin{split}
    \label{xphi.EM1}
    &
    \pifb{d\phi_a}{d\phi_b}
    \\
    &
    \,=\,
        \pifb{d\phi_a}{d\phi_b}
        -
            \pifb{d\phi_a}{d\fx^{\m\sprime}}
            \mem \BB{
                -\tfrac{1}{2}\,
                \fS_{\k\sprime\mem\l'}
                R^{\k\sprime\mem\l\sprime}{}_{\m\sprime\mem\n'\hhnem}(\fx)
            }\,
            \pifb{d\fx^{\n\sprime}}{d\phi_b}
        +
            \cdots
        \,,\\
    &
    \,=\,
        \BB{
            -\tfrac{1}{2}\,
            \fS_{\k\sprime\mem\l'}
            R^{\k\sprime\mem\l\sprime}{}_{\m\sprime\mem\n'\hhnem}(\fx)
        }\, w^{\m\sprime}{}_a\hem w^{\n\sprime}{}_b
    \,,
\end{split}
\end{align}
which is nonvanishing.
Apparently, the gauge algebra is deformed
and is not closed.

To fix this problem, the most general ansatz is
\begin{subequations}
\begin{align}
    \label{GR.ansatz}
    &
    \omega 
    \mem\,=\,\mem
    \omega^\fbullet
    \mem-\mem
    \tfrac{1}{4}\,
        \fS_{\k\sprime\mem\l'}
        R^{\k\sprime\mem\l\sprime}{}_{\m\sprime\mem\n'\hhnem}(\fx)\,
            d\fx^{\m\sprime} \swedge d\fx^{\n\sprime}
    \mem+\mem
    \fom
    \,,
\end{align}
where
\begin{align}
    d\fom \,=\, 0
    \transition{and}
    \lim_{\fb\to0} \fom \,=\, 0
    \,.
\end{align}
\end{subequations}
To automate the closure $d\fom = 0$,
we can take $\fom = d\fth$.
Let $\Pi$ be the pointwise inverse of $\omega$ in \eqref{GR.ansatz}.
We stipulate the abelian spin gauge algebra
as a bootstrap condition:
\begin{align}
    \label{boots.GR}
    0
    \,=\,
        \pb{\phi_a}{\phi_b}
    \,=\,
        \pia{d\phi_a}{d\phi_b}
    \,.
\end{align}

The astute reader will point out that
there exists a rough isomorphism between the current problem
and the former problem,
by recalling \eqref{cov-dc}.
Namely, the ``covariant double copy'' correspondence
identifying spin as a gravitational charge
(local Lorentz charge)
arises between \eqrefs{EM.ansatz}{GR.ansatz},
yet in terms of spurious variables:
\begin{align}
    \label{cov-dc.f}
    \tfrac{1}{2}\,
    qF_{\m\n}(x)\, d\fx^\m \swedge d\fx^\n
    \quad\xleftrightarrow[]{\,\,\,\,\,\,}\quad
    -\tfrac{1}{4}\,
        \fS_{\k\sprime\mem\l'}
        R^{\k\sprime\mem\l\sprime}{}_{\m\sprime\mem\n'\hhnem}(\fx)\,
            d\fx^{\m\sprime} \swedge d\fx^{\n\sprime}
    \,.
\end{align}


Therefore,
an educated guess for the fictitious symplectic potential is a sum of
\begin{align}
\begin{split}
    \label{fth.GR.guess}
    \fth_\text{(s1)}
    \mem\,&=\,\mem 
        \tfrac{1}{2}\,
        \fS_{\k\sprime\mem\l'}
        R^{\k\sprime\mem\l\sprime}{}_{\r\sprime\mem\n'\hnem}(\fx)\,
            \fb^{\r\sprime}\mem d\fx^{\n\sprime}
    \,,\\
    \fth_\text{(s2)}
    \mem\,&=\,\mem 
        -
        \tfrac{1}{4}\,
        \fS_{\k\sprime\mem\l'}
        R^{\k\sprime\mem\l\sprime}{}_{\r\sprime\mem\n'\hnem}(\fx)\,
            \fb^{\r\sprime}\mem d\fb^{\n\sprime} 
        + \cdots
    \,,
\end{split}
\end{align}
and so on.
This directly applies the replacement rule in \eqref{cov-dc.f}
to the electromagnetic fictitious symplectic potential in \eqref{fth.EM}:
\begin{align}
\begin{split}
\label{fth.EM.few}
    \fth_\text{charge}
    \,=\,
        - qF_{\r\n}(\fx)\mem 
            \fb^\r\,
            d\fx^\n
        + \tfrac{1}{2}\,
        \BB{
                qF_{\r\n}(\fx)\mem 
                \fb^\r\,
                    d\fb^\n
            +
                \cdots
        }
        + \cdots
    \,.
\end{split}
\end{align}
For simplicity of our demonstration,
we ignore terms involving derivatives of external fields.

By construction, the exterior derivative of \eqref{fth.GR.guess}
is given by
the sum of
(a)
a group of terms isomorphic to the electromagnetic fictitious symplectic form in \eqref{fom.EM.series}
via \eqref{cov-dc.f}
and
(b)
a group of new contributions
due to the non-constancy of the gravitational charge $\fS_{\k\sprime\mem\l'}$:
\begin{align}
    \label{fom.GR.guess}
    \fom_\text{(s1)}
    \mem\,&=\,\mem 
        \tfrac{1}{2}\,
        D\fS_{\k\sprime\mem\l'} \wedge
        R^{\k\sprime\mem\l\sprime}{}_{\r\sprime\mem\n'\hnem}(\fx)\,
            \fb^{\r\sprime}\mem d\fx^{\n\sprime}
        + \cdots
    \\
    \fom_\text{(s2)}
    \mem\,&=\,\mem 
        -
        \tfrac{1}{4}\,
        D\fS_{\k\sprime\mem\l'} \wedge
        R^{\k\sprime\mem\l\sprime}{}_{\r\sprime\mem\n'\hnem}(\fx)\,
            \fb^{\r\sprime}\mem D\fb^{\n\sprime} 
        + \cdots
    \,.
    \nonumber
\end{align}
These new terms are essentially corrections due to symplecticity
and the nonabelian nature of the Lorentz group.

The perturbation on the Poisson bracket $\pb{\phi_a}{\phi_b}$
due to $\fom_\text{(s1)}$ is
\begin{subequations}
\begin{align}
\begin{split}
    \label{calc.ws1}
    &
    {- \tfrac{1}{2}}\,
        R^{\k\sprime\mem\l\sprime}{}_{\r\sprime\mem\n'\hnem}(\fx)\,
            \fb^{\r\sprime}
    \,\mem
    \pifb{d\phi_a}{D\fS_{\k\sprime\mem\l'}}
    \,
    \pifb{d\fx^{\n\sprime}}{d\phi_b}
    - (a{\,\leftrightarrow\,}b)
    \\
    &\,=\,
        -\tfrac{1}{2}\,
            R^{\k\sprime\mem\l\sprime}{}_{\m\sprime\mem\n'\hnem}(\fx)
            \,
            \fb^{\m\sprime}\mem
            \bigbig{
                w_a \wedge p
            }{}_{\k\sprime\mem\l'}
            \,
            w^{\n\sprime}_b 
        - (a{\,\leftrightarrow\,}b)
    \\
    &
    \,=\,
        p_{\k'}
        R^{\k\sprime}{}_{\l\sprime\mem\m\sprime\mem\n'\hnem}(\fx)
        \,
        \fb^{\l\sprime}\mem
            w^{\m\sprime}{}_a  w^{\n\sprime}{}_b
    \,,
\end{split}
\end{align}
where the last equality uses the algebraic Bianchi identity.
Similarly,
the perturbation on the Poisson bracket $\pb{\phi_a}{\phi_b}$
due to $\fom_\text{(s2)}$ 
is 
\begin{align}
\begin{split}
    \label{calc.ws2}
    &
    \tfrac{1}{4}\,
        R^{\k\sprime\mem\l\sprime}{}_{\r\sprime\mem\n'\hnem}(\fx)\,
            \fb^{\r\sprime}
    \,\mem
    \pifb{d\phi_a}{D\fS_{\k\sprime\mem\l'}}
    \,
    \pifb{D\fb^{\n\sprime}}{d\phi_b}
    - (a{\,\leftrightarrow\,}b)
    \\
    &\,=\,
        \tfrac{1}{4}\,
            R^{\k\sprime\mem\l\sprime}{}_{\m\sprime\mem\n'\hnem}(\fx)
            \,
            \fb^{\m\sprime}\mem
            \bigbig{
                w_a \wedge p
            }{}_{\k\sprime\mem\l'}
            \mem
            w^{\n\sprime}_b 
        - (a{\,\leftrightarrow\,}b)
    \\
    &
    \,=\,
        - \tfrac{1}{2}\,
        p_{\k'}
        R^{\k\sprime}{}_{\l\sprime\mem\m\sprime\mem\n'\hnem}(\fx)
        \,
        \fb^{\l\sprime}\mem
            w^{\m\sprime}{}_a  w^{\n\sprime}{}_b
    \,,
\end{split}
\end{align}
\end{subequations}
where the last equality also arises via the algebraic Bianchi identity.
As a result, the calculations in \eqrefs{calc.ws1}{calc.ws2}
are identical except the coefficients.
Their sum is
\begin{align}
    \label{1a-2a-red}
        + \tfrac{1}{2}\,
        p_{\k'}
        R^{\k\sprime}{}_{\l\sprime\mem\m\sprime\mem\n'\hnem}(\fx)
        \,
        \fb^{\l\sprime}\mem
            w^{\m\sprime}{}_a  w^{\n\sprime}{}_b
    \,.
\end{align}

To cancel out \eqref{1a-2a-red},
we may consider a class of terms that do not directly reference the spurious spin.
A generic ansatz reads
\begin{align}
\begin{split}
    \fth_\text{(m2)}
    \,&=\,
        \mathtt{k}_\text{(m2)}\mem
        \BB{
            p_{\k'}
            R^{\k\sprime}{}_{\l\sprime\mem\m\sprime\mem\n'\hnem}(\fx)
            \,
            \fb^{\l\sprime}\mem
            \fb^{\m\sprime}\mem d\fx^{\n\sprime}
        }
    \,,\\
    \fth_\text{(m3)}
    \,&=\,
        \mathtt{k}_\text{(m3)}\mem
        \BB{
            p_{\k'}
            R^{\k\sprime}{}_{\l\sprime\mem\m\sprime\mem\n'\hnem}(\fx)
            \,
            \fb^{\l\sprime}\mem
            \fb^{\m\sprime}\mem D\fb^{\n\sprime}
        }
        +
        \cdots 
    \,.
\end{split}
\end{align}
These extra fictitious terms
probe the effects of the other gravitational charge,
i.e., mass
(diffeomorphism charge).
Again, we wish to ignore higher-order terms
involving derivatives of the Riemann tensor.


The perturbation on the Poisson bracket $\pb{\phi_a}{\phi_b}$
due to $\fom_\text{(m2)} = d\fth_\text{(m2)}$ is
\begin{subequations}
\begin{align}
\begin{split}
    \label{calc.wm2}
    &
    - \mathtt{k}_\text{(m2)}\,
    \BB{
        2\mem
        p_{\k'}
        R^{\k\sprime}{}_{(\l\sprime\mem\m')\n'\hnem}(\fx)
        \mem\fb^{\l\sprime}
    }
    \,\mem
    \pifb{d\phi_a}{D\fb_{\m\sprime}}
    \,
    \pifb{d\fx^{\n\sprime}}{d\phi_b}
    - (a{\,\leftrightarrow\,}b)
    \\
    &\,=\,
        \mathtt{k}_\text{(m2)}\,
        p_{\k'}
        \BB{
            R^{\k\sprime}{}_{\l\sprime\mem\m\sprime\mem\n'\hnem}(\fx)
            +
            R^{\k\sprime}{}_{\m\sprime\mem\l\sprime\mem\n'\hnem}(\fx)
        }
        \mem\fb^{\l'}
        w^{\m\sprime}{}_a\hem
        w^{\n\sprime}{}_b
    - (a{\,\leftrightarrow\,}b)
    \\
    &\,=\,
        \mathtt{k}_\text{(m2)}\,
        p_{\k'}
        \BB{
            R^{\k\sprime}{}_{\l\sprime\mem\m\sprime\mem\n'\hnem}(\fx)
            +
            \tfrac{1}{2}
            R^{\k\sprime}{}_{\l\sprime\mem\m\sprime\mem\n'\hnem}(\fx)
        }
        \mem\fb^{\l'}
        w^{\m\sprime}{}_a\hem
        w^{\n\sprime}{}_b
    - (a{\,\leftrightarrow\,}b)
    \\
    &
    \,=\,
        3\mem \mathtt{k}_\text{(m2)}\,
        p_{\k'}
            R^{\k\sprime}{}_{\l\sprime\mem\m\sprime\mem\n'\hnem}(\fx)
        \mem\fb^{\l'}
        w^{\m\sprime}{}_a\hem
        w^{\n\sprime}{}_b
    \,,
\end{split}
\end{align}
which uses the algebraic Bianchi identity.
Similarly,
the perturbation on the Poisson bracket $\pb{\phi_a}{\phi_b}$
due to $\fom_\text{(m3)} = d\fth_\text{(m3)}$ is
\begin{align}
    \label{calc.wm3}
    3\mem \mathtt{k}_\text{(m3)}\,
    p_{\k'}
        R^{\k\sprime}{}_{\l\sprime\mem\m\sprime\mem\n'\hnem}(\fx)
    \mem \fb^{\l'}
    w^{\m\sprime}{}_a\hem
    w^{\n\sprime}{}_b
    \,,
\end{align}
\end{subequations}
where we have ignored the derivative term.
Recall from \eqref{wfl.fcov} that
$
    \pifb{d\fx^{\m\sprime}}{d\phi_a}
    =
    \pifb{D\fb^{\m\sprime}}{d\phi_a}
$.
The sum of \eqrefss{1a-2a-red}{calc.wm2}{calc.wm3} gives
\begin{align}
    \label{kcond1}
    \BB{
        \tfrac{1}{2} +
        3\mem \mathtt{k}_\text{(m2)} + 3\mem \mathtt{k}_\text{(m3)}
    }\,
    p_{\k'}
        R^{\k\sprime}{}_{\l\sprime\mem\m\sprime\mem\n'\hnem}(\fx)
    \mem \fb^{\l'}
    w^{\m\sprime}{}_a\hem
    w^{\n\sprime}{}_b
    \,.
\end{align}
Thus, additional information is needed to fix the unknown coefficients
$\mathtt{k}_\text{(m2)}$ and $\mathtt{k}_\text{(m3)}$.

To this end, we stipulate the spin gauge invaraince of the momentum-squared
as another bootstrap condition:
\begin{align}
    \label{boots.GR.psq}
    0
    \,=\,
        \pb{
            \tfrac{1}{2}\mem p^2
        }{\phi_a}
    \,=\,
        p^{\m'}\mem
        \pia{
            Dp_{\m'}
        }{d\phi_a}
    \,.
\end{align}
We have considered four types of fictitious terms so far:
$\fom_\text{(s1)}$,
$\fom_\text{(s2)}$,
$\fom_\text{(m2)}$,
and
$\fom_\text{(m3)}$.
Among these,
only 
$\fom_\text{(s1)}$
and
$\fom_\text{(m2)}$
contribute to the spin gauge flow of $p_\m$.
Computation shows that
the sum of
their contributions to the Poisson bracket
$\pia{
    Dp_{\m'}
}{d\phi_a}
$ is
\begin{align}
    \label{resDp}
    \tfrac{1}{2}\,
        \fS_{\k\sprime\mem\l'}
        R^{\k\sprime\mem\l\sprime}{}_{\m\sprime\mem\n'\hnem}(\fx)\,
            w^{\n\sprime}{}_a
    - 
        p_{\k'} w^{\l\sprime}{}_a\mem
            R^{\k\sprime}{}_{\l\sprime\mem\m\sprime\mem\n'\hnem}(\fx)
        \,
        \fb^{\n\sprime}
    +
        2\mem \mathtt{k}_\text{(m2)}\,
        p_{\k'} w^{\l\sprime}{}_a\mem
            R^{\k\sprime}{}_{(\l\sprime\mem\n')\m'\hnem}(\fx)
        \,
        \fb^{\n\sprime}
\end{align}
Again, the first term in \eqref{resDp},
which directly references the spurious spin $\fS_{\k\sprime\mem\l'}$,
arises by a calculation isomorphic to the electromagnetic case.
The other terms
are specific to gravity.
Contracting \eqref{resDp} with $p^{\m\sprime}$ gives
\begin{align}
\begin{split}
    \label{respDp}
    &
    \tfrac{1}{2}\,
        \fS_{\k\sprime\mem\l'}
        R^{\k\sprime\mem\l\sprime}{}_{\m\sprime\mem\n'\hnem}(\fx)\,
            p^{\m\sprime}\hem
            w^{\n\sprime}{}_a
    - 
        p_{\k'} w^{\l\sprime}{}_a\mem
            R^{\k\sprime}{}_{\l\sprime\mem\m\sprime\mem\n'\hnem}(\fx)
        \,
        p^{\m\sprime}\hem
        \fb^{\n\sprime}
    \\
    &
    -
        \mathtt{k}_\text{(m2)}\,
        p_{\k'} w^{\l\sprime}{}_a\mem
            R^{\k\sprime}{}_{\l\sprime\mem\m\sprime\mem\n'\hnem}(\fx)
        \,
        p^{\m\sprime}\hem
        \fb^{\n\sprime}
    -
        \mathtt{k}_\text{(m2)}\,
        p^{\m\sprime}\mem w^{\l\sprime}{}_a\mem
            R_{\m\sprime\mem\l\sprime\mem\k\sprime\mem\n'\hnem}(\fx)
        \,
        p^{\k\sprime}\hem
        \fb^{\n\sprime}
    \,,
\end{split}
\end{align}
where we have used the index symmetries
of the Riemann tensor.
The role of first term in \eqref{respDp}
is to cancel the contribution from the 
spurious minimal coupling.
The other terms in \eqref{respDp}
cancel each other iff
$\mathtt{k}_\text{(m2)} {\:=\:} {-1/2}$,
in which case
\eqref{kcond1} implies
$\mathtt{k}_\text{(m3)} {\:=\:} {1/3}$.
This completely fixes the unknown coefficients.


In summary, we have established a bottom-up determination
of the fictitious symplectic potential
in the regime where derivatives of the Riemann tensor are ignored,
by imposing the preservation of the spin gauge algebra,
\eqref{boots.GR},
and the spin gauge invariance of the momentum-squared,
\eqref{boots.GR.psq}:\footnote{
    In principle, one can also examine the spin gauge invariance of the physical spin-squared.
}
\begin{align}
\begin{split}
    \label{fthgr}
    \fth
    \,\,=\,\,
    &
        \tfrac{1}{2}\,
        \fS_{\k\sprime\mem\l'}
        R^{\k\sprime\mem\l\sprime}{}_{\r\sprime\mem\n'\hnem}(\fx)\,
            \fb^{\r\sprime}\mem d\fx^{\n\sprime}
        -
        \tfrac{1}{4}\,
        \fS_{\k\sprime\mem\l'}
        R^{\k\sprime\mem\l\sprime}{}_{\r\sprime\mem\n'\hnem}(\fx)\,
            \fb^{\r\sprime}\mem D\fb^{\n\sprime}
    \\
    &
        -\tfrac{1}{2}\,
            p_{\k'}
            R^{\k\sprime}{}_{\l\sprime\mem\m\sprime\mem\n'\hnem}(\fx)
            \,
            \fb^{\l\sprime}\mem
            \fb^{\m\sprime}\mem d\fx^{\n\sprime}
        + \tfrac{1}{3}\,
            p_{\k'}
            R^{\k\sprime}{}_{\l\sprime\mem\m\sprime\mem\n'\hnem}(\fx)
            \,
            \fb^{\l\sprime}\mem
            \fb^{\m\sprime}\mem D\fb^{\n\sprime}
        + \cdots
    \,.
\end{split}
\end{align}
\eqref{fthgr} shows perfect match with the top-down answer
presented in \eqref{vths}:
$1/2 -1 = -1/2$
and
$-1/6 + 1/2 = 1/3$.
The correspondence in \eqref{cov-dc.f}
determines the fictitious terms due to the local Lorentz charge (spin),
which,
through symplecticity constraint $d\fom = 0$ and the nonabelian nature of the Lorentz group,
induces the fictitious terms due to the diffeomorphism charge (mass).

\section{Covariant Geometry of General-Relativistic Phase Spaces}
\label{COVPS}

Eventually,
we can explicitly prove
the postulates and claims
of \Secs{INT>CUPB}{INT>MPD}
for each of the four models.
The structure of our exposition below
shall follow \rcite{csg}.

\subsection{Spinor Oscillator}
\label{COVPS>SO}

The action in \eqref{SOGR.action} describes the symplectic potential
\begin{align}
    \label{SOGR.theta}
    \theta
    \,=\,
        p_{\a\da}\mem e^{\da\a}
        + 
        \frac{1}{2i}\mem
        \BB{\mem
            \bzeta_\da\, \hat{p}^{\da\a} D\zeta_\a 
            - D\bzeta_\da\mem \hat{p}^{\da\a}\hem \zeta_\a
        }
    \,.
\end{align}
The exterior derivative of \eqref{SOGR.theta} 
splits as
$\omega = d\theta =\omega^\bullet + \omega'$.
First,
\begin{align}
    \label{SOGR.omega-cov}
    \omega^\bullet
    \,=\,
        Dp_{\a\da} \swedge e^{\da\a}
        -i\mem \hat{p}^{\da\a}\mem
            D\bzeta_\da \swedge D\zeta_\a
        + 
        D\hat{p}^{\da\a} 
        \swedge
        \frac{1}{2i}\mem
        \BB{\mem
            \bzeta_\da\, D\zeta_\a 
            - D\bzeta_\da\mem \zeta_\a
        }
\end{align}
covariantizes the free theory's symplectic form.
Second, the curvature correction $\omega'$ arises by the square $D^2$ of the covariant exterior derivative:
\begin{align}
\begin{split}
    \label{SOGR.omega'}
    \omega'
    \,&=\,
        \frac{1}{2i}\mem
        \BB{\mem
            \bzeta_\da\, \hat{p}^{\da\a} D^2\zeta_\a 
            - D^2\bzeta_\da\mem \hat{p}^{\da\a}\hem \zeta_\a
        }
    \,,\\
    \,&=\,
        \frac{1}{2i}\mem
        \BB{\mem
            \bzeta_\da\, \hat{p}^{\da\a} \bR_\a{}^\b\hem \zeta_\b
            - \bzeta_\db\mem R^\db{}_\da\mem \hat{p}^{\da\a}\hem \zeta_\a
        }
    \,,\\
    \,&=\,
        \frac{i}{2}\mem
        \BB{\mem
            \bxi^\a \bR_\a{}^\b\mem \zeta_\a
            - \bzeta_\da\mem R^\da{}_\db\mem \xi^\db
        }
    \,,\\
    \,&=\,
        - \frac{1}{2}\, S_{mn}\mem R^{mn}
    \,=\,
        p_m\mem {\star}R^m{}_n\hem \hy^n
    \,.
\end{split}
\end{align}
Any term involving no $D^2$ 
is reproduced by the covariantization $\omega^\bullet$.

\newpage
A complete noncoordinate basis of one-forms
and its dual
are given
on $\P^\SO_{12}$ as
\begin{align}
    \label{SO.bases}
    \bigbig{
        e^m
        ,\mem
        Dp_m
        ,\mem
        D\zeta_\a
        ,\mem
        D\bzeta_\da
    }
    \quad\xleftrightarrow[]{
        \,\,\,\text{dual}\,\,\,\mem
    }\quad
    \bb{
        \tE_m
        ,\mem
        \frac{\partial}{\partial p_m}
        ,\mem
        \frac{\partial}{\partial\zeta_\a}
        ,\mem
        \frac{\partial}{\partial\bzeta_\da}
    }
    \,,
\end{align}
which transform covariantly.
Here, $\tE_m$ is the horizontal lift
\cite{ehresmann1948connexions,Mason:2013sva,gde}
of the spacetime tetrad $E_m = E^\m{}_m(x)\mem \partial_\m$:
\begin{align}
    \label{SO.tE}
    \tE_r
    \,=\,
        E^\r{}_r(x)\,
        \bb{
            \frac{\partial}{\partial x^\r}
            + p_n\mem \gamma^n{}_{m\r}(x)\,
                \frac{\partial}{\partial p_m}  
            - \gamma_\a{}^\b{}_\r(x)\mem \zeta_\b\,
                \frac{\partial}{\partial \zeta_\a}
            - \bzeta_\db\mem \bgamma^\db{}_{\da\r}(x)\,
                \frac{\partial}{\partial \bzeta_\da}
        }
    \,.
\end{align}
\noindent
As a result, the pointwise inverse of \eqref{SOGR.omega-cov}
is explicitly found as
\begin{align}
\begin{split}
    \label{SO.Pib}
    \Pi^\bullet
    \,\,=\,\,
\begin{aligned}[t]
    &
        \tE_m \swedge \frac{\partial}{\partial p_m}
        +
        i\mem \hat{p}_{\a\da}\,
            \frac{\partial}{\partial\zeta_\a}
            \swedge
            \frac{\partial}{\partial\bzeta_\da}
        +
        \frac{1}{-2p^2}\,
            S^{mn}\mem
        \tE_m \swedge \tE_n
    \\
    &
        +
        \frac{
            (\s^{mn}\zeta)_\a
        }{2p^2}\,    
            p_n\mem
        \tE_m \swedge \frac{\partial}{\partial\zeta_\a}
        +
        \frac{
            (\bzeta\hhem\bs^{mn})_\da
        }{2p^2}\,    
            p_n\mem
        \tE_m \swedge \frac{\partial}{\partial\bzeta_\da}
    \,.
\end{aligned}
\end{split}
\end{align}

The universal covariant Poisson bracket relation is verified.
Plugging in the one-forms
$(e^m,D\hy^m,Dp_m)$
to \eqref{SO.Pib}
reproduces \eqref{xyp.cov}.

The minimal covariant symplectic perturbation in \eqref{GR.sp0}
is verified.
\eqref{SOGR.omega'} is the simplest curvature correction such that
$d\omega {\:=\:} dd\theta {\:=\:} 0$ by construction.

Consequently,
we establish the existence of
the universal functions $(x^\m,\hy^m,p_m)$
in the general-relativistic phase space $\P_{12}^\SO$ of the spinor oscillator.

\subsection{Massive Twistor}
\label{COVPS>MT}

The action in
\eqref{MTGR.action}
describes the symplectic potential
\begin{align}
    \label{MTGR.theta}
    \theta
    \,=\,
    	- \lambda_\a{}^I\mem \rambda_{I\da}\, e^{\da\a}
    	+ i\mem y^{\da\a}\mem
    	\BB{
    		\lambda_\a{}^I\hem D\rambda_{I\da}
    		{\,-\,}
    		D\lambda_\a{}^I\mem \rambda_{I\da}
    	}
    \,.
\end{align}
The exterior derivative of \eqref{MTGR.theta} is
$\omega = \omega^\bullet + \omega'$.
First, $\omega^\bullet$
covariantizes the free theory's symplectic form:
\begin{align}
\begin{split}
    \label{MTGR.omega-cov}
    \begin{aligned}[t]
        &
        \rambda_{I\da}\mem
        \bigbig{
            e^{\da\a} \mminus i\hem Dy^{\da\a}
        } \wedge D\lambda_\a{}^I
        - D\rambda_{I\da} \wedge
        \bigbig{
            e^{\da\a} \mplus i\hem Dy^{\da\a}
        }\mem \lambda_\a{}^I
        - 2i\mem D\rambda_{I\da} \wedge y^{\da\a} D\lambda_\a{}^I
        \,.
    \end{aligned}
    \kern-0.5em
\end{split}
\end{align}
Second,
the curvature correction $\omega'$
arises by $D^2$:
\begin{align}
\begin{split}
    \label{MTGR.omega'}
    \omega'
    \,&=\,
        i\mem y^{\da\a}\mem
        	\BB{
        		\lambda_\a{}^I\hem D^2\rambda_{I\da}
        		{\,-\,}
        		D^2\lambda_\a{}^I\mem \rambda_{I\da}
        	}
    \,,\\
    \,&=\,
        i\mem y^{\da\a}\mem
        	\BB{
        		\lambda_\a{}^I\hem \rambda_{I\db}\mem R^\db{}_\da
        		{\,-\,}
        		\bR_\a{}^\b\hem \lambda_\b{}^I\mem \rambda_{I\da}
        	}
    \,,\\
    \,&=\,
        -i\mem 
            p_\wrap{\a\db}\mem R^\db{}_\da\mem y^{\da\a}
        +i\mem p_{\b\da}\mem y^{\da\a}
            \bR_\a{}^\b
    \,=\,
        p_m\mem {\star}R^m{}_n\hem \hy^n
    \,.
\end{split}
\end{align}
Any term involving no $D^2$
is reproduced by the covariantization $\omega^\bullet$.

A complete noncoordinate basis of one-forms and its dual are given
on $\P_{16}^\MT$ as
\begin{align}
    \label{MT.bases}
    \bigbig{
        e^m
        ,\mem
        Dy^m
        ,\mem
        D\lambda_\a{}^I
        ,\mem
        D\rambda_{I\da}
    }
    \quad\xleftrightarrow[]{
        \,\,\,\text{dual}\,\,\,\mem
    }\quad
    \bb{
        \tE_m
        ,\mem
        \frac{\partial}{\partial y^m}
        ,\mem
        \frac{\partial}{\partial\lambda_\a{}^I}
        ,\mem
        \frac{\partial}{\partial\rambda_{I\da}}
    }
    \,,
\end{align}
which transform covariantly.
The horizontal lift is given by
\begin{align}
    \label{MT.tE}
    \tE_r
    \,=\,
        E^\r{}_r(x)\,
        \bb{
            \frac{\partial}{\partial x^\r}
            - \gamma^m{}_{n\r}(x)\mem y^n\,
                \frac{\partial}{\partial y^m}  
            - \gamma_\a{}^\b{}_\r(x)\mem \lambda_\b{}^I\,
                \frac{\partial}{\partial \lambda_\a{}^I}
            - \rambda_{I\db}\mem \bgamma^\db{}_{\da\r}(x)\,
                \frac{\partial}{\partial \rambda_{I\da}}
        }
    \,.
\end{align}
As a result,
the pointwise inverse of \eqref{MTGR.omega-cov}
is explicitly found as
\begin{align}
\begin{split}
    \label{MT.Pib}
    \Pi^\bullet
    \,\,=\,\,
\begin{aligned}[t]
    &
        \frac{1}{2}\,
        (\rambda^{-1})^{\da I}\,
        \frac{\partial}{\partial\lambda_\a{}^I}
        \mem{\mem\wedge}\mem
        \bb{
            \tE_{\a\da}
            + i\mem \frac{\partial}{\partial y^{\da\a}}
        }
    +
        \frac{1}{2}\,
        (\lambda^{-1})_I{}^\a\,
        \frac{\partial}{\partial\rambda_{I\da}}
        \mem{\mem\wedge}\mem
        \bb{
            \tE_{\a\da}
            - i\mem \frac{\partial}{\partial y^{\da\a}}
        }
    \\
    &
        -\frac{i}{2}\, y^{\da\b}\mem (\rambda^{-1}\lambda^{-1})^{\db\a}\,
        \bb{
            \tE_{\a\da}
            - i\mem \frac{\partial}{\partial y^{\da\a}}
        }
        \mem{\mem\wedge}\mem
        \bb{
            \tE_{\a\da}
            + i\mem \frac{\partial}{\partial y^{\da\a}}
        }
    \,.
\end{aligned}
\end{split}
\end{align}

The universal covariant Poisson bracket relation is verified.
Plugging in the one-forms
$(e^m,D\hy^m,Dp_m)$
to \eqref{MT.Pib}
reproduces \eqref{xyp.cov}.


The minimal covariant symplectic perturbation in \eqref{GR.sp0}
is verified.
\eqref{MTGR.omega'} is the simplest curvature correction such that
$d\omega {\:=\:} dd\theta {\:=\:} 0$ by construction.

Consequently,
we establish the existence of
the universal functions $(x^\m,\hy^m,p_m)$
in the general-relativistic phase space $\P_{16}^\MT$ of the massive twistor.

\subsection{Vector Oscillator}
\label{COVPS>VO}

\para{Physical Coupling}

To identify the physical coupling
for the vector oscillator,
we gather the preliminary explorations in 
\eqrefs{VOGR.action}{VOGR.action.correct}
and the technique demonstrated with the spherical top in \Sec{HR}.

At the free theory level,
we have reformulated
the symplectic potential in \eqref{VO.thetac}
into the form of \eqref{VO.theta0.split}
by splitting out a complex gauge component $\fpsi_1$.
We reproduce this reformulation
in the setup of
flat spacetime in noninertial frames:
\begin{subequations}
\label{VOGR.reforms}
\begin{align}
    \label{VOGR.reform'}
    \theta^\circ
    \,&=\,
        p_{\m'}\hem d\fx^{\m\sprime}
        + \frac{i}{2}\mem
        \BB{
            \fba_{\m'} D\hnem\fa^{\m\sprime}
            - D\hnem\fba_{\m'} \fa^{\m\sprime}
        \mem}
    \,,\\
    \label{VOGR.reform}
    \,&=\,
        p_\m\mem dx^\m
        + \frac{i}{2}\mem
        \BB{
            \ba_\m\mem D\a^\m
            - D\ba_\m\mem \a^\m
        }
        + \frac{i}{2}\mem
        \BB{
            \fbpsi_1\mem d\fpsi_1
            - d\fbpsi_1\mem \fpsi_1
        }
    \,.
\end{align}
\end{subequations}
Just like in \eqref{genthetas},
both are healthy in terms of general covariance
but one is spoiled due to spin gauge noninvariance.
Of course, 
the correct answer is \eqref{VOGR.reform}
as discussed in \eqref{VOGR.action.correct}.
The gravitational fields are coupled at the physical center, $x^\m$.

We transplant \eqref{VOGR.reform}
to curved spacetime.
By recalling \eqref{VO.bparam},
we then impose
$\a^\m = \hdelta^\m{}_\n\mem \fa^\n$,
$\fpsi_1 = -i\mem \hp\mdot\hnem\fba$,
and
\eqref{fcen.curved}
with
\begin{align}
    \fb^\m
    \,=\,
        - \frac{1}{|p|}\,\BB{
            \a^\m\mem \fpsi_1
            + 
            \ba^\m\mem \fbpsi_1
        }
    \,.
\end{align}
We also stipulate that the local tensor degrees of freedom
such as $\fa^\m$, $\a^\m$, $p_\m$
are transported by the geodesic parallel propagator.
The result is
\begin{align}
\begin{split}
    \label{VOGR.theta}
    \theta
    \,&=\,
        p_\m\mem dx^\m
        + \frac{i}{2}\mem
        \BB{
            \ba_\m\mem D\a^\m
            - D\ba_\m\mem \a^\m
        }
        + \frac{i}{2}\mem
        \BB{
            \fbpsi_1\mem d\fpsi_1
            - d\fbpsi_1\mem \fpsi_1
        }
    \\[0.3\baselineskip]
    \,&=\,
    \begin{aligned}[t]
        &
        p_{\m'}\hem d\fx^{\m\sprime}
        - p_{\m'}\hem D\fb^{\m\sprime}
        + \fth_\text{mass}
        \\
        &
        + \frac{i}{2}\mem
        \BB{
            \ba_{\m'} D\hnem\a^{\m\sprime}
            - D\hnem\ba_{\m'} \a^{\m\sprime}
        \mem}
        + \fth_\text{spin}
        + \frac{i}{2}\mem
        \BB{
            \fbpsi_1\mem d\fpsi_1
            - d\fbpsi_1\mem \fpsi_1
        }
        \,,
    \end{aligned}
    \\[0.27\baselineskip]
    \,&=\,
        p_{\m'}\hem d\fx^{\m\sprime}
        + \frac{i}{2}\mem
        \BB{
            \fba_{\m'} D\hnem\fa^{\m\sprime}
            - D\hnem\fba_{\m'} \fa^{\m\sprime}
        \mem}
        \,+\,
        \underbrace{
            \fth_\text{mass}
            +
        \mathrlap{\vphantom{\Big|}}   
            \fth_\text{spin}
        }_{\adjustbox{raise=-2pt}{\textfake{
            \scshape\footnotesize Fictitious
        }}}
    \,.
\end{split}
\end{align}
Again, it is important that $\fpsi_1$ is spacetime scalar.


In conclusion,
the symplectic potential
of the general-relativistic vector oscillator,
minimally coupled at the physical center,
develops fictitious terms
in the defining chart
of $\P_{16}^\VO$
in \eqref{VOGR.coords}.
They precisely take the identical form as in 
\eqref{vths}
arising from the tensors 
$\Delt X^{\m\sprime}{}_{\s'}$,
$\Delt Y^{\m\sprime}{}_{\s'}$,
$\acX^{\m\sprime}{}_{\n'\nem\s'}$,
and
$\acY^{\m\sprime}{}_{\n'\nem\s'}$
listed in \eqref{gdelow},
yet with the definition
\begin{align}
    - \fb^{\m\sprime}
    \,=\,
        \frac{1}{p^2}\,\mem
        i\mem\BB{
            \fa^{\m\sprime}
            \fba^{\n\sprime}
            -
            \fba^{\m\sprime}
            \fa^{\n\sprime}
        }\,
        p_{\n\sprime}
    \,.
\end{align}

\para{Symplectic Form}

We shall now compute the symplectic form.
Taking exterior derivative on the unprimed expression in \eqref{VOGR.theta}
gives
$\omega = \omega^\bullet + \omega'$.
First,
\begin{align}
    \label{VOGR.omega-cov}
    \omega^\bullet
    \,&=\,
        Dp_\m \swedge dx^\m
        + i\, D\ba_\m \swedge D\a^\m
        + i\, d\fbpsi_1 \swedge d\fpsi_1
\end{align}
covariantizes 
the alternative representation of
the free theory's symplectic form.
Second, the curvature correction $\omega'$ arises by the square $D^2$ of the covariant exterior derivative:
\begin{align}
\begin{split}
    \label{VOGR.omega'}
    \omega'
    \,&=\,
        \frac{i}{2}\mem
        \BB{
            \ba_\m\mem D^2\a^\m
            - D^2\ba_\m\mem \a^\m
        }
    \,,\\
    \,&=\,
        \frac{i}{2}\mem
        \BB{
            \ba_\m\mem R^\m{}_\n\mem \a^\n
            + \ba_\n\mem R^\n{}_\m\mem \a^\m
        }
    \,=\,
        i\,
            \ba_\m\hem R^\m{}_\n\hem \a^\n
    \,=\,
        p_\m\hem {*\hnem}R^\m{}_\n\hhem \hy^\n
    \,.
\end{split}
\end{align}
In the last line,
we have imported 
the definition of $\hy^\m$ in \eqref{VO.xy}
within $T_x\M$.

\para{Poisson Structure}

\eqref{VOGR.omega-cov} defines a nondegnerate two-form on $\P_{16}^\MT$,
so its pointwise inverse $\Pi^\bullet$ is well-defined.
Based on the isomorphism
between \eqref{VOGR.omega-cov}
and the exterior derivative of \eqref{VO.theta0.split}
(covariantization map with respect to the physical center),
it is not difficult to see that
\begin{align}
\begin{split}
    \label{VOGR.pib}
    \pib{dx^\m}{dx^\n}
    \,=\,
        \frac{1}{-p^2}\,
            S^{\m\n}
    &\,,\quad
    \pib{dx^\m}{Dp_\n}
    \,=\, 
        \delta^\m{}_\n
    \,,\\
    \pib{dx^\m}{D\a^\m}
    \,=\,
        \frac{1}{-p^2}\,
            \a^\m\mem p^\n
    &\,,\quad
    \pib{D\a^\m}{D\ba_\n}
    \,=\,
        -i\mem \hdelta^\m{}_\n
    \,,
\end{split}
\end{align}
where $S^{\m\n} = i\mem (\a^\m \ba^\n - \ba^\m \a^\n)$.

\para{Universality}

We establish the existence of
the universal functions $(x^\m,\hy^m,p_m)$
in the general-relativistic phase space $\P_{16}^\VO$ of the vector oscillator.
Explicitly, define
\begin{align}
\begin{split}
    \label{phys-in-VO}
    x^\m
    \,&=\,
    \delta^\m{}_{\m'}\hem
    \bb{\,
        \fx^{\m\sprime} \hem\hhem-\mem \fb^{\m\sprime}
        \hem\hhem-\mem \frac{1}{2}\, \Gamma^{\m\sprime}{}_{\r'\nem\s'}(\fx)\mem \fb^{\r\sprime}\hem \fb^{\s\sprime}
        + \O(\fb^3)
    \,}
    \,,\\
    \hy^\m
    \,&=\,
        \frac{i}{-p^2}\,
            \bb{
                \delta^\m{}_\l
                + \Gamma^\m{}_{\l\r}(x)\mem \fb^\r
                + \O(\fb^2)
            }\mem \delta^\l{}_{\m'}\mem
            \ve^{\m'\nem\n'\nem\r'\nem\s\sprime} p_{\n'} \fa_{\r'} \fba_{\s'}
    \,,\\[0.12\baselineskip]
    p_\m
    \,&=\,
        p_{\m'}\hem 
        \hem\delta^{\m\sprime}{}_\l
        \mem\bb{
            \delta^\l{}_\m
            - \Gamma^\l{}_{\m\r}(x)\mem \fb^\r
            + \O(\fb^2)
        }
    \,,
\end{split}
\end{align}
and $\hy^m = e^m{}_\m(x)\mem \hy^\m$,
$p_m = p_\m\mem E^\m{}_m(x)$.

The universal covariant Poisson bracket relation is verified.
Plugging in the one-forms
$(e^m,D\hy^m,Dp_m)$
to $\Pi^\bullet$ described in \eqref{VOGR.pib}
reproduces \eqref{xyp.cov}.

The minimal covariant symplectic perturbation in \eqref{GR.sp0}
is verified.
\eqref{VOGR.omega'} is the simplest curvature correction such that
$d\omega {\:=\:} dd\theta {\:=\:} 0$ by construction.


Note that it was necessary to go through the whirlwind trip
from Eqs.\:(\ref{VOGR.reforms}) to (\ref{VOGR.theta}).
If one naively takes \eqref{VOGR.reform}
as the symplectic potential
in curved spacetime,
one finds
$   
        Dp_{\m\sprime}\hem \wedge d\fx^{\m\sprime}
        + i\mem
            D\hnem\fba_{\m\sprime}\hem 
            \wedge
            D\hnem\fa^{\m\sprime}
        + 
            \frac{i}{2}\,
            \fba_{\m'}\hem R^{\m\sprime}{}_{\n'\nem\r'\nem\s'\hnem}(\fx)\mem \fa^{\n\sprime}
            \,
                d\fx^{\r\sprime}
                \swedge
                d\fx^{\s\sprime}
$
for the symplectic form,
which is wrong:
the apparent covariant symplectic potential is given at the fake center, \textit{failing} the postulate in \eqref{GR.sp0}.

The refinement of \eqref{VOGR.action}
arises by taking the action as
$
\int \theta
    - (\k^0\nem/2m)\mem\pr{
        p^2 + m^2
    }\mem d\t
    - \kappa^1\mem p\mdot\hnem\fa
    - \bar{\kappa}^1\mem p\mdot\hnem\fba
    - {\kappa^2} \fa^2/2
    - {\bkappa^2} \fba^2/2
$,
where $\theta$ is the pullback of \eqref{VOGR.theta}.

\subsection{Spherical Top}
\label{COVPS>ST}

\para{Symplectic Form}

The action in
\eqref{STGR.action.correct}
uses the symplectic potential in \eqref{varths},
which is equivalent to \eqref{HRGR.theta}.
The exterior derivative of \eqref{HRGR.theta} is
$\omega = \omega^\bullet + \omega'$.
First,
\begin{align}
\begin{split}
    \label{HRGR.omega-cov}
    \omega^\bullet
    \,=\,
        Dp_\m \wedge dx^\m
        + \frac{1}{2}\,
            DS_{\m\n} \wedge
                (\L^\m{}_a\mem
                D\hhnem\L^{\n a})
        + \frac{1}{2}\,
            S_{\m\n}\mem
                D\L^\m{}_a \wedge
                D\hhnem\L^{\n a}
        + d\phi_a \wedge d\bet^a
\end{split}
\end{align}
covariantizes the free theory's symplectic form
in \eqref{piece.omega}.
Second,
the curvature correction $\omega'$
arises by the square $D^2$ of the covariant exterior derivative:
\begin{align}
\begin{split}
    \label{HRGR.omega'}
    \omega'
    \,&=\,
        \frac{1}{2}\,
            S_{\m\n}\mem
            \BB{
                \L^\m{}_a\mem
                D^2\hhnem\L^{\n a}
            }
    \,=\,
        \frac{1}{2}\,
            S_{\m\n}\mem
            \BB{
                \L^\m{}_a\mem
                R^\n{}_\r\hem \L^{\r a}
            }
    \,,\\
    \,&=\,
        \frac{1}{2}\,
            S_{\m\n}\mem
            \BB{
                R^\n{}_\r\mem
                \hat{\eta}^{\r\m}
            }
    \,=\,
        - \frac{1}{2}\,
            S_{\m\n}\mem R^{\m\n}
    \,=\,
        p_\m\hem {*\hnem}R^\m{}_\n\hhem \hy^\n
    \,.
\end{split}
\end{align}
In the last line,
we have put
$\hy^\m = - {*}S^{\m\n} p_\n / p^2$
within $T_x\M$
by recalling \eqref{ST.xy}.

Again, it is important to use the correct symplectic potential in \eqref{gentheta}.
If one mistakenly uses \eqref{gentheta'}
in curved spacetime,
the symplectic form is found as
$
    Dp_{\m'} \wedge d\fx^{\m\sprime}
    +
    \frac{1}{2}\,
        D\fS_{\m\sprime\mem\n'}
        \wedge
            (\fL^{\m\sprime}{}_A\mem
            D\nem\fL^{\n\sprime A})
    +
    \frac{1}{2}\,
        \fS_{\m\sprime\mem\n'}
            D\fL^{\m\sprime}{}_A
            \wedge
            D\nem\fL^{\n\sprime A}
    -
    \frac{1}{4}\,
        \fS_{\m\sprime\mem\n'}
        R^{\m\sprime\mem\n'\nem\r'\nem\s'\hnem}(\fx)\,
        d\fx^{\r\sprime} \swedge d\fx^{\s\sprime}
$,
which does not include any fictitious terms.
It is left as an exercise to derive the fictitious terms for the symplectic form
by either
applying the methods of \rcite{gde}
to \eqrefs{HRGR.omega-cov}{HRGR.omega'}
or taking exterior derivative on \eqref{varths}.

\para{Poisson Structure}

\eqref{HRGR.omega-cov} defines a nondegnerate two-form on $\P_{20}^\ST$,
so its pointwise inverse $\Pi^\bullet$ is well-defined.
Again, based on the isomorphism
between \eqref{HRGR.omega-cov}
and the free-theory symplectic form in \eqref{piece.omega}
(covariantization map with respect to the physical center),
it is not difficult to see that
\begin{align}
\begin{split}
\label{HRGR.pib}
    \pib{ dx^\mu }{ dx^\nu }
    \,=\, 
        \frac{1}{-p^2}\,
            S^{\m\n}
    &\,,\quad
    \pib{ dx^\m }{ DS_{\r\s} } 
    \,=\, 
        \frac{1}{p^2}\,
            2\mem S^\m{}_\wrap{[\r}\mem p_\wrap{\s]}
    \,,\\
    \pib{ dx^\mu }{ Dp_\nu } 
    \,=\, 
        \delta^\m{}_\n 
    &\,,\quad
    \pib{ DS_{\m\n} }{ DS_{\r\s} }
    \,=\,
        \BB{
            -4\mem \hdelta^{[\k}{}_{[\m}\mem
            \hat{\eta}_{\n][\r}\mem
            \hdelta^{\l]}{}_{\s]}
        }\mem S_{\k\l}
    \,,\\
    \pib{ D\L^\m{}_a }{ dx^\n } 
    \,=\,
        \frac{1}{p^2}\,
            p^\m\hem
            \L^\n{}_a
    &\,,\quad
    \pib{ D\L^\m{}_a }{ DS_{\r\s} } 
    \,=\,
        \BB{
            -2\mem \hdelta^\m{}_{[\r} \hat{\eta}_{\s]\n}
        }\mem
        \L^\n{}_a
    \,.
\end{split}
\end{align}

\para{Universality}

We establish the existence of
the universal functions $(x^\m,\hy^m,p_m)$
in the general-relativistic phase space $\P_{20}^\ST$ of the spherical top.
Their explicit formulae are
\begin{align}
\begin{split}
    \label{phys-in-HR}
    x^\m
    \,&=\,
    \delta^\m{}_{\m'}\hem
    \bb{\,
        \fx^{\m\sprime} \hem\hhem-\mem \fb^{\m\sprime}
        \hem\hhem-\mem \frac{1}{2}\, \Gamma^{\m\sprime}{}_{\r'\nem\s'}(\fx)\mem \fb^{\r\sprime}\hem \fb^{\s\sprime}
        + \O(\fb^3)
    \,}
    \,,\\
    \hy^\m
    \,&=\,
        \frac{1}{-2p^2}\,
            \bb{
                \delta^\m{}_\l
                + \Gamma^\m{}_{\l\r}(x)\mem \fb^\r
                + \O(\fb^2)
            }\mem \delta^\l{}_{\m'}\mem
            \ve^{\m'\nem\n'\nem\r'\nem\s\sprime} 
            p_{\n'}
            \fS_{\r'\nem\s'}
    \,,\\[0.12\baselineskip]
    p_\m
    \,&=\,
        p_{\m'}\hem 
        \hem\delta^{\m\sprime}{}_\l
        \mem\bb{
            \delta^\l{}_\m
            - \Gamma^\l{}_{\m\r}(x)\mem \fb^\r
            + \O(\fb^2)
        }
    \,,
\end{split}
\end{align}
where $\hy^m = e^m{}_\m(x)\mem \hy^\m$,
$p_m = p_\m\mem E^\m{}_m(x)$.
Just as \eqref{phys-in-VO},
\eqref{phys-in-HR} is highly nonlinear.

The universal covariant Poisson bracket relation is verified.
Plugging in the one-forms
$(e^m,D\hy^m,Dp_m)$
to $\Pi^\bullet$ described in \eqref{HRGR.pib}
reproduces \eqref{xyp.cov}.

The minimal covariant symplectic perturbation in \eqref{GR.sp0}
is verified.
\eqref{HRGR.omega'} is the simplest curvature correction such that
$d\omega {\:=\:} dd\theta {\:=\:} 0$ by construction.


\section{Further Applications}
\label{APPL}

\subsection{Dynamical Symmetries and Conserved Charges}
\label{APPL>Q}

This subsection demonstrates how
model-ignorant conserved charges
are directly uncovered from
the universal Poisson and covariant Poisson bracket relations.
For simplicity, we limit our attention to spacetime isometries.

\para{Symmetry in Hamiltonian Mechanics}

Let us first review the definition of symmetry in Hamiltonian mechanics.
Suppose a Hamiltonian system defined by
a symplectic form $\omega$ and a Hamiltonian $H$
on a phase space.
A vector field $V$ implements
an infinitesimal symmetry transformation on
this system iff
both of the following conditions are met:
\begin{subequations}
\label{symdef}
\begin{align}
\label{symdef.symp}
    \text{Symplectic Condition}:
    &\quad
    \pounds_V\hem \omega
    \,=\,
        0
    \,,
    \,\,
    \\
\label{symdef.ham}
    \text{Hamiltonian Condition}:
    &\quad
    \pounds_V H
    \,=\,
        V\act{ H }
    \,=\,
        0
    \,.
    \,\,
\end{align}
\end{subequations}

Due to the Cartan formula
and the closure of $\omega$,
the symplectic condition
translates to $d(\i_V\hem\omega) = 0$,
which implies local existence of a function $Q$
such that
\begin{align}
    \label{symQ}
    \i_V\hem\omega \,=\, -dQ
    \qfq
    V \,=\, \pb{\blank}{Q}
    \,.
\end{align}
Here, $\pb{\blank}{f} = \omega^{-1}(\blank,df)$ denotes the Hamiltonian vector field of $f$.
This shows that symmetries in Hamiltonian mechanics
are realized as Hamiltonian actions on the phase space.
In other words, we have approached the Noether theorem
in the symplectic geometry language:
the function $Q$
is the Noether charge.

On the other hand, the Hamiltonian condition
encodes compatibility with the system's 
time-evolution generator
(or constraints if any).
Plugging in \eqref{symQ},
one finds that it encodes
Poisson commutativity of $Q$ and $H$:
\begin{align}
    \label{symH}
    \pb{H}{Q}
    \,=\,
    V\act{ H }
    \,=\,
        0
    \,.
\end{align}

We shall also note that the above definition
does not invoke the symplectic potential $\theta$.
Famously, the symplectic potential 
involves redundancies $\theta \sim \theta + \Delt\theta$
for any closed one-form $\Delt\theta$ on the phase space.
Thus it cannot provide an invariant definition of symmetry.

Instead, the role of the symplectic potential 
in this discussion
would be
a customary gadget
that provides a convenient formula
for the Noether charge.
By using the freedom $\theta \sim \theta + \Delt\theta$,
one may find a particular $\theta$ such that
$\pounds_V\hhem \theta = 0$.
In this case, the Cartan formula asserts
$\i_V\hem\omega = -d\i_V\hhem \theta$,
leading to a concrete identification of the Noether charge as
\begin{align}
    \label{charge-from-theta}
    \pounds_V\hhem \theta \,=\, 0
    \qiq
        Q \,=\, \i_V\hhem \theta
    \,.
\end{align}
In fact, such a symmetry-preserving symplectic potential
$\theta$ will give rise to a natural, symmetry-adapted phase space action.

\para{Symmetry-Preserving Symplectic Perturbations}

Next, we develop a general theory of symmetries
in the framework of symplectic perturbations.
Suppose $(\ps,\omega^\circ,H)$ defines the free theory of a Hamiltonian system.
Suppose a vector field $V$ on $\ps$
implements an infinitesimal symmetry transformation on $(\ps,\omega^\circ,H)$,
whose Noether charge is $Q^\circ$:
\begin{align}  
\label{sym0}
    \i_V \omega^\circ
    \,=\,
        -dQ^\circ
    \,,\quad
    \pb{H}{Q^\circ}^\circ
    \,=\,
        V\act{ H }
    \,=\,
        0
    \,.
\end{align}

A closed two-form $\omega'$ on $\ps$
is a \textit{symmetry-preserving symplectic perturbation}
iff
\begin{align}
    \label{sym.SPT}
    \pounds_V\hem \omega'
    \,=\,
        0
    \,.
\end{align}
Again, 
through the Cartan formula
and $d\omega' \eqq 0$,
\eqref{sym.SPT}
translates to $d(\i_V\hem\omega') = 0$,
which implies local existence of a function $Q'$
such that
\begin{align}
    \label{sym'Q}
    \i_V\hem\omega' \,=\, -dQ'
    \,.
\end{align} 
Via \eqref{sym0}, \eqref{sym'Q} implies that
\begin{align}
    \label{sym1}
    \i_V\hem\omega \,=\, -dQ
    \,,\quad
    \pb{H}{Q}
    \,=\,
        V\act{
            H
        }
    \,=\,
        0
\end{align}
where $Q = Q^\circ + Q'$ and $\omega = \omega^\circ + \omega'$.

The implication of \eqrefs{sym0}{sym1}
is that
$V$,
the \textit{same} vector field from the free theory $(\omega^\circ,H)$,
implements an infinitesimal symmetry transformation
on the interacting theory $(\omega,H)$ as well,
yet via the modified Noether charge
$Q = Q^\circ + Q'$ in \eqref{sym1}:
\begin{align}
    \label{Vsame0}
    V
    \,=\,
        \pb{\blank}{Q^\circ}^\circ
    \,=\,
        \pb{\blank}{Q}
    \,.
\end{align}

Further, suppose the free theory admits a number of symmetry charges forming an algebra.
The above analysis shows that the symmetry actions on the phase space 
(as the vector fields)
remain the same
under symmetry-preserving symplectic perturbations.
As a result, the symmetry algebra remains the same
in the interacting theory.

\para{Scalar Particle in Electromagnetism}

For a concrete demonstration of the above abstract formalization,
we could revisit the familiar example of the charged scalar particle
described in \Sec{INT>SPT}.

The free theory is defined in \eqref{scalar.omega0},
which supposes
the phase space $T^*\mflat$
with coordinates $x^\m$ and $p_\m$.
The symplectic form is $\omega^\circ = dp_\m \swedge dx^\m$,
while the mass-shell constraint $\phi_0 = \frac{1}{2}\mem (\eta^{\m\n} p_\m p_\n \mplus m^2)$ serves as the Hamiltonian.
A well-known fact is that
\begin{align}
    \label{V0}
    V
    \,=\,
        K^\m(x)\, \frac{\partial}{\partial x^\m}
        - p_\n\mem K^\n{}_{,\m}(x)\, \frac{\partial}{\partial p_\m}
\end{align}
generates a symmetry transformation
on $(T^*\mflat,\omega^\circ,\phi_0)$
iff $K^\m(x)\mem \partial/\partial x^\m$
is a Killing vector of flat spacetime $\mflat \eqq (\R^4,\eta)$.
To elaborate,
the symplectic condition
is trivially satisfied by the ansatz in \eqref{V0},
identifying the Noether charge as
\begin{align}
    Q^\circ
    \,=\,
        p_\m\mem K^\m(x)
    \,.
\end{align}
The Hamiltonian condition
then yields the
Killing equation in flat spacetime,
\begin{align}
    \label{killing0}
    \pounds_K \eta
    \,=\, 0
    \qfq
    K_{\m,\n}(x) \,=\, -K_{\n,\m}(x)
    \,.
\end{align}


Now consider an electromagnetic field strength $F$
on flat spacetime
such that
\begin{align}
    \label{isometry-F}
    \pounds_K F \,=\, 0
    \,.
\end{align}
Via the Cartan formula
and the closure $dF = 0$,
\eqref{isometry-F} implies the local existence of a scalar field $\a$ such that \cite{Hughston:1972qf}
\begin{align}
    \label{symQ.EM}
    \i_K F \,=\, -d\a
    \qfq
    K^\m(x)\mem F_{\m\n}(x)
    \,=\,   
        -\a_{,\n}(x)
    \,.
\end{align}

Crucially, the condition in \eqref{isometry-F}
implies that
$\omega' \eqq qF$
(now viewed as a two-form on $\ps$
via the pullback of the bundle projection)
is a symmetry-preserving symplectic perturbation.
Namely, it is easy to show that $\pounds_V\hhem \omega' = 0$ by $\pounds_K F = 0$.
According to \eqref{sym1},
the correction to the Noether charge  
can be taken as
$Q' = q\hem \a$,
as
$-dQ' = \i_V\hem \omega' = -q\mem d\a$.
Therefore, the exact Noether charge
in the interacting theory is
\begin{align}
    \label{scalar.QEM}
    Q
    \,=\,
        p_\m\mem K^\m(x)
        + \a(x)
    \,.
\end{align}
\eqref{scalar.QEM} is precisely
the conserved charge of the charged scalar probe
\cite{Hughston:1972qf}.

To derive the perhaps more widely-recognized formula
$Q = (\hem{p_\m + qA_\m(x)})\mem K^\m(x)$,
note that the scalar field $\a$ in \eqref{symQ.EM} can be explicitly given as follows
if one is willing to make an explicit gauge choice:
\begin{align}
    \pounds_K A \,=\, 0
    \qiq
    \a(x)
    \,=\,
        q A_\m(x)\mem K^\m(x)
    \,.
\end{align}

\para{Spinning Particles in Electromagnetism}

Now we are ready to derive the \textit{universal} (model-ignorant) conserved charges
of massive spinning particles
in electromagnetism.

Suppose any Hamiltonian formulation of a free massive spinning particle
based on a phase space $(\ps,\omega^\circ)$.
Suppose infinitesimal symmetry $V$ exists for this particle.
Our ansatz for the symmetry action on the physical variables is
\begin{align}
\begin{split}
    \label{V0.act}
    V\act{ x^\m }
    \,&=\,
    \pb{x^\m}{Q^\circ}^\circ
    \,=\,
        K^\m(x)
    \,,\\
    V\act{ \hy^\m }
    \,&=\,
    \pb{\hy^\m}{Q^\circ}^\circ
    \,=\,
        K^\m{}_{,\n}(x)\mem \hy^\n
    \,,\\
    V\act{ p_\m }
    \,&=\,
    \pb{p_\m}{Q^\circ}^\circ
    \,=\,
        - p_\n\mem K^\n{}_{,\m}(x)
    \,.
\end{split}
\end{align}
To validate this ansatz,
we apply the two conditions in \eqref{symdef}.
First, the symplectic condition is solved by
taking the Noether charge as
\begin{align}
    \label{Q0.univ}
    Q^\circ 
    \,=\,
        p_\m\mem K^\m(x)
        - \frac{1}{2}\:
            \ve^{\m\n\r\s} \hy_\r\mem p_\s
            \mem K_{\m,\n}(x)
    \,,
\end{align}
which reproduces \eqref{V0.act} via the
universal Poisson bracket relation in \eqref{xyp}
provided that $K^\m(x)$ exhibits linearity:
$K^\m{}_{,\r\s}(x) = 0$.
Second, the Hamiltonian condition is 
satisfied if $K_{\m,\n}(x) = -K_{\n,\m}(x)$,
as \eqref{V0.act} implies that
the symmetry action on the mass-shell constraint
$\phi_0 \eqq \frac{1}{2}\,(p^2\mplus m^2)$
is $V\act{\phi_0} = -p_\n\mem K^\n{}_{,\m}(x)\mem p^\m$.
Crucially, since $Q^\circ$ in \eqref{Q0.univ} is composed solely of the physical variables $(x^\m,\hy^\m,p_\m)$,
it must Poisson-commute with any gauge constraints
of the model.
Consequently, the Hamiltonian action of \eqref{Q0.univ}
passes both the symplectic and Hamiltonian tests of symmetry in \eqref{symdef}.

Note that this fact could be shown
without writing down
an explicit formula for $V$
(which will depend on the microscopic implementations, i.e., the details of the model).

It follows that
the symplectic perturbation $\omega' \eqq qF$
due to the universal minimal coupling recipe 
in \eqref{EM.sp0}
is symmetry-preserving
as long as the electromagnetic background $F$
exhibits symmetry \`a la \eqref{isometry-F}.
Since \eqref{V0.act} implies
$\i_V dx^\m = V\act{x^\m} = K^\m(x)$,
we have
(with abuse of notation regarding the pullbacks due to bundle projection)
\begin{align}
    \pounds_V \hem\omega'
    \,=\,
    d\mem \i_V \omega'
    \,&=\,
        q\mem d\BB{
            K^\m(x)\mem
                F_{\m\n}(x)\mem dx^\n
        }    
    \,=\,
        q\mem d\mem \i_K F
    \,=\,
        q\mem \pounds_K F
    \,=\,
        0
    \,.
\end{align}
Therefore, the same vector field $V$
still defines a symmetry in the interacting theory.

It remains to find the Noether charge.
Since
$-dQ' = \i_V\hem \omega' = q\mem \i_KF = -q\mem d\a$,
so we can take $Q' = q\mem \a(x)$.
Thus,
we conclude that
minimally coupled 
massive spinning particles
in electromagnetism
universally exhibits the conserved charge
\begin{align}
    \label{Q.EM0}
    Q
    \,=\,
        p_\m\mem K^\m(x)
        - \tfrac{1}{2}\:
            \ve^{\m\n\r\s} \hy_\r\mem p_\s
            \mem K_{\m,\n}(x)
        + q\mem \a(x)
    \,,
\end{align}
where the gauge choice $\pounds_K A = 0$
will take $\a(x) = A_\m(x)\mem K^\m(x)$.

To reiterate,
the Poisson commutativity of
\eqref{Q.EM0}
with the various constraints of the particle
is automatic
since the infinitesimal symmetry transformation $V$ is left unchanged.
Still, it is a nice and interesting exercise to
reconfirm $\pb{H}{Q} = 0$
by direct computation,
by using the exact universal
EoM derived in \eqref{uBMT},
for instance.

Incorporating the multipolar couplings is not difficult.
For instance,
the non-minimal coupling
in \eqref{EM.sp1}
adds
$\theta' = (g/2)\mem q\, {*}F_{\m\n}(x) $ $\hy^\m\mem dx^\n$
on the symplectic potential.
Computation using \eqref{V0.act} shows that
$\pounds_V\hhem \theta' = 0$
by $\pounds_K {*}F = 0$.
Here, $\pounds_K$ preserves the Hodge star
because it preserves the flat metric.
This shows that $\theta'$ is a symmetry-preserving perturbation on the symplectic potential.
Based on our earlier discussion around \eqref{charge-from-theta}, 
this implies that the universal conserved charge
with the generic gyromagnetic ratio $g$ is
\begin{align}
    Q
    \,=\,
        p_\m\mem K^\m(x)
        - \frac{1}{2}\:
            \ve^{\m\n\r\s} \hy_\r\mem p_\s
            \mem K_{\m,\n}(x)
        + q\mem \a(x)
        + \frac{g}{2}\:
            q\, {*}F_{\m\n}(x)\mem \hy^\m\mem K^\n(x)
    \,.
\end{align}

\para{Spinning Particles in Gravity}

Lastly, we concern gravitational backgrounds
with Killing vectors.
To expedite our discussion,
let us directly jump to spinning particles.

Suppose a curved spacetime $\M = (\R^4,g)$
admits a Killing vector $K = K^\m(x)\mem \partial/\partial x^\m$:
\begin{align}
    \label{killingcurved}
    K_{\m;\n}(x)
    \,=\,
        - K_{\n;\m}(x)
    \qiq
    K_{\m;\n;\r}(x)
    \,=\,
        - K_\s(x)\mem R^\s{}_{\r\m\n}(x)
    \,.
\end{align}
Let $(\P,\omega)$ be the phase space of any massive spinning particle put in this curved spacetime,
minimally coupled
so that $\omega = \omega^\bullet + \omega'$
and $\omega' = p_\m\mem {*}R^\m{}_\n\mem \hy^\n$
as postulated in \eqrefs{covsplit.re}{GR.sp0}.
The ansatz for the associated conserved charge is
\begin{align}
    \label{Q.univ}
    Q
    \,=\,
        p_\m\mem K^\m(x)
        - \tfrac{1}{2}\:
            \ve^{\m\n\r\s} \hy_\r\mem p_\s
            \mem K_{\m;\n}(x)
    \,,
\end{align}
which arises by covariantizing \eqref{Q0.univ}.
The differential $dQ \eqq DQ$ of \eqref{Q.univ} splits into two parts as
$dQ = (dQ)^\bullet + (dQ)'$, where
\begin{subequations}
\label{dQ}
\begin{align}
    \label{dQb}
    (dQ)^\bullet
    \,&=\,
        Dp_\m\mem K^\m(x)
        + p_\n\hem K^\n{}_{;\m}(x)\mem dx^\m
        - \tfrac{1}{2}\:
            \ve^{\m\n\r\s}\mem 
                D(\hy_\r\mem p_\s)
            \mem K_{\m;\n}(x)
    \,,\\
    \label{dQ'}
    (dQ)'
    \,&=\,
        dx^\m\mem
        \BB{
            p_\r\mem {*}R^\r{}_{\s\m\n}\mem \hy^\s
        }\hem K^\n(x)
    \,.
\end{align}
\end{subequations}
The first part,
\eqref{dQb},
would directly follow by covariantizing 
the free theory's differential $dQ^\circ$ 
for \eqref{Q0.univ}.
The second part,
\eqref{dQ'},
is a curvature effect.
We have used the well-known identity in \eqref{killingcurved} when deriving \eqref{dQ'}.


With this understanding, the straightforward covariantization 
of the vector field $V$ in \eqref{V0.act}
would be
$\V = \Pi^\bullet(\blank,(dQ)^\bullet)$.
In particular, $\V$ coincides with $V$
in the vanishing curvature limit
such that $(\P,\omega) \to (\ps,\omega^\circ)$.

Our task now is to prove the following proposition
as the generalization of \eqref{Vsame0}
for gravitational interactions:
\begin{align}
    \label{Vsame}
    \V
    \,=\,
        \pib{\blank}{(dQ)^\bullet}
    \,=\,
        \pia{\blank}{dQ}
    \,.
\end{align}
That is, tidal effects do not alter the symmetry action $\V$ on the general-relativistic phase space $\P$.
Note that the first expression of $\V$ in \eqref{Vsame}
completely characterizes the symmetry action on the universal variables 
via covariantization of \eqref{V0.act},
which should be clear from \eqref{dQb}:
\begin{align}
\begin{split}
    \label{V.act}
    \cont{dx^\m}{\V}
    \,&=\,
    \pib{dx^\m}{(dQ)^\bullet}
    \,=\,
        K^\m(x)
    \,,\\
    \cont{D\hy^\m}{\V}
    \,&=\,
    \pib{D\hy^\m}{(dQ)^\bullet}
    \,=\,
        K^\m{}_{;\n}(x)\mem \hy^\n
    \,,\\
    \cont{Dp_\m}{\V}
    \,&=\,
    \pib{Dp_\m}{(dQ)^\bullet}
    \,=\,
        - p_\n\mem K^\n{}_{;\m}(x)
    \,.
\end{split}
\end{align}
The proof proceeds by
computing the difference between the two expressions in \eqref{Vsame}:
\begin{align}
\begin{split}
    \label{deltV-gymnastics}
    \quad
    \Delt\V
    \,:=\,
    {}&{}
        \pia{\blank}{dQ}
        \,-\,
            \pib{\blank}{(dQ)^\bullet}
    \\
    \,=\,
    {}&{}
        \pib{\blank}{(dQ)'}
        - \pib{\blank}{dx^\m}\mem
            \BB{
                p_\r\mem {*}R^\r{}_{\s\m\n}(x)\mem \hy^\s
            }\mem
            \pia{dx^\n}{dQ}
    \,,\\
    \,=\,
    {}&{}
        \pib{\blank}{dx^\m}\mem
        \BB{
            p_\r\mem {*}R^\r{}_{\s\m\n}(x)\mem \hy^\s
        }\hem 
        \BB{
            K^\n(x)
            -
            \pia{dx^\n}{dQ}
        }
    \,,\\
    \,=\,
    {}&{}
    {
        - \pib{\blank}{dx^\m}\mem
        \BB{
            p_\r\mem {*}R^\r{}_{\s\m\n}(x)\mem \hy^\s
        }\mem 
        \cont{dx^\n}{\Delt\V\hem}
    }
    \,.
    \kern-1em
\end{split}
\end{align}
The second line follows by using the identity in \eqref{ppertc}.
The third line arises by plugging in \eqref{dQ'}.
The fourth line is due to the first line of \eqref{V.act}.
Clearly, \eqref{deltV-gymnastics} implies
\begin{align}
    \label{deltV-matrix}
    \bb{\mem
        \delta^\m{}_\s
        + \pib{dx^\m}{dx^\r}\mem
            \bigbig{
                p_\k\mem {*}R^\k{}_{\l\r\s}(x)\mem \hy^\l
            }
    }\mem 
        \cont{dx^\s}{\Delt\V\hem}
    \,=\,
        0
    \,.
\end{align}
The matrix described in \eqref{deltV-matrix}
is, at least for mildly curved spacetimes,
invertible.
Hence we conclude that $\langle\hem{ dx^\s , \Delt\V\hem }\hem\rangle = 0$,
which implies $\Delt\V = 0$ through \eqref{deltV-gymnastics}.

It is left as an exercise to develop this gymnastics 
in the presence of multipolar couplings.
In particular,
the quadrupole coupling in \eqref{GR.sp2} 
adds
$\theta' 
    = - (C_2/2)\,
        p_\m\hem R^\m{}_{\n\r\s}(x)
$ $
        \hy^\n \hy^\r dx^\s
$
on the symplectic potential,
in which case the Noether charge is
\begin{align}
    \label{Q.univ.C2}
    Q
    \,=\,
        p_\m\mem K^\m(x)
        - \frac{1}{2}\:
            \ve^{\m\n\r\s} \hy_\r\mem p_\s
            \mem K_{\m;\n}(x)
        - 
            \frac{C_2}{2!}\,
            p_\m\hem R^\m{}_{\n\r\s}(x)\mem \hy^\n
                \hy^\r\mem K^\s(x)
    \,.
\end{align}

\para{Model-Specific Derivations}

Of course, the above discussions become much more transparent
if one works in the model-specific fashion.
For example, the symmetry action in the spinor oscillator model is
\begin{align}
    \label{V-for-SO}
    \V
    \,&=\,
        K^m(x)\,
            \tE_m
        - p_n\mem K^n{}_{;m}(x)\,
            \frac{\partial}{\partial p_m}
        + \frac{1}{2}\: K_{m;n}(x)\mem 
            (\s^{mn})_\a{}^\b\mem \zeta_\b\:
            \frac{\partial}{\partial \zeta_\a}
    \,,
\end{align}
and contracting \eqref{V-for-SO}
with \eqref{SOGR.theta}
immediately yields \eqref{Q.univ}.

It should be interesting to explore symmetries due to Killing or Killing-Yano tensors
in this universal framework
and understand whether
they obey
\eqrefs{Vsame0}{Vsame}.

\subsection{Post-Minkowskian Expansion}
\label{APPL>PM}

The PM expansion seeks to
solve the dynamics of two gravitationally interacting and radiating
relativistic bodies
while maintaining the global Poincar\'e symmetry
of the flat Minkowski background manifestly.
Its modern driving force has been
scattering amplitudes, effective field theory, and worldline methods
\cite{Neill:2013wsa,Damour:2016gwp,Cheung:2018wkq,Cristofoli:2019neg,Bjerrum-Bohr:2021din,Kalin:2020mvi,Kalin:2020fhe,Mogull:2020sak,Jakobsen:2021smu,Bjerrum-Bohr:2022blt,Vines:2017hyw}.
In this subsection, we discuss 
the PM expansion for spinning bodies
from the perspectives of this work.

\para{Classical Eikonal for Universal Observables}

The ultimate objective of a classical mechanics problem
is to determine the final state from the initial state
by solving the EoM.
Recently, a number of works
\cite{Gonzo:2024zxo,Kim:2024svw,Kim:2025hpn,Alessio:2025flu,Kim:2025olv,Kim:2025gis,Kim:2025sey}
have identified an efficient currency for classical mechanics
that is well-suited for in-in boundary value problems:
the Magnusian.
The Magnusian is the generator of the symplectomorphism
mapping the initial phase space to the final phase space
in Hamiltonian formulation.
Its historical roots can be traced back to 
\rrcite{hunziker1968s,simon1971wave,herbst1974classical,sokolov1979classical,narnhofer1981canonical,thirring1981classical,osborn1980levinson,levinson1949uniqueness},
while
the name Magnusian derives from the fact that
it is computed by the Magnus series \cite{magnus1954exponential}
(to be compared with the Dyson series).
In the scattering context, the Magnusian is referred to as classical eikonal or scattering generator.
In a rigorous fashion,
it is established that
the classical eikonal
defines the classical limit of the log of the S-matrix
\cite{Kim:2025ebl}.
The classical eikonal is a single scalar function on the asymptotic (free) phase space
that encapsulates all classical scattering observables.

In the worldline framework
known as worldline quantum field theory (WQFT)
\cite{Mogull:2020sak,Jakobsen:2021smu},
the diagrammatic computations encode
Berends-Giele recursion for solving EoM;
see \rrcite{Kim:2024grz,Hoogeveen:2025tew} for examples with spinning particles.
The Magnus series formula for classical eikonal
describes nested Poisson/Peierls brackets of
interaction vertices at different times,
producing a weighted sum of WQFT diagrams composed of retarded propagators.

Regarding universality,
any two equivalent spinning particle models
should predict the same relation between 
the initial and final universal observables,
thus defining Magnusians
that generate the same effect on the $(x,\hy,p)$ space.
At intermediate steps during computations, however,
the four models will produce
totally different diagrammatic expansions in WQFT.
A detailed comparison on the efficiency of their WQFT diagrammatics
would be interesting,
say, in terms of the total number of diagrams for an observable.

In fact,
one could also directly study the model-independent part of the time evolution
by focusing on the EoM of $(x,\hy,p)$:
an approach initiated in \rcite{Kim:2024grz}.
This method approaches the universal observables directly
and would be preferred over
model-specific or gauge-dependent methods.
The universality in the classical theory
will be ensured as long as
all interaction vertices are expressed in terms of $(x,\hy,p)$.
A possible technicality is that
the nonlinear nature of the Poisson algebra
(\eqref{M3*} or \eqref{M3})
may induce extra vertices
if the nested bracket formulae are applied in a straightforward fashion.

It is also worth emphasizing that,
for gravitational interactions,
the intermediate time evolution describes
the \textit{general-relativistic} $(x,\hy,p)$ variables in the bulk.
However, the classical eikonal for scattering in asymptotically flat spacetimes
act on the \textit{flat} $(x,\hy,p)$ space,
as it connects between
the initial and final asymptotic (free) phase spaces
(cf. \rcite{sst-asym}).

\para{Model-Specific Features}

We now turn to
the effects of model-specific features
on PM expansion.
First of all,
models with less \textit{redundancy} may be argued as preferable.
In this perspective,
the spinor oscillator model is identified as the most ideal,
while the spherical top is by no means efficient.
It carries the angle (body frame) variables that do not affect the universal observables $(x,\hy,p)$
as long as the interactions are spin-magnitude preserving;
similar comments apply to the vector oscillator and massive twistor models as well.

Second of all,
there is a discrepancy
between \textit{linear} models and \textit{nonlinear} models.
Take the spherical top model, 
for instance,
which describes a sigma model to a nonlinear target space 
(the cotangent bundle of Poincar\'e group).
As a result,
its symplectic potential in \eqref{ST.theta}
is cubic due to the spin term,
so the worldline perturbation theory is already complicated
at the level of \textit{free theory}.
Remarkably,
the massive twistor model
reformulates the spherical top
as a sigma model to a linear target space (${\sim\,} \C^8$)
in which the symplectic potential
takes the canonical, bilinear form as shown in \eqref{MT.theta}
(cf. Darboux theorem \cite{darboux1882probleme}).
Consequently, the diagrammatics of the massive twistor model
is much simpler than the spherical top model.
Similar analysis applies to the spinor and vector oscillators as well;
the former is nonlinear, while the latter is linear.

Third of all, 
when one aims to describe spinning \textit{black holes},
there exists a distinction between
the spinorial models and the vectorial models.
As is glimpsed in \eqrefs{SO.thetac}{incidence}
and will be explicated in \Sec{APPL>BH},
the spinorial models feature descriptions
in which
the so-called spinspacetime coordinates \cite{sst-asym}
can be taken as basic variables,
which can be important for comprehending and implementing the Newman-Janis shift.
For instance, the one-line reformulation of the universal Poisson bracket relation in \eqref{zzpb}
is immediate in the massive twistor model,
although typically obscured.

Last but not least, one shall also pay attention to the Lagrange multipliers.
In the vector oscillator model,
the Lagrange multiplier terms
produce vertex factors that one cannot read off directly from the Lagrangian. 
The WQFT rule of the spinor oscillator model is not affected by constraints, so the bare Feynman rules are sufficient for computing observables. 

\para{Further Comments}

We close this subsection with some comments on \rcite{Hoogeveen:2025tew}, where a fully systematic WQFT expansion based on the vector oscillator model is provided. A similar expansion based on our presentation of the vector oscillator model will eventually produce the same set of observables, but the intermediate steps may look rather different. 

First, we are using symplectic perturbation to implement nonminimal couplings while \rcite{Hoogeveen:2025tew} uses Hamiltonian perturbation to achieve the same effect. The two approaches could be related to by a noncanonical change of variables.
Second, while in our approach the SSC is imposed from the outset nonperturbatively, \rcite{Hoogeveen:2025tew} verifies the preservation of the covariant SSC perturbatively, order by order in spin. 
Finally, \rcite{Hoogeveen:2025tew} makes an impression that the WQFT prefers the Lagrangian (second-order) formulation while the constraint analysis is usually done in the Hamiltonian (first-order) formulation. 
However, as elucidated in Refs.~\cite{Gonzo:2024zxo,Kim:2024svw,Kim:2025hpn,Alessio:2025flu,Kim:2025olv,Kim:2025gis,Kim:2025sey}, 
the Hamiltonian formulation is both conceptually more suitable and computationally more efficient 
than the Lagrangian formulation when it comes to computing the classical eikonal (Magnusian).

\subsection{Post-Newtonian Expansion} 
\label{APPL>PN}

In this subsection,
we explicate the implications of our work for
the PN effective field theory.
The plan is to revisit
the leading order spin-orbit coupling
in a streamlined fashion
and comment on the necessary role of the
fictitious vertices
which the rigorous construction of the interacting spherical top in \Sec{HR} stipulates.
Within this subsection, we switch to the usual notation:
$(\fx^{\m\sprime},\fL^{\m\sprime}{}_A,p_{\m'},\fS_{\m\sprime\mem\n'})
\to
(x^\m,\L^\m{}_A,p_\m,S_{\m\n})
$.

\para{Canonical Gauge}

The PN expansion is conventionally performed in a fixed lab frame,
in which case
it is natural to employ the canonical (Pryce-Newton-Wigner) spin gauge
described in \eqref{PNW}. 
Below, we record more details on the canonical spin gauge.\footnote{
    The comparison between the canonical and covariant spin gauges is detailed in
    Appendix A of \rcite{Lee:2023nkx}.
}

The time axis of the lab frame is characterized by
a unit timelike four-vector $l^\m = (1,\vec{0})$.
The gauge-fixing functions for the translational and rotational sectors are
\begin{align}
    \label{gf-lab}
    \chi^0 
    \,=\, 
        -\frac{1}{m}\: l_\m\mem x^\m  
    \,,\quad 
    \chi^a 
    \,=\,
        l_\mu\mem \Lambda^{\mu a} 
    \,,
\end{align}
respectively.
In particular, 
the spin gauge-fixing condition $\chi^a = 0$ sets that
\begin{align}
    \Lambda^\mu{}_0 
    \,=\,
        l^\mu \,,
    \quad 
    \BB{
        p^\mu + m\mem l^\mu
    }\mem S_{\mu\nu}\mem \Lambda^{\nu}{}_a 
    \,=\,
        0\,.
    \label{lab-constraint}
\end{align}
After the $1+3$ decomposition,
the second equation in \eqref{lab-constraint} describes
\begin{align}
    (E + m)\mem S_{0i} + p^j S_{ji} \,=\, 0
    \qfq 
    S_{0i} 
    \,=\,
        \frac{S_{ij}\hem p^j}{E+m} 
    \,=\,
        - \frac{(\vec{S}\times \vec{p})_i}{E+m} 
    \,,
    \label{S0i-how}
\end{align}
where we have defined
\begin{align}
    S_i 
    \,:=\,
        \frac{1}{2}\: \ve_{ijk}\mem S^{jk}
    \,, 
    \quad 
    E := \sqrt{\vec{p}^2+m^2} \,.
\end{align}
Similarly,
the temporal and spatial components of the vector $b^\mu$ in \eqref{fb} are
\begin{align}
\label{b-vector-PN}
    b^0 \,=\, \frac{1}{m^2}\: S^{0i}\mem p_i \,=\, 0 
    \,,\quad 
    b^i \,=\, \frac{1}{m^2}\: \BB{
        S^{i0} p_0 + S^{ij}\hhem p_j
    }
    \,=\,
        -\frac{(\vec{S}\times \vec{p})^i}{m(E+m)} 
    \,.
\end{align}
It could be also helpful to 
note that the physical spin length vector $\hy^\m$ 
is given by
\cite{Lee:2023nkx} 
\begin{align}
\label{y-vector-PN} 
    - \hy^0 
    \,=\,
        \frac{\vec{p} \md \vec{S}}{m} 
    \,,\quad 
    -\hy^i
    \,=\,
        \frac{S^i}{m}
        \mem+\mem
        \frac{p^i\, (\vec{p}\md\vec{S})}{m^2(E+m)} 
    \,.  
\end{align}

\para{1+3 Decomposition in Free Theory} 

Next, we review the derivation of the ``non-covariant'' action
at the level of free theory.

The gauge-fixing function $\chi^0$ in \eqref{gf-lab}
essentially states that
the worldline parameter
will be taken as the lab time coordinate $x^0 = t$.
By resolving the mass-shell constraint as well,
the symplectic potential of the translational sector
is boiled down to
\begin{align}
    \label{labsplit.px}
    p_\mu\mem dx^\mu 
    \,=\,
        p_i\mem dx^i - \sqrt{\vec{p}^2+m^2}\mem dt 
    \,,
\end{align}
which identifies $\sqrt{\vec{p}^2+m^2}$ as the Hamiltonian.


For the rotational sector,
the $1\mplus3$ decomposition describes
\begin{align}
    \label{labsplit.LS}
    \frac{1}{2}\, 
        S_{\m\n}\mem \L^\m{}_A\mem d\hnem\L^{\n A}
    \,=\,
    \frac{1}{2}\, 
        S_{ij}\mem \L^i{}_A\mem d\hnem\L^{j A}
    \,,
\end{align}
as the canonical spin gauge-fixing in \eqref{lab-constraint} mandates that the body frame can only exhibit $\SO(3)$ rotations transverse to the time axis $l^\m$.
Hence \eqref{labsplit.LS} is immediate
despite the complicated expression for $S_{0i}$ in \eqref{S0i-how}.

With these understandings,
the ``non-covariant'' action is found as
\begin{align}
    \label{Snc}
    \mathcal{S}^\circ \,=\,
    \int dt\,\,
    \bb{
        p_i\mem \dot{x}^i
        \mem-\mem
            \frac{1}{2}\: S_{ij}\mem \Omega^{ij}
        \mem-\mem
            \sqrt{\vec{p}^2+m^2}
    }
    \transition{where}
    \Omega^{ij} 
    \,:=\,
        \Lambda^i{}_a\mem \dot{\Lambda}^{ja}
    \,.
\end{align}
In this subsection, the overdot denotes $d/dt$.
It is instructive to derive the free EoM
from the non-covariant action in \eqref{Snc}:
$\dot{x}^i = p^i\nem/E$,
$\dot{p}_i = 0$,
$\dot{S}_{ij} = 0$,
and $\Omega_{ij} \propto S_{ij}$.

\para{1+3 Decomposition in Interacting Theory}

Now we incorporate the gravitational interactions.
In the PN effective field theory formalism
pioneered by Levi and Steinhoff \cite{Levi:2015msa},
the metric is parametrized by 
the temporal Kaluza-Klein variables
put forward by Kol and Smolkin \cite{Kol:2007bc}:
\begin{subequations}
\begin{align}
    ds^2 
    \:=\:
        - \mathe^{2\phi}\mem
            \BB{
                dt - A_i\mem dx^i
            }^2 
        + \mathe^{-2\phi}\, \sigma_{ij}\mem dx^i\hem dx^j 
    \,.
\end{align}
The effects of $\sigma_{ij}$ can be safely ignored
at the lowest orders in $v^i \sim \dot{x}^i$.
Hence, for the purposes of our discussion, it suffices to take
\begin{align}
    \label{metric-low}
    ds^2 
    \:\approx\:
        - \mathe^{2\phi}\mem
            \BB{
                dt - A_i\mem dx^i
            }^2 
        + \mathe^{-2\phi}\, d\vec{x}^2
    \,,
\end{align}
\end{subequations}
together with the conditions
$\dot{\phi} \approx 0$ and $\dot{A}_i \approx 0$. 

Consequently,
the $1+3$ decomposition of the vierbein will be
\begin{subequations}
\label{decomp.fields}
\begin{align}
    \label{decomp.vier}
    e^0 
    \,=\,
        \mathe^{\phi}\mem
        \BB{
            dt - A_i\mem dx^i
        }
    \,,\quad
    e^i
    \,=\,
        \mathe^{-\phi}\mem dx^i
    \,.
\end{align}
Computation shows that the spin connection is given by
\begin{align}
\begin{split}
    \label{decomp.sc}
    \gamma^0{}_i 
    \,&=\, 
        \mathe^\phi\mem
        (\partial_i \phi)\mem
            e^0 
        - \frac{1}{2}\: 
            \mathe^{3\phi}\mem F_{ij}\mem e^j 
    \,,\\
    \gamma_{ij} 
    \,&=\, 
        \mathe^\phi\mem
        \BB{
            (\partial_i \phi)\mem e_j - (\partial_j \phi)\mem e_i
        }
        - \frac{1}{2}\: 
            \mathe^{3\phi}\mem 
                F_{ij}\mem e^0
    \,,
\end{split} 
\end{align}
\end{subequations}
where $F_{ij} = \partial_i A_j - \partial_j A_i$ denotes the field strength for the ``graviphoton'' field.

Based on \eqref{decomp.vier},
the translational sector 
of the gravity-coupled symplectic potential
is boiled down to
\begin{align}
\begin{split}
    \label{ncs.mass}
    p_m\mem e^m{}_\m(x)\mem dx^\mu 
    \,&\approx\,
            \vec{p} \mdot d\vec{x} 
        \,-\,
        \sqrt{\vec{p}^2+m^2}\:
        (1 + \phi)
            \BB{
                dt - \vec{A} \mdot d\vec{x}
            }
    \,,\\
    \,&\approx\,
        \BB{
            \vec{p} \mdot d\vec{x} 
            - \sqrt{\vec{p}^2+m^2}\: dt
        }
        + \BB{
            - m\mem\phi
            + m\hem \vec{A} \mdot d\vec{x}
        }
    \,,
\end{split}
\end{align}
where we have consistently ignored
terms irrelevant for the leading-order calculation.
\eqref{ncs.mass} identifies
the ``point-particle part'' of the interaction action as
\begin{align}
    \label{vertices-mass}
    \mathcal{S}_\mass
    \,=\,   
        \int dt\,\,
        \bb{
            - m\mem\phi
            + m\hem \vec{A} \mdot \dot{\vec{x}}
        }
    \,.
\end{align}


On the other hand,
the rotational sector of the gravity-coupled symplectic potential
describes
$
    \frac{1}{2}\:
        S_{mn}\mem \Lambda^{m A}\mem D\Lambda^n{}_A 
    \,=\,
    \frac{1}{2}\:
        S_{mn}\mem \Lambda^{m A}\mem d\Lambda^n{}_A 
    - \frac{1}{2}\:
        S_{mn}\mem \gamma^{mn}{}_\r\mem dx^\r
$,
which identifies the ``spin part'' of the interaction action as
\begin{align}
    \mathcal{S}_\spin
    \,=\,
    \int dt\,\,
        \bb{
            \hla{{
                -S_{0i}\mem
                    \gamma^{0i}{}_0
            }}
            \hlx{
                - \frac{1}{2}\: 
                    S_{ij}\mem  
                    \gamma^{ij}{}_k
                    \mem \dot{x}^k
            }
            - \frac{1}{2}\: 
                S_{ij}\mem
                \gamma^{ij}{}_0
        }
    \,.
\end{align}
By inserting \eqref{decomp.sc}
and recalling some basic facts concerning the canonical gauge in \eqref{S0i-how},
we arrive at the desired spin vertex factors:
\begin{align}
    \label{vertices-spin}
    \mathcal{S}_\mathrm{spin} 
    \,=\,
        \int dt\,\, \bb{
            \bb{
                \hla{\frac{1}{2}} \mem+\mem \hlx{1} 
            }\,
                \ve^i{}_{jk}\,
                \partial_i\phi\,
                S^j\mem \dot{x}^k 
            \,+\, 
                \frac{1}{4}\: 
                    \ve^{kij}\hhem S_k\hem F_{ij}
        }
    \,.
\end{align}
Note that the factors $\mathe^{\phi}$ and $\mathe^{3\phi}$ in \eqref{decomp.sc}
can be safely ignored
when deriving this expression.

\para{Leading Spin-Orbit Coupling}

The propagators of PN gravity are given by
\begin{align}
\begin{split}
    \label{PNprops}
    \expval{
        \phi(\vex_1)\mem
        \phi(\vex_2)
    }
    \,=\,
        - \frac{G}{r}
    \,,\quad
    \expval{
        A_i(\vex_1)\mem
        A_j(\vex_2)
    }
    \,=\,
        \frac{4G}{r}\,
            \delta_{ij}
    \,.
\end{split}
\end{align}
From the above propagators
and
the leading-order interaction actions in
\eqrefs{vertices-mass}{vertices-spin},
we obtain the PN effective Hamiltonian
encoding
the leading-order spin-orbit coupling:
\begin{align}
    H_\mathrm{SO}^\mathrm{LO} \,=\,
        -\frac{Gm_2}{r^2}\:
            \vec{S}_1 \cdot \bb{
                \frac{3}{2}\, \vec{v}_1 \mtimes \hat{n} 
                - 2\mem \vec{v}_2 \mtimes \hat{n}
            }
        + (1\mlra2)
    \,.
\end{align}

Compared to the original derivation in \rcite{Levi:2015msa}, our computation is more streamlined in a few ways. 
We need not deal with any acceleration term $(a^i \eqq \ddot{x}^i)$ or a redefinition of the position variable $x^i$. 
We stay in the canonical gauge throughout; there is no need for switching between the two gauge choices during the derivation.
The absence of an acceleration term in the leading spin vertex factors, \eqref{vertices-spin}, keeps affecting higher order terms in the PN expansion. 
For example, the $\vec{a}_1$ term in Eq.\:(6.14) of \rcite{Levi:2015msa} does not appear in our framework. 

\para{Fictitious Vertices in Feynman Rules}

Yet another implication of our work on the PN expansion is that 
the \textit{fictitious} Feynman vertices involving the $\fb$ vector are unavoidable, 
as is clearly stated in \eqref{genGR.theta}.
The fictitious terms
are present in the canonical spin gauge.
The only spin gauge in which the fictitious terms are absent
is the covariant spin gauge.

For a simple example where the fictitious vertices cannot be avoided, consider the computation of the NNLO spin-squared potential in \rcite{Levi:2015ixa}. When extracting the spin-squared vertices from the worldline action, we should include the $\fS \fb$ term and the $\fb \fb$ term in \eqref{fthgr}, as well as the quadrupolar coupling with the constant coefficient $C_2$ in \eqref{GR.sp2}, with the $1\mplus3$ decomposition specified in \eqref{S0i-how} and \eqref{b-vector-PN}. 
If we proceed to the spin-cubic potential, we should include the $\fS \fb \fb$ term and the $\fb^3$ term in \eqref{fthgr}, the fictitious coupling descending from the quadrupolar coupling in \eqref{GR.sp2}, 
and the genuine cubic coupling we are not discussing in this paper. 
The pattern will continue with a proliferation of fictitious vertices at higher orders. 

\subsection{Higher Multipoles and Beyond}
\label{APPL>POLES}

\para{Linear Couplings}

As is glimpsed in \eqrefs{EM.sp1}{GR.sp2},
our universal frameworks in \Secs{INT1}{INT2}
readily generalize to all multipole orders.
For electromagnetism,
an ansatz that conforms to the physical principles in \eqref{bootstrap.EM} is
\begin{align}
    \label{EM.spell}
    \omega
    \,=\,
        \omega^\circ
        \,+\,
        \frac{1}{2}\,
            qF_{\m\n}(x)\,
            dx^\m \swedge dx^\n
        \,+\,
        q\, d\mem\bbsq{\,
        \sum_{\ell=1}^\infty\mem
        \frac{c_\ell}{\ell!}\,\mem
            {*^\ell}\hnem F_{\r_1\s,\r_2\cdots\r_\ell}(x)\,
            \hy^{\r_1}\hy^{\r_2}\cdots\hy^{\r_\ell}\,
            dx^\s
        \mem}
        \vphantom{\bigg|}
    \,,
\end{align}
where $*^\ell$ means to apply the Hodge star $\ell$ times.
\eqref{EM.spell} achieves
the most general linear electromagnetic coupling,
where
the coefficient $c_\ell$ parameterizes
the spin-induced $2^\ell$-pole moment.
Similarly,
an ansatz that conforms to the physical principles in \eqref{bootstrap.GR} is
\begin{align}
    \label{GR.spell}
    \omega
    \,\,=\,\,
        \omega^\bullet
        \,+\,
            p_m\mem {\star}R^m{}_n\mem \hy^n
        \,+\,
        d\mem\bbsq{\,
        \sum_{\ell=2}^\infty\mem
        \frac{C_\ell}{\ell!}\,\mem
            p_m\,
                {\star^\ell}\hnem R^m{}_{r_1r_2s;r_3;\cdots;r_\ell}\hnem(x)\,
            \hy^{r_1}\hy^{r_2}\hy^{r_3}\nem\cdots\hy^{r_\ell}\,
                e^s
        \mem}
        \vphantom{\bigg|}
    \,.
\end{align}
\eqref{GR.spell} is the transcription of the Levi-Steinhoff action \cite{Levi:2015msa}
in the symplectic perturbation framework,
which achieves
the most general linear-in-Riemann gravitational coupling.
The Wilson coefficient $C_\ell$
parameterizes the spin-induced mass $2^\ell$-pole moment.

Note that the Hodge star implements parity invariance.
Thus
$c_{2k}$ and $C_{2k}$ are electric moments,
while
$c_{2k+1}$ and $C_{2k+1}$ are magnetic moments.

\para{Nonlinear Couplings}

The above multipole coefficients $c_\ell$ and $C_\ell$
dictate only the linear couplings
to external fields.
At the nonlinear level,
there are broad possibilities for generalization.
We may want to illuminate two categories of terms,
in particular:
\begin{subequations}
\begin{enumerate}
    \item 
        Ordinary terms
        ($dx$, $e$):
        \begin{align}
        \label{nlv.contact}
        \smash{
            \kern-0.3em
            \frac{q^2}{m^2}\mem
            (\hem{
                \hy^\m F_{\m\r}(x)\hem F^\r{}_\n(x)\mem \hy^\n
            }\hem)
            \mem 
            (\hem{
                p_\k\mem dx^\k
            }\hhem)
            \,,\,\,\,
            p\mem R(x)\hhhem R(x)\mem \hy\hy\hy\hy\mem e
            \,.
            \kern-0.3em
        }
        \end{align} 
    \item 
        Spin-derivative terms
        ($d\hy$, $D\hy$):
        \begin{align}
        \label{nlv.spin}
        \smash{
            qF_{\m\n}(x)\mem \hy^\m\mem d\hy^\n
            \,,\quad
            p_m\hem {*} R^m{}_{nrs}(x)\mem \hy^n \hy^r D\hy^s
            \,,\quad
            p\, {*}R(x)\hem R(x)\mem \hy\hy\hy\hy\mem D\hy
            \,.
        }
        \end{align}
\end{enumerate}
\end{subequations}
\vspace{-4pt}
Here, 
we have enumerated
perturbations on the symplectic potential.

The systematic enumeration of all possible terms
at a given order
is an important problem,
although it goes beyond the present paper's scope.
Generally speaking, this counting will depend on
the field basis choice
(symplectic versus Hamiltonian perturbations),
while computing scattering amplitudes can provide some guidance.
As a reminder, our symplectic perturbation approach
specifies a particular worldline field basis
by fixing the mass-shell constraint in the free-theory form $\frac{1}{2}\mem (\eta^{mn} p_m p_n + m^2)$;
see \App{REVIEW>QMPD} for a related discussion.

The above two classes of terms
play different roles in worldline perturbation theory.
Consider the diagrammatic computation of two-quanta Compton amplitudes,
for instance.
The example terms listed in \eqref{nlv.contact}
contribute
to the amplitude
as \textit{contact vertices}.
In contrast,
the example terms in \eqref{nlv.spin}
contribute essentially
via \textit{exchange channels}
due to propagation of worldline fluctuations,
as the background worldline
will exhibit $\dot{\hy} = 0$.

This point should also justify
the identification of the spin-derivative terms in \eqref{nlv.spin} as ``nonlinear couplings.''
Although
some of them are linear in the curvature tensors,
they are still nonlinear in the sense that 
they do not contribute to the three-point amplitudes with one massless quantum.
Their effects on the classical EoM,
however,
could be just as significant as
the ordinary terms
at the same curvature order.

As shown by \rrcite{probe-nj,njmagic.1,njmagic.11},
spin-derivative terms are \textit{necessary}
for achieving black hole couplings
(see \eqref{rkerr}).
Also, they may not be so fundamentally different than the ordinary terms
in the spinspacetime perspective of \Sec{APPL>BH}.

\para{Definition of Black Hole Coupling}

Physically, the nonlinear couplings will encode various tidal deformability effects, taking different values for each astrophysical object. 

A particularly intriguing case is black holes,
which have attracted considerable attention in the current literature.
The determination of the exact black hole coupling to all orders
has been a long-standing problem.
The central question reads
\begin{center}
    \vphantom{.}\llap{``\:}\textit{What defines black holes}
    among all massive \\ spinning objects
    in the point-particle effective theory?\rlap{\:''}
\end{center}
The problem stands unambiguously solved
at least 
in the following subsectors.
\begin{enumerate}
    \item 
        In \textsc{Linearized Gravity}:
        The Kerr black hole is defined by
        the \textit{unity of multipole coefficients},
        $C_\ell = 1$ for all $\ell = 2,3,4,\cdots$
        \cite{%
            Hansen:1974zz,Newman:1965tw-janis,Hernandez:1967zza,Thorne:1980ru%
        }
        (cf. \rrcite{ahh2017,Guevara:2018wpp,Guevara:2019fsj,chkl2019,aho2020}).
    \item
        In \textsc{Self-Dual Gravity}:
        The Kerr black hole is defined by
        a \textit{superintegrability}
        in type-D backgrounds
        with Killing-Yano tensors
        \cite{probe-nj}
        (cf. \rrcite{Johansson:2019dnu,Aoude:2020onz,Lazopoulos:2021mna}).
\end{enumerate}
We have reviewed this development in \App{NJB}
as a preliminary for \Sec{APPL>BH}.
If a worldline model passes the above tests (a) and (b),
then it qualifies as an effective description of the Kerr black hole
according to our best theoretical understanding 
at the current moment.

Criterion (a) is necessary for reproducing the Newman-Janis \cite{Newman:1965tw-janis} property
in linearized gravity.
The Newman-Janis shift
describes that 
the Kerr black hole is
in some sense
a Schwarzschild black hole
displaced into ``complex spacetime''
along an imaginary direction
set by its spin length pseudovector.
Criterion (b) arises via a nonlinear generalization of this property
in curved backgrounds \cite{probe-nj}.

The problem of constructing worldline models for black holes
may have been approached in model-specific manners
so far:
spherical top model \cite{gmoov},
massive twistor model \cite{njmagic.1,njmagic.11},
fermionic model \cite{bonocore2025higher},
etc.
Meanwhile, \rcite{probe-nj} has explicated
the model-independent nature of the problem 
by introducing a probe counterpart of the Newman-Janis \cite{Newman:1965tw-janis} algorithm
in terms of the universal variables $(x,\hy,p)$
and providing
the definition of black holes in terms of global symmetries
(criterion (b) above).

In this paper, we hope to
illuminate this model-independent aspect
again.
For a quick demonstration,
we will concern the electromagnetic analog of the Kerr black hole:
The zero-gravitation limit 
$G_\mathrm{N} \too 0$
of the Kerr-Newman solution,
dubbed the {\Kerr} solution
\cite{aho2020,Lynden-Bell:2002dvr}.
The {\Kerr} solution
exhibits
the unity of multipole coefficients 
\cite{Newman:1965tw-janis,Newman:1973yu,Lynden-Bell:2002dvr}:
$c_\ell \eqq 1$ for all $\ell \eqq 1,2,3,\cdots$.
Its dynamics as a probe
enjoys an analogous superintegrability.

\subsection{Black Holes}
\label{APPL>BH}

\para{Hidden Complex Geometry of Universal Phase Space}

Recall \Sec{FREE>PART},
where we reviewed the fact \cite{sst-asym} that
the universal Poisson bracket relation in \eqref{M3}
holds as a direct implication of
Poincar\`e symmetry.

In fact,
we can now disclose to the reader that
the universal Poisson bracket relation
admits a remarkably simpler presentation \cite{sst-asym}.
Consider the complex combination,
\begin{align}
    \label{zcoords}
    z^\m
    \,=\,
        x^\m \mem+\hem i\hy^\m
    \,.
\end{align}
Straightforward algebra shows that \eqref{M3}
is equivalent to
\begin{subequations}
\label{M3-in-zp}
\begin{align}
    \label{zp-canonical}
    \pb{z^\m}{z^\n}^\circ
    \,\,&=\,\,
        0
    \,,\quad
    \pb{z^\m}{p_\n}^\circ
    \,=\,
        \delta^\m{}_\n
    \,,\quad
    \pb{p_\m}{p_\n}^\circ
    \,=\,
        0
    \,,\\
    \label{zzpb-vec}
    \pb{z^\m}{\bz^\n}^\circ
    \,\,&=\,\,
        \frac{1}{p^2}\,
        \BB{
            \delta^\m{}_\r\mem \delta^\n{}_\s
            + \delta^\m{}_\s\mem \delta^\n{}_\r
            + i\mem \ve^{\m\n}{}_{\r\s}
        \nem}\,
            (z^\r {\mem-\,} \bz^\r)\, p^\s
    \,,
\end{align}
\end{subequations}
which fits on two lines.
Many components are set to zero.
This rewriting
not only provides a succinct summary
of \eqref{M3}
but also reveals that
the the components of $z^\m$ in \eqref{zcoords}
are Poisson-commutative,
despite the peculiar noncommutativity of 
the physical center
coordinates $x^\m$
in \eqref{M3}.

Following \cite{sst-asym},
we refer to $z^\m$ as \textit{spinspacetime coordinates},
as they unify spacetime $x^\m$ and spin $\hy^\m$
as real and imaginary parts.
Mathematically,
they are the holomorphic coordinates
for the complexified Minkowski space $\mflat^\C = (\C^4,\eta^\C)$
whose real section is the flat spacetime $\mflat = (\R^4,\eta)$.
The property that
\begin{align}
    \label{zigzag}
    \pb{z^\m}{z^\n}^\circ
    \,=\,0
    \,,\quad
    \pb{z^\m}{\bz^\n}^\circ
    \,\neq\,0
    \,,\quad
    \pb{\bz^\m}{\bz^\n}^\circ
    \,=\, 0
\end{align}
is referred to as the zig-zag structure,
as the nicknames ``zig'' and ``zag''
are used for referring to
``holomorphic''\:($z$) and ``anti-holomorphic''\:($\bz$),
respectively
\cite{sst-asym}.\footnote{
    This terminology is inspired by Penrose
    \cite{penr04-zigzag}.
}

It should be evident that
spinspacetime and its zig-zag structure exist \textit{universally} in
every special-relativistic massive spinning system.
In fact, 
glimpses toward spinspacetime
had occurred in \eqrefss{SO.thetac}{Gpformula}{incidence}:
the models with less redundancies.

\para{Inherent Association Between Holomorphy and Self-Duality}

The idea of spinspacetime 
traces back to
Newman and Winicour \cite{newman1974curiosity}.\footnote{
    See also  
    \rrcite{newman1974collection,newman1988remarkable,Newman:1973afx,Newman:2004ba,Newman:1973yu,Newman:2002mk,ko1981theory,grg207flaherty},
    which portray
    Newman's serious take on spinspacetime.
}
The observation was that
the self-dual part of the total angular momentum in \eqref{JD} is
\begin{align}
    \label{OSD}
    J
    \,=\,
        (x \wedge p)
        + {*}(\hy \wedge p)
    \qiq
    J^+
    \,=\,
        (z \wedge p)^+
    \,.
\end{align}
Here, the $+$ superscript
signifies self-dual projection by
$\frac{1}{2}\mem ( 1 \mminus\mem i\mem{*} )$.
The Hodge duality ${*}$ in \eqref{OSD} turns into the imaginary unit $+i$ in the self-dual sector,
so holomorphic coordinates $z^\m = x^\m + i\hy^\m$
arise from the self-dual angular momentum $J^+$.
In the same way, anti-holomorphic coordinates $\bz^\m = x^\m - i\hy^\m$
arise from the anti-self-dual angular momentum $J^-$.
Importantly, an inherent association arises between 
\textit{holomorphy and self-duality}
in spinspacetime,
consistently with the fact that $\hy^\m$ is a \textit{pseudo}vector.


Newman and Winicour \cite{newman1974curiosity}'s discussion on spinspacetime,
however,
does not examine
its Poisson structure
in \eqref{M3-in-zp}.
The modern reboot of the spinspacetime program \cite{sst-asym}
proposes that one should pay attention to the Poisson structure of spinspacetime
and its zig-zag property,
as they are physical, universal structures.

For instance,
it is instructive to employ the spinor notation,
in which case \eqref{M3-in-zp} is boiled down to a further succinct form:
\begin{align}
    \label{zzpb}
    \pb{z^{\da\a}}{p_\wrap{\b\db}}^\circ
    \,=\,
        \delta^\da{}_\wrap{\db}\mem \delta_\wrap{\b}{}^\a
    \,,\quad
    \pb{z^{\da\a}}{\bz^{\db\b}}^\circ
    \,=\,
        \frac{1}{-p^2}\,
            (z \mminus \bz)^{\da\b}\hem p^{\db\a}
    \,.
\end{align}
\eqref{zzpb} is a one-line summary of the entirety of universal Poisson brackets in \eqref{M3}.
\newpage

The chiral nature of the zig-zag bracket
$\pb{z^{\da\a}}{\bz^{\db\b}}^\circ$ in \eqref{zzpb}
reflects the aforementioned association between 
holomorphy and self-duality,
which
dictates the precession behavior of spinning particles under influence of external fields
and derives the probe-level Newman-Janis shift
from an ideal spin precession behavior
\cite{ambikerr1,sst-asym}.

\begin{figure}[t]
    \centering
    \includegraphics[scale=1.5, valign=c]{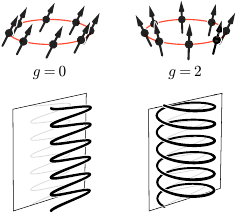}
    \caption{
        Spinspacetime offers a new angle on spin precession.
        The Dirac gyromagnetic ratio $g \eqq 2$
        can be geometrically characterized
        as the synchronization of
        orbital and spin precessions in external fields.
        When seen from spinspacetime,
        such a motion appears as an isotropic spiral trajectory,
        pointing to the equal blending of all $dx, d\hy$ components in $dz \swedge dz$.
        In contrast, the traditional minimal coupling $g \eqq 0$
        translates to a flattened sinusoidal curve,
        which points to 
        the reality of $dx \swedge dx$.
        Spinspacetime identifies $g \eqq 2$
        as the simplest coupling.
    }
    \label{g=0-g=2}
\end{figure}

\para{The Simplest Symplectic Perturbation from Spinspacetime}

Spinspacetime opens a new chapter in relativity, offering a unique perspective on spinning-particle mechanics in four dimensions. It is now not only \textit{space} and \textit{time} but also \textit{spin} that are unified. 
This invites a different way of thinking:
spin is the imaginary part of spacetime.
Spacetime and spin may not be fundamentally distinct.
\fref{g=0-g=2} invites this shift.

An electromagnetic field on spacetime
takes the form $\frac{1}{2}\, F_{\m\n}(x)\mem dx^\m \swedge dx^\n$.
Spacetime $\mflat$ is embedded in spinspacetime $\mflat^\C$ as the real section.
Spacetime fields would somehow permeate into spinspacetime.
This will describe
$dz \swedge dz$,
$dz \swedge d\bz$,
or
$d\bz \swedge d\bz$
components.

For 
charged scalar particles,
electromagnetic fields
are symplectic perturbations
(Sec. \ref{INT>SPT}).
For 
charged spinning particles,
\textit{spinspacetime fields}
will be
symplectic perturbations.

For scalar particles, 
$\omega' = \frac{1}{2}\, qF_{\m\n}(x)\mem dx^\m \swedge dx^\n$
is said to be the minimal coupling
because it is the simplest symplectic perturbation
that one can possibly write down.
For spinning particles,
the simplest symplectic perturbation on \textit{spinspacetime}
will declare the true minimal coupling.

The spacetime field
splits into self-dual and anti-self-dual parts as
\begin{align}
    \label{Fsplit}
    \frac{1}{2}\, F_{\m\n}(x)\mem dx^\m \swedge dx^\n
    \,=\,
        \frac{1}{2}\, F^+_{\m\n}(x)\mem dx^\m \swedge dx^\n
    +
        \frac{1}{2}\, F^-_{\m\n}(x)\mem dx^\m \swedge dx^\n
    \,.
\end{align}
When studying how spacetime fields permeate into spinspacetime,
it can be helpful to examine their self-dual and anti-self-dual parts separately.

Recalling the earlier association,
suppose the self-dual field
is \textit{holomorphic}
on spinspacetime:\footnote{
    Note that the argument of $F^+_{\m\n}(z)$ here had to be also holomorphic,
    due to the principle of symplecticity (closure $d\omega' = 0$).
}
\begin{align}
    \label{sp.heaven}
    \omega'
    \,=\,
        \frac{1}{2}\, 
            qF^+_{\m\n}(z)\mem dz^\m \swedge dz^\n
    \,.
\end{align}
Holomorphic symplectic perturbations are
the simplest symplectic perturbations
on spinspacetime,
in the sense that it minimally deforms the universal Poisson brackets in \eqref{zigzag}.
Holomorphic symplectic perturbations
preserve the Poisson-commutativity of holomorphic coordinates,
due to the very zig-zag structure in \eqref{zigzag}
(recall \eqref{ppert}):
\begin{align}
    \label{zz=0}
    \pb{z^\m}{z^\n}
    \,\,=\,\,
        \pb{z^\m}{z^\n}^\circ
        \mem-\mem
        \pb{z^\m}{z^\r}^\circ\,
            \hem qF^+_{\r\s}(z)
        \,\pb{z^\s}{z^\n}
    \,\,=\,\,
        0
    \,.
\end{align}
In this manner, only the zag-zag bracket
is deformed by the zig symplectic perturbation:
\begin{align}
    \label{zigzag'}
    \pb{z^\m}{z^\n}
    \,=\,0
    \,,\quad
    \pb{z^\m}{\bz^\n}
    \,\neq\,0
    \,,\quad
    \pb{\bz^\m}{\bz^\n}
    \,\neq\, 0
    \,.
\end{align}

We should understand the physics due to 
the simplest symplectic perturbation in \eqref{sp.heaven}.
By straightforward application of
the methodology in \Sec{INT>BMT},
the EoM are derived from the universal Poisson brackets in \eqref{M3-in-zp}---%
equivalently \eqref{zzpb}, if one likes to work with spinors---%
that
\begin{align}
    \label{eom.heavenly.bz}
    \dot{z}^\m
    \,=\,
        \hp^\m
    \,,\quad
    \dot{\bz}^\m
    \,=\,
        \hp^\m
        - 
        \frac{2iq}{m}\, F^+{}^\m{}_\n(z)\mem \hy^\n
    \,,\quad
    \dot{p}^\m
    \,=\,
        \frac{q}{m}\, F^+{}^\m{}_\n(z)\mem p^\n
    \,.
\end{align}
We see that
the EoM for the zig coordinates are
completely ignorant of the external field,
due to the very Poisson-commutativity shown in \eqref{zz=0}:
$\dot{z}^\m = \hp^\m$.

\eqref{eom.heavenly.bz} translates to
\begin{align}
    \label{eom.heavenly}
    \dot{z}^\m
    \,=\,
        \hp^\m
    \,,\quad
    \dot{\hy}^\m
    \,=\,
        \frac{q}{m}\, F^+{}^\m{}_\n(z)\mem \hy^\n
    \,,\quad
    \dot{p}^\m
    \,=\,
        \frac{q}{m}\, F^+{}^\m{}_\n(z)\mem p^\n
    \,,
\end{align}
which describes that
the orbital ($\hp^\m$) and spin ($\hy^\m$) precessions are perfectly synchronized
by the cyclotron angular frequency 
$(q/m)\mem F^+{}^\m{}_\n(z)$.
Recalling the preliminary exploration in \fref{g=0-g=2},
this synchronization implies the gyromagnetic ratio $g \eqq 2$, which is $c_1 \eqq 1$.\footnote{
    If self-dual fields were anti-holomorphic in spinspacetime,
    then the particle exhibits $g = -2$,
    which deviates from the black hole behavior.
    See also \rcite{sst-asym} for 
    a physical interpretation of
    the synchronization of orbital and spin precessions
    for gravitational interactions
    (``spinning equivalence principle'').
}
In fact,
we will see shortly that
all multipole coefficients are unity:
$c_\ell \eqq 1$ for all $\ell \eqq 1,2,3,\cdots$.


\para{The Simplest Real Symplectic Perturbation from Spinspacetime}

The symplectic perturbation in \eqref{sp.heaven}
is formal,
giving rise to a complexified Poisson structure as in \eqref{zigzag'}.
For a real 
dynamics,
we add the missing complex conjugate
so that the anti-self-dual field is restored:
\begin{align}
    \label{sp.rkerr}
    \omega'
    \,&=\,
        \frac{1}{2}\:
            qF^+_{\m\n}(z)\mem dz^\m \swedge dz^\n
        +
        \frac{1}{2}\:
            qF^-_{\m\n}(\bz)\mem d\bz^\m \swedge d\bz^\n
    \,.
\end{align}
By methods in differential geometry \cite{gde},
it follows that \eqref{sp.rkerr}
describes
\cite{njmagic.1,njmagic.11}\footnote{
    To elaborate,
    the potential one-form for 
    \eqref{sp.rkerr} can be represented as
    $\theta' = \cos(\pounds_N)\mem (A_\m(x)\mem dx^\m) + \sinc(\pounds_N)\mem ({*}F_{\m\n}(x)\mem \hy^\m\mem dx^\n)$
    via $F^\pm = (1 \mp i\mem {*})F /2$,
    where $N$ is a vector field such that
    $\i_N dx^\m = \hy^\m$,
    $\i_N d\hy^\m = 0$,
    and
    $\i_N dp_\m = 0$.
}
\begin{align}
\begin{split}
    \label{rkerr}
    \omega'
    \,=\,
    {}&{}
        \frac{1}{2}\,
            qF_{\m\n}(x)\,
            dx^\m \swedge dx^\n
        \,+\,
        q\, d\mem\bbsq{\,
        \sum_{\ell=1}^\infty\mem
        \frac{1}{\ell!}\,\mem
            {*^\ell}\hnem F_{\r_1\s,\r_2\cdots\r_\ell}(x)\,
            \hy^{\r_1}\hy^{\r_2}\cdots\hy^{\r_\ell}\,
            dx^\s
        \mem}
    \,,\\[-0.1\baselineskip]
    {}&{}
        \,+\,
        q\, d\mem\bbsq{\,
        \sum_{\ell=1}^\infty\mem
        \frac{(\ell{\,-\mem}1)}{\ell!}\,\mem
            {*^\ell}\hnem F_{\r_1\s,\r_2\cdots\r_{\ell-1}}(x)\,
            \hy^{\r_1}\hy^{\r_2}\cdots\hy^{\r_{\ell-1}}\,
            d\hy^\s
        \mem}
    \,.
\end{split}
\end{align}

Compare \eqref{rkerr} with \eqref{EM.spell}.
For the linear couplings,
we identify the unity $c_\ell \eqq 1$ of multipole moments.
Thus \eqref{rkerr} passes the first test of spinning black holes,
criterion (a).
Furthermore,
\eqref{rkerr} also describes
nonlinear couplings
via
\textit{spin-derivative terms},
which are necessary for passing the second test of spinning black holes.\footnote{
    For instance,
    it can be diagrammatically shown \cite{njmagic.11}
    that the resulting same-helicity Compton amplitudes
    exhibit spin exponentiation for all multiplicities.
    Thus \eqref{sp.rkerr}
    is the symplectic perturbation
    of the {\Kerr} spinning particle
    up to mixed-chirality terms
    (earthly deformations \cite{probe-nj})
    that vanish in the self-dual limit,
    such as $(q^2/m^2)\mem (\hy\mem F^+(x)\hem F^-(x)\mem\hy)\mem (p\mem dx)$.
}

The message here is that the simplest symplectic perturbation from spinspacetime
directly yields
the simplest (black hole) coupling.
This fact is obscured in typical spacetime-based approaches;
it is a unique insight that spinspacetime can provide.
Also, it should be clear that
this derivation of a black hole coupling
is model independent.

The traditional minimal coupling, $c_\ell \eqq 0$,
arises by
the simplest symplectic perturbation from the standards of spacetime:
$\frac{1}{2}\, qF_{\m\n}(x)\mem dx^\m \swedge dx^\n$.
The spinspacetime Poisson brackets in \eqref{M3-in-zp} shows that
this grossly complicates the dynamics of the particle,
as $dx \swedge dx$
contains all the
$dz \swedge dz$,
$dz \swedge d\bz$,
and
$d\bz \swedge d\bz$
components.

\para{Model-Independent Derivation of {\Kerr} EoM}

Via \eqrefs{ppert}{M3-in-zp},
it is easy to 
derive the complete {\Kerr} EoM 
due to the real symplectic perturbation in \eqref{sp.rkerr}:
\begin{align}
\begin{split}
    \label{eom.earthly}
    \dot{p}_\m
    \,&=\,
        qF^+_{\m\n}(z)\mem \dot{z}^\m
        + qF^-_{\m\n}(\bz)\mem \dot{\bz}^\m
    \,,\\
    \dot{z}^\m
    \,&=\,
        \hp^\m
        - \frac{2iq}{m^2}\,
        \BB{
            \hy^\m p^\n + p^\m \hy^\n
            -i\mem \ve^{\m\n\r\s} \hy_\r p_\s
        }\mem
        F^-_{\n\k}(\bz)\mem \dot{\bz}^\k
    \,.
\end{split}
\end{align}
The EoM for $\bz^\m$ follow by complex conjugation
since the dynamics is real.
The above equations have been known from the massive twistor model \cite{Kim:2024grz}.
Here, we have reproduced them in the model-agnostic fashion.
Note that \eqref{eom.earthly} describes \textit{all-orders-in-spin} EoM.

It is left as an exercise to
deduce the EoM for $x^\m$ and $y^\m$
by taking real and imaginary parts.
One may use an identity
$
\ve^{\m\n\r\s} \phi^\pm_{\n\k}\mem u^\k
= \mp\mem 3\hhem i\mem \phi^\pm{}^{[\r\s} u^{\m]}
$
to remove the epsilon tensors
if wanted,
where $\phi^+_{\m\n}$\,($\phi^-_{\m\n}$) is self-dual\:(anti-self-dual).


Overall, spinspacetime is a powerful tool
for making predictions on the all-orders-in-spin dynamics of spinning objects.

As demonstrated in \rcite{ambikerr1},
spinspacetime is also useful for
deriving the all-orders-in-spin EoM of particles with generic multipole moments.
In this case,
one perturbs away from the black hole coupling by
redefining the Wilson coefficients on the complex worldline.

\para{Dynamical Newman-Janis Shift of Conserved Charges}

As a glimpse toward the applications of the above EoM,
we provide a reproduction of
the fact \cite{njmagic.11} that 
an idealized version of the spinning black hole binary problem
exhibits superintegrability
via exact hidden symmetries.
We follow the model-independent presentation due to \rcite{probe-nj}.
This discussion confirms the criterion (b),
confirming that \eqref{rkerr} qualifies as an effective point-particle description of {\Kerr}.

The dream of a theoretical relativist
is to understand the exact dynamics of the
Kerr-Kerr black hole binary system.
However, the problem is too hard.
As a first simplification,
we suppose that one of the black holes is significantly heavier than the other:
the zeroth self-force limit.
As a second simplification,
we suppose the electromagnetic analog:
the {\Kerr}-{\Kerr} binary problem.

As a result, we study the motion of a {\Kerr} probe
in the background of the {\Kerr} solution.
Since Newman and Janis \cite{Newman:1965tw-janis},
it has been available that the field strength of the {\Kerr} solution takes the form
\begin{align}
    \label{bkg.F}
    F_{\m\n}(x)
    \,=\,
        \frac{Q}{8\pi|\vex \mminus i\vea|^3}\:
            {*}Y^+_{\m\n}(x)
    \mem+\mem
        \frac{Q}{8\pi|\vex \mplus i\vea|^3}\:
            {*}Y^-_{\m\n}(x)
    \,,
\end{align}
where $Q$ is the charge parameter,
$a^\m$ is a spacelike pseudovector encoding the ring radius,
and $u^\m$ is a timelike unit vector defining the stationary direction.
We have denoted
\begin{align}
    {*}Y^\pm_{\m\n}(x)
    \,=\,
        \bigbig{
            u \wedge (x\mp ia)
        }\hnem{}^\pm_{\m\n}
    \,,
\end{align}
which
encodes the position three-vector
from imaginary centers $\pm ia$.
As disclosed by \rcite{nja},
these centers are precisely
self-dual and anti-self-dual dyons.

The problem is still too hard.
With both self-dual and anti-self-dual fields,
the geometric series expansion of the Poisson bivector 
does not truncate at a finite order,
so \eqref{eom.earthly} describes time derivatives on its right-hand sides
unlike as in \eqref{eom.heavenly.bz}.

Hence we simplify the problem again by dropping
the anti-self-dual part of the field configuration in \eqref{bkg.F}:
\begin{align}
    \label{bkg.Fp}
    F_{\m\n}(x)
    \,=\,
        \frac{Q}{8\pi|\vex \mminus i\vea|^3}\:
            {*}Y^+_{\m\n}(x)
    \qiq
    {*}F_{\m\n}(x)
    \,=\,
        +i\mem F_{\m\n}(x)
    \,.
\end{align}
This extracts the self-dual part of the {\Kerr} background in the precise sense;
see \rcite{sdtn} for the gauge potential.
It is not difficult to see that \eqref{bkg.Fp} describes
\begin{align}
    \vec{E}(\vex)
    \,=\,
        i\hem \vec{B}(\vex)
    \,=\,
        \frac{Q}{8\pi}\mem
        \frac{\vex \mminus i\vea}{|\vex \mminus i\vea|^3}
    \,,
\end{align}
which is the electromagnetic field of a dyon
with electric charge $Q/2$ and magnetic charge $-i\hem Q/2$:
a self-dual dyon.
The magnetic dipole moment of {\Kerr} arises precisely like
a \textit{static} magnet made of monopoles:
$Q\vea = (-iQ/2)(2i\vea)$.
See \rcite{nja} for more details.


In short, we study
the motion of a {\Kerr} probe
in the background of a self-dual dyon.
The relevance to the original problem
is that the self-dual dyon
is a part of {\Kerr}.

In the self-dual dyon background,
we concern the combination
\begin{align}
    Y_{\m\n}(x)
    \,=\,
        Y^+_{\m\n}(x)
        \mem+\mem
        Y^-_{\m\n}(x)
    \,=\,
        \ve_{\m\n\r\s}\mem x^\r\hem u^\s
        - (u \wedge a)_{\m\n}
    \,.
\end{align}
It turns out that this defines a Killing-Yano tensor \cite{yano1952some} such that
\begin{align}
    Y_{\m\n,\r}(x)
    \,=\,
        \ve_{\m\n\r\s}\mem u^\s
    \,=\,
        Y_{[\m\n,\r]}(x)
    \,,\quad
    F^\m{}_\r(x)\mem Y^\r{}_\n(x)
    \,=\,
        Y^\m{}_\r(x)\mem F^\r{}_\n(x)
    \,.
\end{align}
The first equation is the definition statement,
while the second equation describes a commuting property with the field strength.

In this background,
the {\Kerr} probe
obeys
the complexified EoM in \eqref{eom.heavenly}:
\begin{align}
    \label{eom.heavenly.sdd}
    \dot{z}^\m
    \,=\,
        \hp^\m
    \,,\quad
    \dot{p}^\m
    \,=\,
        \frac{q}{m}\, F^\m{}_\n(z)\mem p^\n
    \,,\quad
    \dot{\hy}^\m
    \,=\,
        \frac{q}{m}\, F^\m{}_\n(z)\mem \hy^\n
    \,,
\end{align}
Notably, the first two equations here
are identical to those of the charged scalar particle
up to the complexification $x \to z$,
which originated from the Poisson commutativity of holomorphic spinspacetime in \eqref{zz=0}.
Thus the well-known discourse on the scalar particle \cite{Hughston:1972qf} 
is straightforwardly recycled as
\begin{align}
    \frac{d}{d\t}\mem\BB{
        Y^\m{}_\n(z)\mem p^\n
    }
    \,&=\,
        Y^\m{}_\n(z)\mem F^\n{}_\r(z)\mem p^\r
        + Y^\m{}_{\n,\r}(z)\mem p^\n\mem \hp^\r
    \,=\,
        F^\m{}_\n(z)\mem Y^\n{}_\r(z)\mem p^\r
    \,,
\end{align}
showing that the precessions of
$p$ and $Y(z)\mem p$ are synchronized.

Crucially,
the precession of the spin length pseudovector $\hy$ is also synchronized with $p$.
Now we find three vectors singing in unison,
$p$, $Y(z)\mem p$, and $\hy$.
It is then immediate that
\begin{align}
    \label{rkerr.RC}
    R \,=\, p_\m\mem Y^\m{}_\n(z)\mem \hy^\n
    \,,\quad
    C \,=\, - p_\m\mem Y^\m{}_\r(z)\mem Y^\r{}_\n(z)\mem p^\n
\end{align}
are conserved quantities:
the all-orders-in-spin completions of
the R\"udiger \cite{Rudiger:1981uu,Rudiger:1984er} and Carter \cite{Carter:1968ks} constants.

Moreover, the self-dual dyon background exhibits isometries $\R \times \SO(3)$.
Again, we immediately deduce the {\Kerr} probe's conserved charge 
by direct dynamical Newman-Janis shift of the Coulomb probe's conserved charge in \eqref{scalar.QEM}:
\begin{align}
    \label{rkerr.Q}
    Q
    \,=\,
        p_\m\mem K^\m(z) + \a(z)
    \,.
\end{align}
The symmetry algebra between these Newman-Janis shifted Killing charges
are the same as in the non-spinning case,
crucially because of the Poisson commutativity
$\pb{z^\m}{z^\n} = 0$
that we have had to emphasize.

Using
\eqrefs{rkerr.RC}{rkerr.Q},
one can show that
the dynamics is exactly solvable (integrable).
See \rrcite{probe-nj,njmagic.11} to grasp more details for the superintegrability.

In summary, we have learned that
the self-dual sector of the {\Kerr}-{\Kerr} binary problem is exactly solvable,
based on
direct dynamical Newman-Janis shifts of the non-spinning probe's conserved charges
as well as
the synchronization of orbital and spin precessions.

This is yet an idealized discussion.
How should one approach the original problem of the {\Kerr}-{\Kerr} binary system, then?

A reasonable pathway may be perturbing away from the self-dual sector
where exact all-orders expressions for the conserved charges are known:
the googly agenda \cite{probe-nj}.

\section{Summary}
\label{FIN}

\noindent
\begin{enumerate}[leftmargin=20pt,label={}\arabic*{.}]
    \item 
        The study of relativistic massive spinning particle models boasts a long history
        and has seen renewed attention
        in the context of point-particle effective theory of compact astrophysical objects.
        An extensive zoo of models has emerged,
        featuring different implementations of spin
        with varying degrees of redundancy.
        
\end{enumerate}

\begin{enumerate}[leftmargin=20pt,label={[}\arabic*{]}]
\setcounter{enumi}{1}
    \item
        Poincar\'e symmetry implies
        the existence of universal variables $(x,\hy,p)$
        in any Hamiltonian formulation of a free massive spinning particle,
        satisfying the universal Poisson bracket relation.
        The space of these variables admits
        ten- or eight-dimensional symplectic submanifolds
        tracing back to Souriau's elementary realization of spin.
        
    \item
        The existence of the universal variables $(x,\hy,p)$
        and their Poisson algebra
        is concretely substantiated in four example models:
        vector oscillator, spinor oscillator, spherical top, and massive twistor.
        The vector oscillator describes the spin phase space $T^*S^2$
        and admits the spinor oscillator as a reformulation with no spin redundancy.
        The spherical top describes the spin phase space $T^*\SO(3)$
        and admits the massive twistor as a reformulation with less spin redundancy.

    \item
        The hierarchy
        $T^*\SO(3) \supset T^*S^2 \supset S^2$
        of spin phase spaces
        arises by
        explicit identifications of the constraints
        in the symplectic reduction and Dirac bracket frameworks.

    \item
        The BMT equations are direct consequences of the universal Poisson bracket relation.
        Universal recipes for the minimal and dipolar couplings
        are identified
        in the symplectic perturbation framework.
        
\end{enumerate}

\begin{enumerate}[leftmargin=20pt,label={(}\arabic*{)}]
\setcounter{enumi}{5}
    \item 
        The equivalence principle asserts
        the existence of
        universal variables $(x,\hy,p)$
        in any Hamiltonian formulation of a massive spinning particle in general relativity, 
        which satisfy the universal covariant Poisson bracket relation.
        The MPD and QMPD equations are direct consequences of this covariant Poisson bracket relation.
        Universal recipes for the minimal and quadrupolar couplings
        are identified
        in the covariant symplectic perturbation framework.

    \item
        To ensure that
        physical observables are ignorant of one's choice of SSC,
        the spherical top model must be coupled to external fields
        in the spin gauge-invariant manner.
        The spin gauge-invariant variables,
        such as the general-relativistic physical center coordinates,
        describe highly nonlinear formulae
        when expressed in terms of the defining variables.
        Noncovariant SSCs necessitate adding infinite towers of fictitious interaction terms in the Lagrangian.
        The spin gauge algebra and flow
        are preserved by the interactions.

    \item
        The existence of the general-relativistic universal variables $(x,\hy,p)$
        and their Poisson algebra
        is explicitly proven in the four example models.

    \item
        Model-independent conserved charges
        are direct consequences of the universal Poisson and covariant Poisson bracket relations.

        In the context of PM expansion,
        equivalent models produce classical eikonals (Magnusians)
        that generate the same effect on the $(x,\hy,p)$ space.
        Models with less redundancy and linear phase space
        can be preferred in general.
        Spinorial models might be optimal for studying all-orders-in-spin dynamics of black holes.
        When reading off the Feynman rules in noncovariant SSCs,
        it is important to include the extra vertices due to the fictitious terms.
        Such fictitious vertices cannot be neglected
        in the PN effective theory computation of the NNLO spin-squared potential,
        for example.

        From the universal variables,
        one builds
        interaction Lagrangians encoding
        not only linear couplings to all multipole orders
        but also nonlinear couplings.
        Nonlinear couplings are incorporated
        in the forms of explicit contact terms or spin-derivative terms.

        The universal Poisson bracket relation admits a radically simple reformulation in terms of the complex spinspacetime coordinates $z = x \hem\mplus i\hy$
        which exhibits Poisson-commutativity $\pb{z}{z}^\circ = 0$.
        The simplicity of black hole couplings
        and
        the Newman-Janis shift
        are attributed to this universal complex geometry
        instead of model-specific arguments.
        The universal derivation of {\Kerr} EoM is given.
        An exactly solvable sector of the {\Kerr}-{\Kerr} binary problem is identified
        in an idealized setup
        via direct dynamical Newman-Janis shifts of conserved charges.
        
\end{enumerate}

\vfill
\begin{center}
\hrule
\bigskip
\begin{minipage}{0.932\linewidth}
{\noindent\large\bfseries Acknowledgements.}
\noindent
    We are grateful to Alessandro Georgoudis, Jung-Wook Kim, David Kosower, Kanghoon Lee, and Euihun Joung for discussions. 
    JHK would like to thank Thibault Damour for 
    stimulating conversations on spinning particle phase spaces
    and sharing his insightful and critical views.
    JHK is supported by the Department of Energy (Grant No.\:DE-SC0011632) and by the Walter Burke Institute for Theoretical Physics.
    The work of 
    SL is supported by National Research Foundation of Korea (NRF) grants,
    NRF-2023-K2A9A1A0609593811 and NRF RS-2024-00351197, as well as KIAS grant PG006002. 
\end{minipage}
\vspace{1pt}
\bigskip
\hrule
\end{center}

\newpage
\appendix 
\section{Symplectic Quotient and Dirac bracket}
\label{DIRAC}

Our expositions in the main text
rely quite heavily on
the languages of symplectic quotient and Dirac bracket.
Therefore it is the duty of this appendix
to provide a physicist-friendly introduction to those concepts.

\para{What is Symplectic Quotient?}

In the mathematics literature,
the idea of symplectic quotient
was systematized
by Marsden and Weinstein
\cite{marsden1974reduction},
and also Meyer \cite{meyer1973symmetries};
see \rcite{marsden2001comments} for a historical account.
Suppose a Lie group $G$ acts on
a symplectic manifold $(\P,\omega)$
by Hamiltonian actions.
The momentum map \cite{souriau1970structure}
$\phi : \P \too \g^*$
is a dual Lie algebra valued function
defining the generators of the $G$-action.
The definition of the symplectic quotient reads
\cite{marsden1974reduction}
\begin{align}
    \label{marsden}
    \P \ss G
    \,=\,
        \phi^{-1}(0)
        / G
    \,.
\end{align}
With \eqref{marsden}, the following facts hold \cite{marsden1974reduction,meyer1973symmetries}.
First,
$\pi : \phi^{-1}(0) \to \P \ss G$
describes a principal $G$-bundle.
Second,
$\P \ss G$ is a symplectic manifold 
by a unique symplectic form,
whose pullback by $\pi$
is the pullback of $\omega$
by the embedding $\phi^{-1}(0) \to \P$.

What does \eqref{marsden} mean, explicitly?
To this end, let $X^I$ be local coordinates on $\P$,
where $I,J,K,L,\cdots \in \{1,{\cdots},\dim\P\}$.
The momentum map is essentially
a set of functions $\phi_a(X)$,
where $a,b,c,d,\cdots \in \{1,{\cdots},\dim\g\}$
are the Lie algebra indices.
As is suggested by the very name ``momentum map,''
$\phi_a(X)$ are literally the generalized momenta for the $G$-action.
Namely,
the Hamiltonian vector fields of $\phi_a(X)$
generate the $G$-transformations on $\P$,
so $\pb{\phi_a}{\phi_b} = \phi_c\mem f^c{}_{ab}$.
The formula in \eqref{marsden} then states that
\begin{align}
    \label{marsden.unpacked}
    \P \ss G
    \,\,=\,\,
        \Big\{\mem\mem{
            X^I \nem\inn \P
        \mem\,\Big|\,\hem\mem
            \phi_a(X) = 0
        }\mem\,\Big\}
        \hem\Big/
            G
    \,,
\end{align}
which represents the ubiquitous pattern
described in the main text:
see \eqrefss{quotient}{ST.quotient14}{MT.quotient14},
for instance.
That is,
$\P \ss G$ is the constrained phase space in $\P$
that arises by
fixing the generalized momenta as $\phi_a = 0$
\textit{and}
performing a quotient by the group action generated by $\phi_a$.
The dimension of the symplectic quotient is
\begin{align}
    \label{hits-twice}
    \dim( \P \ss G )
    \,\,=\,\,
        \dim \P \mem-\mem 2\mem \dim G
    \,.
\end{align}
Clearly, the dimension of a symplectic quotient
is always even.

\para{What is Dirac Bracket?}
Meanwhile, the theory of constrained Hamiltonian mechanics
was developed by 
Dirac \cite{Dirac:1950pj,dirac1964lectures}
and is well-reviewed in the book \cite{Henneaux:1992ig}.

The gist of the Dirac bracket
is really simple.
Let $(\P_{2n},\omega_{2n})$ be a $2n$-dimensional symplectic manifold.
Suppose an embedding $i : \P_{2m} \to \P_{2n}$ of a $2m$-dimensional submanifold,
defined as the common zero locus of 
functions $\Psi^\a$ on $\P_{2n}$
where $\a,\b,\cdots \inn \{1,{\cdots},2(n \mminus m)\}$.
If the pullback $\omega_{2m} := i^* \omega_{2n}$
is nondegenerate in $\P_{2m}$,
then $(\P_{2m},\omega_{2m})$ is a symplectic manifold.
In turn, $\omega_{2m}$ can be inverted \textit{within} $\P_{2m}$
to define a Poisson bivector $\Pi_{2m}$.
The \textit{Dirac bivector}
is the pushforward
$i_* \Pi_{2m}$.
By construction,
this Dirac bivector
defines a Poisson structure on $\P_{2n}$ of rank $2m$
whose null directions are characterized by the functions $\Psi^\a$:
$(i_* \Pi_{2m})(\blank,d\Psi^\a) = 0$.
The Dirac bracket is 
$\pb{f}{g}_{2m} := (i_* \Pi_{2m})(df,dg)$
for $f,g \in \Cinfty(\P_{2n})$.


Explicitly, the formula for the Dirac bracket reads
\cite{Dirac:1950pj,dirac1964lectures,Henneaux:1992ig}.
\begin{align}
    \label{db}
    \pb{f}{g}_{2m}
    \mem\,=\,\mem
        \pb{f}{g}_{2n}
        \,-\,
        \pb{f}{\Psi^\a}_{2n}
        \,
            (C^{-1})_{\a\b}
        \,
        \pb{\Psi^\b}{g}_{2n}
    \,,
\end{align}
where we assume that $(C^{-1})_{\a\b}$ exists and is well-defined as the inverse of
\begin{align}
    \label{Cmatrix}
    C^{\a\b}
    \,:=\,
        \pb{\Psi^\a}{\Psi^\b}_{2n}
    \,,
\end{align}
so $C^{\a\c} (C^{-1})_{\c\b} = \delta^\a{}_\b$.
It should be clear that the right-hand side of \eqref{db} is invariant
under any linear transformation on the constraints $\Psi^\a$.

A derivation of the Dirac bracket
that can be physically insightful
is the following
(see also Appendix A of \rcite{Kim:2021rda}).
Recall how Poisson bracket arises from the variational principle.
Suppose a symplectic manifold $(\P,\omega)$.
A generic phase space action
takes the form
\begin{align}
    \label{psa.g}
    \int d\t\,\,
    \bb{
        \theta_I(X)\mem \dot{X}^I
        - g(X)
    }
    \,,
\end{align}
which defines a sigma model $\R \too (\P,\omega)$.
Here,
$\theta_I(X)\mem \dot{X}^I\mem d\t$ is the pullback
of a symplectic potential $\theta$
such that $d\theta = \omega$,
while $g \in \Cinfty(\P)$.
By discarding a boundary term,
the bulk variation of \eqref{psa.g} is found as
\begin{align}
    \label{vpsa.g}
    \int d\t\,\,
        \dX^I\,
        \bb{
            \omega_{IJ}(X)\mem \dot{X}^J
            - g_{,I}(X)
        }
    \,,
\end{align}
from which one derives the Hamiltonian EoM.
The resulting time evolution of a function $f \inn \Cinfty(\P)$
is given by
\begin{align}
    \label{pbfg}
    \dot{f}
    \,=\,
        f_{,I}\mem \dot{X}^I
    \,=\,
        f_{,I}\mem \Pi^{IJ}\hem g_{,J}
    \,=\,
        \pb{f}{g}
    \,.
\end{align}
Here, $\Pi$ denotes the pointwise inverse of $\omega$,
so $\omega_{IK}\mem \Pi^{KJ} = \delta_I{}^J$.

The action in \eqref{psa.g}
can be thought of as describing
a particle in $\P$.
We can constrain this particle on a submanifold $\Psi^\a = 0$
in terms of Lagrange multipliers $\k_\a(\t)$:
\begin{align}
    \label{psa.gc}
    \int d\t\,\,
    \bb{
        \theta_I(X)\mem \dot{X}^I
        - g(X)
        - \Psi^\a(X)\, \k^\a
    }
    \,.
\end{align}
The bulk variation of \eqref{psa.gc} is
\begin{align}
    \label{vpsa.gc}
    \int d\t\,\,
        \dX^I\,
        \bb{
            \omega_{IJ}(X)\mem \dot{X}^J
            - g_{,I}(X)
            - \Psi^\a{}_{,I}(X)\, \k^\a
        }
        - \Psi^\a(X)\, \d\k_\a
    \,,
\end{align}
from which the EoM are found as
\begin{align}
    \label{heom.gc}
    \dot{X}^I
    \,=\,
        \Pi^{IJ}\hem g_{,J}
        + \Pi^{IJ}\hem \Psi^\a{}_{,J}\mem \k_\a
    \,=\,
        \pb{X^I}{g}
        + \pb{X^I}{\Psi^\a}\mem \k_\a
    \,,\quad
    \Psi^\a
    \,=\,
        0
    \,.
\end{align}
As usual, the Lagrange multipliers are determined by imposing $\dot{\Psi}^\a = 0$:
\begin{align}
    \label{ksol}
    0 \,=\, \dot{\Psi}^\a
    \,=\,
        \pb{\Psi^\a}{g}
        + \pb{\Psi^\a}{\Psi^\b}\mem \k_\b
    \qiq
    \k_\a
    \,=\,
        - (C^{-1})_{\a\b}\, \pb{\Psi^\b}{g}
    \,.
\end{align}
By plugging in \eqref{ksol} back to \eqref{heom.gc},
one finds
\begin{align}
    \label{dbfg}
    \dot{f}
    \,=\,
        f_{,I}\mem \dot{X}^I
    \,=\,
        \pb{f}{g}
        + \pb{f}{\Psi^\a}\mem \k_\a
    \,=\,
        \pb{f}{g}
        - \pb{f}{\Psi^\a}\,
            (C^{-1})_{\a\b}\, 
        \pb{\Psi^\b}{g}
    \,.
\end{align}
\eqref{dbfg}
reproduces the right-hand side in \eqref{db},
upon restoring the $2n$ subscripts.


\para{Successive Reductions}

Now suppose a nested embedding of symplectic submanifolds,
\begin{align}
    (\P_{2r},\omega_{2r})
        \:\:\xrightarrow[]{\:\:\:{
            i'
        }\:\:\:}\:\:
    (\P_{2m},\omega_{2m})
        \:\:\xrightarrow[]{\:\:\:{
            i
        }\:\:\:}\:\:
    (\P_{2n},\omega_{2n})
    \,.
\end{align}
Suppose $\P_{2m}$ is defined in $\P_{2n}$ by
$\Psi^\a = 0$,
while
$\P_{2n}$ is defined in $\P_{2n}$ by
$\Psi^\a = 0$ and $\Psi^{\a'} = 0$.
Here, $\a,\b,\cdots$ run through $2(n\mminus m)$ integers
while $\a',\b',\cdots$ run through $2(n'\mminus m')$ integers.

The rank-$2m$ Dirac bivector on $\P_{2n}$ is
$i_* \Pi_{2m}$.
The rank-$2r$ Dirac bivector on $\P_{2n}$ is
$i_* i'_* \Pi_{2r}$.
Suppose
one wants to find
the rank-$2r$ Dirac bracket
while the rank-$2m$ Dirac bracket is already known.
It can be shown that
\begin{align}
    \pb{f}{g}_{2r}
    \mem\,=\,\mem
        \pb{f}{g}_{2m}
        \,-\,
        \pb{f}{\Psi^{\a\sprime\mem}}_{2m}
        \,
            (C'^{-1})_{\a\sprime\mem\b'}
        \,
        \pb{\Psi^{\b\sprime}}{g}_{2m}
    \,,
\end{align}
where $(C'^{-1})_{\a\sprime\mem\b'}$ is the inverse of
$C'^{\a\sprime\mem\b'} := \pb{\Psi^{\a\sprime}}{\Psi^{\b\sprime\mem}}_{2m}$.
This ``shortcut''
was used in \Sec{FOUR}.

\para{Relation between Symplectic Quotient and Dirac Bracket}

Finally,
we explicate the concrete link between 
symplectic quotient and Dirac bracket:
\begin{align}
    \text{Symplectic Quotient}
    \quad\xleftrightarrow[]{\quad}\quad
    \text{Dirac Bracket from First-Class Constraints}
    \,.
\end{align}
Below, we study the symplectic quotient $\P_{2n} \ss G$
with $\dim \P_{2n} = 2n$
and $\dim G = n \mminus m$.

First of all,
recall that the momentum map
describes generalized momenta $\phi_a$
that represent the Lie algebra $\g$
under the Poisson bracket:
$\pb{\phi_a}{\phi_b} = \phi_c\mem f^c{}_{ab}$.
In the Dirac terminology
\cite{Dirac:1950pj,dirac1964lectures,Henneaux:1992ig},
this means that $\phi_a = 0$ are first-class constraints:
$\pb{\phi_a}{\phi_b}$ vanish on the common zero locus of $\phi_a$.

The symplectic quotient in \eqref{marsden.unpacked}
is explicitly realized as a symplectic submanifold in $\P$
if one picks a particular slice on the $G$-orbit.
Concretely, this is done by a well-behaved set of functions $\chi^a$.
Without loss of generality, we may assume $\pb{\chi^a}{\phi_b} = \delta^a{}_b$
on account of the well-behavedness.

Now take $\Psi^\a = (\chi^a,\phi_a)$
as a set of $2(n \mminus m)$ constraints.
In the Dirac terminology
\cite{Dirac:1950pj,dirac1964lectures,Henneaux:1992ig},
these $\Psi^\a$ are second-class constraints:
the matrix $C^{\a\b} = \pb{\Psi^\a}{\Psi^\b}$ does not vanish on the common zero locus of $\Psi^\a$
and exhibits full rank
due to the well-behavedness.

Therefore, the formula in \eqref{db}
readily applies to define the rank-$2m$ Dirac bracket:
\begin{align}
    \label{dbq}
    \pb{f}{g}_{2m}
    \mem\,=\,\mem
        \pb{f}{g}_{2n}
        \,-\,
        \pb{f}{\phi_a}_{2n}
        \,
        \pb{\chi^a}{g}_{2n}
        \,+\,
        \pb{f}{\chi^a}_{2n}
        \,
        \pb{\phi_a}{g}_{2n}
    \,.
\end{align}
\eqref{dbq} describes the pushforward
of a rank-$2m$ Poisson bivector
on the explicit realization of $\P_{2n} \ss G$.
This concretely shows 
how the symplectic quotient $\P_{2n} \ss G$
is a symplectic manifold,
stated earlier as a mathematical fact.

If $G$ is a gauge group,
then $\phi_a$ are the \textit{gauge generators}
while $\chi^a$ are the \textit{gauge-fixing functions}.
The gauge generators $\phi_a$ must form first-class constraints
for the closure of the gauge algebra,
while $(\chi^a,\phi_a)$ must form second-class constraints
to produce a well-defined symplectic structure on the symplectic quotient
as well as a well-defined Dirac bracket on the original phase space.

We clarify that
this paper does not concern the cases
in which gauge redundancies are formulated by
presymplectic manifolds.
The symplectic reduction and Dirac bracket
procedures
in this paper
always start from a nondegenerate symplectic manifold.

\newpage

\section{Regge Trajectory}
\label{REGGE}

In the context of relativistic spinning particle mechanics,
the Regge trajectory \cite{Hanson:1974qy} is a positive function of the spin-squared,
$M(\xi) {\:>\:} 0$,
which is real-analytic in the domain $\xi {\:>\:} 0$.
Physically, it describes how the rotational kinetic energy contributes to the rest mass
via the mass-energy equivalence.
On a related note,
it is useful to denote
\begin{align}
    \label{inertia}
    I(\xi)
    \,:=\,
        \frac{1}{2M'(\xi)}
    \,.
\end{align}
It is not difficult to see that
$I(\xi)$ in \eqref{inertia} describes the rotational inertia
by considering
the nonrelativistic limit \cite{Kim:2021rda}.

This appendix 
explicitly demonstrates that
all of the four models
in \fref{overview}
allow for a generic Regge trajectory
in the mass-shell constraint:
\begin{align}
    \label{regge}
    \phi_0
    \,=\,
        \frac{1}{2}\mem \BB{
            p^2 + M^2(-p^2\hy^2)
        \nem}
    \,.
\end{align}
Below, we enumerate
the Dirac brackets
that imposes the generic mass-shell constraint in \eqref{regge}
in each model.
It suffices to suppose free theory.
For a covariant gauge fixing,
the gauge-fixing function
$\chi^0$
will be always taken as proportional to the dilatation charge.\footnote{
    It is realized in 
    \rcite{sst-asym} that
    the dilatation charge
    can be used as a proper time
    for the massive particle
    (up to customary rescalings by mass).
    A fruitful application is provided in its appendix
    with a discussion on Poincar\'e charges for binary systems.
}
We also describe the free-theory time evolution,
which clarifies the role of the Regge trajectory.

\subsection{Vector Oscillator}

The Dirac brackets on $\ps_{12}^\VO$
are shown in \eqref{VO.db12}.

\para{Dirac Brackets}

We impose the constraint pair $(\chi^0,\phi_0)$
on $\ps_{12}^\VO$
to reach a $10$-dimensional symplectic quotient
$\widetilde{\ps}^\VO_{10}$
(which differs from $\ps^\VO_{10}$ in \eqref{VO.series}).
The resulting Dirac bracket describes
\begin{align}
\label{VO.db10}
\begin{split}
    \pb{\fx^\mu}{\fx^\nu}_{10}
    &\,=\, 
        \frac{ J^{\m\n} }{M^2} 
    \,,\quad
    \pb{\fx^\mu}{p_\nu}_{10} 
    \,=\, 
        \hat{\delta}^\m{}_\n 
    \,,
    \\
    \pb{\fa^\mu}{\fba_\nu}_{10} 
    &\,=\, 
    -i\, \BB{\hat{\delta}^\m{}_\n  - \frac{\fba^\m \fa_\n}{\fba\cdot \fa} }
    \,,\\
    \pb{ \fa^\m }{ \fx^\n }_{10} 
    \,&=\,
        - \frac{1}{M^2}\,
            p^\m\hem
            \fa^\n{}
        + \frac{1}{MI}\,
            \fS^\m{}_\r\hem \fa^\r{}\,
            \fx^\n
    \,,\\
    \pb{\fa^\m }{p_\n}_{10}
    \,&=\,
        - \frac{1}{MI}\,
            \fS^\m{}_\r\hem \fa^\r{}\,
            p_\n
    \,,
\end{split}
\end{align}
where we have denoted
$M = M(-p^2\hy^2)$ and $I = I(-p^2\hy^2)$
to avoid clutter.

\para{Time Evolution}

The free-theory time evolution
due to the generic Hamiltonian constraint in \eqref{regge} can be found by
using the Dirac brackets in \eqref{VO.db12}:
$\dot{f} = \k^0\mem \pb{f}{\phi_0}_{12}$.
For $\k^0 = 1/M$,
we obtain
\begin{align}
    \label{VO.regge}
    \dot{\fx}^\m
    \,=\,
        \hat{p}^\m
    \,,\quad
    \dot{p}_\m
    \,=\,
        0
    \,,\quad
    \dot{\fa}^\m
    \,=\,
        -i\,\frac{\fba\mdot\fa}{I}\,
            \fa^\m
    \,.
\end{align}
This shows that
the sole effect of the Regge trajectory
on the time evolution
is a $\U(1)$ phase rotation of $\fa^\m$
by the angular frequency
\begin{align}
    \label{VO.angular}
    \Omega
    \,:=\,
        \frac{\fba\mdot\fa}{I}
    \qiq
    \fba\mdot\fa
    \,=\,
        I\mem \Omega
    \,.
\end{align}
Recalling \eqref{VO.constraints},
we see that $\fba\mdot\fa$
computes the spin magnitude in the vector oscillator model.
\eqref{VO.angular} identifies the familiar relation
between spin magnitude and angular velocity,
such that $I$ indeed describes the rotational inertia.

\subsection{Spinor Oscillator}

The Poisson brackets on $\ps_{12}^\SO$
are shown in \eqref{SO.pb}.

\para{Dirac Brackets}

We impose the constraint pair $(\chi^0,\phi_0)$
on $\ps_{12}^\SO$
to reach a $10$-dimensional symplectic quotient
$\ps^\SO_{\widetilde{10}}$
(which differs from $\ps^\SO_{10}$ in \eqref{SO.series}).
The resulting Dirac bracket describes
\begin{align}
\begin{split}
    &
    \pb{x^\m}{x^\n}_{\,\widetilde{\,\mathclap{10}\,}\,}
    \,=\,
        \frac{J^{\m\n}}{M^2}\,
    \,,\quad
    \pb{x^\m}{p_\n}_{\,\widetilde{\,\mathclap{10}\,}\,}
    \,=\,
        \hat{\delta}^\m{}_\n
    \,,\quad
    \pb{\zeta_\a}{\bzeta_\da}_{\,\widetilde{\,\mathclap{10}\,}\,}
    \,=\,
        i\mem \hat{p}_{\a\da}
    \,,\\
    &
    \pb{\zeta_\a}{x^\m}_{\,\widetilde{\,\mathclap{10}\,}\,}
    \,=\,
        \frac{
            (\sigma^{\m\n}\zeta)_\a
        }{2M^2}\,    
            p_\n
            + \frac{
                S_\a{}^\b \zeta_\b 
            }{MI}\,
            x^\m
    \,,\quad
    \pb{\zeta_\a}{p_\m}_{\,\widetilde{\,\mathclap{10}\,}\,}
    \,=\, 
            - \frac{
                S_\a{}^\b \zeta_\b
            }{MI}\,
            p_\m
    \,,
\end{split}
\end{align}
where 
$S_{\a\b} = \frac{i}{4}\mem (\zeta_\a \xi_\b + \xi_\a \zeta_\b)$
describes \eqref{SO.S} in the spinor notation.

\para{Time Evolution}

The free-theory time evolution
due to the generic Hamiltonian constraint in \eqref{regge} is found by
using the Poisson brackets in \eqref{SO.pb}:
$\dot{f} = \k^0\mem \pb{f}{\phi_0}_{12}$.
For $\k^0 = 1/M$,
we obtain
\begin{align}
\label{SO.regge}
    \dot{x}^\m
    \,=\,
        \hat{p}^\m
    \,,\quad
    \dot{p}_\m
    \,=\,
        0
    \,,\quad
    \dot{\zeta}_\a
    \,=\,
        \frac{1}{2i} \,
        \bb{
            -\frac{1}{2I}\,
                \bar{\zeta}_\db\mem \hat{p}^{\db \b} \zeta_\b
        }\,
            \zeta_\a
    \,.
\end{align}
This shows that
the sole effect of the Regge trajectory
on the time evolution
is a $\U(1)$ phase rotation of $\zeta_\a$
by the angular frequency
\begin{align}
    \label{SO.angular}
    \Omega
    \,:=\,
        -\frac{1}{2I}\,
            \bar{\zeta}_\db\mem \hat{p}^{\db \b} \zeta_\b
    \qiq
    \frac{1}{2}\,
        \bxi^\a \zeta_\a 
    \,=\,
        I\mem \Omega
    \,,
\end{align}
where we have used \eqref{Econj}.
Recalling \eqref{SO.constraints},
we see that $\bxi^\a \zeta_\a / 2$
computes the spin magnitude in the spinor oscillator model.
Therefore,
we again identify the familiar relation between
spin magnitude and angular velocity
such that $I$ is the rotational inertia.

Note that we peel off $1/2i$ in \eqref{SO.regge}
(instead of $1/i$ in \eqref{VO.regge})
because the phase of a spinor rotates in a half speed:
the double cover.

\subsection{Spherical Top}

\para{Dirac Brackets}

First,
imposing
the constraints
$\phi_a {\:=\:} 0$ and $\chi^a {\:=\:} 0$
identifies
the $14$-dimensional physical phase space
$\ps_{14}^\ST$:
Its Poisson structure
is characterized by the Dirac bracket
whose nonzero components are
\begin{align}
\begin{split}
\label{ST.db14}
    \pb{\fx^\mu}{\fx^\nu}_{14}
    \,=\, 
        \frac{\fS^{\m\n} }{-p^2} 
    &\,,\quad
    \pb{\fx^\m}{\fS_{\r\s}}_{14} 
    \,=\, 
        \frac{1}{p^2}\,
            2\mem \fS^\m{}_\wrap{[\r}\mem p_\wrap{\s]}
    \,,\\
    \pb{\fx^\mu}{p_\nu}_{14} 
    \,=\, 
        \delta^\m{}_\n 
    &\,,\quad
    \pb{ \fL^\m{}_a }{ \fS_{\r\s} }_{14} 
    \,=\,
        \BB{
            -2\mem \hat{\delta}^\m{}_{[\r} \hat{\eta}_{\s]\n}
        }\mem
        \fL^\n{}_a
    \,,\\
    \pb{ \fL^\m{}_a }{ \fx^\n }_{14} 
    \,=\,
        \frac{1}{p^2}\,
            p^\m\hem
            \fL^\n{}_a
    &\,,\quad
    \pb{\fS_{\m\n}}{\fS_{\r\s}}_{14}
    \,=\,
        \BB{
            -4\mem \hat{\delta}^{[\k}{}_{[\m}\mem
            \hat{\eta}_{\n][\r}\mem
            \hat{\delta}^{\l]}{}_{\s]}
        }\mem \fS_{\k\l}
    \,,
\end{split}
\end{align}
where $\hat{\delta}^\m{}_\n {\::=\:} \delta^\m{}_\n - p^\m p_\n/p^2$
and $\hat{\eta}_{\m\n} {\::=\:} \eta_{\m\n} - p_\m p_\n /p^2$.

Second,
imposing
the constraints
$\phi_0 {\:=\:} 0$ and $\chi^0 {\:=\:} 0$
identifies
the $12$-dimensional symplectic submanifold
$\ps_{12}^\ST$. 
Its Poisson structure
is characterized by the Dirac bracket
whose nonzero components are
\begin{subequations}
\label{ST.db12}
\begin{align}
\begin{split}
    \pb{\fx^\mu}{\fx^\nu}_{12}
    \,=\, 
        \frac{ J^{\m\n} }{M^2} 
    &\,,\quad
    \pb{\fx^\m}{\fS_{\r\s}}_{12} 
    \,=\, 
        - \frac{1}{M^2}\,
            2\mem \fS^\m{}_\wrap{[\r}\mem p_\wrap{\s]}
    \,,\\
    \pb{\fx^\mu}{p_\nu}_{12} 
    \,=\, 
        \hat{\delta}^\m{}_\n 
    &\,,\quad
    \pb{ \fL^\m{}_a }{ \fS_{\r\s} }_{12} 
    \,=\,
        \BB{
            -2\mem \hat{\delta}^\m{}_{[\r} \hat{\eta}_{\s]\n}
        }\mem
        \fL^\n{}_a
    \,,\\
    &\phantom{{}\,,\quad{}}
    \pb{\fS_{\m\n}}{\fS_{\r\s}}_{12}
    \,=\,
        \BB{
            -4\mem \hat{\delta}^{[\k}{}_{[\m}\mem
            \hat{\eta}_{\n][\r}\mem
            \hat{\delta}^{\l]}{}_{\s]}
        }\mem \fS_{\k\l}
    \,,
\end{split}
\end{align}
and
\begin{align}
\begin{split}
    \pb{ \fL^\m{}_a }{ \fx^\n }_{12} 
    \,&=\,
        - \frac{1}{M^2}\,
            p^\m\hem
            \fL^\n{}_a
        + \frac{1}{MI}\,
            \fS^\m{}_\r\hem \fL^\r{}_a\,
            \fx^\n
    \,,\\
    \pb{\fL^\m{}_a}{p_\n}_{12}
    \,&=\,
        - \frac{1}{MI}\,
            \fS^\m{}_\r\hem \fL^\r{}_a\,
            p_\n
    \,,
\end{split}
\end{align}
\end{subequations}
where $M = M(-p^2\hy^2)$ and $I = I(-p^2\hy^2) := (\hhem 2M'\hnem(-p^2\hy^2))^{-1}$
are \textit{functions} of $-p^2\hy^2$.
By taking the nonrelativistic limit,
one sees that $I$ describes the rotational inertia.

It suffices to consider the spatial part $\fL^\m{}_a$ of the body frame
in the above Dirac brackets,
as the boost components are removed.
In the same way,
it is more helpful to examine
the pseudovector $\hy^\m$
rather than the bulky six-component variable $\fS_{\m\n}$.
Straightforward algebra shows that
\eqref{ST.db14}
is rewritten in the
$(\fx^\m,\fL^\m{}_a,p_\m,\hy^\m)$ basis as
\label{ST.db14.re}
\begin{align}
\begin{split}
    \pb{\fx^\m}{\fx^\n}_{14}
    \,&=\,
        \frac{1}{-p^2}\,
            \ve^{\m\n\r\s} \hy_\r\mem p_\s
    \,=\,
    \pb{\hy^\m}{\hy^\n}_{14}
    \,,\\
    \pb{\fx^\m}{\hy^\n}_{14}
    \,&=\,
        \frac{1}{-p^2}\,
            \BB{
                \hy^\m p^\n + p^\m \hy^\n
            }
    \,,\\[0.12\baselineskip]
    \pb{\fx^\m}{p_\n}_{14}
    \,&=\,
        \delta^\m{}_\n 
    \,,\\[0.12\baselineskip]
    \pb{\fL^\m{}_a}{\fx^\n}_{14}
    \,&=\,
        \frac{1}{p^2}\,
            p^\m\hem \fL^\n{}_a
    \,,\\
    \pb{\fL^\m{}_a}{\hy^\n}_{14}
    \,&=\,
        \frac{1}{p^2}\,
            \ve^{\m\n}{}_{\r\s}\,
            p^\r\hem \fL^\s{}_a
\end{split}
\end{align}
Similarly,
\eqref{ST.db12}
is rewritten in the
$(\fx^\m,\fL^\m{}_a,p_\m,\hy^\m)$ basis as
\begin{subequations}
\begin{align}
\begin{split}
    \pb{\fx^\m}{\fx^\n}_{12}
    \,=\,
        \frac{J^{\m\n}}{M^2} 
    &\,,\quad
    \pb{\fx^\m}{\hy^\n}_{12}
    \,=\,
        \frac{1}{M^2}\,
            \hy^\m p^\n 
    \,,\\
    \pb{\fx^\m}{p_\n}_{12}
    \,=\,
        \hat{\delta}^\m{}_\n
    &\,,\quad
    \pb{\hy^\m}{\hy^\n}_{12}      
    \,=\,
        \frac{1}{M^2}\,
            \ve^{\m\n\r\s} \hy_\r\mem p_\s 
    \,,
\end{split}
\end{align}
and
\begin{align}
\begin{split}
\label{ST.db12.re}
    \pb{\fL^\m{}_a}{\fx^\n}_{12}
    \,&=\,
        -\frac{1}{M^2}\,
            p^\m\hem \fL^\n{}_a
        + \frac{1}{MI}\,
            (\ve^\m{}_{\r\k\l}\mem 
            \fL^\r{}_a\mem
            \hy^\k p^\l)
            \,
            \fx^\n
    \,,\\
    \pb{\fL^\m{}_a}{\hy^\n}_{12}
    \,&=\,
        -\frac{1}{M^2}\,
            \ve^{\m\n}{}_{\r\s}\,
            p^\r\hem \fL^\s{}_a
    \,,\\
    \pb{\fL^\m{}_a}{p_\n}_{12}
    \,&=\,
        - \frac{1}{MI}\,
            (\ve^\m{}_{\r\k\l}\mem
            \fL^\r{}_a\mem
            \hy^\k p^\l)
            \,
            p_\n
    \,.
\end{split}
\end{align}
\end{subequations}

\para{Time Evolution}

The free-theory time evolution
due to \eqref{regge} follows by
using the Dirac brackets in \eqref{ST.db14@}:
$\dot{f} = \k^0\mem \pb{f}{\phi_0}_{14}$.
For $\k^0 = 1/M$,
we obtain
\begin{align}
    \label{ST.regge}
    \dot{\fx}^\m
    \,=\,
        \hat{p}^\m
    \,,\quad
    \dot{p}_\m
    \,=\,
        0
    \,,\quad
    \fL^\m{}_a\mem \dot{\fL}{}^{\n a}
    \,=\,
        \frac{1}{I}\: \fS^{\m\n} 
    \,.
\end{align}
This identifies
a proportionality between
spin and angular velocity
such that $I$ describes an isotropic rotational inertia
(hence the name \textit{spherical} top):
\begin{align}
    \Omega^{\m\n}
    \,:=\,
        \fL^\m{}_a\mem \dot{\fL}{}^{\n a}
    \qiq
    \fS^{\m\n}
    \,=\,
        I\mem \Omega^{\m\n}
    \,.
\end{align}

\subsection{Massive Twistor}

\para{Dirac Brackets}

The Poisson structure of $\ps_{14}^\mathrm{MT}$ in the hybrid basis
$(x^\m,\hy^\m,\lambda_\a{}^I)$ is specified by the following nonzero components, 
\begin{align}
\begin{split}
\label{MT.db14}
    \pb{x^\m}{x^\n}_{14}
    \,&=\,
        \frac{1}{-p^2}\,
            \ve^{\m\n\r\s} \hy_\r\mem p_\s
    \,=\,
    \pb{\hy^\m}{\hy^\n}_{14}
    \,,\\
    \pb{x^\m}{\hy^\n}_{14}
    \,&=\,
        \frac{1}{-p^2}\,
            \BB{
                \hy^\m p^\n + p^\m \hy^\n
            }
    \,,\\[0.12\baselineskip]
    \pb{\lambda_\a{}^I}{x^{\db\b}}_{14}
    \,&=\,
        \frac{1}{2}\,
            \delta_\a{}^\b\mem (\rambda^{-1})^{\db I} 
    \,,\\
    \pb{\lambda_\a{}^I}{\hy^{\db\b}}_{14}
    \,&=\,
        \frac{i}{2}\,
        \bb{
            \delta_\a{}^\b\mem (\rambda^{-1})^{\db I}
            - \frac{1}{2}\, \lambda_\a{}^I\mem (\rambda^{-1}\lambda^{-1})^{\db\b}
        }
    \,,
\end{split}
\end{align}
Imposing the constraints $\phi_0=0$ and $\chi^0=0$ leads us to $\ps_{12}^\mathrm{MT}$ with 
\begin{subequations}
\begin{align}
    \pb{x^\m}{x^\n}_{12}
    \,=\,
        \frac{J^{\m\n}}{M^2} 
    \,,\quad
    \pb{\hy^\m}{\hy^\n}_{12}      
    \,=\,
 \frac{1}{M^2}\,
            \ve^{\m\n\r\s} \hy_\r\mem p_\s 
    \,,\quad
    \pb{x^\m}{\hy^\n}_{12}
    \,=\,
        \frac{1}{M^2}\,
                \hy^\m p^\n 
    \,,
\end{align}
and
\begin{align}
\begin{split}
\label{MT.db12} 
    \pb{\lambda_\a{}^I}{x^{\db\b}}_{12}
    \,&=\,
        \frac{1}{2}\,
        \BB{
            \delta_\a{}^\b\mem (\rambda^{-1})^{\db I}
            - \frac{1}{2}\, \lambda_\a{}^I\mem (\rambda^{-1}\lambda^{-1})^{\db\b}
        } 
        - \frac{i}{2MI}\,
            (\lambda W)_\a{}^I\,
            x^{\db\b}
    \,,\\
    \pb{\lambda_\a{}^I}{y^{\db\b}}_{12}
    \,&=\,
        \frac{i}{2}\, 
            \delta_\a{}^\b\mem (\rambda^{-1})^{\db I} 
        - \frac{i}{2MI}\,
            (\lambda W)_\a{}^I\,
            y^{\db\b}
    \,,\\
    \pb{\lambda_\a{}^I}{\lambda_\b{}^J}_{12}
    \,&=\,
        \frac{i}{4MI}\,
        \BB{
            (\lambda W)_\a{}^I\,
            \lambda_\b{}^J
            - \lambda_\a{}^I\,
            (\lambda W)_\b{}^J
        }
    \,,\\
    \pb{\lambda_\a{}^I}{\rambda_\wrap{J\db}}_{12}
    \,&=\,
        \frac{i}{4MI}\,
        \BB{
            (\lambda W)_\a{}^I\,
            \rambda_\wrap{J\db}
            + \lambda_\a{}^I\,
            (W \rambda)_\wrap{J\db}
        }
    \,.
\end{split}
\end{align}
\end{subequations}
Here,
$W_I{}^J$
is an $\SU(2)$ matrix
encoding the body-frame components $W^a$ of the spin pseudovector:
\begin{align}
    W^a
    \,:=\,
        - |p|
        \, (\hy \mdot \Lambda^a)
    \,,\quad
    W_I{}^J
    \,=\,
        W^a\mem (\s_a)_I{}^J
    \,.
\end{align}

\para{Time Evolution}

The free-theory time evolution due \eqref{regge}
follows by
using the Dirac brackets in \eqref{MT.db14}:
$\dot{f} = \k^0\mem \pb{f}{\phi_0}_{14}$.
For $\k^0 = 1/M$,
we obtain
\begin{align}
    \label{MT.regge}
    \dot{x}^\m
    \,=\,
        \hat{p}^\m
    \,,\quad
    \dot{y}^\m
    \,=\,
        0
    \,,\quad
    \dot{\lambda}_\a{}^I 
    \,=\, 
        \frac{1}{2iI}\:
            \lambda_\a{}^J\:
                W^a\mem
                (\s_a)_I{}^J
    \,.
\end{align}
In this case, the isotropic rotational inertia $I$
is best identified in the internal frame:
\begin{align}
    \frac{1}{2i}\:
        \Omega^a\,
        (\s_a)_I{}^J
    \,:=\,
        (\hlambda^{-1})_I{}^\a\, \dot{\hlambda}_\a{}^J
    \qiq
    W^a
    \,=\,
        I\mem \Omega^a
    \,.
\end{align}
In fact,
\eqref{MT.regge} can be shown to be equivalent to \eqref{ST.regge} 
via the map described in \eqref{ST.solve}.
Note also that $\frac{1}{2i}\mem (\s_a)_I{}^J$
are the $\SU(2)$ generators
in the anti-Hermitian convention,
which should explain the $1/2i$ factors above:
$\comm{\s_a/2i}{\s_b/2i} = \ve^c{}_{ab}\mem (\s_c/2i)$.

\section{Review of Spinning-Particle Equations of Motion}
\label{REVIEW}

\subsection{BMT Equations}
\label{REVIEW>BMT}

The original statement of
the BMT equations reads \cite{Bargmann:1959gz}
\begin{align}
\begin{split}
\label{OBMT}
    \ddot{x}^\m
    \,&=\,
        \frac{q}{m}\,
            F^\m{}_\n(x)\mem \dot{x}^\n
    \,,\quad
    \dot{\hy}^\m
    \,=\,
        \frac{q}{m}\,
        \bbsq{    
            \frac{g}{2}\:
                F^\m{}_\n(x)
            - \bb{1 - \frac{g}{2}}\,
                \dot{x}^\m \dot{x}_\k\mem
                F^\k{}_\n(x)
        }\, \hy^\n
    \,,
\end{split}
\end{align}
where a proper time parametrization is assumed
such that $\dot{x}^2 = -1$.

In this paper,
we advocated a refinement of \eqref{OBMT}
into \eqref{BMT},
which we reproduce below for the reader's sake:
\begin{subequations}
\label{BMT(re)}
\begin{align}
\label{BMT(re).x}
    \dot{x}^\m
    \,&=\,
        \frac{p^\m}{m}
        \mem-\mem
        \frac{q}{m}\,
        \bbsq{    
            \frac{g}{2}\,
                {*}\hnem F^\m{}_\n(x)
            + \bb{1 - \frac{g}{2}}\,
                \hdelta^\m{}_\k\,
                {*}\hnem F^\k{}_\n(x)
        }\, \hy^\n
    + \cdots
    \,,\\
\label{BMT(re).y}
    \dot{\hy}^\m
    \,&=\,
        \frac{q}{m}\,
        \bbsq{    
            F^\m{}_\n(x)
            - \bb{1 - \frac{g}{2}}\,
                \hdelta^\m{}_\k\mem
                F^\k{}_\n(x)
        }\, \hy^\n
    + \cdots
    \,,\\[0.12\baselineskip]
\label{BMT(re).p}
    \dot{p}^\m
    \,&=\,
        \frac{q}{m}\,
            F^\m{}_\n(x)\mem p^\n
    + \cdots
    \,.
\end{align}
\end{subequations}
The ellipses signify that
terms of $\O(q^{1+n})$ or $\O(\partial^n\nem F)$
are discarded from the right-hand sides
for $n \geq 1$.

We shall emphasize again that
the unit timelike vector $p^\m\nem/m$
\textit{differs} from the four-velocity $\dot{x}^\m$
in this refinement.
This mismatch is crucial for
the existence of a well-defined Hamiltonian system
giving rise to \eqref{BMT(re)}
up to the $\O(q^2)$ and $\O(\hy^2)$ truncations,
as is established in \Sec{INT>BMT}.

To elaborate, a necessary condition for the existence of a proper Hamiltonian formulation is that
the following holds for any choice of classical observables $f$ and $g$:
\begin{align}
    \label{consistency}
    \frac{d}{d\t}\,
        \pb{f}{g}
    \,=\,
        \pb{\dot{f}}{g} + \pb{f}{\dot{g}}
    \,.
\end{align}
This describes a consistency condition
about the time derivative,
mandated by 
(a) the Jacobi identity of the Poisson bracket
and 
(b) the fact that time evolution is a Hamiltonian action.
Crucially, 
the very identity in
\eqref{consistency} will \textit{fail}
if one sets $\dot{x}^\m = p^\m\nem/m$ in \eqref{BMT(re).x}!

\newpage

Mathematically, \eqref{consistency} states the existence of a Poisson manifold equipped with a Hamiltonian.
In the context of symplectic realizations,
\eqref{consistency} encodes the principle of \textit{symplecticity},
which states that the theory is Lagrangian
while ensuring the conservation of classical probability.
See the appendix of \rcite{csg}
for a detailed discussion.

We shall end with stating the exact equation
that fulfills the consistency condition in \eqref{consistency},
without any truncations.
The symplectic perturbation in \eqref{EM.sp1}
unpacks to
\begin{align}
    \label{EM.sp1(unpacked)}
    \omega'
    \,\,=\,\,
        \frac{1}{2}\,
            \BB{
                qF_{\m\n}(x)
                + c_1q\,
                    {*}F_{\m\n,\r}(x)\, \hy^\r
            }\,
            dx^\m \swedge dx^\n
        \,+\,
        c_1q\,
            {*}F_{\m\n}(x)\, 
            d\hy^\m \swedge dx^\n
    \,,
\end{align}
where we have assumed that
the background is on-shell:
${*}F_{[\m\n,\r]}(x) = 0$.
We do not ignore the derivative term in \eqref{EM.sp1(unpacked)}.
The exact EoM arises via \eqref{ppert}:
\begin{subequations}
\label{EXACTBMT}
\begin{align}
\label{EXACTBMT.x}
    \dot{x}^\m
    \,&=\,
    \begin{aligned}[t]
        \frac{p^\m}{m}
        &-
        \frac{q}{m^2}\mem
        \BB{
            \ve^{\m\r\k\l} \hy_\k\mem p_\l
        }\mem
        F_{\r\s}(x)\mem \dot{x}^\s
        \\
        &
        -
        \frac{c_1q}{m^2}\mem
        \BB{
            F^\m{}_{\n,\r}(x)\mem \hy^\n \hy^\r
            \mem p_\k \dot{x}^\k
            - F^\m{}_{\n,\r}(x)\mem p^\n \hy^\r
            \mem \hy_\k\mem \dot{x}^\k
            + \dot{x}^\m\mem
                F_{\r\s,\k}(x)\mem \hy^\r p^\s \hy^\k
        }
        \\
        &
        - \frac{c_1q}{m^2}\mem
        \BB{
            \hy^\m\mem p^\r
            + p^\m\mem \hy^\r
        }\mem
            {*}F_{\m\n}(x)\mem \dot{x}^\n
        \\
        &
        + \frac{c_1q}{m^2}\mem
        \BB{ 
            -
            F^\m{}_\n(x)\mem \hy^\n
            \mem
            p_\r\mem \dot{\hy}^\r
            +
                \dot{\hy}^\m\mem
                (F_{\r\s}(x)\mem p^\r \hy^\s)
        }
        \,,
    \end{aligned}
    \\
\label{EXACTBMT.y}
    \dot{\hy}^\m
    \,&=\,
    \begin{aligned}[t]
        &
        \frac{q}{m^2}\mem
        \BB{
            \hy^\m\mem p^\r
            + p^\m\mem \hy^\r
        }\mem
        \BB{
            F_{\r\s}(x)
            + c_1\mem
                {*}F_{\r\s,\k}(x)\mem \hy^\k
        }\mem \dot{x}^\s
        \\
        &
        + \frac{c_1q}{m^2}\mem
        \BB{
            \hy^\m\mem p^\r
            + p^\m\mem \hy^\r
        }\mem
            {*}F_{\m\n}(x)\mem \dot{\hy}^\n
        \\
        &
        + \frac{c_1q}{m^2}\mem
        \BB{
            F^\m{}_\n(x)\mem p^\n
            \mem
            \hy_\r\mem \dot{x}^\r
            -
            F^\m{}_\n(x)\mem \hy^\n
            \mem
            p_\r\mem \dot{x}^\r
            +
                \dot{x}^\m\mem
                (F_{\r\s}(x)\mem p^\r \hy^\s)
        }
        \,,
    \end{aligned}
    \\[0.12\baselineskip]
\label{EXACTBMT.p}
    \dot{p}_\m
    \,&=\,
        q\mem \BB{
            F_{\m\n}(x)
            + c_1\mem
                {*}F_{\m\n,\r}(x)\mem \hy^\r
        }\mem \dot{x}^\n
        + c_1q\,
            {*}F_{\m\n}(x)\mem \dot{\hy}^\n
    \,.
\end{align}
\end{subequations}
By iterating \eqref{EXACTBMT} and truncating, one reproduces \eqref{BMT(re)}.

\subsection{MPD Equations}
\label{REVIEW>MPD}

The original statement of the MPD equations reads
\cite{Mathisson:1937zz,Papapetrou:1951pa,Dixon:1970zza}
\begin{subequations}
\begin{align}
    \label{OMPD}
    \frac{Dp_\m}{d\t}
    \,&=\,
        -\frac{1}{2}\,
        S_{\k\l}\hem
        R^{\k\l}{}_{\m\n}(x)\mem 
        \dot{x}^\n
    \,,\quad
    \frac{DS^{\m\n}}{d\t}
    \,=\,
        -\BB{
            \dot{x}^\m p^\n
            - p^\m \dot{x}^\n
        }
    \,.
\end{align}
The conventional practice
determines
the EoM for $x^\m$ by imposing a SSC.
For the covariant SSC
(referred to as Tulcjzew-Dixon SSC \cite{tulczyjew1959motion,dixon1964covariant} in this context),
$S^{\m\n} p_\n = 0$,
one finds
\begin{align}
    \label{OMPD.x}
    \dot{x}^\m
    \,=\,
        \frac{p^\m}{m}
        \mem-\mem
        \frac{1}{2p^2}\,
        S^{\m\r}\mem
            S^{\k\l}\hem
            R_{\k\l\r\s}(x)\mem
            \dot{x}^\s
    \,.
\end{align}
\end{subequations}
To see this, note that
\eqrefs{OMPD}{OMPD.x} together implies
\begin{align}
    \dot{x}^\m
    \,=\,
        \frac{p^\m}{m}
        + \frac{1}{p^2}\,
            S^{\m\r} \frac{Dp_\r}{d\t}
    \,,\quad
    \frac{DS^{\m\n}}{d\t}
    +
        S^{\m\r}\mem \frac{Dp_\r}{d\t}
        \mem 
        \frac{p^\n}{p^2}
    -
        \frac{p^\m}{p^2}\mem
        S^{\n\r}\mem \frac{Dp_\r}{d\t}
    \,=\,
        0
    \,,
\end{align}
so $(DS^{\m\n}\nem/d\t)\mem p_\n + S^{\m\n}\hem (Dp_\n/d\t) = 0$ is assured.

We point out that
\eqrefs{OMPD}{OMPD.x}
can be rewritten
in terms of the pseudovector parametrization of spin
$S^{\m\n} = \ve^{\m\n\r\s} \hy_\r p_\s$,
which explicitly imposes the covariant SSC:
\begin{subequations}
\label{MPD(resolved)}
\begin{align}
\begin{split}
    \dot{x}^\m
    \,&=\,
    \begin{aligned}[t]
    &
        \frac{p^\m}{m}
        - \hdelta^\m{}_\k\mem R^\k{}_{\n\r\s}(x)\mem \hy^\n \hy^\r \dot{x}^\s
        + R^\m{}_{\n\r\s}(x)\mem 
            \hp^\n \hp^\r \dot{x}^\s
        \hem\hy^2
        + \hy^\m\mem
            \hp_\k\hem
            R^\k{}_{\n\r\s}(x)\mem
            \hy^\n \hp^\r \dot{x}^\s
        \,,
        \kern-0.5em
    \end{aligned}
\end{split}
    \\
    \kern-0.2em
    \frac{D\hy^\m}{d\t}
    \,&=\,
        \hp^\m\mem \hp_\k\mem
            {*}R^\k{}_{\l\r\s}(x)\mem
            \hy^\l \hy^\r \dot{x}^\s
        +
        \hy^\m\mem
            \hp_\k\mem
            {*}R^\k{}_{\l\r\s}(x)\mem
            \hy^\l \hp^\r \dot{x}^\s
    \,,\\
    \kern-0.2em
    \frac{Dp_\m}{d\t}
    \,&=\,
        p_\k\mem
        {*}R^\k{}_{\l\m\n}(x)\mem 
        \hy^\l\mem
        \dot{x}^\n
    \,.
\end{align}
\end{subequations}
Here, we have made uses 
${*}R^\r{}_{\m\r\n} = 0$
and $R^\r{}_{\m\r\n} = 0$.
By iterating \eqref{MPD(resolved)} and truncating, one reproduces
\eqref{MPD} stated in the main text.

\subsection{QMPD Equations}
\label{REVIEW>QMPD}

The quadrupolar extension of the MPD equations
has been envisioned in the literature
since
\rrcite{khriplovich1989particle,Khriplovich:1997ni,Yee:1993ya},
where
the quadrupolar coupling coefficient $C_2$
is named the gravimagnetic ratio $\k$.
More recent works include
\rrcite{Steinhoff:2009tk,Harte:2011ku,Vines:2016unv,Compere:2023alp,Ramond:2026fpi}.

In this paper,
we presented a transparent derivation of the QMPD equations
in terms of the covariant symplectic perturbation $\omega'$ in
\eqref{GR.sp2}.
When expanded out, it is
\begin{align}
    \label{GR.sp2(re)}
    \nonumber
    &
    \frac{1}{2}\:
        p_\m\mem
        \bb{
            {*\hnem}R^\m{}_{\n\r\s}(x)
            - \frac{C_2}{2}\:
                R^\m{}_{\n\r\s;\k}(x)\mem \hy^\k
        }\,
            \hy^\n\mem
            dx^\r \swedge dx^\s
        - \frac{C_2}{2}\:
            p_\m\hem R^\m{}_{\n\r\s}(x)\mem \hy^\n\hem
            D\hy^\r \wedge dx^\s
    \\
    &
    \hlg{
        - \frac{C_2}{4}\:
            D\hnem S_{\k\l}
            \wedge
            {*}R^{\k\l}{}_{\r\s}\:
                \hy^\r\mem dx^\s
    }
    \,,
\end{align}
where we have intentionally used both $D\hy$ and $DS$.
The first line in \eqref{GR.sp2(re)}
corresponds to \eqref{EM.sp1(unpacked)}
in light of
the \textit{abelian} double copy mapping
identified in \eqref{cov-dc}.
However, as we experienced in \Sec{HR} when deriving the fictitious terms,
the abelian double copy mapping
misses the fact that $S$,
as the (local Lorentz) gravitational charge,
is a variable.
Hence the second line in \eqref{GR.sp2(re)} arises.

In fact, the covariant symplectic perturbation for
the \textit{nonabelian} BMT equations in gauge theory backgrounds
(the spinning generalization of the Wong's equations \cite{Wong:1970fu})
will precisely
parallel the covariant symplectic perturbation for
the QMPD equations in \eqref{GR.sp2(re)},
via the nonabelian double copy mapping 
$q_a F^a{}_{\m\n}(x) \leftrightarrow -\frac{1}{2}\, S_{\k\l}\mem R^{\k\l}{}_{\m\n}(x)$
explicated in \rcite{csg} 
(see also Appendix A of \rcite{ambikerr1}).

By using \eqref{GR.sp2(re)} and the universal covariant Poisson bracket relation in \eqref{xyp.cov},
we find the exact and complete version of the QMPD equations:
\begin{subequations}
\label{EXACTQMPD}
\begin{align}
\label{EXACTQMPD.x}
    \dot{x}^\m
    \,&=\,
    \begin{aligned}[t]
        \frac{p^\m}{m} 
        &+
        \frac{1}{m^2}\mem
            R^\m{}_{\n\r\s}(x)\mem \hy^\n \hy^\r p^\s
            \mem p_\k \dot{x}^\k
        \\
        &-
        \frac{1}{m^2}\mem
        \BB{
            R^\m{}_{\n\r\s}(x)\mem p^\n \hy^\r p^\s
            \mem \hy_\k \dot{x}^\k
            + \dot{x}^\m\mem
                p_\zeta R^\zeta{}_{\xi\k\l}(x)\mem 
                \hy^\xi \hy^\k p^\l
        }
        \\
        &+
        \frac{C_2}{2m^2}\mem
        \BB{
            \ve^{\m\r\k\l} \hy_\k\mem p_\l
        }\mem
        \mem p_\zeta \hy^\xi\mem
        R^\zeta{}_\xi{}_{\r\s;\t}(x)\mem \hy^\t\mem \dot{x}^\s
        \\
        &
        + \frac{C_2}{2m^2}\mem
        \BB{
            \hy^\m\mem p^\r
            + p^\m\mem \hy^\r
        }\mem
        \mem p_\zeta \hy^\xi\mem
            R^\zeta{}_\xi{}_{\m\n}(x)\mem \dot{x}^\n
        \\
        &
        + \frac{C_2}{2m^2}\mem
        \mem p_\zeta \hy^\xi\mem
        \BB{
            -
            {*}R^\zeta{}_\xi{}^\m{}_\n(x)\mem \hy^\n
            \mem
            p_\r\mem \frac{D\hy^\r}{d\t}
            +
                \frac{D\hy^\m}{d\t}\mem
                ({*}R^\zeta{}_\xi{}_{\r\s}(x)\mem p^\r \hy^\s)
        }
    \end{aligned}
\end{align}
\begin{align*}
        &
        \hlg{
            + \frac{C_2}{2}\:
                \hdelta^\m{}_\k\mem
                R^\k{}_{\n\r\s}(x)\mem \hy^\n
                \hy^\r \dot{x}^\s
        }
        \\
        &
        \hlg{
            - \frac{C_2}{2m^2}
                \mem\BB{
                    {*}R^\m{}_{\n\k\l}(x)\mem p^\n\mem \hy^2
                    + \hy^\m\mem 
                        p_\zeta\mem {*}R^\zeta{}_{\xi\k\l}(x)\mem \hy^\xi
                }\mem
                \frac{D}{d\t}\mem ( \hy^\k p^\l )
        }
        \,,
\end{align*}
\vspace{-1.5\baselineskip}
\begin{align}
\label{EXACTQMPD.y}
    \frac{D\hy^\m}{d\t}
    \,&=\,
    \begin{aligned}[t]
        &
        \frac{1}{m^2}\mem
        \BB{
            \hy^\m\mem p^\r
            + p^\m\mem \hy^\r
        }\mem
        \mem p_\zeta \hy^\xi\mem
        \bb{
            {*}R^\zeta{}_\xi{}_{\r\s}(x)
            - \frac{C_2}{2}\mem
                R^\zeta{}_\xi{}_{\r\s,\k}(x)\mem \hy^\k
        }\mem \dot{x}^\s
        \\
        &
        - \frac{C_2}{2m^2}\mem
        \BB{
            \hy^\m\mem p^\r
            + p^\m\mem \hy^\r
        }\mem
        \mem p_\zeta \hy^\xi\mem
            R^\zeta{}_\xi{}_{\m\n}(x)\mem \frac{D\hy^\n}{d\t}
        \\
        &
        + \frac{C_2}{2m^2}\mem
        \mem p_\zeta \hy^\xi\mem
        \BB{
            {*}R^\zeta{}_\xi{}^\m{}_\n(x)\mem p^\n
            \mem
            \hy_\r\mem \dot{x}^\r
            -
            {*}R^\zeta{}_\xi{}^\m{}_\n(x)\mem \hy^\n
            \mem
            p_\r\mem \dot{x}^\r
            +
                \dot{x}^\m\mem
                ({*}R^\zeta{}_\xi{}_{\r\s}(x)\mem p^\r \hy^\s)
        }
        \kern-3em
        \\
        &
        \hlg{
            - \frac{C_2}{2}\:
                \hdelta^\m{}_\k\mem
                {*}R^\k{}_{\n\r\s}(x)\mem \hy^\n
                \hy^\r \dot{x}^\s
        }
        \hlg{
            + \frac{C_2}{2m^2}
                \mem\BB{
                    \hy^\m\mem 
                        p_\zeta\mem R^\zeta{}_{\xi\k\l}(x)\mem \hy^\xi
                }\mem
                \frac{D}{d\t}\mem ( \hy^\k p^\l )
        }
        \,,
    \end{aligned}
    \\[0.12\baselineskip]
\label{EXACTQMPD.p}
    \frac{Dp_\m}{d\t}
    \,&=\,
    \begin{aligned}[t]
        &
        p_\zeta \hy^\xi\mem
        \bb{
            {*}R^\zeta{}_\xi{}_{\m\n}(x)
            - \frac{C_2}{2}\mem
                R^\zeta{}_\xi{}_{\m\n,\r}(x)\mem \hy^\r
        }\mem \dot{x}^\n
        - \frac{C_2}{2}\,
            R^\zeta{}_\xi{}_{\m\n}(x)\mem \frac{D\hy^\n}{d\t}
        \\
        &
        \hlg{
            - \frac{C_2}{2}\:
                p_\k\mem {*}R^\k{}_{\zeta\m\n}(x)\mem
                R^\n{}_{\xi\r\s}(x)\mem
                \hy^\zeta \hy^\xi
                \, \hy^\r \dot{x}^\s
            + \frac{C_2}{2}\,
                R_{\m\n\k\l}(x)\mem \hy^\n
                \mem \frac{D}{d\t}\mem (\hy^\k p^\l)
        }
    \,.
    \end{aligned}
\end{align}
\end{subequations}
Here, we have made uses 
${*}R^\r{}_{\m\r\n} = 0$
and $R^\r{}_{\m\r\n} = 0$.
By iterating \eqref{EXACTQMPD} and truncating,
one reproduces \eqref{QMPD} stated in the main text.

For the reader's sake, we reproduce \eqref{QMPD} below:
\begin{align}
    &
    \dot{x}^\m
    \,=\,
        \frac{p^\m}{m}
        \mem-\mem
        \frac{1}{m}\mem
        \bbsq{\mem
            \frac{C_2}{2}\,
            \BB{
                R^\m{}_{\n\r\s}(x)\mem 
                \hy^\r p^\s
            }
            +
            \bigbig{1-C_2}\,
            \hdelta^\m{}_\k\mem
            \BB{
                R^\k{}_{\n\r\s}(x)\mem 
                \hy^\r p^\s
            }
        }\mem
        \hy^\n
        + \cdots
    \nonumber
    \,,\\
    &
    \frac{D\hy^\m}{d\t}
    \,=\, -
        \frac{1}{m}\mem
        \bbsq{
            \BB{
                {*}\hnem R^\m{}_{\n\r\s}(x)\mem 
                \hy^\r p^\s
            }
            - 
            \bigbig{1-C_2}\,
            \hdelta^\m{}_\k\,
            \BB{
                {*}\hnem R^\k{}_{\n\r\s}(x)\mem 
                \hy^\r p^\s
            }
        }\mem
        \hy^\n
        + \cdots
    \nonumber
    \,,\\[0.12\baselineskip]
    &
    \frac{Dp^\m}{d\t}
    \,=\, -
        \frac{1}{m}\mem
        \BB{
            {*}\hnem R^\m{}_{\n\r\s}(x)\mem
            \hy^\r p^\s
        }\mem 
        p^\n
        + \cdots
    \,,
    \label{QMPD(re)}
\end{align}
where the ellipses signify that terms of $\O(R^{1+n})$, $\O(\hy^{2+n})$, or $\O(D^n\nem R)$ are discarded from the right-hand sides
for $n \geq 1$.

It remains to clarify whether our QMPD equations
agree with the equations stated in the literature
\cite{Steinhoff:2009tk,Harte:2011ku,Vines:2016unv,Compere:2023alp,Ramond:2026fpi}.
This is a nontrivial process,
as our symplectic perturbation framework
fixes the mass-shell constraint
whereas typical literature
exhibits inclinations toward the so-called ``dynamical mass'' approach.
We have checked that
our QMPD equations are equivalent to those in \rcite{Compere:2023alp}
via 
a redefinition of momentum, 
a redefinition of mass, 
and 
a redefinition of the worldline parameter $\tau$.

To elaborate,
recall from \Sec{COVPS} that
the symplectic potential of 
a massive spinning particle model
can be represented in the form
$p_\m\mem dx^\m + (\text{spin term}) + (\text{gauge terms})$
when minimally coupled,
where the content number of gauge terms 
vary over models.
When the quadrupolar coupling coefficient $C_2$ 
is introduced via \eqref{GR.sp2},
it becomes
\begin{subequations}
\begin{align}
    \label{theta-C2-cov}
    \theta
    \,=\,
        p_\m\mem dx^\m
        - 
        \frac{C_2}{2}\,
        p_\m\hem R^\m{}_{\n\r\s}(x)\mem \hy^\n\hem
            \hy^\r\mem dx^\s
        + \cdots
    \,.
\end{align}
Therefore, \textit{without} altering the spin and gauge terms,
one may rewrite \eqref{theta-C2-cov} as
\begin{align}
    \label{theta-C2-can}
    \theta
    \,=\,
        P_\m\mem dx^\m
        + \cdots
    \transition{where}
    P_\s
    \,=\,
        p_\m\mem
        \bb{
            \delta^\m{}_\s
            - 
            \frac{C_2}{2}\,
                R^\m{}_{\n\r\s}(x)\mem \hy^\n \hy^\r
        }
    \,.
\end{align}
\end{subequations}
While
\eqref{theta-C2-can} returns the symplectic potential
to its basic (minimally coupled) form,
the mass-shell constraint gets deformed as
(Hamiltonian perturbation)
\begin{align}
\begin{split}
    \phi_0
    \,&=\,
        \frac{1}{2}\,\BB{
            p^2 + m^2
        }
    \,=\,
        \frac{1}{2}\,\BB{
            P^2 
            + m^2
            + C_2\,
                R_{\m\n\r\s}(x)\mem 
                P^\m \hy^\n \hy^\r P^\s
            + \O(R^2\hy^4)
        }
    \,.
\end{split}
\end{align}
This reproduces \rcite{Compere:2023alp}'s dynamical mass function,
\begin{align}
    \M^2
    \,=\,
        m^2 
        + C_2\,
            R_{\m\n\r\s}(x)\mem 
            P^\m \hy^\n \hy^\r P^\s
        + \O(S^3)
    \,.
\end{align}

Evidently, this manipulation
counts heavily on
the fact that the quadrupolar coupling in \eqref{theta-C2-cov} is a simple $dx$-type term.
It seems unlikely that the dynamical mass function 
(Hamiltonian perturbation)
approach will succeed in achieving the exact 
root-Kerr \cite{gmoov,Kim:2024grz,njmagic.1}, Kerr \cite{njmagic.1}, and Kerr-Newman \cite{njmagic.11} 
black hole couplings
realized in the symplectic perturbation framework,
at least in simple ways
since they crucially involve spin-derivative terms
as explained in \Sec{APPL>BH}.

\section{Definition of Black Hole Coupling}
\label{NJB}

In this appendix, we aim to review the developments
around the question,
\begin{center}
    \vphantom{.}\llap{``\:}\textit{What defines black holes}
    among all massive \\ spinning objects
    in the point-particle effective theory?\rlap{\:''}
\end{center}
We will only cover the ``infrared'' approaches within the point-particle effective theory,
although a large volume of the current literature
pursues
the determination of black hole coupling
from the ultraviolet (i.e., matching with general relativity).

\smallskip
\begin{center}
\begin{minipage}{0.7\linewidth}
\vphantom{.}\llap{1.\: }
    In \textit{linearized gravity},
    the Kerr black hole is defined by
    the \textit{unity of multipole coefficients}
    (i.e., \textit{three-point spin exponentiation}).
\end{minipage}
\end{center}

This result is due to the classic works
\cite{%
    Hansen:1974zz,Geroch:1970cd,%
    Newman:1965tw-janis,janis1965structure,%
    Hernandez:1967zza,%
    Thorne:1980ru%
}
in the '60s and '70s:
the Kerr solution
is shown to exhibit
$C_\ell \eqq 1$ for all $\ell \eqq 1,2,3,\cdots$.

According to the traditional definition,
this seems like a grossly non-minimal coupling.
However, a modern analysis
\cite{ahh2017,Guevara:2018wpp,Guevara:2019fsj,chkl2019,aho2020}
shows that it is the simplest coupling
in terms of 
a large simplification of
the three-point graviton scattering amplitudes
(cf. \rcite{Holstein:2006wi}).
This simplicity is represented as
a complexification of the impact parameter
that adds spin length as an imaginary component
and is dubbed the \textit{three-point spin exponentiation}.

To elaborate,
these three-point amplitudes
describe the process of the black hole (as a point-particle source) sourcing a single massless quanta.
Thus,
they are synonymous to the linearized part of the Kerr metric
\cite{Duff:1973zz,Neill:2013wsa,Vines:2017hyw}.
In this manner,
it is shown that
the three-point spin exponentiation
is the Newman-Janis shift 
in linearized gravity \cite{aho2020}.

The Newman-Janis shift
refers to the property 
discovered by Newman and Janis \cite{Newman:1965tw-janis} in 1965
that the Kerr black hole is secretly
a Schwarzschild black hole
displaced into ``complex spacetime''
in some rough sense,
in which case the spin length (as ring radius) describes the imaginary direction.
This trick was the very method
that enabled the historical discovery of the Kerr-Newman solution \cite{Newman:1965my-kerrmetric}.
The aspects of Newman-Janis shift
in linearized gravity
have been well-studied from the modern scattering amplitudes perspective by \rrcite{ahh2017,Guevara:2018wpp,Guevara:2019fsj,chkl2019,aho2020},
while
its nonlinear aspect has been recently elucidated by \rcite{nja}.

\begin{center}
\begin{minipage}{0.7\linewidth}
\vphantom{.}\llap{2.\: }
    In \textit{self-dual gravity},
    the Kerr black hole is defined by
    the \textit{n-point spin exponentiation}.
\end{minipage}
\end{center}

At the nonlinear orders,
it has been unclear
what principle in the point-particle effective theory 
defines spinning black holes.
Meanwhile,
explorations via scattering amplitudes
investigated the $n$-point Compton scattering process
in which the Kerr black hole
receives $(n \mminus 2)$ gravitons,
where $n \geq 3$.
This research showed that
the $n$-point gravitational Compton amplitudes
can exhibit the spin exponentiation property
as a theoretically allowed possibility,
iff all the helicities of the gravitons are the same
\cite{Johansson:2019dnu,Aoude:2020onz,Lazopoulos:2021mna}.\footnote{
    The mixed-helicity Compton amplitudes are plagued by a contact term issue,
    first noticed in \rcite{ahh2017}.
}

Therefore, by extending the definition of the simplicity of scattering amplitudes,
it has been believed that
this ``$n$-point spin exponentiation''
for all $n = 3,4,5,6,\cdots$
may define the black hole coupling at all nonlinear orders
within the \textit{same-helicity} sector
as a beautiful possibility,
yet without a necessary physical reason.

This scattering amplitudes statement
is equivalent to the proposition that
the point-particle effective action of the Kerr black hole
is unique
(up to worldline field redefintions)
in background self-dual spacetimes.
A self-dual spacetime 
describes a formal complexified limit of spacetime
in which the Riemann tensor becomes self-dual:
${*}R_{\m\n\r\s} = {+i\hem R_{\m\n\r\s}}$.
A well-established fact is that
the same-helicity gravitational Compton amplitudes
due to incoming positive-helicity gravitons
describe
scattering processes in 
self-dual background spacetimes;
see Figure 1 in \rcite{Mason:2009afn}
and also 
\rrcite{bialynicki1981note,ashtekar1986note}.

This unique effective action has been rigorously established in \rrcite{probe-nj,njmagic.1}.
The lesson is that
the $n$-point spin exponentiation
for positive-helicity gravitons
amounts to
a nonlinear generalization of
Newman-Janis shift
for black hole probes
in self-dual spacetimes.

In sum,
three-point spin exponentiation
is Newman-Janis shift 
in linearized gravity,
while
$n$-point spin exponentiation
is nonlinear Newman-Janis shift
in self-dual gravity.

It has been recently shown \cite{probe-nj} that
hidden symmetries (superintegrability) in the self-dual sector
implies the spin exponentiation property of same-helicity Compton amplitudes to all multiplicities
\cite{probe-nj}.\footnote{
    \rcite{probe-nj}
    also provided a physical interpretation of
    the $n$-point spin exponentiation:
    the helicity selection rule \cite{Adamo:2023fbj} exhibited by
    the self-dual Taub-Newman-Unti-Tamburino solution,
    which consists a part in the Kerr black hole
    \cite{nja}.
}
This result has promoted
the $n$-point spin exponentiation
from a mere beauty statement
to a physical assertion based on global symmetries.

\begin{center}
\begin{minipage}{0.7\linewidth}
\vphantom{.}\llap{3.\: }
    In \textit{full gravity},
    the Kerr black hole is defined by
    \textit{(?)}.
\end{minipage}
\end{center}

The unique characterization in the self-dual sector
does not pinpoint the complete black hole couplings
in full nonlinear gravity
due to mixed-helicity interactions.
The extent of this ambiguity
is explicitly examined in
the analysis of \rcite{probe-nj}.

So far, the construction of the exact worldline effective theory of the Kerr black hole 
may have been 
based on model-specific perspectives.
An early sketch was given in \rcite{gmoov}
by extending
the Levi-Steinhoff action
in the spherical top model,
which identified the Newman-Janis shift 
as an important clue.
This work provided valuable intuitions and insights,
although
the subtleties of spin gauge redundancy,
geodesic deviation, and integrability of complex structure
are not 
clearly addressed or resolved
so that
the explicit expressions are limited to the linear-in-Riemann order.
In the meantime, \rcite{ambikerr1} provided an approach based on
the K\"ahler geometry of massive twistor space,
unearthing a unique geometrical perspective on the Newman-Janis shift
that was also demonstrated in \rcite{sst-asym}.
A recent work \cite{bonocore2025higher}
seemed to ponder on the relevance of
supersymmetry as a model-specific feature of a fermionic model.

Eventually,
\rrcite{njmagic.1,njmagic.11}
have established
rigorous constructions of
the exact worldline actions for
Kerr and Kerr-Newman black holes
in the massive twistor model,
providing 
their explicit formulae
to all orders
via the systematic formalism of \rcite{gde}.
As clearly portrayed in 
\rcite{njmagic.1},
this construction
envisions
a curved generalization of
the magical feature of twistor particle theory
that spin is literally an imaginary
deviation in terms of the complexified incidence relation \cite{Shirafuji:1983zd,penrose:maccallum,newman1974curiosity}:
\eqref{incidence}.
The dynamical Newman-Janis shift
is implemented by
a solid mathematical framework known as
adapted complex structure
\cite{guillemin1992grauert,halverscheid2002complexifications,hall2011adapted},
refining \rcite{gmoov}'s earlier sketch.

While 
\rrcite{njmagic.1,njmagic.11}
paved a top-down pathway
to the problem
from massive twistor theory,
\rcite{probe-nj} provided a complementary, bottom-up view
that emphasizes model independence.
\rcite{probe-nj}
defines
an EoM-level implementation of
the Newman-Janis algorithm
in terms of the universal variables $(x,\hy,p)$
of the spinning probes.
A corollary is
the aforementioned hidden symmetry in the self-dual sector.
Notably,
the conserved charges of the spinning probe
are directly obtained
by dynamically Newman-Janis shifting
the conserved charges of the scalar probe.

In this universal manner,
\rcite{probe-nj} also suggests
a natural proposal for uplifting 
the self-dual sector couplings
to the full nonlinear sector
in a rather model-independent fashion.
In this case, the principle that defines black holes
is ``orbit-spin duality,''
a generalization of an old idea due to Newman and Winicour \cite{newman1974curiosity}
in curved backgrounds
\cite{sst-asym,probe-nj,njmagic.1}.
A dedicated article will appear soon \cite{sodual}
for an elaboration on this point.


\let\c\oldc
\let\i\oldi
\let\o\oslash
\bibliographystyle{utphys-modified}
\bibliography{biblio}

\providecommand{\href}[2]{#2}\begingroup\raggedright\begin{thebibliography}{100}

\bibitem{Frenkel:1926zz}
J.~Frenkel, ``{Die Elektrodynamik des rotierenden Elektrons},''
  \href{http://dx.doi.org/10.1007/BF01397099}{{\em Z. Phys.} {\bfseries 37}
  (1926) 243--262}.

\bibitem{Thomas:1927yu}
L.~H. Thomas, ``{The Kinematics of an electron with an axis},''
  \href{http://dx.doi.org/10.1080/14786440108564170}{{\em Phil. Mag. Ser. 7}
  {\bfseries 3} (1927) 1--21}.

\bibitem{Mathisson:1937zz}
M.~Mathisson, ``{Neue mechanik materieller systemes},'' {\em Acta Phys. Polon.}
  {\bfseries 6} (1937) 163--200.

\bibitem{Papapetrou:1951pa}
A.~Papapetrou, ``{Spinning test particles in general relativity. 1.},''
  \href{http://dx.doi.org/10.1098/rspa.1951.0200}{{\em Proc. Roy. Soc. Lond. A}
  {\bfseries 209} (1951) 248--258}.

\bibitem{Dixon:1970zza}
W.~G. Dixon, ``{Dynamics of extended bodies in general relativity. I. Momentum
  and angular momentum},'' \href{http://dx.doi.org/10.1098/rspa.1970.0020}{{\em
  Proc. Roy. Soc. Lond. A} {\bfseries 314} (1970) 499--527}.

\bibitem{Hanson:1974qy}
A.~J. Hanson and T.~Regge, ``{The Relativistic Spherical Top},''
  \href{http://dx.doi.org/10.1016/0003-4916(74)90046-3}{{\em Annals Phys.}
  {\bfseries 87} (1974) 498}.

\bibitem{Bailey:1975fe}
I.~Bailey and W.~Israel, ``{Lagrangian Dynamics of Spinning Particles and
  Polarized Media in General Relativity},''
  \href{http://dx.doi.org/10.1007/BF01609434}{{\em Commun. Math. Phys.}
  {\bfseries 42} (1975) 65--82}.

\bibitem{souriau1970structure}
J.-M. Souriau, {\em Structure des syst{\`e}mes dynamiques: ma{\^\i}trises de
  math{\'e}matiques}.
\newblock Collection Dunod universit{\'e}. Dunod, 1970.

\bibitem{souriau:1970b}
J.-M. Souriau, ``{Sur le mouvement des particules {\`a} spin en relativit{\'e}
  g{\'e}n{\'e}rale},'' {\em C. R. Acad. Sc. Paris, S{\'e}rie A} {\bfseries 271}
  (1970) 751--753.

\bibitem{Souriau:1974ahp}
J.-M. Souriau, ``Mod\`ele de particule \`a spin dans le champ
  \'electromagn\'etique et gravitationnel,'' {\em Annales de l'institut Henri
  Poincar\'e. Section A, Physique Th\'eorique} {\bfseries 20} no.~4, (1974)
  315--364. \url{https://www.numdam.org/item/AIHPA_1974__20_4_315_0/}.

\bibitem{Kunzle:1972uk}
H.~P. K{\"u}nzle, ``{Canonical dynamics of spinning particles in gravitational
  and electromagnetic fields},''
  \href{http://dx.doi.org/10.1063/1.1666045}{{\em J. Math. Phys.} {\bfseries
  13} (1972) 739--744}.

\bibitem{Grassberger:1977tn}
P.~Grassberger, ``{Classical Charged Particles with Spin},''
  \href{http://dx.doi.org/10.1088/0305-4470/11/7/009}{{\em J. Phys. A}
  {\bfseries 11} (1978) 1221}.

\bibitem{Rempel:2015foa}
T.~Rempel and L.~Freidel, ``{Interaction Vertex for Classical Spinning
  Particles},'' \href{http://dx.doi.org/10.1103/PhysRevD.94.044011}{{\em Phys.
  Rev. D} {\bfseries 94} no.~4, (2016) 044011},
  \href{http://arxiv.org/abs/1507.05826}{{\ttfamily arXiv:1507.05826
  [hep-th]}}.

\bibitem{Deriglazov:2015bqa}
A.~A. Deriglazov and W.~Guzm{\'a}n~Ramirez, ``{Lagrangian formulation for
  Mathisson-Papapetrou-Tulczyjew-Dixon (MPTD) equations},''
  \href{http://dx.doi.org/10.1103/PhysRevD.92.124017}{{\em Phys. Rev. D}
  {\bfseries 92} (2015) 124017},
  \href{http://arxiv.org/abs/1509.04926}{{\ttfamily arXiv:1509.04926 [gr-qc]}}.

\bibitem{Basile:2023vyg}
T.~Basile, E.~Joung, and T.~Oh, ``{Manifestly covariant worldline actions from
  coadjoint orbits. Part I. Generalities and vectorial descriptions},''
  \href{http://dx.doi.org/10.1007/JHEP01(2024)018}{{\em JHEP} {\bfseries 01}
  (2024) 018}, \href{http://arxiv.org/abs/2307.13644}{{\ttfamily
  arXiv:2307.13644 [hep-th]}}.

\bibitem{Haddad:2024ebn}
K.~Haddad, G.~U. Jakobsen, G.~Mogull, and J.~Plefka, ``{Spinning bodies in
  general relativity from bosonic worldline oscillators},''
  \href{http://dx.doi.org/10.1007/JHEP02(2025)019}{{\em JHEP} {\bfseries 02}
  (2025) 019}, \href{http://arxiv.org/abs/2411.08176}{{\ttfamily
  arXiv:2411.08176 [hep-th]}}.

\bibitem{Berezin:1976eg}
F.~A. Berezin and M.~S. Marinov, ``{Particle Spin Dynamics as the Grassmann
  Variant of Classical Mechanics},''
  \href{http://dx.doi.org/10.1016/0003-4916(77)90335-9}{{\em Annals Phys.}
  {\bfseries 104} (1977) 336}.

\bibitem{Casalbuoni:1975hx}
R.~Casalbuoni, ``{Relativity and Supersymmetries},''
  \href{http://dx.doi.org/10.1016/0370-2693(76)90044-7}{{\em Phys. Lett. B}
  {\bfseries 62} (1976) 49--50}.

\bibitem{Casalbuoni:1976tz}
R.~Casalbuoni, ``{The Classical Mechanics for Bose-Fermi Systems},''
  \href{http://dx.doi.org/10.1007/BF02729860}{{\em Nuovo Cim. A} {\bfseries 33}
  (1976) 389}.

\bibitem{Brink:1976sz}
L.~Brink, S.~Deser, B.~Zumino, P.~Di~Vecchia, and P.~S. Howe, ``{Local
  Supersymmetry for Spinning Particles},''
  \href{http://dx.doi.org/10.1016/0370-2693(76)90115-5}{{\em Phys. Lett. B}
  {\bfseries 64} (1976) 435}. [Erratum: Phys.Lett.B 68, 488 (1977)].

\bibitem{Brink:1976uf}
L.~Brink, P.~Di~Vecchia, and P.~S. Howe, ``{A Lagrangian Formulation of the
  Classical and Quantum Dynamics of Spinning Particles},''
  \href{http://dx.doi.org/10.1016/0550-3213(77)90364-9}{{\em Nucl. Phys. B}
  {\bfseries 118} (1977) 76--94}.

\bibitem{woodhouse1997geometric}
N.~M.~J. Woodhouse, {\em Geometric quantization}.
\newblock Oxford University Press, 2~ed., 1997.

\bibitem{Lyakhovich:1996we}
S.~L. Lyakhovich, A.~Y. Segal, and A.~A. Sharapov, ``{A Universal model of D =
  4 spinning particle},''
  \href{http://dx.doi.org/10.1103/PhysRevD.54.5223}{{\em Phys. Rev. D}
  {\bfseries 54} (1996) 5223--5238},
  \href{http://arxiv.org/abs/hep-th/9603174}{{\ttfamily arXiv:hep-th/9603174}}.

\bibitem{Rempel:2016jbn}
T.~Rempel and L.~Freidel, ``{A Classical and Spinorial Description of the
  Relativistic Spinning Particle},''
  \href{http://arxiv.org/abs/1612.00551}{{\ttfamily arXiv:1612.00551
  [hep-th]}}.

\bibitem{Zima:1995db}
V.~G. Zima and S.~Fedoruk, ``{Spinor (super)particle with a commuting index
  spinor},'' {\em JETP Lett.} {\bfseries 61} (1995) 251--256.

\bibitem{Mezincescu:2015apa}
L.~Mezincescu, A.~J. Routh, and P.~K. Townsend, ``{Twistors and the massive
  spinning particle},''
  \href{http://dx.doi.org/10.1088/1751-8113/49/2/025401}{{\em J. Phys. A}
  {\bfseries 49} no.~2, (2016) 025401},
  \href{http://arxiv.org/abs/1508.05350}{{\ttfamily arXiv:1508.05350
  [hep-th]}}.

\bibitem{Penrose:1974di}
R.~Penrose, ``Twistors and particles: An outline,'' in {\em {Feldafing
  Conference of the Max-Planck Inst. on Quantum Theory and the Structure of
  Space-time}}, pp.~129--145.
\newblock 1974.

\bibitem{Perjes:1974ra}
Z.~Perj\'es, ``Twistor variables of relativistic mechanics,''
  \href{http://dx.doi.org/10.1103/PhysRevD.11.2031}{{\em Phys. Rev. D}
  {\bfseries 11} (1975) 2031--2041}.

\bibitem{tod1977some}
P.~Tod, ``Some symplectic forms arising in twistor theory,'' {\em Reports on
  Mathematical Physics} {\bfseries 11} no.~3, (1977) 339--346.

\bibitem{Bette:1989zt}
A.~Bette, ``{On Lorentz particle with spin. A twistorial approach},''
  \href{http://dx.doi.org/10.1016/0034-4877(89)90029-3}{{\em Rept. Math. Phys.}
  {\bfseries 28} (1989) 133--140}.

\bibitem{Bette:2004ip}
A.~Bette, J.~A. de~Azcarraga, J.~Lukierski, and C.~Miquel-Espanya, ``{Massive
  relativistic particle model with spin and electric charge from two-twistor
  dynamics},'' \href{http://dx.doi.org/10.1016/j.physletb.2004.06.051}{{\em
  Phys. Lett. B} {\bfseries 595} (2004) 491--497},
  \href{http://arxiv.org/abs/hep-th/0405166}{{\ttfamily arXiv:hep-th/0405166}}.

\bibitem{Fedoruk:2003td}
S.~Fedoruk and V.~G. Zima, ``{Bitwistor formulation of massive spinning
  particle},'' \href{http://arxiv.org/abs/hep-th/0308154}{{\ttfamily
  arXiv:hep-th/0308154}}.

\bibitem{Fedoruk:2007dd}
S.~Fedoruk, A.~Frydryszak, J.~Lukierski, and C.~Miquel-Espanya, ``{Massive
  particle model with spin from a hybrid (spacetime-twistorial) phase space
  geometry and its quantization},'' in {\em {22nd Max Born Symposium on
  Quantum, Super and Twistors: A Conference in Honor of Jerzy Lukierski on His
  70th Birthday}}.
\newblock 2, 2007.
\newblock \href{http://arxiv.org/abs/hep-th/0702050}{{\ttfamily
  arXiv:hep-th/0702050}}.

\bibitem{Fedoruk:2014vqa}
S.~Fedoruk and J.~Lukierski, ``{Massive twistor particle with spin generated by
  Souriau{\textendash}Wess{\textendash}Zumino term and its quantization},''
  \href{http://dx.doi.org/10.1016/j.physletb.2014.04.059}{{\em Phys. Lett. B}
  {\bfseries 733} (2014) 309--315},
  \href{http://arxiv.org/abs/1403.4127}{{\ttfamily arXiv:1403.4127 [hep-th]}}.

\bibitem{deAzcarraga:2014hda}
J.~A. de~Azcarraga, S.~Fedoruk, J.~M. Izquierdo, and J.~Lukierski,
  ``{Two-twistor particle models and free massive higher spin fields},''
  \href{http://dx.doi.org/10.1007/JHEP04(2015)010}{{\em JHEP} {\bfseries 04}
  (2015) 010}, \href{http://arxiv.org/abs/1409.7169}{{\ttfamily arXiv:1409.7169
  [hep-th]}}.

\bibitem{Deguchi:2015iuw}
S.~Deguchi and S.~Okano, ``{Gauged twistor formulation of a massive spinning
  particle in four dimensions},''
  \href{http://dx.doi.org/10.1103/PhysRevD.93.045016}{{\em Phys. Rev. D}
  {\bfseries 93} no.~4, (2016) 045016},
  \href{http://arxiv.org/abs/1512.07740}{{\ttfamily arXiv:1512.07740
  [hep-th]}}. [Erratum: Phys.Rev.D 93, 089906 (2016)].

\bibitem{Kim:2021rda}
J.-H. Kim, J.-W. Kim, and S.~Lee, ``{The relativistic spherical top as a
  massive twistor},'' \href{http://dx.doi.org/10.1088/1751-8121/ac11be}{{\em J.
  Phys. A} {\bfseries 54} no.~33, (2021) 335203},
  \href{http://arxiv.org/abs/2102.07063}{{\ttfamily arXiv:2102.07063
  [hep-th]}}.

\bibitem{ambikerr1}
J.-H. Kim and S.~Lee, ``{Symplectic Perturbation Theory in Massive Ambitwistor
  Space: A Zig-Zag Theory of Massive Spinning Particles},''
  \href{http://arxiv.org/abs/2301.06203}{{\ttfamily arXiv:2301.06203
  [hep-th]}}.

\bibitem{Frydryszak:1996mu}
A.~Frydryszak, ``{Lagrangian models of the particles with spin: The First
  seventy years},'' \href{http://arxiv.org/abs/hep-th/9601020}{{\ttfamily
  arXiv:hep-th/9601020}}.

\bibitem{Rivas:2002}
M.~Rivas, \href{http://dx.doi.org/10.1007/0-306-47133-7}{{\em {Kinematical
  Theory of Spinning Particles}}}.
\newblock Springer Dordrecht, 2002.

\bibitem{Plyushchay:1990eq}
M.~S. Plyushchay, ``{Lagrangian formulation for the massless (super)particles
  in (super)twistor approach},''
  \href{http://dx.doi.org/10.1016/0370-2693(90)90421-2}{{\em Phys. Lett. B}
  {\bfseries 240} (1990) 133--136}.

\bibitem{Plyushchay:1992ga}
M.~S. Plyushchay, ``{Pseudoclassical description of the massive spinning
  particle in $d$-dimensions},''
  \href{http://dx.doi.org/10.1142/S0217732393000969}{{\em Mod. Phys. Lett. A}
  {\bfseries 8} (1993) 937--945}.

\bibitem{Bars:2005ze}
I.~Bars and M.~Picon, ``{Single twistor description of massless, massive, AdS,
  and other interacting particles},''
  \href{http://dx.doi.org/10.1103/PhysRevD.73.064002}{{\em Phys. Rev. D}
  {\bfseries 73} (2006) 064002},
  \href{http://arxiv.org/abs/hep-th/0512091}{{\ttfamily arXiv:hep-th/0512091}}.

\bibitem{Bars:2005ax}
I.~Bars and M.~Picon, ``{Twistor transform in d dimensions and a unifying role
  for twistors},'' \href{http://dx.doi.org/10.1103/PhysRevD.73.064033}{{\em
  Phys. Rev. D} {\bfseries 73} (2006) 064033},
  \href{http://arxiv.org/abs/hep-th/0512348}{{\ttfamily arXiv:hep-th/0512348}}.

\bibitem{Deriglazov:2017jub}
A.~A. Deriglazov and W.~Guzm{\'a}n~Ramirez, ``{Recent progress on the
  description of relativistic spin: vector model of spinning particle and
  rotating body with gravimagnetic moment in General Relativity},''
  \href{http://dx.doi.org/10.1155/2017/7397159}{{\em Adv. Math. Phys.}
  {\bfseries 2017} (2017) 7397159},
  \href{http://arxiv.org/abs/1710.07135}{{\ttfamily arXiv:1710.07135 [gr-qc]}}.

\bibitem{Jakobsen:2023tvm}
G.~U. Jakobsen, ``{Spin and Susceptibility Effects of Electromagnetic
  Self-Force in Effective Field Theory},''
  \href{http://dx.doi.org/10.1103/PhysRevLett.132.151601}{{\em Phys. Rev.
  Lett.} {\bfseries 132} no.~15, (2024) 151601},
  \href{http://arxiv.org/abs/2311.04151}{{\ttfamily arXiv:2311.04151
  [hep-th]}}.

\bibitem{Damour:2024mzo}
T.~Damour and P.~Iglesias-Zemmour, ``{Editorial note to: On the motion of
  spinning particles in general relativity by Jean-Marie Souriau},''
  \href{http://dx.doi.org/10.1007/s10714-024-03294-w}{{\em Gen. Rel. Grav.}
  {\bfseries 56} no.~10, (2024) 127},
  \href{http://arxiv.org/abs/2401.10013}{{\ttfamily arXiv:2401.10013 [gr-qc]}}.

\bibitem{sst-asym}
J.-H. Kim, ``{Asymptotic spinspacetime},''
  \href{http://dx.doi.org/10.1103/PhysRevD.111.105011}{{\em Phys. Rev. D}
  {\bfseries 111} no.~10, (2025) 105011},
  \href{http://arxiv.org/abs/2309.11886}{{\ttfamily arXiv:2309.11886
  [hep-th]}}.

\bibitem{Kim:2024grz}
J.-H. Kim, J.-W. Kim, and S.~Lee, ``{Massive twistor worldline in
  electromagnetic fields},''
  \href{http://dx.doi.org/10.1007/JHEP08(2024)080}{{\em JHEP} {\bfseries 08}
  (2024) 080}, \href{http://arxiv.org/abs/2405.17056}{{\ttfamily
  arXiv:2405.17056 [hep-th]}}.

\bibitem{Bargmann:1959gz}
V.~Bargmann, L.~Michel, and V.~L. Telegdi, ``{Precession of the polarization of
  particles moving in a homogeneous electromagnetic field},''
  \href{http://dx.doi.org/10.1103/PhysRevLett.2.435}{{\em Phys. Rev. Lett.}
  {\bfseries 2} (1959) 435--436}.

\bibitem{kirillov1975elements}
A.~A. Kirillov, {\em Elements of the Theory of Representations}.
\newblock Berlin, Heidelberg, New York: Springer 1975, Chap. 15, 1975.

\bibitem{kostant1970orbits}
B.~Kostant, ``Orbits and quantization theory,'' in {\em Proc. Int. Cong. Math.,
  Nice}, vol.~2, p.~395.
\newblock 1970.

\bibitem{bacry1967space}
H.~Bacry, ``Space-time and degrees of freedom of the elementary particle,''
  {\em Communications in Mathematical Physics} {\bfseries 5} no.~2, (1967)
  97--105.

\bibitem{arens1971classical}
R.~Arens, ``Classical {Lorentz} invariant particles,'' {\em Journal of
  Mathematical Physics} {\bfseries 12} no.~12, (1971) 2415--2422.

\bibitem{Carinena:1989uw}
J.~F. Carinena, J.~M. Gracia-Bondia, and J.~C. Varilly, ``{Relativistic Quantum
  Kinematics in the Moyal Representation},''
  \href{http://dx.doi.org/10.1088/0305-4470/23/6/015}{{\em J. Phys. A}
  {\bfseries 23} (1990) 901}.

\bibitem{pryce1948mass}
M.~H.~L. Pryce, ``The mass-centre in the restricted theory of relativity and
  its connexion with the quantum theory of elementary particles,'' {\em
  Proceedings of the Royal Society of London. Series A. Mathematical and
  Physical Sciences} {\bfseries 195} no.~1040, (1948) 62--81.

\bibitem{Newton:1949cq}
T.~D. {Newton} and E.~P. Wigner, ``Localized states for elementary systems,''
  \href{http://dx.doi.org/10.1103/RevModPhys.21.400}{{\em Rev. Mod. Phys.}
  {\bfseries 21} (1949) 400--406}.

\bibitem{moller1949dynamique}
C.~M{\o}ller, ``Sur la dynamique des syst{\`e}mes ayant un moment angulaire
  interne,'' {\em Annales de l'institut Henri Poincar{\'e}} {\bfseries 11}
  no.~5, (1949) 251--278.

\bibitem{moller1949definition}
C.~M{\o}ller, ``On the definition of the centre of gravity of an arbitrary
  closed system in the theory of relativity,'' {\em Communications of the
  Dublin Institute for Advanced Studies} no.~5, (1949) 3--42.

\bibitem{fleming1965covariant}
G.~N. Fleming, ``Covariant position operators, spin, and locality,'' {\em
  Physical Review} {\bfseries 137} no.~1B, (1965) B188.

\bibitem{Dixon:1970zz}
W.~G. Dixon, ``{Dynamics of extended bodies in general relativity. II. Moments
  of the charge-current vector},''
  \href{http://dx.doi.org/10.1098/rspa.1970.0191}{{\em Proc. Roy. Soc. Lond. A}
  {\bfseries 319} (1970) 509--547}.

\bibitem{Costa:2014nta}
L.~F.~O. Costa and J.~Nat\'ario, ``{Center of mass, spin supplementary
  conditions, and the momentum of spinning particles},''
  \href{http://dx.doi.org/10.1007/978-3-319-18335-0_6}{{\em Fund. Theor. Phys.}
  {\bfseries 179} (2015) 215--258},
  \href{http://arxiv.org/abs/1410.6443}{{\ttfamily arXiv:1410.6443 [gr-qc]}}.

\bibitem{Mikoczi:2016fiy}
B.~Mik{\'o}czi, ``{Spin supplementary conditions for spinning compact
  binaries},'' \href{http://dx.doi.org/10.1103/PhysRevD.95.064023}{{\em Phys.
  Rev. D} {\bfseries 95} no.~6, (2017) 064023},
  \href{http://arxiv.org/abs/1609.01536}{{\ttfamily arXiv:1609.01536 [gr-qc]}}.

\bibitem{Witzany:2018ahb}
V.~Witzany, J.~Steinhoff, and G.~Lukes-Gerakopoulos, ``{{Hamilton}ians and
  canonical coordinates for spinning particles in curved space-time},''
  \href{http://dx.doi.org/10.1088/1361-6382/ab002f}{{\em Class. Quant. Grav.}
  {\bfseries 36} no.~7, (2019) 075003},
  \href{http://arxiv.org/abs/1808.06582}{{\ttfamily arXiv:1808.06582 [gr-qc]}}.

\bibitem{Steinhoff:2015ksa}
J.~Steinhoff, ``{Spin gauge symmetry in the action principle for classical
  relativistic particles},'' \href{http://arxiv.org/abs/1501.04951}{{\ttfamily
  arXiv:1501.04951 [gr-qc]}}.

\bibitem{Steinhoff:2009tk}
J.~Steinhoff and D.~Puetzfeld, ``{Multipolar equations of motion for extended
  test bodies in General Relativity},''
  \href{http://dx.doi.org/10.1103/PhysRevD.81.044019}{{\em Phys. Rev. D}
  {\bfseries 81} (2010) 044019},
  \href{http://arxiv.org/abs/0909.3756}{{\ttfamily arXiv:0909.3756 [gr-qc]}}.

\bibitem{Harte:2011ku}
A.~I. Harte, ``{Mechanics of extended masses in general relativity},''
  \href{http://dx.doi.org/10.1088/0264-9381/29/5/055012}{{\em Class. Quant.
  Grav.} {\bfseries 29} (2012) 055012},
  \href{http://arxiv.org/abs/1103.0543}{{\ttfamily arXiv:1103.0543 [gr-qc]}}.

\bibitem{Vines:2016unv}
J.~Vines, D.~Kunst, J.~Steinhoff, and T.~Hinderer, ``{Canonical {Hamilton}ian
  for an extended test body in curved spacetime: To quadratic order in spin},''
  \href{http://dx.doi.org/10.1103/PhysRevD.104.029902}{{\em Phys. Rev. D}
  {\bfseries 93} no.~10, (2016) 103008},
  \href{http://arxiv.org/abs/1601.07529}{{\ttfamily arXiv:1601.07529 [gr-qc]}}.
  [Erratum: Phys.Rev.D 104, 029902 (2021)].

\bibitem{Compere:2023alp}
G.~Comp{\`e}re, A.~Druart, and J.~Vines, ``{Generalized Carter constant for
  quadrupolar test bodies in Kerr spacetime},''
  \href{http://dx.doi.org/10.21468/SciPostPhys.15.6.226}{{\em SciPost Phys.}
  {\bfseries 15} no.~6, (2023) 226},
  \href{http://arxiv.org/abs/2302.14549}{{\ttfamily arXiv:2302.14549 [gr-qc]}}.

\bibitem{Ramond:2026fpi}
P.~Ramond, S.~Isoyama, and A.~Druart, ``{Symplectic mechanics of relativistic
  spinning compact bodies. III. quadratic-in-spin integrability in Type-D
  Einstein spacetimes: persistence and breakdown},''
  \href{http://arxiv.org/abs/2601.06416}{{\ttfamily arXiv:2601.06416 [gr-qc]}}.

\bibitem{khriplovich1989particle}
I.~Khriplovich, ``Particle with internal angular momentum in a gravitational
  field,'' {\em Soviet Physics-JETP (English Translation)} {\bfseries 69}
  no.~2, (1989) 217--219.

\bibitem{Yee:1993ya}
K.~Yee and M.~Bander, ``{Equations of motion for spinning particles in external
  electromagnetic and gravitational fields},''
  \href{http://dx.doi.org/10.1103/PhysRevD.48.2797}{{\em Phys. Rev. D}
  {\bfseries 48} (1993) 2797--2799},
  \href{http://arxiv.org/abs/hep-th/9302117}{{\ttfamily arXiv:hep-th/9302117}}.

\bibitem{Khriplovich:1997ni}
I.~B. Khriplovich and A.~A. Pomeransky, ``{Equations of motion of spinning
  relativistic particle in external fields},''
  \href{http://dx.doi.org/10.1134/1.558554}{{\em J. Exp. Theor. Phys.}
  {\bfseries 86} (1998) 839--849},
  \href{http://arxiv.org/abs/gr-qc/9710098}{{\ttfamily arXiv:gr-qc/9710098}}.

\bibitem{Thorne:1984mz}
K.~S. Thorne and J.~B. Hartle, ``{Laws of motion and precession for black holes
  and other bodies},'' \href{http://dx.doi.org/10.1103/PhysRevD.31.1815}{{\em
  Phys. Rev. D} {\bfseries 31} (1984) 1815--1837}.

\bibitem{dyson1990feynman}
F.~J. Dyson, ``{Feynman’s} proof of the maxwell equations,'' {\em American
  Journal of Physics} {\bfseries 58} no.~3, (1990) 209--211.

\bibitem{lee1990feynman}
C.~Lee, ``{The Feynman-Dyson proof of the gauge field equations},'' {\em
  Physics letters A} {\bfseries 148} no.~3-4, (1990) 146--148.

\bibitem{tanimura1992relativistic}
S.~Tanimura, ``{Relativistic generalization and extension to the non-Abelian
  gauge theory of Feynman's proof of the Maxwell equations},'' {\em Annals of
  Physics} {\bfseries 220} no.~2, (1992) 229--247.

\bibitem{stern1993deformed}
A.~Stern and I.~Yakushin, ``{Deformed Wong particles},'' {\em Physical Review
  D} {\bfseries 48} no.~10, (1993) 4974.

\bibitem{chou1994dynamical}
C.~Chou, ``{Dynamical equations of spinning particles. {Feynman's} proof},''
  {\em Physics Letters B} {\bfseries 323} no.~2, (1994) 147--152.

\bibitem{berard1999dirac}
A.~B{\'e}rard, Y.~Grandati, and H.~Mohrbach, ``{Dirac} monopole with {Feynman}
  brackets,'' {\em Physics Letters A} {\bfseries 254} no.~3-4, (1999) 133--136.

\bibitem{csg}
J.-H. Kim, ``{Covariant Symplectic Geometry of Classical Particles},''
  \href{http://arxiv.org/abs/2603.21934}{{\ttfamily arXiv:2603.21934
  [hep-th]}}.

\bibitem{dewitt1965dynamical}
B.~S. DeWitt, {\em Dynamical theory of groups and fields}.
\newblock Gordon and Breach, 1965.

\bibitem{DeWitt:2011nnj}
B.~DeWitt, \href{http://dx.doi.org/10.1007/978-3-540-36911-0}{{\em {Bryce
  DeWitt's Lectures on Gravitation}}}, vol.~826.
\newblock Springer, 2011.

\bibitem{vanHolten:1992bu}
J.~W. van Holten and R.~H. Rietdijk, ``{Symmetries and motions in manifolds},''
  \href{http://dx.doi.org/10.1016/0393-0440(93)90079-T}{{\em J. Geom. Phys.}
  {\bfseries 11} (1993) 559},
  \href{http://arxiv.org/abs/hep-th/9205074}{{\ttfamily arXiv:hep-th/9205074}}.

\bibitem{Gibbons:1993ap}
G.~W. Gibbons, R.~H. Rietdijk, and J.~W. van Holten, ``{SUSY in the sky},''
  \href{http://dx.doi.org/10.1016/0550-3213(93)90472-2}{{\em Nucl. Phys. B}
  {\bfseries 404} (1993) 42--64},
  \href{http://arxiv.org/abs/hep-th/9303112}{{\ttfamily arXiv:hep-th/9303112}}.

\bibitem{dAmbrosi:2015ndl}
G.~d'Ambrosi, S.~Satish~Kumar, and J.~W. van Holten, ``{Covariant {Hamilton}ian
  spin dynamics in curved space{\textendash}time},''
  \href{http://dx.doi.org/10.1016/j.physletb.2015.03.007}{{\em Phys. Lett. B}
  {\bfseries 743} (2015) 478--483},
  \href{http://arxiv.org/abs/1501.04879}{{\ttfamily arXiv:1501.04879 [gr-qc]}}.

\bibitem{tulczyjew1959motion}
W.~Tulczyjew, ``Motion of multipole particles in general relativity theory,''
  {\em Acta Phys. Pol} {\bfseries 18} no.~393, (1959) 94.

\bibitem{dixon1964covariant}
W.~G. Dixon, ``A covariant multipole formalism for extended test bodies in
  general relativity,'' {\em Il Nuovo Cimento (1955-1965)} {\bfseries 34}
  no.~2, (1964) 317--339.

\bibitem{gde}
J.-H. Kim, ``{Geodesic deviation to all orders via a tangent bundle
  formalism},'' \href{http://arxiv.org/abs/2509.23600}{{\ttfamily
  arXiv:2509.23600 [gr-qc]}}.

\bibitem{Ruse:1931ht}
H.~S. Ruse, ``{Taylor's theorem in the tensor calculus},''
  \href{http://dx.doi.org/10.1112/plms/s2-32.1.87}{{\em Proc. Lond. Math. Soc.}
  {\bfseries 32} (1931) 87}.

\bibitem{Synge:1931zz}
J.~L. Synge, ``{A characteristic function in Riemannian space and its
  applications to the solution of geodesic triangles},''
  \href{http://dx.doi.org/10.1112/plms/s2-32.1.241}{{\em Proc. Lond. Math.
  Soc.} {\bfseries 32} (1931) 241}.

\bibitem{Poisson:2011nh}
E.~Poisson, A.~Pound, and I.~Vega, ``{The Motion of point particles in curved
  spacetime},'' \href{http://dx.doi.org/10.12942/lrr-2011-7}{{\em Living Rev.
  Rel.} {\bfseries 14} (2011) 7},
  \href{http://arxiv.org/abs/1102.0529}{{\ttfamily arXiv:1102.0529 [gr-qc]}}.

\bibitem{Vines:2014oba}
J.~Vines, ``{Geodesic deviation at higher orders via covariant bitensors},''
  \href{http://dx.doi.org/10.1007/s10714-015-1901-9}{{\em Gen. Rel. Grav.}
  {\bfseries 47} no.~5, (2015) 59},
  \href{http://arxiv.org/abs/1407.6992}{{\ttfamily arXiv:1407.6992 [gr-qc]}}.

\bibitem{Goldberger:2004jt}
W.~D. Goldberger and I.~Z. Rothstein, ``{An Effective field theory of gravity
  for extended objects},''
  \href{http://dx.doi.org/10.1103/PhysRevD.73.104029}{{\em Phys. Rev. D}
  {\bfseries 73} (2006) 104029},
  \href{http://arxiv.org/abs/hep-th/0409156}{{\ttfamily arXiv:hep-th/0409156}}.

\bibitem{Porto:2005ac}
R.~A. Porto, ``{Post-{Newton}ian corrections to the motion of spinning bodies
  in NRGR},'' \href{http://dx.doi.org/10.1103/PhysRevD.73.104031}{{\em Phys.
  Rev. D} {\bfseries 73} (2006) 104031},
  \href{http://arxiv.org/abs/gr-qc/0511061}{{\ttfamily arXiv:gr-qc/0511061}}.

\bibitem{Levi:2015msa}
M.~Levi and J.~Steinhoff, ``{Spinning gravitating objects in the effective
  field theory in the post-{Newton}ian scheme},''
  \href{http://dx.doi.org/10.1007/JHEP09(2015)219}{{\em JHEP} {\bfseries 09}
  (2015) 219}, \href{http://arxiv.org/abs/1501.04956}{{\ttfamily
  arXiv:1501.04956 [gr-qc]}}.

\bibitem{Porto:2016pyg}
R.~A. Porto, ``{The effective field theorist\textquoteright{}s approach to
  gravitational dynamics},''
  \href{http://dx.doi.org/10.1016/j.physrep.2016.04.003}{{\em Phys. Rept.}
  {\bfseries 633} (2016) 1--104},
  \href{http://arxiv.org/abs/1601.04914}{{\ttfamily arXiv:1601.04914
  [hep-th]}}.

\bibitem{Levi:2018nxp}
M.~Levi, ``{Effective Field Theories of Post-{Newton}ian Gravity: A
  comprehensive review},''
  \href{http://dx.doi.org/10.1088/1361-6633/ab12bc}{{\em Rept. Prog. Phys.}
  {\bfseries 83} no.~7, (2020) 075901},
  \href{http://arxiv.org/abs/1807.01699}{{\ttfamily arXiv:1807.01699
  [hep-th]}}.

\bibitem{Will:2011nz}
C.~M. Will, ``{On the unreasonable effectiveness of the post-{Newton}ian
  approximation in gravitational physics},''
  \href{http://dx.doi.org/10.1073/pnas.1103127108}{{\em Proc. Nat. Acad. Sci.}
  {\bfseries 108} (2011) 5938},
  \href{http://arxiv.org/abs/1102.5192}{{\ttfamily arXiv:1102.5192 [gr-qc]}}.

\bibitem{Blanchet:2013haa}
L.~Blanchet, ``Gravitational radiation from post-{Newton}ian sources and
  inspiralling compact binaries,''
  \href{http://dx.doi.org/10.12942/lrr-2014-2}{{\em Living Rev. Rel.}
  {\bfseries 17} (2014) 2},
  \href{http://arxiv.org/abs/gr-qc/0202016}{{\ttfamily arXiv:gr-qc/0202016
  [gr-qc]}}.

\bibitem{Damour:2016gwp}
T.~Damour, ``Gravitational scattering, post-{Minkowski}an approximation and
  effective-one-body theory,''
  \href{http://dx.doi.org/10.1103/PhysRevD.94.104015}{{\em Phys. Rev. D}
  {\bfseries 94} no.~10, (2016) 104015},
  \href{http://arxiv.org/abs/1609.00354}{{\ttfamily arXiv:1609.00354 [gr-qc]}}.

\bibitem{Cheung:2018wkq}
C.~Cheung, I.~Z. Rothstein, and M.~P. Solon, ``From scattering amplitudes to
  classical potentials in the post-{Minkowski}an expansion,''
  \href{http://dx.doi.org/10.1103/PhysRevLett.121.251101}{{\em Phys. Rev.
  Lett.} {\bfseries 121} no.~25, (2018) 251101},
  \href{http://arxiv.org/abs/1808.02489}{{\ttfamily arXiv:1808.02489
  [hep-th]}}.

\bibitem{Cristofoli:2019neg}
A.~Cristofoli, N.~E.~J. Bjerrum-Bohr, P.~H. Damgaard, and P.~Vanhove, ``On
  post-{Minkowski}an {Hamilton}ians in general relativity,''
  \href{http://dx.doi.org/10.1103/PhysRevD.100.084040}{{\em Phys. Rev. D}
  {\bfseries 100} no.~8, (2019) 084040},
  \href{http://arxiv.org/abs/1906.01579}{{\ttfamily arXiv:1906.01579
  [hep-th]}}.

\bibitem{Bjerrum-Bohr:2021din}
N.~E.~J. Bjerrum-Bohr, P.~H. Damgaard, L.~Plante, and P.~Vanhove, ``The
  amplitude for classical gravitational scattering at third post-{Minkowski}an
  order,'' \href{http://dx.doi.org/10.1007/JHEP08(2021)172}{{\em JHEP}
  {\bfseries 08} (2021) 172}, \href{http://arxiv.org/abs/2105.05218}{{\ttfamily
  arXiv:2105.05218 [hep-th]}}.

\bibitem{Kalin:2020mvi}
G.~Kalin and R.~A. Porto, ``From boundary data to bound states,''
  \href{http://dx.doi.org/10.1007/JHEP01(2020)072}{{\em JHEP} {\bfseries 01}
  (2020) 072}, \href{http://arxiv.org/abs/1910.03008}{{\ttfamily
  arXiv:1910.03008 [hep-th]}}.

\bibitem{Kalin:2020fhe}
G.~Kalin, Z.~Liu, and R.~A. Porto, ``Conservative dynamics of binary systems to
  third post-{Minkowski}an order from the effective field theory approach,''
  \href{http://dx.doi.org/10.1103/PhysRevLett.125.261103}{{\em Phys. Rev.
  Lett.} {\bfseries 125} (2020) 261103},
  \href{http://arxiv.org/abs/2007.04977}{{\ttfamily arXiv:2007.04977
  [hep-th]}}.

\bibitem{Mogull:2020sak}
G.~Mogull, J.~Plefka, and J.~Steinhoff, ``{Classical black hole scattering from
  a worldline quantum field theory},''
  \href{http://dx.doi.org/10.1007/JHEP02(2021)048}{{\em JHEP} {\bfseries 02}
  (2021) 048}, \href{http://arxiv.org/abs/2010.02865}{{\ttfamily
  arXiv:2010.02865 [hep-th]}}.

\bibitem{Jakobsen:2021smu}
G.~U. Jakobsen, G.~Mogull, J.~Plefka, and J.~Steinhoff, ``Classical
  gravitational bremsstrahlung from a worldline quantum field theory,''
  \href{http://dx.doi.org/10.1103/PhysRevLett.126.201103}{{\em Phys. Rev.
  Lett.} {\bfseries 126} (2021) 201103},
  \href{http://arxiv.org/abs/2101.12688}{{\ttfamily arXiv:2101.12688}}.

\bibitem{Bjerrum-Bohr:2022blt}
N.~E.~J. Bjerrum-Bohr, P.~H. Damgaard, L.~Plante, and P.~Vanhove, ``{The SAGEX
  review on scattering amplitudes Chapter 13: Post-{Minkowski}an expansion from
  scattering amplitudes},''
  \href{http://dx.doi.org/10.1088/1751-8121/ac7a78}{{\em J. Phys. A} {\bfseries
  55} no.~44, (2022) 443014}, \href{http://arxiv.org/abs/2203.13024}{{\ttfamily
  arXiv:2203.13024 [hep-th]}}.

\bibitem{Vines:2017hyw}
J.~Vines, ``Scattering of two spinning black holes in post-{Minkowski}an
  gravity, to all orders in spin, and effective-one-body mappings,''
  \href{http://dx.doi.org/10.1088/1361-6382/aaa3a8}{{\em Class. Quant. Grav.}
  {\bfseries 35} no.~8, (2018) 084002},
  \href{http://arxiv.org/abs/1709.06016}{{\ttfamily arXiv:1709.06016 [gr-qc]}}.

\bibitem{Cheung:2023lnj}
C.~Cheung, J.~Parra-Martinez, I.~Z. Rothstein, N.~Shah, and J.~Wilson-Gerow,
  ``{Effective Field Theory for Extreme Mass Ratio Binaries},''
  \href{http://dx.doi.org/10.1103/PhysRevLett.132.091402}{{\em Phys. Rev.
  Lett.} {\bfseries 132} no.~9, (2024) 091402},
  \href{http://arxiv.org/abs/2308.14832}{{\ttfamily arXiv:2308.14832
  [hep-th]}}.

\bibitem{Cui:2025bgu}
Q.~Cui, W.-B. Han, and Z.~Pan, ``{Secondary spins of extreme mass-ratio
  inspirals: A probe to the formation channels},''
  \href{http://dx.doi.org/10.1103/PhysRevD.111.103044}{{\em Phys. Rev. D}
  {\bfseries 111} no.~10, (2025) 103044},
  \href{http://arxiv.org/abs/2502.00856}{{\ttfamily arXiv:2502.00856
  [astro-ph.HE]}}.

\bibitem{Skoupy:2025nie}
V.~Skoup{\'y}, G.~A. Piovano, and V.~Witzany, ``{Spherical inspirals of
  spinning bodies into Kerr black holes},''
  \href{http://dx.doi.org/10.1103/x9yy-c2jq}{{\em Phys. Rev. D} {\bfseries 112}
  no.~12, (2025) 124054}, \href{http://arxiv.org/abs/2506.20726}{{\ttfamily
  arXiv:2506.20726 [gr-qc]}}.

\bibitem{Chen:2025ncm}
Y.-P. Chen, T.~Hsieh, and D.-S. Lee, ``{Motion of spinning particles in the
  Kerr-Newman black hole exterior. I. Periodic orbits},''
  \href{http://arxiv.org/abs/2510.05603}{{\ttfamily arXiv:2510.05603 [gr-qc]}}.

\bibitem{gmoov}
A.~Guevara, B.~Maybee, A.~Ochirov, D.~O'connell, and J.~Vines, ``{A worldsheet
  for Kerr},'' \href{http://dx.doi.org/10.1007/JHEP03(2021)201}{{\em JHEP}
  {\bfseries 03} (2021) 201}, \href{http://arxiv.org/abs/2012.11570}{{\ttfamily
  arXiv:2012.11570 [hep-th]}}.

\bibitem{probe-nj}
J.-H. Kim, ``{Note on the Kerr Spinning-Particle Equations of Motion},''
  \href{http://arxiv.org/abs/2512.23697}{{\ttfamily arXiv:2512.23697 [gr-qc]}}.

\bibitem{njmagic.1}
J.-H. Kim, ``{The Kerr two-twistor particle},''
  \href{http://arxiv.org/abs/2602.19495}{{\ttfamily arXiv:2602.19495 [gr-qc]}}.

\bibitem{njmagic.11}
J.-H. Kim, ``{The Kerr-Newman two-twistor particle},''
  \href{http://arxiv.org/abs/2603.07537}{{\ttfamily arXiv:2603.07537 [gr-qc]}}.

\bibitem{Compere:2021kjz}
G.~Comp{\`e}re and A.~Druart, ``{Complete set of quasi-conserved quantities for
  spinning particles around Kerr},''
  \href{http://dx.doi.org/10.21468/SciPostPhys.12.1.012}{{\em SciPost Phys.}
  {\bfseries 12} no.~1, (2022) 012},
  \href{http://arxiv.org/abs/2105.12454}{{\ttfamily arXiv:2105.12454 [gr-qc]}}.

\bibitem{Gonzo:2024zxo}
R.~Gonzo and C.~Shi, ``{Scattering and Bound Observables for Spinning Particles
  in Kerr Spacetime with Generic Spin Orientations},''
  \href{http://dx.doi.org/10.1103/PhysRevLett.133.221401}{{\em Phys. Rev.
  Lett.} {\bfseries 133} no.~22, (2024) 221401},
  \href{http://arxiv.org/abs/2405.09687}{{\ttfamily arXiv:2405.09687
  [hep-th]}}.

\bibitem{Ramond:2024ozy}
P.~Ramond, ``{On the integrability of extended test body dynamics around black
  holes},'' \href{http://dx.doi.org/10.1088/1361-6382/adb197}{{\em Class.
  Quant. Grav.} {\bfseries 42} no.~6, (2025) 065019},
  \href{http://arxiv.org/abs/2402.02670}{{\ttfamily arXiv:2402.02670 [gr-qc]}}.

\bibitem{Akpinar:2025tct}
D.~Akpinar, G.~R. Brown, R.~Gonzo, and M.~Zeng, ``{Unexpected Symmetries of
  Kerr Black Hole Scattering},''
  \href{http://arxiv.org/abs/2508.10761}{{\ttfamily arXiv:2508.10761
  [hep-th]}}.

\bibitem{Newman:1965tw-janis}
E.~T. Newman and A.~I. Janis, ``{Note on the {Kerr} spinning particle
  metric},'' \href{http://dx.doi.org/10.1063/1.1704350}{{\em J. Math. Phys.}
  {\bfseries 6} (1965) 915--917}.

\bibitem{marsden1974reduction}
J.~Marsden and A.~Weinstein, ``Reduction of symplectic manifolds with
  symmetry,'' {\em Reports on mathematical physics} {\bfseries 5} no.~1, (1974)
  121--130.

\bibitem{meyer1973symmetries}
K.~R. Meyer, ``Symmetries and integrals in mechanics,'' in {\em Dynamical
  systems}, pp.~259--272.
\newblock Elsevier, 1973.

\bibitem{Dirac:1950pj}
P.~A.~M. Dirac, ``{Generalized {Hamilton}ian dynamics},''
  \href{http://dx.doi.org/10.4153/CJM-1950-012-1}{{\em Can. J. Math.}
  {\bfseries 2} (1950) 129--148}.

\bibitem{dirac1964lectures}
P.~A.~M. Dirac, {\em Lectures on Quantum Mechanics}.
\newblock Yeshiva University Press, New York, 1964.

\bibitem{Henneaux:1992ig}
M.~Henneaux and C.~Teitelboim, {\em {Quantization of gauge systems}}.
\newblock Princeton University Press, 1992.

\bibitem{jordan1935zusammenhang}
P.~Jordan, ``Der zusammenhang der symmetrischen und linearen gruppen und das
  mehrk{\"o}rperproblem,'' {\em Zeitschrift f{\"u}r Physik} {\bfseries 94}
  no.~7, (1935) 531--535.

\bibitem{Schwinger1952NYO3071}
J.~Schwinger, \href{http://dx.doi.org/10.2172/4389568}{``On angular
  momentum,''} U.S. Atomic Energy Commission Technical Report NYO-3071, Harvard
  University and Nuclear Development Associates, Inc., Cambridge, MA, USA,
  Jan., 1952.
\newblock \url{https://www.osti.gov/biblio/4389568}.
\newblock Unpublished report; distributed by the U.S. Atomic Energy Commission
  (predecessor of the U.S. Department of Energy).

\bibitem{penrose:1985spinors1}
R.~Penrose and W.~Rindler,
  \href{http://dx.doi.org/10.1017/CBO9780511564048}{{\em {Spinors and
  space-time: Volume 1, Two-spinor calculus and relativistic fields}}}.
\newblock Cambridge Monographs on Mathematical Physics. Cambridge Univ. Press,
  Cambridge, UK, 4, 2011.

\bibitem{Lee:2023nkx}
H.~Lee and S.~Lee, ``{Poincar{\'e} invariance of spinning binary dynamics in
  the post-{Minkowski}an {Hamilton}ian approach},''
  \href{http://dx.doi.org/10.1088/1361-6382/ad0992}{{\em Class. Quant. Grav.}
  {\bfseries 40} no.~24, (2023) 245004},
  \href{http://arxiv.org/abs/2305.10739}{{\ttfamily arXiv:2305.10739 [gr-qc]}}.

\bibitem{ahh2017}
N.~Arkani-Hamed, T.-C. Huang, and Y.-t. Huang, ``{Scattering amplitudes for all
  masses and spins},'' \href{http://dx.doi.org/10.1007/JHEP11(2021)070}{{\em
  JHEP} {\bfseries 11} (2021) 070},
  \href{http://arxiv.org/abs/1709.04891}{{\ttfamily arXiv:1709.04891
  [hep-th]}}.

\bibitem{penrose:maccallum}
R.~Penrose and M.~A.~H. MacCallum, ``{Twistor theory: An Approach to the
  quantization of fields and space-time},''
  \href{http://dx.doi.org/10.1016/0370-1573(73)90008-2}{{\em Phys. Rept.}
  {\bfseries 6} (1972) 241--316}.

\bibitem{penrose1967twistoralgebra}
R.~Penrose, ``Twistor algebra,'' {\em Journal of Mathematical Physics}
  {\bfseries 8} no.~2, (1967) 345--366.

\bibitem{newman1974curiosity}
E.~T. Newman and J.~Winicour, ``{A curiosity concerning angular momentum},''
  \href{http://dx.doi.org/10.1063/1.1666761}{{\em J. Math. Phys.} {\bfseries
  15} (1974) 1113--1115}.

\bibitem{Shirafuji:1983zd}
T.~Shirafuji, ``{Lagrangian mechanics of massless particles With spin},''
  \href{http://dx.doi.org/10.1143/PTP.70.18}{{\em Prog. Theor. Phys.}
  {\bfseries 70} (1983) 18--35}.

\bibitem{Pauli:1927qhd}
W.~Pauli, ``{Zur Quantenmechanik des magnetischen Elektrons},''
  \href{http://dx.doi.org/10.1007/bf01397326}{{\em Zeitschrift f{\"u}r Physik}
  {\bfseries 43} no.~9-10, (1927) 601--623}.

\bibitem{Kuzenko:1994ju}
S.~M. Kuzenko, S.~L. Lyakhovich, and A.~Y. Segal, ``{A Geometric model of
  arbitrary spin massive particle},''
  \href{http://dx.doi.org/10.1142/S0217751X95000735}{{\em Int. J. Mod. Phys. A}
  {\bfseries 10} (1995) 1529--1552},
  \href{http://arxiv.org/abs/hep-th/9403196}{{\ttfamily arXiv:hep-th/9403196}}.

\bibitem{shankar2012principles}
R.~Shankar, {\em Principles of quantum mechanics}.
\newblock Springer Science \& Business Media, 2012.
\newblock Chapter 21.3: Spin and Fermion Path Integrals.

\bibitem{kirillov2004lectures}
A.~A. Kirillov, {\em Lectures on the Orbit Method}, vol.~64 of {\em Graduate
  Studies in Mathematics}.
\newblock American Mathematical Society, Providence, RI, 2004.

\bibitem{torrence1973gauge}
R.~Torrence and W.~Tulczyjew, ``Gauge invariant canonical mechanics for charged
  particles,'' {\em Journal of Mathematical Physics} {\bfseries 14} no.~12,
  (1973) 1725--1732.

\bibitem{guillemin1978equations}
V.~Guillemin and S.~Sternberg, ``Equations of motion of a classical particle in
  a yang-mills field and the principle of general covariance,'' {\em Hadronic
  J.} {\bfseries 1} no.~1, (1978) 1.

\bibitem{sternberg1978classical}
S.~Sternberg and T.~Ungar, ``{Classical and Prequantized Mechanics Without
  Lagrangians Or {Hamilton}ians},'' {\em Hadronic J.} {\bfseries 1} (1978)
  33--76.

\bibitem{guillemin1990symplectic}
V.~Guillemin and S.~Sternberg, {\em Symplectic techniques in physics}.
\newblock Cambridge university press, 1990.

\bibitem{Kim:2025sey}
J.-H. Kim, ``{Manifest symplecticity in classical scattering},''
  \href{http://arxiv.org/abs/2511.07387}{{\ttfamily arXiv:2511.07387
  [hep-th]}}.

\bibitem{ehresmann1948connexions}
C.~Ehresmann, ``Les connexions infinit{\'e}simales dans un espace fibr{\'e}
  diff{\'e}rentiable,'' {\em S{\'e}minaire Bourbaki} {\bfseries 1} (1948)
  153--168.

\bibitem{Mason:2013sva}
L.~Mason and D.~Skinner, ``{Ambitwistor strings and the scattering
  equations},'' \href{http://dx.doi.org/10.1007/JHEP07(2014)048}{{\em JHEP}
  {\bfseries 07} (2014) 048}, \href{http://arxiv.org/abs/1311.2564}{{\ttfamily
  arXiv:1311.2564 [hep-th]}}.

\bibitem{Hansen:1974zz}
R.~O. Hansen, ``{Multipole moments of stationary space-times},''
  \href{http://dx.doi.org/10.1063/1.1666501}{{\em J. Math. Phys.} {\bfseries
  15} (1974) 46--52}.

\bibitem{Hernandez:1967zza}
W.~C. Hernandez, ``{Material Sources for the Kerr Metric},''
  \href{http://dx.doi.org/10.1103/PhysRev.159.1070}{{\em Phys. Rev.} {\bfseries
  159} (1967) 1070--1072}.

\bibitem{Thorne:1980ru}
K.~S. Thorne, ``{Multipole Expansions of Gravitational Radiation},''
  \href{http://dx.doi.org/10.1103/RevModPhys.52.299}{{\em Rev. Mod. Phys.}
  {\bfseries 52} (1980) 299--339}.

\bibitem{Dixon:1974xoz}
W.~G. Dixon, ``{Dynamics of extended bodies in general relativity III.
  Equations of motion},'' \href{http://dx.doi.org/10.1098/rsta.1974.0046}{{\em
  Phil. Trans. Roy. Soc. Lond. A} {\bfseries 277} no.~1264, (1974) 59--119}.

\bibitem{stuckelberg1938wechselwirkungskrafte}
E.~C. St{\"u}ckelberg, ``Die wechselwirkungskr{\"a}fte in der elektrodynamik
  und in der feldtheorie der kr{\"a}fte,'' {\em Helv. Phys. Acta} {\bfseries
  11} (1938) 225.

\bibitem{Hughston:1972qf}
L.~P. Hughston, R.~Penrose, P.~Sommers, and M.~Walker, ``{On a quadratic first
  integral for the charged particle orbits in the charged kerr solution},''
  \href{http://dx.doi.org/10.1007/BF01645517}{{\em Commun. Math. Phys.}
  {\bfseries 27} (1972) 303--308}.

\bibitem{Neill:2013wsa}
D.~Neill and I.~Z. Rothstein, ``{Classical Space-Times from the S Matrix},''
  \href{http://dx.doi.org/10.1016/j.nuclphysb.2013.09.007}{{\em Nucl. Phys. B}
  {\bfseries 877} (2013) 177--189},
  \href{http://arxiv.org/abs/1304.7263}{{\ttfamily arXiv:1304.7263 [hep-th]}}.

\bibitem{Kim:2024svw}
J.-H. Kim, J.-W. Kim, S.~Kim, and S.~Lee, ``{Classical eikonal from Magnus
  expansion},'' \href{http://dx.doi.org/10.1007/JHEP01(2025)111}{{\em JHEP}
  {\bfseries 01} (2025) 111}, \href{http://arxiv.org/abs/2410.22988}{{\ttfamily
  arXiv:2410.22988 [hep-th]}}.

\bibitem{Kim:2025hpn}
J.-W. Kim, ``{Radiation eikonal for post-{Minkowski}an observables},''
  \href{http://dx.doi.org/10.1103/PhysRevD.111.L121702}{{\em Phys. Rev. D}
  {\bfseries 111} no.~12, (2025) L121702},
  \href{http://arxiv.org/abs/2501.07372}{{\ttfamily arXiv:2501.07372
  [hep-th]}}.

\bibitem{Alessio:2025flu}
F.~Alessio, R.~Gonzo, and C.~Shi, ``{Dirac brackets for classical radiative
  observables},'' \href{http://arxiv.org/abs/2506.03249}{{\ttfamily
  arXiv:2506.03249 [hep-th]}}.

\bibitem{Kim:2025olv}
S.~Kim, H.~Lee, and S.~Lee, ``{Classical eikonal in relativistic scattering},''
  \href{http://dx.doi.org/10.1007/JHEP11(2025)032}{{\em JHEP} {\bfseries 11}
  (2025) 032}, \href{http://arxiv.org/abs/2509.01922}{{\ttfamily
  arXiv:2509.01922 [hep-th]}}.

\bibitem{Kim:2025gis}
J.-W. Kim, R.~Patil, T.~Scheopner, and J.~Steinhoff, ``{Magnusian: Relating the
  Eikonal Phase, the On-Shell Action, and the Scattering Generator},''
  \href{http://arxiv.org/abs/2511.05649}{{\ttfamily arXiv:2511.05649
  [hep-th]}}.

\bibitem{hunziker1968s}
W.~Hunziker, ``{The S-matrix in classical mechanics},'' {\em Communications in
  Mathematical Physics} {\bfseries 8} no.~4, (1968) 282--299.

\bibitem{simon1971wave}
B.~Simon, ``Wave operators for classical particle scattering,'' {\em
  Communications in Mathematical Physics} {\bfseries 23} no.~1, (1971) 37--48.

\bibitem{herbst1974classical}
I.~W. Herbst, ``Classical scattering with long range forces,'' {\em
  Communications in Mathematical Physics} {\bfseries 35} no.~3, (1974)
  193--214.

\bibitem{sokolov1979classical}
S.~Sokolov, ``Classical analogues of the moeller operators, of the pearson
  example and of the birmann-kato invariance principle,'' {\em Il Nuovo Cimento
  A (1965-1970)} {\bfseries 52} no.~1, (1979) 1--22.

\bibitem{narnhofer1981canonical}
H.~Narnhofer and W.~Thirring, ``Canonical scattering transformation in
  classical mechanics,'' {\em Physical Review A} {\bfseries 23} no.~4, (1981)
  1688.

\bibitem{thirring1981classical}
W.~Thirring, ``Classical scattering theory,'' in {\em New Developments in
  Mathematical Physics}, pp.~3--28.
\newblock Springer, 1981.

\bibitem{osborn1980levinson}
T.~Osborn, R.~Froese, and S.~Howes, ``Levinson's theorems in classical
  scattering,'' {\em Physical Review A} {\bfseries 22} no.~1, (1980) 101.

\bibitem{levinson1949uniqueness}
N.~Levinson, ``On the uniqueness of the potential in a schrodinger equation for
  a given asymptotic phase,'' {\em Kgl. Danske Videnskab Selskab. Mat. Fys.
  Medd.} {\bfseries 25} no.~9, (1949) 1.

\bibitem{magnus1954exponential}
W.~Magnus, ``On the exponential solution of differential equations for a linear
  operator,'' {\em Communications on pure and applied mathematics} {\bfseries
  7} no.~4, (1954) 649--673.

\bibitem{Kim:2025ebl}
J.-H. Kim, ``{Phase Space Formulation of S-matrix},''
  \href{http://arxiv.org/abs/2512.23100}{{\ttfamily arXiv:2512.23100
  [hep-th]}}.

\bibitem{Hoogeveen:2025tew}
J.~Hoogeveen, G.~U. Jakobsen, and J.~Plefka, ``{Spinning the probe in Kerr with
  WQFT},'' \href{http://dx.doi.org/10.1007/JHEP10(2025)201}{{\em JHEP}
  {\bfseries 10} (2025) 201}, \href{http://arxiv.org/abs/2506.14626}{{\ttfamily
  arXiv:2506.14626 [hep-th]}}.

\bibitem{darboux1882probleme}
G.~Darboux, ``Sur le probleme de pfaff,'' {\em Bulletin des sciences
  math{\'e}matiques et astronomiques} {\bfseries 6} no.~1, (1882) 14--36.

\bibitem{Kol:2007bc}
B.~Kol and M.~Smolkin, ``{Non-Relativistic Gravitation: From {Newton} to
  Einstein and Back},''
  \href{http://dx.doi.org/10.1088/0264-9381/25/14/145011}{{\em Class. Quant.
  Grav.} {\bfseries 25} (2008) 145011},
  \href{http://arxiv.org/abs/0712.4116}{{\ttfamily arXiv:0712.4116 [hep-th]}}.

\bibitem{Levi:2015ixa}
M.~Levi and J.~Steinhoff, ``{Next-to-next-to-leading order gravitational
  spin-squared potential via the effective field theory for spinning objects in
  the post-{Newton}ian scheme},''
  \href{http://dx.doi.org/10.1088/1475-7516/2016/01/008}{{\em JCAP} {\bfseries
  01} (2016) 008}, \href{http://arxiv.org/abs/1506.05794}{{\ttfamily
  arXiv:1506.05794 [gr-qc]}}.

\bibitem{Guevara:2018wpp}
A.~Guevara, A.~Ochirov, and J.~Vines, ``{Scattering of Spinning Black Holes
  from Exponentiated Soft Factors},''
  \href{http://dx.doi.org/10.1007/JHEP09(2019)056}{{\em JHEP} {\bfseries 09}
  (2019) 056}, \href{http://arxiv.org/abs/1812.06895}{{\ttfamily
  arXiv:1812.06895 [hep-th]}}.

\bibitem{Guevara:2019fsj}
A.~Guevara, A.~Ochirov, and J.~Vines, ``{Black-hole scattering with general
  spin directions from minimal-coupling amplitudes},''
  \href{http://dx.doi.org/10.1103/PhysRevD.100.104024}{{\em Phys. Rev. D}
  {\bfseries 100} no.~10, (2019) 104024},
  \href{http://arxiv.org/abs/1906.10071}{{\ttfamily arXiv:1906.10071
  [hep-th]}}.

\bibitem{chkl2019}
M.-Z. Chung, Y.-T. Huang, J.-W. Kim, and S.~Lee, ``{The simplest massive
  S-matrix: from minimal coupling to Black Holes},''
  \href{http://dx.doi.org/10.1007/JHEP04(2019)156}{{\em JHEP} {\bfseries 04}
  (2019) 156}, \href{http://arxiv.org/abs/1812.08752}{{\ttfamily
  arXiv:1812.08752 [hep-th]}}.

\bibitem{aho2020}
N.~Arkani-Hamed, Y.-t. Huang, and D.~O'Connell, ``{Kerr black holes as
  elementary particles},''
  \href{http://dx.doi.org/10.1007/JHEP01(2020)046}{{\em JHEP} {\bfseries 01}
  (2020) 046}, \href{http://arxiv.org/abs/1906.10100}{{\ttfamily
  arXiv:1906.10100 [hep-th]}}.

\bibitem{Johansson:2019dnu}
H.~Johansson and A.~Ochirov, ``{Double copy for massive quantum particles with
  spin},'' \href{http://dx.doi.org/10.1007/JHEP09(2019)040}{{\em JHEP}
  {\bfseries 09} (2019) 040}, \href{http://arxiv.org/abs/1906.12292}{{\ttfamily
  arXiv:1906.12292 [hep-th]}}.

\bibitem{Aoude:2020onz}
R.~Aoude, K.~Haddad, and A.~Helset, ``{On-shell heavy particle effective
  theories},'' \href{http://dx.doi.org/10.1007/JHEP05(2020)051}{{\em JHEP}
  {\bfseries 05} (2020) 051}, \href{http://arxiv.org/abs/2001.09164}{{\ttfamily
  arXiv:2001.09164 [hep-th]}}.

\bibitem{Lazopoulos:2021mna}
A.~Lazopoulos, A.~Ochirov, and C.~Shi, ``{All-multiplicity amplitudes with four
  massive quarks and identical-helicity gluons},''
  \href{http://dx.doi.org/10.1007/JHEP03(2022)009}{{\em JHEP} {\bfseries 03}
  (2022) 009}, \href{http://arxiv.org/abs/2111.06847}{{\ttfamily
  arXiv:2111.06847 [hep-th]}}.

\bibitem{bonocore2025higher}
D.~Bonocore, A.~Kulesza, and J.~Pirsch, ``Higher-spin effects in black hole and
  neutron star binary dynamics: worldline supersymmetry beyond minimal
  coupling,'' {\em Physical Review Letters} {\bfseries 135} no.~21, (2025)
  211404.

\bibitem{Lynden-Bell:2002dvr}
D.~Lynden-Bell, ``A magic electromagnetic field,'' {\em Stellar astrophysical
  fluid dynamics} (2003) 369--375,
  \href{http://arxiv.org/abs/astro-ph/0207064}{{\ttfamily
  arXiv:astro-ph/0207064}}.

\bibitem{Newman:1973yu}
E.~T. Newman, ``{{Maxwell}'s equations and complex {{Minkowski}} space},''
  \href{http://dx.doi.org/10.1063/1.1666160}{{\em J. Math. Phys.} {\bfseries
  14} (1973) 102--103}.

\bibitem{penr04-zigzag}
R.~Penrose, {\em The road to reality: A complete guide to the laws of the
  universe}.
\newblock Random house, 2005.
\newblock Chapter 25.2: The zigzag picture of the electron, 628--633.

\bibitem{newman1974collection}
{Newman, Ezra T}, ``{Complex space-time {\&} some curious consequences},'' in
  {\em {Feldafing Conference of the Max-Planck Inst. on Quantum Theory and the
  Structure of Space-time}}, pp.~117--127.
\newblock 1974.

\bibitem{newman1988remarkable}
E.~T. Newman, ``The remarkable efficacy of complex methods in general
  relativity,'' in {\em Highlights in Gravitation and Cosmology}, pp.~67--78.
\newblock Cambridge University Press, 1988.

\bibitem{Newman:1973afx}
E.~T. Newman, ``{Complex coordinate transformations and the Schwarzschild-Kerr
  metrics},'' \href{http://dx.doi.org/10.1063/1.1666393}{{\em J. Math. Phys.}
  {\bfseries 14} no.~6, (1973) 774}.

\bibitem{Newman:2004ba}
E.~T. Newman, ``{Maxwell fields and shear free null geodesic congruences},''
  \href{http://dx.doi.org/10.1088/0264-9381/21/13/007}{{\em Class. Quant.
  Grav.} {\bfseries 21} (2004) 3197--3222},
  \href{http://arxiv.org/abs/gr-qc/0402056}{{\ttfamily arXiv:gr-qc/0402056}}.

\bibitem{Newman:2002mk}
E.~T. Newman, ``{On a classical, geometric origin of magnetic moments, spin
  angular momentum and the {Dirac} gyromagnetic ratio},''
  \href{http://dx.doi.org/10.1103/PhysRevD.65.104005}{{\em Phys. Rev. D}
  {\bfseries 65} (2002) 104005},
  \href{http://arxiv.org/abs/gr-qc/0201055}{{\ttfamily arXiv:gr-qc/0201055}}.

\bibitem{ko1981theory}
M.~Ko, M.~Ludvigsen, E.~Newman, and K.~Tod, ``The theory of
  $\mathscr{H}$-space,'' {\em Physics Reports} {\bfseries 71} no.~2, (1981)
  51--139.

\bibitem{grg207flaherty}
E.~J. Flaherty~Jr., ``Complex variables in relativity,'' in {\em General
  Relativity and Gravitation: One Hundred Years After the Birth of Albert
  Einstein}, vol.~2, pp.~207--239, International Society on General Relativity
  and Gravitation.
\newblock 1980.

\bibitem{nja}
J.-H. Kim, ``{Newman-Janis Algorithm from Taub-NUT Instantons},''
  \href{http://arxiv.org/abs/2412.19611}{{\ttfamily arXiv:2412.19611 [gr-qc]}}.

\bibitem{sdtn}
J.-H. Kim, ``{Single Kerr-Schild metric for Taub-NUT instanton},''
  \href{http://dx.doi.org/10.1103/PhysRevD.111.L021703}{{\em Phys. Rev. D}
  {\bfseries 111} no.~2, (2025) L021703},
  \href{http://arxiv.org/abs/2405.09518}{{\ttfamily arXiv:2405.09518
  [hep-th]}}.

\bibitem{yano1952some}
K.~Yano, ``Some remarks on tensor fields and curvature,'' {\em Annals of
  Mathematics} {\bfseries 55} no.~2, (1952) 328--347.

\bibitem{Rudiger:1981uu}
R.~R{\"u}diger, ``{The Dirac Equation and Spinning Particles in General
  Relativity},'' \href{http://dx.doi.org/10.1098/rspa.1981.0132}{{\em Proc.
  Roy. Soc. Lond. A} {\bfseries 377} (1981) 417--424}.

\bibitem{Rudiger:1984er}
R.~R{\"u}diger, ``{Separable Systems for the Dirac Equation in Curved
  Space-times},'' \href{http://dx.doi.org/10.1063/1.526169}{{\em J. Math.
  Phys.} {\bfseries 25} (1984) 649--654}.

\bibitem{Carter:1968ks}
B.~Carter, ``{{Hamilton}-Jacobi and Schrodinger separable solutions of
  Einstein's equations},'' \href{http://dx.doi.org/10.1007/BF03399503}{{\em
  Commun. Math. Phys.} {\bfseries 10} no.~4, (1968) 280--310}.

\bibitem{marsden2001comments}
J.~E. Marsden and A.~Weinstein, ``Comments on the history, theory, and
  applications of symplectic reduction,'' in {\em Quantization of singular
  symplectic quotients}, pp.~1--19.
\newblock Springer, 2001.

\bibitem{Wong:1970fu}
S.~K. Wong, ``{Field and particle equations for the classical Yang-Mills field
  and particles with isotopic spin},''
  \href{http://dx.doi.org/10.1007/BF02892134}{{\em Nuovo Cim. A} {\bfseries 65}
  (1970) 689--694}.

\bibitem{Geroch:1970cd}
R.~P. Geroch, ``{Multipole moments. II. Curved space},''
  \href{http://dx.doi.org/10.1063/1.1665427}{{\em J. Math. Phys.} {\bfseries
  11} (1970) 2580--2588}.

\bibitem{janis1965structure}
A.~I. Janis and E.~T. Newman, ``Structure of gravitational sources,'' {\em
  Journal of Mathematical Physics} {\bfseries 6} no.~6, (1965) 902--914.

\bibitem{Holstein:2006wi}
B.~R. Holstein, ``{How large is the `natural' magnetic moment?},''
  \href{http://dx.doi.org/10.1119/1.2345655}{{\em Am. J. Phys.} {\bfseries 74}
  (2006) 1104--1111}, \href{http://arxiv.org/abs/hep-ph/0607187}{{\ttfamily
  arXiv:hep-ph/0607187}}.

\bibitem{Duff:1973zz}
M.~J. Duff, ``{Quantum Tree Graphs and the Schwarzschild Solution},''
  \href{http://dx.doi.org/10.1103/PhysRevD.7.2317}{{\em Phys. Rev. D}
  {\bfseries 7} (1973) 2317--2326}.

\bibitem{Newman:1965my-kerrmetric}
E.~T. Newman, R.~Couch, K.~Chinnapared, A.~Exton, A.~Prakash, and R.~Torrence,
  ``{Metric of a rotating, charged mass},''
  \href{http://dx.doi.org/10.1063/1.1704351}{{\em J. Math. Phys.} {\bfseries 6}
  (1965) 918--919}.

\bibitem{Mason:2009afn}
L.~J. Mason and D.~Skinner, ``{Gravity, Twistors and the MHV Formalism},''
  \href{http://dx.doi.org/10.1007/s00220-009-0972-4}{{\em Commun. Math. Phys.}
  {\bfseries 294} (2010) 827--862},
  \href{http://arxiv.org/abs/0808.3907}{{\ttfamily arXiv:0808.3907 [hep-th]}}.

\bibitem{bialynicki1981note}
I.~Bialynicki-Birula, E.~Newman, J.~Porter, J.~Winicour, B.~Lukacs, Z.~Perjes,
  and A.~Sebestyen, ``A note on helicity,'' {\em Journal of Mathematical
  Physics} {\bfseries 22} no.~11, (1981) 2530--2532.

\bibitem{ashtekar1986note}
A.~Ashtekar, ``A note on helicity and self-duality,'' {\em Journal of
  mathematical physics} {\bfseries 27} no.~3, (1986) 824--827.

\bibitem{Adamo:2023fbj}
T.~Adamo, G.~Bogna, L.~Mason, and A.~Sharma, ``{Scattering on self-dual
  Taub-NUT},'' \href{http://dx.doi.org/10.1088/1361-6382/ad12ee}{{\em Class.
  Quant. Grav.} {\bfseries 41} no.~1, (2024) 015030},
  \href{http://arxiv.org/abs/2309.03834}{{\ttfamily arXiv:2309.03834
  [hep-th]}}.

\bibitem{guillemin1992grauert}
V.~Guillemin and M.~Stenzel, ``{Grauert tubes and the homogeneous Monge-Ampere
  equation. II},'' {\em Journal of Differential Geometry} {\bfseries 35} no.~3,
  (1992) 627--641.

\bibitem{halverscheid2002complexifications}
S.~Halverscheid, ``Complexifications of geodesic flows and adapted complex
  structures,'' {\em Reports on Mathematical Physics} {\bfseries 50} no.~3,
  (2002) 329--338.

\bibitem{hall2011adapted}
B.~C. Hall and W.~D. Kirwin, ``Adapted complex structures and the geodesic
  flow,'' {\em Mathematische Annalen} {\bfseries 350} no.~2, (2011) 455--474.

\bibitem{sodual}
J.-H. Kim, ``{Kerr effective action from orbit-spin duality}.'' To appear.

\end{thebibliography}\endgroup

\end{document}